\documentstyle[a4,epsbox]{report}

\begin{document}

%======================================%
%<<<<<<<<<<< DEFINITIONS >>>>>>>>>>>>>>%
%======================================%

%%% Math fonts %%%
\def\A{{\cal A}}
\def\B{{\cal B}}
\def\F{{\cal F}}
\def\H{{\cal H}}
\def\I{{\cal I}}
\def\P{{\cal P}}
\def\T{{\cal T}}
\def\U{{\cal U}}
\def\Tr{{\rm Tr}}
\newcommand{\bm}[1]{%boldmath
\mbox{\boldmath$ #1 $}}
%%%%%%%%%%%%%%%%%%%

%%% Enviroments %%%%
\newtheorem{theorem}{Theorem}
\newtheorem{lemma}[theorem]{Lemma}
\newtheorem{definition}[theorem]{Definition}
\newenvironment{proof}{\par\noindent {\bf Proof}\par}{\par\medskip}
\newtheorem{prop}{Proposition}
\def\QED{$\Box$}

%======================================%
%<<<<<<<<<<<< TITLE PAGE >>>>>>>>>>>>>>%
%======================================%

\title{The origin of black hole entropy}
\author{
Shinji Mukohyama\\
Yukawa Institute for Theoretical Physics, 
Kyoto University\\
Kyoto 606-8502, Japan
}
\date{
Doctoral thesis submitted to\\
Department of Physics, Kyoto University\\
December 1998}

\maketitle

%======================================%
%<<<<<<<<<<<<< ABSTRACT >>>>>>>>>>>>>>>% 
%======================================%

\begin{abstract}

In this thesis properties and the origin of black hole entropy are
investigated from various points of view. First, laws of black hole
thermodynamics are reviewed. In particular, the first and generalized
second laws are investigated in detail. It is in these laws that the
black hole entropy plays key roles. Next, three candidates for the
origin of the black hole entropy are analyzed: the D-brane
statistical-mechanics, the brick wall model, and the entanglement
thermodynamics. Finally, discussions are given on semiclassical
consistencies of the brick wall model and the entanglement
thermodynamics and on the information loss problem. 

\end{abstract}

\tableofcontents

\pagestyle{headings}

%%%%%%%%%%%%%%%%%%%%%%%%%%%%%%%%%%%%%%%%%%%%%%%%%%%%%%%%%%%%%%%%%%%%
%%%%%%%%%%%%%%%%%%%%%%%%%%%%%%%%%%%%%%%%%%%%%%%%%%%%%%%%%%%%%%%%%%%%
% CHAPTER 1
%%%%%%%%%%%%%%%%%%%%%%%%%%%%%%%%%%%%%%%%%%%%%%%%%%%%%%%%%%%%%%%%%%%%
%%%%%%%%%%%%%%%%%%%%%%%%%%%%%%%%%%%%%%%%%%%%%%%%%%%%%%%%%%%%%%%%%%%%
\chapter{Introduction}
	\label{chap:intro}

Thermodynamics describes behavior of coarse-grained or averaged
quantities of a system with a large number of physical degrees of
freedom. The behavior is traced by a small number of parameters. 
Mathematically, the microscopic description of thermodynamics,
statistical mechanics, is grounded by the ergodic hypothesis. 
On the other hand, in the theory of black holes, the no hair
theorem~\cite{Heusler_BHUT} allows us to describe a stationary black 
hole by a small number of parameters. 
Since the cosmic censorship conjecture~\cite{Penrose1969&1976}
combined with the singularity theorem~\cite{Hawking&Ellis_LSS}
predicts inevitable occurrence of black holes, it seems that the no
hair theorem plays the same role as the ergodic hypothesis plays in
thermodynamics.

In fact, it is well known that black holes have many properties
analogous to those of thermodynamics. Those are as a whole 
called {\it black hole thermodynamics}. In particular, four laws of
black holes combined with the generalized second law make up a main 
framework of the black hole thermodynamics. 
In these laws, {\it black hole entropy} defined as follows plays an
important role. 
%============< EQUATION >==============%
%
\begin{equation}
  S_{BH} = \frac{k_B c^3}{4\hbar G}A,
\label{eqn:1-0:BHformula}
\end{equation}
%======================================%
where $A$ is area of black hole horizon. Moreover, it was suggested
that the black hole entropy, or the horizon area, is an adiabatic
invariant~\cite{Bekenstein1974&1998} and that it can be used as a
potential function in catastrophe theory to judge stability of black
hole solutions~\cite{MTT1994}. The formula (\ref{eqn:1-0:BHformula})
for black hole entropy is often called {\it Bekenstein-Hawking
formula} since the concept of black hole entropy was first introduced 
by Bekenstein~\cite{Bekenstein1973} as a quantity proportional to the
horizon area and the proportionality coefficient was fixed by
Hawking's discovery of thermal radiation from a black
hole~\cite{Hawking1975} (see arguments below). He
showed that a black hole radiates thermal radiation with temperature
given by 
%============< EQUATION >==============%
%
\begin{equation}
 k_B T_{BH} = \frac{\hbar\kappa}{2\pi c}, 
\label{eqn:1-0:HawkingT}
\end{equation}
%======================================%
where $\kappa$ is the surface gravity of a background black hole. This 
thermal radiation and its temperature are called 
{\it Hawking radiation} and {\it Hawking temperature}, respectively.

Let us recall basic properties of the  black-hole thermodynamics by
taking a simple example. We consider a one-parameter family of
Schwarzschild black holes parameterized by the mass $M_{BH}$. 
We assume that a relation analogous to the
first law of thermodynamics holds for a black-hole system. In the
present example, there is only one parameter $M_{BH}$ characterizing a
black hole. Therefore, this relation should be of the simplest form
%============< EQUATION >==============%
%
\begin{equation}
 \delta E_{BH} = T_{BH}\delta S_{BH},
\label{eqn:2-0:1st-law}
\end{equation}
%======================================%
where $E_{BH}$, $S_{BH}$ and $T_{BH}$ are quantities that are
identified with the energy, the entropy and the temperature of a black
hole, respectively. The relation Eq.(\ref{eqn:2-0:1st-law}) is called 
{\it the first law} of the black-hole
thermodynamics~\cite{BCH1973}. Thus, if two of the 
quantities $E_{BH}$, $S_{BH}$ and $T_{BH}$ are given,
$Eq.(\ref{eqn:2-0:1st-law})$ determines the remaining quantity.

In the present example, $M_{BH}$ is the only parameter characterizing
the family of black holes. Therefore the simplest combination which
yields the dimension of energy is 
%============< EQUATION >==============%
%
\begin{equation}
 E_{BH} \equiv M_{BH}c^2.
\label{eqn:2-0:E-BH}
\end{equation}
%======================================%
This is the energy of the black hole.

There is also a natural choice for
$T_{BH}$~\cite{Hawking1975}. Hawking showed 
that a black hole with surface gravity $\kappa$ emits thermal
radiation of a quantum matter field (which plays the role of a
thermometer) at temperature given by (\ref{eqn:1-0:HawkingT}).
Moreover, as shown in section \ref{sec:GSL}, 
if a matter field in a thermal-equilibrium state at some
temperature is scattered by a black hole, then it always becomes
closer to the thermal-equilibrium state at the Hawking temperature 
(\ref{eqn:1-0:HawkingT})~\cite{Panangaden&Wald1977,Mukohyama1997a}.
Thus it is natural to define the temperature of a Schwarzschild black
hole with mass $M_{BH}$ by 
%============< EQUATION >==============%
%
\begin{equation}
 k_BT_{BH} = \frac{\hbar c^3}{8\pi GM_{BH}},
\label{eqn:2-0:T-BH}
\end{equation}
%======================================%
since $\kappa =c^4/4GM_{BH}$~\cite{Wald_GR}.

From Eqs.(\ref{eqn:2-0:1st-law})-(\ref{eqn:2-0:T-BH}), we get an
expression for  $S_{BH}$ as 
%============< EQUATION >==============%
%
\begin{equation}
 S_{BH} = \frac{k_B c^3}{4\hbar G}A+C,	
\label{eqn:2-0:S-BH}
\end{equation}
%======================================%
where $A \equiv 16\pi G^2 M_{BH}^2 / c^4$ is the area of 
the event horizon and $C$ is some constant. Since a value of $C$ is
not essential in our discussions, we shall set hereafter
%============< EQUATION >==============%
%
\begin{equation}
 C = 0.
\label{eqn:2-0:constants}
\end{equation}
%======================================%
Note that Eq.~(\ref{eqn:2-0:S-BH}) with (\ref{eqn:2-0:constants}) is a 
special case of the Bekenstein-Hawking formula
(\ref{eqn:1-0:BHformula}).

It is well-known that classically the area of the event horizon does
not decrease in time (the area law~\cite{Hawking1971} or {\it the
second law} of black hole)  just as the ordinary thermodynamical
entropy. The Bekenstein-Hawking formula (\ref{eqn:1-0:BHformula})
looks reasonable in this 
sense. Indeed this observation was the original motivation for the
introduction of the black-hole entropy~\cite{Bekenstein1973}.
Moreover, when quantum effects are taken into account, it is believed
that a sum of the black hole entropy and matter entropy does not
decrease ({\it the generalized second law}).

{\it The zeroth law} of black hole thermodynamics states that surface 
gravity of a Killing horizon is constant throughout the horizon. 
This supports our choice of the black hole temperature. 
(For a proof of the zeroth law, see
Refs.~\cite{BCH1973,Wald_GR,Racz&Wald1992}.)

At this stage we would like to point out that for a black hole the
third law does not hold in the sense of Planck: 
$S_{BH}\to\infty$ as $T_{BH}\to 0$ for the family of Schwarzschild
black holes, irrespective of the choice of the
value for $C$ as is seen from Eqs.(\ref{eqn:2-0:T-BH}) and
(\ref{eqn:2-0:S-BH}). 
Rather, {\it the third law} does hold in the sense of Nernst: 
it is impossible by any process, no matter how idealized, to reduce
$\kappa$ to zero in a finite sequence of operations~\cite{BCH1973}.
(See Ref.~\cite{Israel1986} for a precise expression and a proof.)

Thermodynamics has a well-established microscopic description: the
quantum statistical mechanics. In the thermodynamical description,
information on each microscopic degree of freedom is lost, and only
macroscopic variables are concerned. However, the number of all
microscopic degrees of freedom is implemented in a macroscopic
variable: entropy $S$ is related to the number of all consistent
microscopic states $N$ as 
%============< EQUATION >==============%
%
\begin{equation}
 S = k_B\ln N.
\end{equation}
%======================================%
In analogy, it is expected that there might be a microscopic
description of the black hole thermodynamics, too. In particular, 
it is widely believed that the black hole entropy might be related to 
a number of microscopic states. Since the microscopic description
seems to require a quantum theory of gravity, detailed investigations
of the black hole entropy should contribute a lot toward construction 
of the theory of quantum gravity. This is one among the several
reasons why the origin of the black hole entropy needs to be
understood at the fundamental level.

Another strong motivation to investigate the black hole entropy is the 
so-called information loss problem. Hawking argued that, if a black
hole is formed by gravitational collapse, then evolution of
quantum fields becomes non-unitary because of evaporation of the black
hole due to the Hawking radiation~\cite{Hawking1976}. This means that
some information is lost in the process of the black hole
evaporation. Moreover, this suggests that the conventional
field-theoretical approach may be useless for the purpose of
construction of the theory of quantum gravity since the field theory
is based on the unitarity. Hence, the 
evaporation of a black hole makes people, who wish to construct a
unitary theory of quantum gravity, be in difficulties: if the
evaporation of a black hole would actually occur and lead to the
information loss, then they would be obliged to give up the
unitarity. Thus, we have to clarify whether information is lost or not
due to the Hawking radiation in order 
to take a step forward. This problem is called the {\it information
loss problem}. On the other hand, since entropy is strongly connected
with information in the theory of information, it is natural to expect
that the black hole entropy might be related to some
information. Therefore, investigations of the origin of the black hole
entropy seem to provide important insight toward the information
loss problem.

In these senses, the origin of the black hole entropy is one of the
most important issues at the present stage of black hole physics.

Recently a microscopic derivation of the black hole entropy was given in 
superstring theory~\cite{GSW_superstring,Polchinski_superstring} by
using the so-called D-brane technology~\cite{PCJ1996}. In this
approach, as will be shown in section~\ref{sec:Dbrane}, 
the black hole entropy is identified with the logarithm of
the number of states of massless strings attached to D-branes, with
D-brane configuration and total momentum of the strings along a
compactified direction fixed to be consistent with the corresponding
black hole~\cite{Strominger&Vafa1996,Maldacena1996}. The  
analysis along this line was extended to the so-called 
M-theory~\cite{Schwarz1997}. In particular, by using a conjectured 
correspondence (the Matrix theory) between the M-theory in the 
infinite momentum frame and a $10$-dimensional 
$U(N)$ supersymmetric Yang-Mills theory dimensionally reduced to 
$(0+1)$-dimension with $N\to\infty$~\cite{BFSS1997}, the black hole
entropy was calculated by means of the Yang-Mills theory. The result
gives the correct Bekenstein-Hawking entropy for BPS black holes and
their low lying excitations~\cite{Li&Martinec1997ab}. Moreover, in
Ref.~\cite{BFKS1998ab} the 
black hole entropy of a Schwarzschild black hole was derived in the
Matrix theory up to a constant of order unity. 
On the other hand, in loop quantum gravity~\cite{Rovelli1997}, black
hole entropy was identified with the logarithm of the number of
different spin-network states for a fixed eigenvalue of the area
operator~\cite{Ashtekar&Krasnov1998}. The result coincides with
the Bekenstein-Hawking entropy up to a constant of order unity.

The derivations in these candidate theories of quantum gravity depend
strongly on details of the theories. In this  
sense, the success of the derivations can be considered as non-trivial 
consistency checks of the theories. However, it is believed 
that proportionality of the black hole entropy to horizon area 
is more universal and does not depend on details of the theory. 
Hence, one should be able to give a statistical or thermodynamical 
derivation of the black hole entropy, which does not depend on details 
of theory, while we are proceeding with theory-dependent derivations of 
it by using the well-established candidate theories of quantum gravity.

There were many attempts to explain the origin of the black hole entropy
besides the above theory-dependent approaches. 
(See Ref.~\cite{Frolov&Fursaev1998} for an up-to-date review.)
For example, in Euclidean gravity the black hole entropy is associated
with the topology of an instanton which corresponds to a black 
hole~\cite{Gibbons&Hawking1977,Hawking&Horowitz1995,Brown&York1993,
BTZ1994}~\footnote{
The $1$-loop correction to the black hole entropy was also calculated
and compared with the brick wall model and the conical singularity 
method~\cite{FFZ1996a}.
}; 
Wald~\cite{Wald1993} defined the black hole entropy as a Noether charge 
associated with a bifurcating Killing horizon~\footnote{
Relations to the approach by Euclidean gravity was investigated in 
Ref.~\cite{Iyer&Wald1995}. 
} (See section \ref{sec:1st_statics}.);
'tHooft~\cite{tHooft1985} identified the black hole entropy with the 
statistical entropy of a thermal gas of quantum particles with a 
mirror-like boundary just outside the horizon (This model is called
the brick wall model and is analyzed in detail in
section~\ref{sec:brick_wall});  
Pretorius et al.~\cite{PVI1998} identified the black hole entropy with 
the thermodynamical entropy of a shell in thermal equilibrium with 
acceleration radiation due to the shell's gravity in the limit that
the shell forms a black hole.

There remains another strong candidate for the statistical origin of 
the black hole entropy, called entanglement 
entropy~\cite{BKLS1986,Srednicki1993,Frolov&Novikov1993}. 
It is a statistical entropy measuring the information loss due to a 
spatial division of a system~\cite{BKLS1986}. 
The entanglement entropy is based only on the spatial division, and can 
be defined independently of the theory, although explicit calculations 
in the literature are dependent on the model employed. Moreover, as
will be explained in section~\ref{sec:entanglement}, it is expected
independently of the details of the theory that the entanglement
entropy is proportional to the area of the boundary of the spatial
division. In this sense, the entanglement entropy is considered to be
a strong candidate for the statistical origin of the black hole
entropy.

In chapter \ref{chap:BHlaws}, laws of black hole thermodynamics are
reviewed. In particular the first laws of black hole statics and
dynamics, and the generalized second law are studied in detail. 
The first law derived in the above simple arguments on Schwarzschild 
black holes relates changes of physical quantities of stationary black
holes corresponding to a variation in a space of stationary black hole 
solutions. In this sense we can call it {\it the first law of black
hole statics}. Historically, the first law of black hole statics is
derived in Ref.~\cite{BCH1973} in general relativity, and is extended
by Wald to a general covariant theory of gravity~\cite{Wald1993}.
In section \ref{sec:1st_statics} we re-analyze the first law
of Wald in detail following Ref.~\cite{Mukohyama1998b}. 
In section \ref{sec:1st_dynamics} we consider a generalization of the
first law to a purely dynamical
situation~\cite{HMA1998,Mukohyama&Hayward1998}. 
We call the dynamical version {\it the first law of black hole
dynamics}. 
In section \ref{sec:GSL} a proof of the generalized second law is
given for a quasi-stationary black hole~\cite{Mukohyama1997a}.

In chapter \ref{chap:BHentropy}, three candidates for the origin of 
the black hole entropy are analyzed in detail.
In section \ref{sec:Dbrane} a microscopic derivation of black hole
entropy by the D-brane technology is shown. 
We consider a $5$-dimensional black hole solution in the low energy
effective theory of Type IIB superstring. This black hole is, in fact, 
a black brane in $10$-dimensional sense and can be interpreted as a 
configuration of D-branes wrapped on $T^5=T^4\times S^1$.
We calculate statistical-mechanical entropy and temperature of open
strings on the D-branes and compare them with the Bekenstein-Hawking
entropy and the Hawking temperature of the original $5$-dimensional
black hole~\cite{Mukohyama1996}.  
In section \ref{sec:brick_wall} we re-examine the brick wall model in
detail and solve problems concerning this
model~\cite{Mukohyama&Israel1998}. 
In section \ref{sec:entanglement} we construct a thermodynamics
(entanglement thermodynamics~\cite{MSK1997,MSK1998}) which includes
the entanglement entropy as the entropy variable, for a massless
scalar field in Minkowski, Schwarzschild and Reissner-Nordstr{\"o}m
spacetimes to understand the statistical origin of black-hole
thermodynamics.  
In section \ref{sec:interpretation} 
a new interpretation of entanglement entropy is 
proposed~\cite{Mukohyama1998a}.

Chapter \ref{chap:summary} is devoted to a summary of this thesis,
discussions on semiclassical consistencies and the information loss 
problem, and speculations.

%%%%%%%%%%%%%%%%%%%%%%%%%%%%%%%%%%%%%%%%%%%%%%%%%%%%%%%%%%%%%%%%%%%%
%%%%%%%%%%%%%%%%%%%%%%%%%%%%%%%%%%%%%%%%%%%%%%%%%%%%%%%%%%%%%%%%%%%%
% CHAPTER 2
%%%%%%%%%%%%%%%%%%%%%%%%%%%%%%%%%%%%%%%%%%%%%%%%%%%%%%%%%%%%%%%%%%%%
%%%%%%%%%%%%%%%%%%%%%%%%%%%%%%%%%%%%%%%%%%%%%%%%%%%%%%%%%%%%%%%%%%%%
\chapter{Laws of black hole thermodynamics}
	\label{chap:BHlaws}

%%%%%%%%%%%%%%%%%%%%%%%%%%%%%%%%%%%%%%%%
%%%%%%%%%%%% SECTION 2-1 %%%%%%%%%%%%%%%
%%%%%%%%%%%%%%%%%%%%%%%%%%%%%%%%%%%%%%%%
\section{The first law of black hole statics}
	\label{sec:1st_statics}

In Ref.~\cite{Wald1993}, the first law of black hole mechanics was
derived not only in general relativity but also in a general covariant
theory of gravity for stationary variations around a stationary black
hole. It is formulated as a relation among variations of those
quantities such as energy, angular momentum and entropy, each of which
is defined in terms of a Noether charge.
The first law was extended to non-stationary variations around a
stationary black hole in Ref.~\cite{Iyer&Wald1994}.

The Noether charge form of the first law has many advantages over
the original first law of Ref.~\cite{BCH1973}.
For example, it gives a general method to calculate stationary black
hole entropy in general covariant theories of
gravity~\cite{Iyer&Wald1994};
it connects various Euclidean methods for computing black 
hole entropy~\cite{Iyer&Wald1995}; 
it suggests a possibility of defining  entropy of non-stationary black 
holes~\cite{Iyer&Wald1994,JKM1994};  etc.

However, in their derivation there are several issues to 
be discussed in more detail. 
\begin{enumerate}
 \renewcommand{\labelenumi}{(\alph{enumi})}
 \item	In Ref.~\cite{Wald1993}, unperturbed and perturbed stationary 
	black holes are identified so that horizon generator Killing 
	fields with unit surface gravity coincide in a neighborhood of 
	the horizons and that stationary Killing fields and axial
	Killing fields coincide in a neighborhood of infinity. This
	corresponds to taking a certain 
	gauge condition in linear perturbation theory. 
	For a complete understanding of the first law, we have to 
	clarify whether such a gauge condition can be 
	imposed or not. If it can, then we wish to know whether such
	a gauge condition is necessary. 
	Note that, on the contrary, the original derivation in general 
	relativity by Bardeen, Carter and Hawking~\cite{BCH1973} is 
	based on a gauge condition such that the stationary Killing 
	fields and the axial Killing fields coincide everywhere on a 
	spacelike hypersurface whose boundary is a union of a horizon 
	cross section and spatial infinity.
 \item	In Ref.~\cite{Iyer&Wald1994}, the first law is extended to 
	non-stationary perturbations around a stationary black hole. 
	In the derivation, change of black hole entropy is calculated
	on a ($n-2$)-surface, which is a bifurcation surface for 
	an unperturbed black hole, but which is not a cross section 
	of an event (nor apparent) horizon for a perturbed 
	non-stationary black hole in general. 
	Does this mean that black hole entropy would be assigned to a 
	surface which is not a horizon cross section for a 
	non-stationary black hole? 
	It seems more natural to assign black hole entropy to 
	a horizon cross section also for a non-stationary black hole. 
\end{enumerate}
In this section these two issues are discussed and it is concluded
that there are no difficulties in the derivation of the Noether charge 
form of the first law for both stationary and non-stationary 
perturbations about a stationary black hole. 
In its course, we give an alternative derivation of the first law
based on a variation in which a horizon generator Killing field with
unit surface gravity is fixed.

In subsection \ref{subsec:gauge_cond} gauge conditions are analyzed. 
In subsection \ref{subsec:1st_stationary} the first law of black holes
is derived for stationary variations around a stationary black hole. 
In subsection \ref{subsec:1st_nonstationary} the derivation is
extended to non-stationary variations around a stationary black hole.

%======================================%
%<<<<<<   SUBSECTION 2-1-1    >>>>>>>>>%
%======================================%
\subsection{Gauge conditions}
	\label{subsec:gauge_cond}

Consider a stationary black hole in $n$-dimensions, which has a
bifurcating Killing horizon. 
Let $\xi^a$ be a generator Killing field of the Killing horizon, which 
is normalized as $\xi^a=t^a+\Omega_H^{(\mu)}\varphi_{(\mu)}^a$, 
and $\Sigma$ be the bifurcation surface. Here, $t^a$ is the stationary 
Killing field with unit norm at infinity, 
$\{\varphi_{(\mu)}^a\}$ ($\mu=1,2,\cdots$) is a family of axial Killing 
fields, and $\{\Omega_H^{(\mu)}\}$ is a family of constants (angular 
velocities).
Let $\kappa$ be the surface gravity corresponding to $\xi^a$:
%============< EQUATION >==============%
%
\begin{equation}
 \xi^b\nabla_b\xi^a = \kappa\xi^a
\end{equation}
%======================================%
on the horizon.

Now let us show that it is not possible in general to impose a gauge 
condition such that $\delta\xi^a=0$ in a neighborhood of the
bifurcation surface. 
For this purpose we shall temporarily assume that $\delta\xi^a=0$ and 
show a contradiction.

On $\Sigma$, the covariant derivative of $\xi^a$ is given by
%============< EQUATION >==============%
%
\begin{equation}
 \nabla_b\xi^a = \kappa{\bf\epsilon}_b^{\ a},
\end{equation}
%======================================%
where ${\bf\epsilon}_{ab}$ is binormal to $\Sigma$. 
However, the variation of the l.h.s. is zero:
%============< EQUATION >==============%
%
\begin{equation}
 \delta(\nabla_b\xi^a) = \delta\Gamma^a_{bc}\xi^c =0
 	\label{eqn:2-1:delta-d-xi-0}
\end{equation}
%======================================%
since $\xi^a=0$, where $\delta\Gamma^a_{bc}$ is given by
%============< EQUATION >==============%
%
\begin{equation}
 \delta\Gamma^a_{bc} = 
	\frac{1}{2}g^{ad}(\nabla_c\delta g_{db}
	+\nabla_b\delta g_{dc}-\nabla_d\delta g_{bc}).
\end{equation}
%======================================%
Hence, 
%============< EQUATION >==============%
%
\begin{equation}
 \delta{\bf\epsilon}_b^{\ a} = 
 	-\frac{\delta\kappa}{\kappa}{\bf\epsilon}_b^{\ a}.
\end{equation}
%======================================%
Substituting this into the identity 
$\delta({\bf\epsilon}_b^{\ a}{\bf\epsilon}^b_{\ a})=0$, we obtain
%============< EQUATION >==============%
%
\begin{equation}
 0 = \delta({\bf\epsilon}_b^{\ a}{\bf\epsilon}^b_{\ a}) = 
	-\frac{4\delta\kappa}{\kappa}.\label{eqn:2-1:delta-kappa-0}
\end{equation}
%======================================%

Thus, the assumption $\delta\xi^a=0$ leads to $\delta\kappa=0$, 
which implies, for example, that $\delta M=0$ for the vacuum general 
relativity in a static case, where $M$ is mass of Schwarzschild black
holes. This peculiar behavior can be understood as appearance of a
coordinate singularity at the bifurcation surface of a coordinate
fixed by the gauge condition $\delta\xi^a=0$ since in the above
argument finiteness of $\delta\Gamma^a_{bc}$ has been assumed
implicitly. Therefore, it is impossible to impose the condition
$\delta\xi^a=0$ in a neighborhood of the bifurcation surface whenever 
$\delta\kappa\ne 0$.

As mentioned above, the original derivation of the first law in 
Ref.~\cite{BCH1973} adopt the gauge condition 
$\delta t^a=\delta\varphi^a=0$. This leads to $\delta\xi^a=0$ when 
$\delta\Omega=0$ (for example, when we consider static black holes).
Of course, in Ref.~\cite{BCH1973}, a general horizon cross section 
(not necessary a bifurcation surface) is considered as a surface on 
which black hole entropy is calculated. 
Hence, the above argument arises no difficulties unless the cross 
section is taken to be the bifurcation surface. 
The derivation in Ref.~\cite{BCH1973} suffers from the above argument
if and only if black hole entropy is estimated on the bifurcation 
surface.

On the other hand, arguments like the above do not lead to any 
contradiction if we adopt a gauge condition such that $\tilde{\xi}^a$ 
is fixed in a neighborhood of the bifurcation surface under
variations, where $\tilde{\xi}^a=\xi^a/\kappa$ is a horizon generator
Killing field with unit surface gravity. 
Moreover, it is concluded that, if we intend
to fix a horizon generator Killing field, then it must have the same 
value of surface gravity for unperturbed and perturbed black holes. 
Hence, the gauge condition $\delta\tilde{\xi}^a=0$ in a neighborhood
of the bifurcation 
surface adopted in Ref.~\cite{Wald1993,Iyer&Wald1994} is very natural 
one.

In fact, it is always possible to identify unperturbed 
and perturbed stationary black holes so that the Killing horizons
and the generator Killing fields with unit surface gravity
coincide. 
As stated in Ref.~\cite{Wald1993}, such an identification can be done 
at least in a neighborhood of the horizon by using the general formula 
for Kruskal-type coordinates $(U,V)$ given in Ref.~\cite{Racz&Wald1992}. 
(The identified Killing horizon is given by $U=0$ and $V=0$. The
identified Killing field with unit surface gravity is given by 
$\tilde{\xi}^a=U(\partial/\partial U)^a-V(\partial/\partial V)^a$.)

The purpose of the next subsection is to discuss the remaining
gauge condition $\delta t^a=\delta\varphi^a=0$ at infinity. 
It is evident that this gauge condition at infinity can be imposed by
identifying the perturbed and unperturbed spacetimes suitably. 
So, our question now is whether this gauge condition is necessary or
not. 
For this purpose we temporarily adopt a gauge condition such that
$\tilde{\xi}^a$ is fixed everywhere on a hypersurface connecting the
bifurcation surface and spatial infinity. 
In deriving the first law in this gauge condition, the gauge condition
$\delta t^a=\delta\varphi^a=0$ at infinity is found to be 
necessary for a proper interpretation of the first law. 
On the other hand, as shown in subsection \ref{subsec:1st_nonstationary},
it is not necessary to fix $\tilde{\xi}^a$ in a neighborhood of  the bifurcation surface, 
strictly speaking. 
Hence, it can be concluded that the minimal set of gauge conditions
necessary for the derivation of the first law is that $t^a$ and
$\varphi^a$ are fixed at spatial infinity.

%======================================%
%<<<<<<   SUBSECTION 2-1-2    >>>>>>>>>%
%======================================%
\subsection{The first law for stationary black holes}
	\label{subsec:1st_stationary}

Before deriving the first law, we review basic ingredients of the 
formalism.

We consider a classical theory of gravity in $n$-dimensions 
with arbitrary matter fields, which is described by a diffeomorphism 
invariant Lagrangian $n$-form ${\bf L}(\phi)$, where $\phi$ denotes 
dynamical fields~\cite{Iyer&Wald1994}.

%<<<<<<<<<< Noether charge >>>>>>>>>>>>%

The Noether current ($n-1$)-form ${\bf j}[V]$ for a vector field
$V^a$ is defined by 
%============< EQUATION >==============%
%
\begin{equation}
 {\bf j}[V] \equiv 
 	{\bf\Theta}(\phi,{\cal L}_V\phi)-V\cdot{\bf L}(\phi),
 	\label{eqn:2-1:j-def}
\end{equation}
%======================================%
where the ($n-1$)-form ${\bf\Theta}(\phi,\delta\phi)$ is defined by 
%============< EQUATION >==============%
%
\begin{equation}
 \delta{\bf L}(\phi) = 
 	{\bf E}(\phi)\delta\phi + d{\bf\Theta(\phi,\delta\phi)}.
 	\label{eqn:2-1:Theta-def}
\end{equation}
%======================================%
It is easily shown that the Noether current is closed as 
%============< EQUATION >==============%
%
\begin{equation}
 d{\bf j}[V] = -{\bf E}(\phi){\cal L}_V\phi = 0,
\end{equation}
%======================================%
where we have used the equations of motion ${\bf E}(\phi)=0$.
Hence, by using the machinery developed in Ref.~\cite{Wald1990}, we 
obtain the Noether charge ($n-2$)-form ${\bf Q}[V]$ such that
%============< EQUATION >==============%
%
\begin{equation}
 {\bf j}[V] = d{\bf Q}[V]. \label{eqn:2-1:def-Q}
\end{equation}
%======================================%

%<<<<<<<<<< Hamiltonian >>>>>>>>>>>>>>>%

Hereafter, we assume that in an asymptotically flat spacetime there exists 
an ($n-1$)-form ${\bf B}$ such that 
%============< EQUATION >==============%
%
\begin{equation}
 \int_{\infty}V\cdot\delta{\bf B}(\phi) = 
	\int_{\infty}V\cdot{\bf\Theta}(\phi,\delta\phi),
\end{equation}
%======================================%
where the integral is taken over an $(n-2)$-dimensional sphere at
infinity. 
By using ${\bf B}$, we can write a Hamiltonian $H[V]$ corresponding
to evolution by $V^a$ as follows~\cite{Wald1993}.
%============< EQUATION >==============%
%
\begin{equation}
 H[V] \equiv \int_{\infty}({\bf Q}[V]-V\cdot{\bf B}).
\end{equation}
%======================================%

%<<< The symplectic current density >>>%

The symplectic current density 
${\bf\omega}(\phi,\delta_1\phi,\delta_2\phi)$ is defined by 
%============< EQUATION >==============%
%
\begin{equation}
 {\bf\omega}(\phi,\delta_1\phi,\delta_2\phi) \equiv 
 	\delta_1[{\bf\Theta}(\phi,\delta_2\phi)] -
 	\delta_2[{\bf\Theta}(\phi,\delta_1\phi)]
\end{equation}
%======================================%
and is linear both in $\delta_1\phi$ and its derivatives, and
$\delta_2\phi$ and its derivatives~\cite{Lee&Wald1990}.

%<<<<<<<<<<<<< Variation >>>>>>>>>>>>>>%

Now we define a space of solutions in which we take a variation to derive 
the first law.

Let $\tilde{\xi}^a$ be a fixed vector field, which vanishes on a
($n-2$)-surface $\Sigma$.
(Note that $\tilde{\xi}^a$ and $\Sigma$ can be defined without
referring to any dynamical fields, eg. the metric $g_{ab}$.)
In the following arguments, we consider a space of stationary, 
asymptotically flat solutions of the equations of motion 
${\bf E}(\phi)=0$, 
each of which satisfies the following three conditions.
(a) There exists a bifurcating Killing horizon with the bifurcation
surface $\Sigma$.
(b) $\tilde{\xi}^a$ is a generator Killing field of the Killing horizon. 
(c) Surface gravity corresponding to $\tilde{\xi}^a$ is $1$:
%============< EQUATION >==============%
%
\begin{equation}
 \tilde{\xi}^b\nabla_b\tilde{\xi}^a = \tilde{\xi}^a
\end{equation}
%======================================%
on the Killing horizon.

For each element in this space, there exist constants $\kappa$ and
$\Omega_H^{(\mu)}$ ($\mu=1,2,\cdots$) such that 
%============< EQUATION >==============%
%
\begin{equation}
 \kappa\tilde{\xi}^a = t^a + \Omega_H^{(\mu)}\varphi_{(\mu)}^a,
	\label{eqn:2-1:xi-t-varphi}
\end{equation}
%======================================%
where $t^a$ is the stationary Killing field with 
unit norm at infinity, $\{\varphi_{(\mu)}^a\}$ ($\mu=1,2,\cdots$) is 
a family of axial Killing fields. Hence, $\kappa$
is surface gravity and $\Omega_H^{(\mu)}$ are angular velocities of 
the horizon.

Note that, by definition, the vector field $\tilde{\xi}^a$ is fixed
under a variation of dynamical fields. We express this explicitly by
denoting the variation by $\tilde{\delta}$:
%============< EQUATION >==============%
%
\begin{equation}
 \tilde{\delta}\tilde{\xi}^a = 0. \label{eqn:2-1:deltaxi=0}
\end{equation}
%======================================%

%<<<<<<<<<< The 1st law    >>>>>>>>>>>>%

We now derive the first law of black hole mechanics.

First, by taking a variation of the definition (\ref{eqn:2-1:j-def}) for 
${\bf j}[\tilde{\xi}]$ and using (\ref{eqn:2-1:deltaxi=0}) and
(\ref{eqn:2-1:Theta-def}), we obtain
%============< EQUATION >==============%
%
\begin{eqnarray}
 \tilde{\delta} {\bf j}[\tilde{\xi}] & = &
 	\tilde{\delta}\left({\bf\Theta}
	(\phi,{\cal L}_{\tilde{\xi}}\phi)\right)
 	- \tilde{\xi}\cdot\left({\bf E}(\phi)\tilde{\delta}\phi 
	+ d{\bf\Theta}(\phi,\tilde{\delta}\phi)\right)
 	\nonumber\\
 	& = & 
 	{\bf\omega}
	(\phi,\tilde{\delta}\phi,{\cal L}_{\tilde{\xi}}\phi)
 	+ d\left(\tilde{\xi}\cdot{\bf\Theta}
	(\phi,\tilde{\delta}\phi)\right). 
	\label{eqn:2-1:delta-j}
\end{eqnarray}
%======================================%
Here we have used the equations of motion ${\bf E}(\phi)=0$ and the 
following identity for an arbitrary vector $V^a$ and an arbitrary
differential form ${\bf\Lambda}$ to obtain the last line.
%============< EQUATION >==============%
%
\begin{equation}
 {\cal L}_V{\bf\Lambda} = V\cdot d{\bf\Lambda} + d(V\cdot{\bf\Lambda}).
 \label{eqn:2-1:identity}
\end{equation}
%======================================%
Since 
${\bf\omega}(\phi,\tilde{\delta}\phi,{\cal L}_{\tilde{\xi}}\phi)$ is
linear in ${\cal L}_{\tilde{\xi}}\phi$ and its derivatives, we obtain 
%============< EQUATION >==============%
%
\begin{equation}
 d(\tilde{\delta}{\bf Q}[\tilde{\xi}]) = 
 	d\left(\tilde{\xi}\cdot
	{\bf\Theta}(\phi,\tilde{\delta}\phi)\right)
 	\label{eqn:2-1:d-delta-Q}
\end{equation}
%======================================%
by using ${\cal L}_{\tilde{\xi}}\phi=0$ and Eq.~(\ref{eqn:2-1:def-Q}).

Next we ingrate Eq.~(\ref{eqn:2-1:d-delta-Q}) over an asymptotically flat 
spacelike hypersurface ${\cal C}$, which is parallel to
$\varphi^a_{(\mu)}$ at infinity and the interior boundary of which is
$\Sigma$. Since $\tilde{\xi}^a=0$ on $\Sigma$, we obtain 
%============< EQUATION >==============%
%
\begin{equation}
 \tilde{\delta}\int_{\Sigma}{\bf Q}[\tilde{\xi}] = 
	\tilde{\delta}H[\tilde{\xi}].
	\label{eqn:2-1:pre-1st-law}
\end{equation}
%======================================%

Finally we rewrite the r.h.s. and the l.h.s. of
(\ref{eqn:2-1:pre-1st-law}) in a form useful to be estimated at
infinity and the horizon, respectively.

A relation among variations of $\kappa$, $\Omega_H^{(\mu)}$, $t^a$
and $\varphi^a_{(\mu)}$ is obtained by substituting
(\ref{eqn:2-1:xi-t-varphi}) in (\ref{eqn:2-1:deltaxi=0}). 
%============< EQUATION >==============%
%
\begin{equation}
 t^a\tilde{\delta}\left(\frac{1}{\kappa}\right) 
	+ \varphi^a_{(\mu)}\tilde{\delta}
	\left(\frac{\Omega_H^{(\mu)}}{\kappa}\right) =
 - \frac{1}{\kappa}\tilde{\delta}t^a
	- \frac{\Omega_H^{(\mu)}}{\kappa}
	\tilde{\delta}\varphi^a_{(\mu)}.
	\label{eqn:2-1:delta-Omega-t-varphi}
\end{equation}
%======================================%
By using this relation and the fact that  $H[V]$ is linear in the
vector field $V$, we can rewrite the r.h.s. of (\ref{eqn:2-1:pre-1st-law}) 
as follows.   
%============< EQUATION >==============%
%
\begin{eqnarray}
 \tilde{\delta}  H[\tilde{\xi}] & = &
 	\frac{1}{\kappa}(\tilde{\delta} H[t] -H[\tilde{\delta} t]) +
 	\frac{\Omega_H^{(\mu)}}{\kappa}
	(\tilde{\delta} H[\varphi_{(\mu)}] -
 	H[\tilde{\delta} \varphi_{(\mu)}]) \nonumber\\
 & = &  \frac{1}{\kappa}\delta_{\infty}H[t] +
 	\frac{\Omega_H^{(\mu)}}{\kappa}
	\delta_{\infty}H[\varphi_{(\mu)}],
\end{eqnarray}
%======================================%
where the variation $\delta_{\infty}$ is defined for linear 
functionals $F[t]$ and $G_{(\mu)}[\varphi_{(\mu)}]$ so that 
%============< EQUATION >==============%
%
\begin{eqnarray}
 \delta_{\infty}F[t] & = & 
 	\tilde{\delta} F[t] -F[\tilde{\delta} t],\nonumber\\
 \delta_{\infty}G_{(\mu)}[\varphi_{(\mu)}] & = & 
	\tilde{\delta} G_{(\mu)}[\varphi_{(\mu)}] -
	G_{(\mu)}[\tilde{\delta} \varphi_{(\mu)}].
	\label{eqn:2-1:def-delta-infty}
\end{eqnarray}
%======================================%
This newly introduced variation corresponds to a variation at 
infinity such that $t^a$ and $\varphi^a$ are fixed:
%============< EQUATION >==============%
%
\begin{equation}
 \delta_{\infty}t^a =\delta_{\infty}\varphi^a_{(\mu)} = 0.
\end{equation}
%======================================%

In Ref.~\cite{Iyer&Wald1994} a useful expression of the Noether 
charge was given as follows.
%============< EQUATION >==============%
%
\begin{equation}
 {\bf Q}[V] = {\bf W}_c(\phi)V^c + 
	{\bf X}^{cd}(\phi)\nabla_{[c}V_{d]}
	+{\bf Y}(\phi,{\cal L}_V\phi) + d{\bf Z}(\phi,V),
 	\label{eqn:2-1:Noether}
\end{equation}
%======================================%
where ${\bf W}_c$, ${\bf X}^{cd}$, ${\bf Y}$ and ${\bf Z}$ are locally 
constructed covariant quantities. In particular, 
${\bf Y}(\phi,{\cal L}_V\phi)$ is linear in 
${\cal L}_V\phi$ and its derivatives, and ${\bf X}^{cd}$ is given by 
%============< EQUATION >==============%
%
\begin{equation}
 \left({\bf X}^{cd}(\phi)\right)_{c_3\cdots c_n} = 
 	-E_R^{abcd}{\bf\epsilon}_{abc_3\cdots c_n}.
\end{equation}
%======================================%
Here $E_R^{abcd}$ is the would-be equations of motion 
form~\cite{Iyer&Wald1994} for the Riemann tensor $R_{abcd}$ and 
${\bf\epsilon}_{abc_3\cdots c_n}$ is the volume $n$-form.

By using the form of ${\bf Q}$ we can rewrite the integral in the
l.h.s. of (\ref{eqn:2-1:pre-1st-law}) as 
%============< EQUATION >==============%
%
\begin{equation}
 \int_{\Sigma}{\bf Q}[\tilde{\xi}] = 
	\int_{\Sigma}{\bf X}^{cd}(\phi)\nabla_{[c}\tilde{\xi}_{d]},
	\label{eqn:2-1:intQ}
\end{equation}
%======================================%
where we have used the Killing equation 
${\cal L}_{\tilde{\xi}}\phi=0$ and the fact that $\tilde{\xi}^a=0$ on
$\Sigma$.

Using the relation
%============< EQUATION >==============%
%
\begin{equation}
 \nabla_c\tilde{\xi}_d = {\bf\epsilon}_{cd}
\end{equation}
%======================================%
on $\Sigma$, for any stationary solutions we can eliminate explicit
dependence of Eq.~(\ref{eqn:2-1:intQ}) on $\tilde{\xi}$, 
where ${\bf\epsilon}_{cd}$ is the binormal to 
$\Sigma$. Hence, at least within the space of stationary solutions, we
can take the variation $\tilde{\delta}$ of the integral without any
difficulties.

Thus, we obtain the first law for stationary black holes by
rewriting Eq.~(\ref{eqn:2-1:pre-1st-law}) as
%============< EQUATION >==============%
%
\begin{equation}
 \frac{\kappa}{2\pi}\tilde{\delta}S = 
	\delta_{\infty}{\cal E} - 
	\Omega_H^{(\mu)}\delta_{\infty}{\cal J}_{(\mu)},
	\label{eqn:2-1:1st-law}
\end{equation}
%======================================%
where entropy $S$ is defined by 
%============< EQUATION >==============%
%
\begin{equation}
 S \equiv 2\pi\int_{\Sigma}{\bf X}^{cd}(\phi){\bf\epsilon}_{cd},
 	\label{eqn:2-1:S1}
\end{equation}
%======================================%
and energy ${\cal E}$ and angular momenta ${\cal J}_{(\mu)}$ are
defined by 
%============< EQUATION >==============%
%
\begin{eqnarray}
 {\cal E} & \equiv & H[t] = 
 	\int_{\infty}({\bf Q}[t]-t\cdot{\bf B}),\nonumber\\
 {\cal J}_{(\mu)} & \equiv & -H[\varphi_{(\mu)}] = 
 	-\int_{\infty}{\bf Q}[\varphi_{(\mu)}].
	\label{eqn:2-1:def-E-J}
\end{eqnarray}
%======================================%

Note that, in the r.h.s. of Eq.~(\ref{eqn:2-1:1st-law}), variations of 
${\cal E}$ and ${\cal J}_{(\mu)}$ are taken with $t^a$ and 
$\varphi_{(\mu)}^a$ fixed. This condition is explicitly implemented 
by the definition (\ref{eqn:2-1:def-delta-infty}) of $\delta_{\infty}$ 
and is necessary for a proper interpretation of the first law.

%<<<<  Another expression of S    >>>>>%

We conclude this subsection by giving another expression of the 
entropy.

Since $\tilde{\xi}^a$ is a generator Killing field of the Killing 
horizon, we have ${\cal L}_{\tilde{\xi}}\phi=0$ and the pull-back of 
$\tilde{\xi}\cdot{\bf L}(\phi)$ to the horizon vanishes. 
Hence, the definition (\ref{eqn:2-1:j-def}) says that the pull-back of 
${\bf j}[\tilde{\xi}]$ to the horizon is zero~\cite{JKM1994}.   
Thus, the integral of ${\bf Q}[\tilde{\xi}]$ is independent of the 
choice of the horizon cross section.

Moreover, it can be shown that the integral in (\ref{eqn:2-1:S1}) does 
not change even if we replace the integration surface $\Sigma$ by an
{\it arbitrary} horizon cross section
$\Sigma'$~\cite{JKM1994}. Therefore we obtain 
%============< EQUATION >==============%
%
\begin{equation}
 S = 2\pi\int_{\Sigma'}{\bf X}^{cd}(\phi){\bf\epsilon}'_{cd},
\end{equation}
%======================================%
where ${\bf\epsilon}'_{cd}$ denotes the binormal to $\Sigma'$.

%======================================%
%<<<<<<   SUBSECTION 2-1-3    >>>>>>>>>%
%======================================%
\subsection{Non-stationary perturbation}
	\label{subsec:1st_nonstationary}
	
In this subsection, we shall derive the first law for a non-stationary 
perturbation about a stationary black hole with a bifurcating
Killing horizon. 
Unfortunately, for non-stationary perturbations, $\delta\kappa$ and 
$\delta\Omega_H^{(\mu)}$ do not have meaning of perturbations of
surface gravity and angular velocity of the Killing horizon, even if
they are defined. 
However, since the first law (\ref{eqn:2-1:1st-law}) does not refer to
$\delta\kappa$ and $\delta\Omega_H^{(\mu)}$ but only to the
unperturbed values of $\kappa$ and $\Omega_H^{(\mu)}$, we expect that 
the first law holds also for non-stationary perturbations. 
In the following, we shall show that it does hold.

%<<<<<<<<<<<<< Variation >>>>>>>>>>>>>>%

First, we specify a space of solutions in which we take a variation.

Let $\tilde{\xi}_0^a$ be a fixed vector field, which vanishes on an
fixed ($n-2$)-surface $\Sigma$.
In this subsection, we consider a space of asymptotically flat
solutions of the field equation ${\bf E}(\phi)=0$, for each of which
$\tilde{\xi}_0^a$ is an asymptotic Killing field.

For each solution in this space, there exist constants $\kappa$ and
$\Omega_H^{(\mu)}$ ($\mu=1,2,\cdots$) such that at spatial infinity 
%============< EQUATION >==============%
%
\begin{equation}
 \kappa\tilde{\xi}_0^a = t^a + \Omega_H^{(\mu)}\varphi_{(\mu)}^a,
\end{equation}
%======================================%
where $t^a$ is a timelike asymptotic Killing field with unit norm at 
infinity, $\{\varphi_{(\mu)}^a\}$ ($\mu=1,2,\cdots$) is a family of
axial asymptotic Killing fields orthogonal to $t^a$ at infinity and 
$\{\Omega_H^{(\mu)}\}$ is a family of constants. 
Note that the constants $\kappa$ and $\Omega_H^{(\mu)}$ do not have 
meaning of surface gravity and angular velocities unless we consider a
stationary solution.
Moreover, in general, $\tilde{\xi}_0^a$ and $\Sigma$ have no meaning but 
an asymptotic Killing field and a fixed ($n-2$)-surface, respectively.

Note that, by definition, the vector field $\tilde{\xi}_0^a$ is fixed
under the variation. We denote the variation by $\tilde{\delta}$: 
%============< EQUATION >==============%
%
\begin{equation}
 \tilde{\delta}\tilde{\xi}_0^a = 0.
\end{equation}
%======================================%
On the contrary, $t^a$, $\varphi_{(\mu)}^a$, $\kappa$ and
$\Omega_H^{(\mu)}$ are not fixed under the variation since definitions
of them refer to dynamical fields, which are varied. Their variations
are related by (\ref{eqn:2-1:delta-Omega-t-varphi}).

Suppose that an element $\phi_0$ of the space of solutions
satisfies the following three conditions.
(a') $\phi_0$ is a stationary solution with a bifurcating Killing
horizon with the bifurcation surface $\Sigma$. 
(b') $\tilde{\xi}_0^a$ is a generator Killing field of the Killing
horizon of $\phi_0$. 
(c') Surface gravity of $\phi_0$ corresponding to $\tilde{\xi}_0^a$ is
$1$: 
%============< EQUATION >==============%
%
\begin{equation}
 \tilde{\xi}_0^b\nabla_b\tilde{\xi}_0^a = \tilde{\xi}_0^a
\end{equation}
%======================================%
on the Killing horizon.

%<<<<<<<<<< The 1st law    >>>>>>>>>>>>%

Now we derive the first law for the {\it non-stationary} perturbation 
$\tilde{\delta}\phi$ about the stationary solution $\phi_0$.

First, we mention that the validity of Eq.~(\ref{eqn:2-1:pre-1st-law}) in
the previous section depends on the following three facts.
(i) The equations of motion ${\bf E}(\phi)=0$ hold for both
unperturbed and perturbed fields. 
(Unless they hold also for perturbed fields, $\tilde{\delta}{\bf j}$
can not be rewritten as $d(\tilde{\delta}{\bf Q})$.)
(ii) $\tilde{\xi}^a$ (corresponding to $\tilde{\xi}_0^a$) is a Killing 
field of the unperturbed solution.
(iii) $\tilde{\xi}^a=0$ (corresponding to $\tilde{\xi}_0^a=0$) on 
$\Sigma$ for unperturbed solution.

These three are satisfied for the unperturbed solution $\phi_0$ and
the non-stationary variation $\tilde{\delta}\phi$ around 
$\phi_0$, too. Thus, Eq.~(\ref{eqn:2-1:pre-1st-law}) is
valid, provided that $\tilde{\xi}^a$ is replaced by $\tilde{\xi}_0^a$.

Since $\tilde{\delta}t^a$, $\tilde{\delta}\varphi^a_{(\mu)}$,
$\tilde{\delta}\kappa$  and $\tilde{\delta}\Omega_H^{(\mu)}$ are
related by  Eq.~(\ref{eqn:2-1:delta-Omega-t-varphi}), we can transform the
r.h.s. of (\ref{eqn:2-1:pre-1st-law}) to obtain
%============< EQUATION >==============%
%
\begin{equation}
 \kappa\tilde{\delta}\int_{\Sigma}{\bf Q}[\tilde{\xi}_0] =
	\delta_{\infty}{\cal E} - 
	\Omega_H^{(\mu)}\delta_{\infty}{\cal J}_{(\mu)},
	\label{eqn:2-1:pre-1st-law2}
\end{equation}
%======================================%
where, as in the previous section, energy ${\cal E}$ and angular
momenta ${\cal J}_{(\mu)}$ are defined by (\ref{eqn:2-1:def-E-J}), and 
the variation $\delta_{\infty}$ is defined at infinity so that $t^a$
and $\varphi_{(\mu)}^a$ are fixed. 
Here note that $\kappa$ and $\Omega_H^{(\mu)}$ are surface gravity and 
angular velocities, respectively, for $\phi_0$.

Up to this point we have not yet used explicitly the fact that 
$\tilde{\xi}_0^a=0$ on $\Sigma$ for the perturbed solution, 
although we have used it implicitly. 
By using it explicitly, we can rewrite the l.h.s. of
(\ref{eqn:2-1:pre-1st-law2}) in a useful form. The result is 
%============< EQUATION >==============%
%
\begin{equation}
 \tilde{\delta}\int_{\Sigma}{\bf Q}[\tilde{\xi}_0] =
        \frac{1}{2\pi}\tilde{\delta}S,
\end{equation}
%======================================%
where $S$ is defined by (\ref{eqn:2-1:S1}). 
(For explicit manipulations, see the proof of Theorem 6.1 of
Iyer-Wald~\cite{Iyer&Wald1994}.)

Finally, we obtain the first law (\ref{eqn:2-1:1st-law}) for 
non-stationary perturbations $\tilde{\delta}\phi$ about a stationary 
black hole solution $\phi_0$.

%<<<<<<  Perturbed entropy  >>>>>>>>>>>%

Now we comment on entropy for the perturbed, non-stationary black 
hole.

As stated above, the ($n-2$)-surface $\Sigma$ has no meaning for the 
perturbed solution. (It is nothing but a surface on which
$\tilde{\xi}_0^a$ vanishes.) 
In general, it does not lie on the event (or 
apparent) horizon for the perturbed solution. Hence, entropy 
evaluated on $\Sigma$ may not coincide with that on a cross 
section of the perturbed horizon, provided that 
the entropy is defined as $2\pi$ times an integral of 
${\bf Q}[\tilde{\xi}_0]$ for both $(n-2)$-surfaces. 
Note that it is in general impossible to make gauge transformation so 
that $\Sigma$ lie on a horizon cross section, if entropy 
(eg. a quarter of area in general relativity) on $\Sigma$ is 
different from entropy on a horizon cross section. 
The difference is given by $2\pi$ times an integral of the Noether 
current ${\bf j}[\tilde{\xi}_0]$ over a hypersurface whose boundary is a 
union of $\Sigma$ and a cross section of the perturbed horizon. 
Since it is natural to assign black hole entropy to the horizon 
cross section~\cite{Iyer&Wald1994}, it might be expected that there 
appears an extra term corresponding to the integral of 
${\bf j}[\tilde{\xi}_0]$ in the first law.

However, as shown in the next paragraph, the integral of 
${\bf j}[\tilde{\xi}_0]$ vanishes to first order in 
$\tilde{\delta}\phi$~\footnote{
The author thanks Professor R. M. Wald for helpful comments on this
point. 
}.
Thus, $\tilde{\delta}S$ evaluated on $\Sigma$ gives the correct 
variation of entropy defined on the horizon to first order in
$\tilde{\delta}\phi$. 
This means that the extra term does not appear and that the first law
of Ref.~\cite{Iyer&Wald1994} derived in this section for
non-stationary perturbation about a stationary black hole 
is the correct formula.

Let us show the above statement. 
Since $\tilde{\xi}_0^a=0$ on $\Sigma$ and 
${\cal L}_{\tilde{\xi}_0}\phi_0=0$, the Noether current 
${\bf j}[\tilde{\xi}_0]$ vanishes on $\Sigma$ for the unperturbed 
solution by the definition~(\ref{eqn:2-1:j-def}). Hence, for the perturbed 
solution, the Noether current is at least first order in
$\tilde{\delta}\phi$ on $\Sigma$. On the other hand, deviation of a 
horizon cross section from $\Sigma$ is at least first order. 
Therefore, the integral of ${\bf j}[\tilde{\xi}_0]$ over a 
hypersurface connecting $\Sigma$ and the perturbed horizon cross
section is at least second order in $\tilde{\delta}\phi$.

Finally, let us apply the first law of this subsection to a stationary 
perturbation. The result is the same as that derived in the previous
subsection. 
It is evident that the gauge condition used in this subsection is
weaker than that used in the previous subsection. In fact,
$\tilde{\xi}^a$ ($\ne\tilde{\xi}_0^a$ for a perturbed solution) is not
fixed in the former condition. 
Hence, it can be concluded that the minimal set of gauge conditions 
necessary for the derivation of the first law is that $t^a$ and 
$\varphi_{(\mu)}^a$ are fixed at spatial infinity.

%%%%%%%%%%%%%%%%%%%%%%%%%%%%%%%%%%%%%%%%
%%%%%%%%%%%% SECTION 2-2 %%%%%%%%%%%%%%%
%%%%%%%%%%%%%%%%%%%%%%%%%%%%%%%%%%%%%%%%
\section{The first law of black hole dynamics}
\label{sec:1st_dynamics}

As will be seen in section \ref{sec:GSL}, the first law of black
hole statics is used in a (quasi-stationary but) dynamical 
situation to prove the generalized second
law~\cite{Frolov&Page1993,Mukohyama1997a}, which is a natural 
generalization of both the second law (or area law~\cite{Hawking1971})
of black hole and the second law of usual thermodynamics. In the
proof, by assuming quasi-stationarity, the use of the first law of
statics can be justified to relate a small change from an initial
stationary black hole to final stationary one. However, if we intend
to extend the proof of the generalized second law to finite changes
between two stationary black holes or a purely dynamical situation,
the first law of black hole statics can not be used.

So, we want to extend the first law to a dynamical situation and call
it a first law of black hole dynamics (BHD). 
It will be discussed in subsection \ref{subsec:remark_GSL} that the
generalized second law might be extended to not quasi-stationary
situations by using the first law of black hole dynamics.

In this section, for simplicity, we consider general relativity
only. 
In subsection \ref{subsec:dynamic-S} we consider two non-statistical
definitions of entropy for dynamic (non-stationary) black holes in
spherical symmetry. 
The first is analogous to the original Clausius definition
of thermodynamic entropy:
there is a first law containing an energy-supply term
which equals surface gravity times a total differential.
The second is Wald's Noether-charge method,
adapted to dynamic black holes by using the Kodama flow.
Both definitions give the same answer for Einstein gravity:
one-quarter the area of the trapping horizon~\cite{HMA1998}.
In subsection \ref{subsec:quasilocal_1st}, the first law of BHD is
derived without assuming any symmetry and any asymptotic
conditions~\cite{Mukohyama&Hayward1998}. In the derivation, a  
definition of dynamical surface gravity is proposed.

%======================================%
%<<<<<<   SUBSECTION 2-2-1    >>>>>>>>>%
%======================================%
\subsection{Dynamic black hole entropy in general relativity with
spherical symmetry}
\label{subsec:dynamic-S}

It is generally thought that
black-hole entropy should have a statistical origin,
presumably in a quantum theory of gravity.
This is, of course, due to the definition of entropy in statistical
mechanics. However, it should be remembered that
the original concept of entropy was not statistical~\cite{Truesdell_thermo}.
The original argument of Clausius was that, in a cyclic reversible process,
the total heat supply $\delta Q$ divided by temperature $\vartheta$
should vanish.
Thus in any reversible process, $\delta Q/\vartheta$
should be the total differential $dS$ of a state function $S$, the entropy.
Moreover, in irreversible processes,
there should be a second law $dS\ge\delta Q/\vartheta$.
The heat supply also occurs in a first law $dU=\delta Q+\delta W$,
where $U$ is the internal energy and $\delta W$ the work being done.
These are basic laws of thermodynamics as stated in typical textbooks and
originally formulated by Clausius before the invention of statistical
mechanics. 
In this subsection, we argue that there is a similar concept
of entropy for dynamic black holes,
suggested by the mathematical structure of the first law:
it contains an energy-supply term
which equals surface gravity times a total differential.

%<<<<<<  Kodama vector  >>>>>>>>>>>%
\subsubsection{Kodama vector and the first law of black hole dynamics
in spherical symmetry}

The relevant quantities and equations in spherical symmetry
may be summarized as follows.
The area $A$ or areal radius $r=\sqrt{A/4\pi}$ of the spheres of symmetry
determines the 1-form
\begin{equation}
k={*}dr
\end{equation}
where $d$ is the exterior derivative and $*$ is the Hodge operator
of the two-dimensional space normal to the spheres of symmetry.
Henceforth, 1-forms and their vector duals
with respect to the space-time metric will not be distinguished.
Then $k$ is the divergence-free vector introduced 
by Kodama~\cite{Kodama1980},
which generates a preferred flow of time
and is a dynamic analogue of a stationary Killing
vector~\cite{Hayward1998a}. The active gravitational energy or mass is 
\begin{equation}
E=(1-dr\cdot dr)r/2
	\label{eqn:2-2:MSenergy}
\end{equation}
where the dot denotes contraction.
Misner and Sharp~\cite{Misner&Sharp1964} originally defined $E$
and, thus, is called the Misner-Sharp energy. (Ref.~\cite{Hayward1996} 
described its physical properties.) The dynamic surface gravity
\begin{equation}
\kappa={*}dk/2	\label{eqn:2-2:kappa_sph}
\end{equation}
was defined in Ref.\cite{Hayward1998a}
by analogy with the standard definition of stationary surface gravity.
This reference also introduced two invariants of the energy tensor $T$:
the energy density (work density)
\begin{equation}
w=-\hbox{tr}\,T/2
\end{equation}
and the energy flux (localized Bondi flux)
\begin{equation}
\psi=T\cdot dr+wdr
\end{equation}
where tr denotes the the two-dimensional normal trace.
One may say that $(A,k)$ are the basic kinematic quantities,
$(E,\kappa)$ the gravitational quantities
and $(w,\psi)$ the relevant matter quantities.
Instead of $\psi$ one may also use
the divergence-free energy-momentum
vector~\cite{Hayward1998a,Kodama1980,Hayward1996} 
\begin{equation}
j={*}\psi+wk.
\end{equation}
Finally, the relevant components of the Einstein equation are
\cite{Hayward1998a} 
\begin{equation}
E=r^2\kappa+4\pi r^3w
\end{equation}
and\cite{Hayward1996}
\begin{equation}
Aj={*}dE.
\end{equation}
The latter may be rewritten as
\begin{equation}
dE=A\psi+wdV
\end{equation}
where $V={4\over3}\pi r^3$ is the areal volume.
This is the unified first law of Ref.\cite{Hayward1998a}.
One may regard $wdV$ as a type of work and $A\psi$ as an energy supply,
analogous to heat supply $\delta Q=\oint q$,
where $q$ is the heat flux.
The energy supply can be written as
\begin{equation}
A\psi={\kappa dA\over{8\pi}}+rd\left({E\over{r}}\right).
\end{equation}
The second term vanishes when projected along a black-hole horizon,
defined as in Refs.\cite{Hayward1998a,Hayward1994,Hayward1996} by a
trapping horizon: 
a hypersurface where $dr$ is null,
so that $E=r/2$.
This also occurs for any hypersurface on which $E/r$ is constant,
thereby covering any smooth space-time.
The key point is that the first term is the product of surface gravity
$\kappa$ and a total differential.
Identifying $\kappa/2\pi$ as a temperature,
this total differential therefore determines a {\it Clausius entropy}
$A/4$. Note that this stems from a purely mathematical property 
of the energy supply occurring in the first law.
The restriction to spherical symmetry will be removed 
in subsection \ref{subsec:quasilocal_1st}.

%<<<<<<  Wald-Kodama entropy  >>>>>>>>>>>%
\subsubsection{Wald-Kodama entropy}

Wald~\cite{Wald1993} also gave a definition of entropy
\begin{equation}
 \kappa S=2\pi\oint Q[\xi]
\end{equation}
where $Q$ henceforth denotes a Noether charge 2-form
obtained from a certain type of Lagrangian.
For Einstein gravity,
$Q_{ab}=-\epsilon_{abcd}\nabla^c\xi^d/16\pi$~\cite{Iyer&Wald1994}, 
where $\epsilon$ is the space-time volume form, 
$\xi$ is a generating vector for the diffeomorphisms
which is taken to be the horizon generating Killing vector of a
stationary black hole, and $\kappa$ is the surface gravity
corresponding to $\xi$. 
In spherical symmetry, we propose using the Kodama vector $k$ for
$\xi$ to give an alternative definition of the entropy of dynamic
black holes. We call it {\it Wald-Kodama entropy}. 
This prescription effectively corresponds to also replacing
$\xi$ by $k$ in $S_2$ of Jacobsen et al.~\cite{JKM1994}. 
Then from the above expression for $Q$ by Wald's method, 
\begin{equation}
\oint Q[k]={A\kappa\over{8\pi}},
\end{equation}
where the integral is over a sphere of symmetry.
Thus the Wald-Kodama entropy is
\begin{equation}
S=A/4.
\end{equation}

%<<<<<<  Speclations  >>>>>>>>>>>%
\subsubsection{Agreement and speculations}

So, the Wald-Kodama entropy agrees with the Clausius entropy. 
The motivations also seem similar,
since Wald's construction involved a first law of black-hole statics
based on perturbations $\delta$ of a stationary solution.
This agreement suggests that some combination of the two methods may
be useful in general, assuming neither stationarity nor Einstein
gravity.

It is also interesting that
both methods formally hold not just on a black-hole horizon,
but anywhere in the space-time, as do all the equations displayed above.
Whether any surface in any space-time should have an entropy
related to its area is arguable,
but this does concur with the entanglement entropy approach, which is
discussed in section \ref{sec:entanglement}.

%======================================%
%<<<<<<   SUBSECTION 2-2-2    >>>>>>>>>%
%======================================%
\subsection{Qusi-local first law of black hole dynamics in general
relativity} 
\label{subsec:quasilocal_1st}

%<<<<<<<<<   Def. of BH      >> >>>>>>>%
\subsubsection{A general definition of a black hole}

Here we would like to treat a dynamical, not necessarily
asymptotically flat spacetime. Even for such a general situation,
there is a definition of a black hole. Namely, a {\it black hole} is
defined as a future outer trapping horizon~\cite{Hayward1994}. The
{\it future outer trapping horizon} is the closure of a three-surface
foliated by marginal surfaces on which $\theta_-<0$ and ${\cal
L}_-\theta_+<0$, where the {\it marginal surface} is a 
spatial two-surface on which one of two null expansions (which we have 
denoted by $\theta_+$) vanishes. Here $\theta_-$ is another null
expansion and ${\cal L}_-$ is a Lie derivative w.r.t. a null vector
defined below. 
For the purpose of this subsection, we only need the fact that 
$\theta_+\theta_-=0$ along the horizon. Hence, the first law we shall 
obtain remains to hold for a general trapping horizon, i.e. a hypersurface 
foliated by marginal surfaces. We mention here that a trapping horizon 
can be regarded as a black hole, a white hole and a wormhole when it is 
future outer, past outer and temporal outer,
respectively~\cite{Hayward1998b}.

%<<<<<<<< Double-null formalism >>>>>>>%
\subsubsection{The double-null formalism}

To investigate behavior of the trapping horizon, the so-called
double-null formalism, or the $(2+2)$ decomposition, of general
relativity is useful. Among several
$(2+2)$-formalisms~\cite{GHP1973&etc,Hayward1993}, we adopt one 
based on Lie derivatives w.r.t null vectors 
developed by Hayward~\cite{Hayward1993}. 
Let us review basic ingredients
of the formalism. Suppose that a four-dimensional spacetime manifold 
$(M,g)$ is foliated (at least locally) by two families of null
hypersurfaces $\Sigma^{\pm}$, each of which is parameterized
by a scalar $\xi^{\pm}$, respectively. 
The null character is described by $g^{-1}(n^{\pm},n^{\pm})=0$, 
where $n^{\pm}=-d\xi^{\pm}$ are normal $1$-forms to $\Sigma^{\pm}$.
The relative normalization of the null normals defines 
a function $f$ as $g^{-1}(n^+,n^-)=-e^f$.
The intersections of $\Sigma^+(\xi^+)$ and
$\Sigma^-(\xi^-)$ define a two-parameter family of two-dimensional
spacelike surfaces $S(\xi^+,\xi^-)$. Hence, by introducing an intrinsic
coordinate system ($\theta^1$,$\theta^2$) of the $2$-surfaces, 
the foliation is described by the imbedding 
$x = x(\xi^+,\xi^-;\theta^1,\theta^2)$.

For the imbedding, the intrinsic metric on the $2$-surfaces is found 
to be $h=g+e^{-f}(n^+\otimes n^- +n^-\otimes n^+)$. 
Correspondingly, the vectors $u_{\pm}=\partial/\partial\xi^{\pm}$ 
have 'shift vectors' $s_{\pm}=\perp u_{\pm}$, where $\perp$ indicates 
projection by $h$.
The $4$-dimensional metric is written in terms of ($h$,$f$,$s_{\pm}$) 
as
%============< EQUATION >==============%
%
\begin{equation}
 g = \left(
	\begin{array}{ccc}
	h(s_+,s_+) &	h(s_+,s_-)-e^{-f} &	h(s_+) \\
	h(s_-,s_+)-e^{-f} &	h(s_-,s_-) &	h(s_-) \\
	h(s_+) &	h(s_-) &	h 
	\end{array}
	\right).
\end{equation}
%======================================%
Geometrical quantities such as {\it expansions} $\theta_{\pm}$, 
{\it shears} $\sigma_{\pm}$ and 
the {\it twist} $\omega$ are defined by 
$\theta_{\pm}=*{\cal L}_{\pm}*1$, 
$\sigma_{\pm}=\perp{\cal L}_{\pm}h-\theta_{\pm}h$ and 
$\omega=e^f [l_-,l_+]/2$, where
$*$ denotes the Hodge-dual operator of $h$,
$l_{\pm}=u_{\pm}-s_{\pm}=e^{-f}g^{-1}(n^{\mp})$ 
are null normal vectors to $\Sigma^{\pm}$, 
and ${\cal L}_{\pm}$ denotes the Lie derivative along $l_{\pm}$, 
respectively. 
It is possible to write down the Einstein tensor in terms of these
geometrical quantities. The component useful for our purpose is
$G_{+-}=G(l_+,l_-)$, which is given by
%============< EQUATION >==============%
%
\begin{equation}
 2e^fG_{+-} = 
	{}^{(2)}R + 
	e^f ({\cal L}_+\theta_- +{\cal L}_-\theta_++2\theta_+\theta_-)
	-2\left[ h(\omega,\omega) + \frac{1}{4}h^{\sharp}(df,df)
	\right] + {\cal D}^2f.
\label{eqn:2-2:G}
\end{equation}
%======================================%
Here $h^{\sharp}=g^{-1}hg^{-1}$ is $h$ raised by $g^{-1}$, 
${\cal D}^2$ and ${}^{(2)}R$ are the two-dimensional Laplacian  
and the Ricci scalar both associated with the metric $h$.

%<<<<<<<<< Hawking energy>>>>>>>>>>>>>>%
\subsubsection{Hawking energy}

Before deriving the first law we have to define energy and surface
gravity in a quasi-local way. In spherical symmetry there 
is a widely accepted energy: the Misner-Sharp (MS) 
energy (\ref{eqn:2-2:MSenergy})~\cite{Misner&Sharp1964}. In
the previous subsection the MS energy is used to derive the first law
of BHD in spherical symmetry~\cite{Hayward1998a}. In this subsection
we adopt the Hawking energy~\cite{Hawking1968}, which reduces to 
the MS energy in spherical symmetry. It is defined by
%============< EQUATION >==============%
%
\begin{equation}
 E(\xi^+,\xi^-) = \frac{r}{16\pi}\int_{S(\xi^+,\xi^-)} 
        d^2\theta\sqrt{h} 
        \left[{}^{(2)}R + e^{f}\theta_+\theta_-\right],
        \label{eqn:2-2:energy}
\end{equation}
%======================================%
where $h$ is the determinant of the two-dimensional metric $h_{ab}$
and the area radius $r$ is defined by 
%============< EQUATION >==============%
%
\begin{equation}
 r = \sqrt{A/4\pi},
 A = \int_{S(\xi^+,\xi^-)} d^2\theta\sqrt{h}.
\end{equation}
%======================================%

%<<<<<< Dynamical surface gravity >>>>>%
\subsubsection{A proposal of dynamical surface gravity}

In Ref.~\cite{Hayward1998a}, a definition of dynamical surface gravity
was proposed in spherical symmetry as (\ref{eqn:2-2:kappa_sph}). A
natural generalization to a non spherically-symmetric case
is 
%============< EQUATION >==============%
%
\begin{equation}
 \kappa (\xi^+,\xi^-) = \frac{-1}{16\pi r}
	\int_{S(\xi^+,\xi^-)}d^2\theta\sqrt{h} \ 
        e^{f}
	({\cal L}_+\theta_-+{\cal L}_-\theta_++\theta_+\theta_-). 
	\label{eqn:2-2:kappa}
\end{equation}
%======================================%
This is the most simple generalization in the sense that 
it includes neither the shear $\sigma_{Aab}$ nor 
the twist $\omega^a$.

Note that this definition of surface gravity and the definition
(\ref{eqn:2-2:energy}) of the Hawking energy are both
quasi-local in the sense originally introduced by
Penrose~\cite{Penrose1982}. Hence we call the corresponding first 
law, which we shall derive below, {\it the quasi-local first law} of
BHD.

%<<<<<<<<< Dynamical 1st law >>>>>>>>>>%
\subsubsection{The dynamical first law}

We now derive the quasi-local first law of BHD for the Hawking energy 
(\ref{eqn:2-2:energy}) and the surface gravity defined by 
(\ref{eqn:2-2:kappa}). 
It is easy to show that 
%============< EQUATION >==============%
%
\begin{equation}
 dE - \frac{\kappa}{8\pi} dA = 
        wAdr + r d\left(\frac{E}{r}\right),\label{eqn:2-2:pre-1st-law}
\end{equation}
%======================================%
where $w$ is defined by
%============< EQUATION >==============%
%
\begin{equation}
 w = \frac{1}{A}
        \left(\frac{E}{r} 
        -\kappa r\right).
\end{equation}
%======================================%
Here note that '$d$' in Eq.~(\ref{eqn:2-2:pre-1st-law}) is not a variation 
in a space of stationary solutions of the Einstein equation as in the
first law of black hole statistics, but is the differentiation w.r.t.
the parameters $\xi^{\pm}$ of the spacetime foliation. 
(For example, $dE=d\xi^+\partial_+E+d\xi^-\partial_-E$.) 
We mention that Eq.~(\ref{eqn:2-2:pre-1st-law}) holds independently of the 
definitions of $E$ and $\kappa$ while the following arguments depend 
on the definitions.

Since the Gauss-Bonnet theorem says that 
%============< EQUATION >==============%
%
\[
 \int_{S(\xi^+,\xi^-)} d^2\theta\sqrt{h} \ {}^{(2)}R = 
        8\pi (1-\gamma ),
\]
%======================================%
where $\gamma$ is the genus or number of handles of $S(\xi^+,\xi^-)$, 
the energy divided by area radius is given by $E/r=(1-\gamma )/2$ on a 
marginal surface and is a constant. Thus, 
%============< EQUATION >==============%
%
\begin{equation}
 E' = \frac{\kappa}{8\pi} A' + wAr',
\end{equation}
%======================================%
where the prime denotes the derivative along the trapping horizon. 
This is the quasi-local first law of BHD. Note that this also holds
along any hypersurface foliated by 2-surfaces on which $E/r$ is
constant.

%<<<<<<<<<< Work term >>>>>>>>>>>>>>>>>%
\subsubsection{The work term}

By using Eq.~(\ref{eqn:2-2:G}) it is easy to show that $w$ is written as 
follows.
%============< EQUATION >==============%
%
\begin{equation}
 w = \rho_m + \rho_j, 
\end{equation}
%======================================%
where the averaged matter energy density $\rho_m$ and the effective
angular energy density $\rho_j$ are defined by 
%============< EQUATION >==============%
%
\begin{eqnarray}
 \rho_m & = & \frac{1}{8\pi A}
        \int_{S(\xi^+,\xi^-)}d^2\theta\sqrt{h}  
        e^{f}G_{+-},\nonumber\\
 \rho_j & = & \frac{1}{8\pi A}
        \int_{S(\xi^+,\xi^-)}d^2\theta\sqrt{h}\left[
        h(\omega,\omega) + \frac{1}{4}h^{\sharp}(df,df) \right].
\end{eqnarray}
%======================================%
The Einstein equation $G=8\pi T$ says that $\rho_m$ is $e^fT(l_+,l_-)$
averaged over the $2$-surface. It seems that $\rho_j$ represents
effective energy density due to angular momentum. 

The term $wAr'$ should be a work term done along the horizon. For 
example, for an electromagnetic field, the term $\rho_mAr'$ reduces to 
the electromagnetic work done along the
horizon~\cite{Hayward1998a}. It seems that the term $\rho_jAr'$ is a
work associated with angular momentum of the trapping horizon.

%<<<<<<<< Comments >>>>>>>>%
\subsubsection{Comments}

In this subsection the quasi-local first law of black hole dynamics
has been derived without assuming any symmetry and any asymptotic
condition.  In the derivation we have given a new
definition of dynamical surface gravity. In spherical symmetry it 
reduces to that defined in Ref.~\cite{Hayward1998a}.

By using the quasi-local first law derived in this subsection, it
might be possible to extend a proof of the generalized second law to
not quasi-stationary situations. (See subsection
\ref{subsec:remark_GSL}.)

Besides the first law derived in this subsection, there exist the 
second law~\cite{Hayward1994} and the third law~\cite{Israel1986} for 
the trapping horizon (or apparent horizon). It seems that by using 
these laws we can formulate black hole thermodynamics consistently as 
trapping horizon dynamics. However, for this purpose, 
there is an important open question:
we have to associate temperature of quantum fields with the 
trapping horizon. 
All we can say here is that the temperature may be given by 
$\hbar\kappa/2\pi$, where $\kappa$ is the surface gravity introduced 
here.

The final comment is in order. The surface gravity $\kappa(\xi^+,\xi^-)$ 
is an invariant of a double-null foliation at the surface. 
Since a non-null trapping horizon locally determines a unique 
double-null foliation, the surface gravity is definitely an 
invariant of the trapping horizon if the horizon is not null.
On the other hand, the null case is ambiguous because of the freedom to 
rescale the other null direction.  
Fixing this would require some kind of limiting argument that might be 
effectively a zeroth law.
Therefore, we have to impose an auxiliary condition 
for the surface gravity $\kappa(\xi^+,\xi^-)$ to work well 
when the trapping horizon is null.
Since surface gravity seems to be related to temperature of quantum 
fields as stated above, it will be valuable to investigate the auxiliary 
condition in detail~\cite{Mukohyama&Hayward1998}.

%%%%%%%%%%%%%%%%%%%%%%%%%%%%%%%%%%%%%%%%
%%%%%%%%%%%% SECTION 2-3 %%%%%%%%%%%%%%%
%%%%%%%%%%%%%%%%%%%%%%%%%%%%%%%%%%%%%%%%

\section{The generalized second law}
	\label{sec:GSL}

The generalized second law of black hole thermodynamics is a statement 
that a sum of black hole entropy and thermodynamic entropy of matter
fields outside the horizon does not decrease
\cite{Zurek&Thorne1985,Wald1992,Frolov&Page1993,Mukohyama1997a}, where
the black hole 
entropy is defined as a quarter of the area of the horizon. Namely it
says that an entropy of the whole system does not decrease. It
interests us in a quite physical sense since it links a world inside a 
black hole and our thermodynamic world. In particular it gives a
physical meaning to black hole entropy indirectly since it concerns
the sum of black hole entropy and ordinary thermodynamic entropy, and
since physical meaning of the latter is well-known by statistical
mechanics. 

Frolov and Page \cite{Frolov&Page1993} proved the generalized second
law for a quasi-stationary eternal black hole by assuming that a state
of matter fields on the past horizon is thermal one and that a set of
radiation modes on the past horizon and a set of radiation modes on
the past null infinity are quantum mechanically uncorrelated. The
assumption is reasonable for the eternal case since a black hole emit
a thermal radiation (the Hawking radiation). When we attempt to apply
their proof to a non-eternal black hole which is formed by
gravitational collapse, we might expect that things would go well by
simply replacing the past horizon with a null surface at a moment of a
formation of a horizon ($v=v_0$ surface in {\it Figure}
\ref{fig:modes}). However, it is not the case since the above
assumption does not hold in this case. The reason is that on a 
background describing gravitational collapse the thermal radiation is
observed not at the moment of the horizon formation but at the future
null infinity and 
that any modes on the future null infinity have correlation with modes
on the past null infinity located after the horizon formation. The
correlation can be seen in the equation (\ref{eqn:2-3:multi}) of this
section explicitly. Thus, their proof does not hold for the case in
which a black hole is formed by gravitational collapse. Since a black
hole is thought to be formed by gravitational collapse in
astrophysical situation, we want to prove the generalized second law
in this case.

The rest of this section is organized as follows. In
subsection \ref{subsec:scalar_field} we consider a real massless
scalar field in a background of a gravitational collapse to show that 
a thermal state with special values of temperature and chemical
potential evolves to a thermal state with the same temperature and the
same chemical potential. These special values are determined by the
background geometry. In subsection \ref{subsec:proof_GSL}, 
first, the generalized second law is rewritten as an inequality which
states that there is a non-decreasing functional of a density matrix
of matter fields. After that, we give a theorem which shows an
inequality between functionals of density matrices. Finally, we apply
it to the scalar field investigated in subsection
\ref{subsec:scalar_field} to prove the generalized second law for the
quasi-stationary background. In subsection \ref{subsec:remark_GSL} we
summarize this section.

%======================================%
%<<<<<<   SUBSECTION 2-3-1    >>>>>>>>>%
%======================================%
\subsection{A massless scalar field in black hole background}
	\label{subsec:scalar_field}

\begin{figure}
 \begin{center}
  \epsfile{file=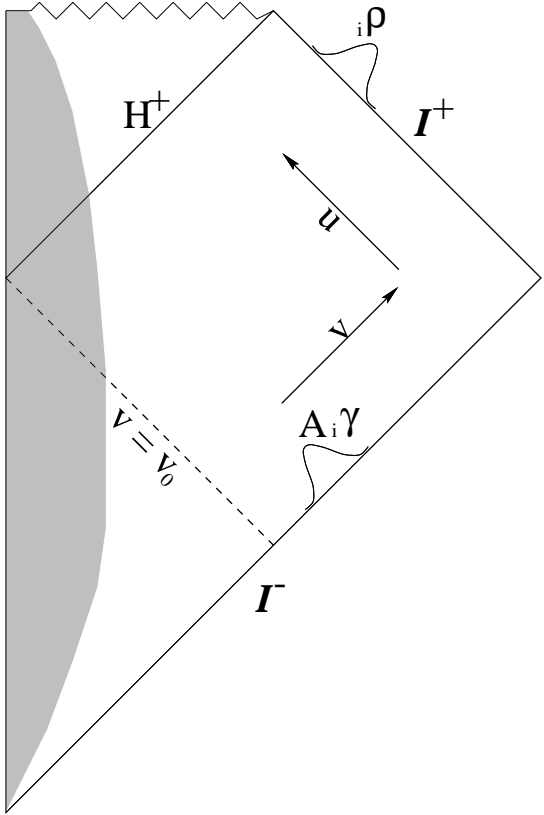,scale=0.7}
 \end{center}
 \caption{
A conformal diagram of a background geometry which describes a 
gravitational collapse. $\I^-$ and $\I^+$ are the past null infinity
and the future null infinity, respectively and $H^+$ is the future
event horizon. 
Shaded region represents collapsing materials which forms the black
hole. Besides the collapsing matter, we consider a real massless
scalar field and investigate a scattering problem by the black hole
after its formation ($v>v_0$). Thus we specify possible initial states 
at $\I^-$ to those states which are excited from the vacuum by only
modes whose support is within $v>v_0$ (elements of
$\F_{\I^-(v>v_0)}$),
and possible mixed states constructed from them (elements of $\P$). In
the diagram, $A\ _i\gamma$ ($i=1,2,\cdots$) is a mode function
corresponding to a wave packet whose peak is at $v>v_0$ on $\I^-$, 
$\!_i\rho$ ($i=1,2,\cdots$) is a mode function corresponding to a wave 
packet on $\I^+$.
}
\label{fig:modes}
\end{figure}

In this subsection we consider a real massless scalar field in a
curved background which describes formation of a quasi-stationary
black hole. Let us denote the past null infinity by $\I^-$, the future
null infinity by $\I^+$ and the future event horizon by
$H^+$. Introduce the usual null coordinates $u$ and $v$, and suppose
that the formation of the event horizon $H^+$ is at $v=v_0$ 
(see {\it Figure}~\ref{fig:modes}). 
On $\I^-$ and $\I^+$, by virtue of the
asymptotic flatness, there is a natural definition of Hilbert spaces
$\H_{\I^-}$ and $\H_{\I^+}$ of mode functions with positive
frequencies \cite{Wald1975&1976}. The Hilbert spaces $\F(\H_{\I^\pm})$
of all asymptotic states are defined as follows with a suitable
completion (symmetric Fock spaces): 
\[
 \F(\H_{\I^\pm}) \equiv 
	\bm{C} \oplus \H_{\I^\pm} \oplus 
	\left(\H_{\I^\pm}\otimes\H_{\I^\pm}\right)_{sym} 
	\oplus \cdots,	
\]
where $(\cdots)_{sym}$ denotes the symmetrization
($(\xi\otimes\eta )_{sym}=
	\frac{1}{2}(\xi\otimes\eta +\eta\otimes\xi )$, etc.). 
Physically, $\bm{C}$ denotes the vacuum state, $\H_{\I^\pm}$ one
particle states, $\left(\H_{\I^\pm}\otimes\H_{\I^\pm}\right)_{sym}$
two particle states, etc.. We suppose that all our observables are 
operators on $\F(\H_{\I^\pm})$ since we observe a radiation of the
scalar field radiated by the black hole at places far away from it. In
this sense $\F(\H_{\I^+})$ are quite physical. Next let us consider
how to set an initial state of the scalar field. We want to see a
response of the scalar field on the quasi-stationary black hole
background which is formed by gravitational collapse of other
materials (a dust, a fluid, etc.). Hence as the initial state at
$\I^-$ we consider a state such that it includes no excitations of
modes located before the formation of the horizon (no excitation at
$v<v_0$). A space of all such states is a subspace of $\F(\H_{\I^-})$,
and we denote it by $\F_{\I^-(v>v_0)}$. We like to derive a thermal
property of a scattering process of the scalar field by the
quasi-stationary black hole. Hence we consider density matrices on
$\F_{\I^-(v>v_0)}$ and $\F(\H_{\I^+})$. Denote a space of all density
matrices on $\F_{\I^-(v>v_0)}$ by $\P$ and a space of all density
matrices on $\F(\H_{\I^+})$ by $\tilde{\P}$. 

Let us discuss the evolution of a state at $\I^-$ to future. Since
$\I^{+}$ is not a Cauchy surface because of the existence 
of $H^+$, $\F(\H_{\I^-})$ is mapped not to $\F(\H_{\I^+})$ but to
$\F(\H_{\I^+})\otimes\F(\H_{H^+})$ by a unitary evolution, where
$\H_{H^+}$ is a Hilbert space of mode functions on the horizon with a
positive frequency, and $\F(\H_{H^+})$ is a Hilbert
space of all states on $H^+$ defined as a symmetric Fock
space (see the definition of $\F(\H_{\I^\pm})$). Although there is no
natural principle to determine positive frequency modes
(equivalently, there is no natural definition of the particle concept) 
on $H^+$, how to define $\H_{H^+}$ does not affect the result since 
we shall trace out the degrees of freedom of $\F(\H_{H^+})$ (see
(\ref{eqn:2-3:def-T})).  To describe the evolution of a quantum state of
the scalar field from $\F(\H_{\I^-})$ to $\F(\H_{\I^+})\otimes\F(\H_{H^+})$
an S-matrix is introduced \cite{Wald1975&1976}. For a given initial
state $|\psi\rangle$ in $\F(\H_{\I^-})$, the corresponding final state
in $\F(\H_{\I^+})\otimes\F(\H_{H^+})$ is $S|\psi\rangle$. Then the
corresponding evolution from $\F_{\I^-(v>v_0)}$ to $\F(\H_{\I^+})$ is
obtained by restricting $S$ to $\F_{\I^-(v>v_0)}$, and we denote it by
$S$, too. The S-matrix elements was given by
Wald~\cite{Wald1975&1976}.

%<<<<<<   Definition    >>>>>>>>>%
\subsubsection{Superscattering matrix $T$}
      \label{subsec:def-T} 
Suppose that the initial state of the scalar field is $|\phi\rangle$
($\in\F(\H_{\I^-})$) and that the corresponding final state is
observed at $\I^+$ (see the argument after the definition of
$\F(\H_{\I^\pm})$). Formally the observation corresponds to a
calculation of a matrix element 
$\langle\phi |S^{\dagger}OS|\phi\rangle$, where $S$ is the
S-matrix which describes the evolution of the scalar field from
$\F_{\I^-(v>v_0)}$ to $\F(\H_{\I^+})\otimes\F(\H_{H^+})$ and $O$ is a
self-adjoint operator on $\F(\H_{\I^+})$ corresponding to a quantity
we want to observe. The matrix element can be rewritten in the
following convenient fashion: 
\[
 \langle\phi |S^{\dagger}OS|\phi\rangle =
 \bm{Tr}_{\I^+}\left[ O\rho_{red}\right],
\]
where 
\[
 \rho_{red} =
	\bm{Tr}_{H^+}\left[ S|\phi\rangle\langle\phi 
		|S^{\dagger}\right],
\]
$\bm{Tr}_{\I^+}$ and $\bm{Tr}_{H^+}$ denote partial trace over
$\F(\H_{\I^+})$ and $\F(\H_{H^+})$, respectively. In viewing this
expression we are lead to an interpretation that the corresponding 
final state at $\I^+$ is represented by the reduced density matrix
$\rho_{red}$. Next we generalize this argument to a wider class of 
initial states, which includes all mixed states. For this case an
initial state is represented not by an element of $\F_{\I^-(v>v_0)}$
but by an element of $\P$ (a density matrix on
$\F_{\I^-(v>v_0)}$). Its evolution to $\I^+$ is represented by the
so-called superscattering matrix $T$ defined as follows: let 
$\rho$ ($\in\P$) be an initial density matrix then the corresponding
final density matrix $T(\rho )$ ($\in\tilde{\P}$) is
\begin{equation}
 T(\rho ) = \bm{Tr}_{H^+}\left[ S\rho S^{\dagger}\right].
		\label{eqn:2-3:def-T}
\end{equation}
Note that $T$ is a linear map from $\P$ into $\tilde{\P}$. 

%============< FIGURE >==============%
%              modes.ps
%======================================%

%<<<<<<   Thermodynamic property    >>>>>>>>>%
\subsubsection{Thermodynamic property of $T$ }

Let us calculate a conditional probability defined as follows: 
\begin{equation}
 P(\{n_{\!_{i}\rho}\}|\{n_{\!_{i}\gamma}\}) \equiv 
     \langle \{n_{\!_{i}\rho}\}| 
     T\left( |\{n_{\!_{i}\gamma}\}\rangle
     \langle \{n_{\!_{i}\gamma}\}|\right)|\{n_{\!_{i}\rho}\}\rangle,
                \label{eqn:2-3:def-P}
\end{equation}
where
\begin{eqnarray}
 |\{n_{\!_{i}\gamma}\}\rangle & \equiv & 
      \left[\prod_{i}\frac{1}{\sqrt{n_{\!_{i}\gamma}!}}
      \left(a^{\dagger}(A\ _{i}\gamma)\right)^{n_{\!_{i}\gamma}}
      \right]|0\rangle, \nonumber\\
 |\{n_{\!_{i}\rho}\}\rangle & \equiv & 
      \left[\prod_{i}\frac{1}{\sqrt{n_{\!_{i}\rho}!}}
      \left(a^{\dagger}(\!_{i}\rho)\right)^{n_{\!_{i}\rho}}
      \right] |0\rangle.    \label{eqn:2-3:def-basis}
\end{eqnarray}
$|\{n_{\!_{i}\gamma}\}\rangle$ is a state in $\F_{\I^-(v>v_0)}$
characterized by a set of integers $n_{\!_{i}\gamma}$
($i=1,2,\cdots$) and $|\{n_{\!_{i}\rho}\}\rangle$ is a state in
$\F(\H_{\I^+})$ characterized by a set of integers $n_{\!_{i}\rho}$
($i=1,2,\cdots$). Therefore
$P(\{n_{\!_{i}\rho}\}|\{n_{\!_{i}\gamma}\})$ is a conditional
probability for the final state to be $|\{n_{\!_{i}\rho}\}\rangle$  when
the initial state is specified to be $|\{n_{\!_{i}\gamma}\}\rangle$. In
the expressions, $A$ is a part of a representation of a Bogoliubov
transformation~\cite{Wald1975&1976}, which represents a map from
$\H_{\I^+}\oplus\H_{H^+}$ to $\H_{\I^-}$, and $\!_{i}\gamma$ is 
a unit vector in $\H_{\I^+}\oplus\H_{H^+}$ such that $A\ _{i}\gamma$
corresponds to a wave packet whose peak is located at a point on
$\I^-$ later than the formation of the horizon ($v>v_0$)
\cite{Wald1975&1976}. On the other hand, $\!_{i}\rho$ is a unit vector
in $\H_{\I^+}$ and corresponds to a wave packet on $\I^+$
\cite{Wald1975&1976} (see {\it Figure} \ref{fig:modes}). The
probability (\ref{eqn:2-3:def-P}) is a generalization of $P(k|j)$
investigated by Panangaden and Wald \cite{Panangaden&Wald1977}. 
(Our $P(\{n_{\!_{i}\rho}\}|\{n_{\!_{i}\gamma}\})$ reduces to $P(k|j)$
of Panangaden-Wald when $n_{\!_{i_0}\gamma}=j$, $n_{\!_{i_0}\rho}=k$
and $n_{\!_{i}\gamma}=n_{\!_{i}\rho}=0$ for all $i$ other than
$i_0$. Here $i_0$ is an arbitrary fixed value of $i$.)
Evidently, our conditional probability
$P(\{n_{\!_{i}\rho}\}|\{n_{\!_{i}\gamma}\})$ includes more abundant
information 
\footnote{
All the information about the response of the scalar field is included
in 
$T_{\{ n_{\!_{i}\rho}\} \{ n'_{\!_{i}\rho}\} }
	^{\{ n_{\!_{i}\gamma}\} \{ n'_{\!_{i}\gamma}\} }$ 
defined in Lemma \ref{lemma:off-diagonal}.
}
about a response of the scalar field than $P(k|j)$. In fact, any
initial states on $\I^-$, which include no excitation 
before the formation of the horizon ($v<v_0$.), can be represented by
using the basis $\left\{|\{n_{\!_{i}\gamma}\}\rangle\right\}$ and any
final states on $\I^+$ can be expressed by the basis
$\left\{|\{n_{\!_{i}\rho}\}\rangle\right\}$, i.e. a set of all 
$|\{n_{\!_{i}\gamma}\}\rangle$ generates $\F(\H_{\I^-(v>v_0)})$ and a
set of all $|\{n_{\!_{i}\rho}\}\rangle$ generates
$\F(\H_{\I^+})$. This is the very reason why we have generalized
$P(k|j)$ to $P(\{n_{\!_{i}\rho}\}|\{n_{\!_{i}\gamma}\})$.

By using the S-matrix elements given in \cite{Wald1975&1976}, the
conditional probability is rewritten as follows (see appendix
\ref{app:probability} for its derivation): 
\begin{eqnarray}
 & &P(\{ n_{\!_{i}\rho}\} |\{ n_{\!_{i}\gamma}\} )\nonumber\\
 & & = \prod_{i}\left[ (1-x_i)x_i^{2n_{\!_{i}\rho}}
      \left( 1-|R_{i}|^{2}\right)^{n_{\!_{i}\gamma}+n_{\!_{i}\rho}}
		\right.\nonumber	\\
 & &  \times\sum_{l_{i}=0}^{\min (n_{\!_{i}\gamma},n_{\!_{i}\rho})}
      \sum_{m_{i}=0}^{\min (n_{\!_{i}\gamma},n_{\!_{i}\rho})}
      \frac{\left[ -|R_{i}|^{2}/(1-|R_{i}|^{2})
      \right]^{l_{i}+m_{i}}n_{\!_{i}\gamma}!n_{\!_{i}\rho}!}
      {l_{i}!(n_{\!_{i}\gamma}-l_{i})!(n_{\!_{i}\rho}-l_{i})!
       m_{i}!(n_{\!_{i}\gamma}-m_{i})!(n_{\!_{i}\rho}-m_{i})!} 
		\nonumber\\
 & &  \left.\times\sum_{n_i=n_{\!_{i}\rho}-\min (l_{i},m_{i})}
      ^{\infty}\frac{n_{i}!(n_{i}-n_{\!_{i}\rho}+n_{\!_{i}\gamma})!}
      {(n_{i}-n_{\!_{i}\rho}+l_{i})!(n_{i}-n_{\!_{i}\rho}+m_{i})!}
      (x_{i}^{2}|R_{i}|^{2})^{n_{i}-n_{\!_{i}\rho}}\right],
	\label{eqn:2-3:probability}
\end{eqnarray}
where $R_i$ is a reflection coefficient for the mode specified by the
integer $i$ on the Schwarzschild metric (see appendix
\ref{app:probability}) and $x_i$ is a constant defined by 
$x_{i}=\exp (-\pi (\omega_{i}-\Omega_{BH}m_{i})/\kappa )$. In the
expression, $\omega_{i}$ and $m_{i}$ are the frequency and the
azimuthal angular momentum quantum number of the mode specified by the
integer $i$, $\Omega_{BH}$ and $\kappa$ are the angular velocity and
the surface gravity of the black hole. 

Now, the expression in the squared bracket in (\ref{eqn:2-3:probability})
appears also in the calculation of $P(k|j)$. Using the result of
\cite{Panangaden&Wald1977}, it is easily shown that 
\begin{equation}
 P(\{ n_{\!_{i}\rho}\} |\{ n_{\!_{i}\gamma}\} ) =
      \prod_{i}\left[K_{i}\sum_{s_{i}=0}
      ^{\min (n_{\!_{i}\rho},n_{\!_{i}\gamma})}
      \frac{(n_{\!_{i}\rho}+n_{\!_{i}\gamma}-s_{i})!v_{i}^{s_{i}}}
      {s_{i}!(n_{\!_{i}\rho}-s_{i})!(n_{\!_{i}\gamma}-s_{i})!}
      \right],   \label{eqn:2-3:multi}
\end{equation}
where 
\begin{eqnarray*}
 K_{i} & = & \frac{(1-x_i)x_{i}^{2n_{\!_{i}\rho}}\left( 
      1-|R_{i}|^{2}\right)^{n_{\!_{i}\gamma}+n_{\!_{i}\rho}}}
      {\left( 1-|R_{i}|^{2}x_{i}^{2}\right)
      ^{n_{\!_{i}\gamma}+n_{\!_{i}\rho}+1}},  \\
 v_{i} & = & \frac{\left( |R_{i}|^{2}-x_{i}^{2}\right)
      \left( 1-|R_{i}|^{2}x_{i}^{2}\right)}
      {\left( 1-|R_{i}|^{2}\right)^{2}x_{i}^{2}}.
\end{eqnarray*}

This is a generalization of the result of \cite{Panangaden&Wald1977},
and the following lemma is easily derived by using this expression. 

%
%=====================================%
%============<  LEMMA  >==============%
%
\begin{lemma}\label{lemma:balance}
For the conditional probability defined by (\ref{eqn:2-3:def-P}) the
following equality holds:
\begin{eqnarray}
 & & P(\{ n_{\!_{i}\rho}=k_{i}\} |\{ n_{\!_{i}\gamma}=j_{i}\} )
      e^{-\beta_{BH}\sum_{i}j_{i}
      (\omega_{i}-\Omega_{BH}m_{i})}	\nonumber\\
 & & =
 P(\{ n_{\!_{i}\rho}=j_{i}\} |\{ n_{\!_{i}\gamma}=k_{i}\} )
      e^{-\beta_{BH}\sum_{i}k_{i}
      (\omega_{i}-\Omega_{BH}m_{i})},\label{eqn:2-3:balance}
\end{eqnarray}
where $\omega_{i}$ and $m_{i}$ are the frequency and the azimuthal angular
momentum quantum number of the mode specified by $i$, $\Omega_{BH}$ is 
the angular velocity of the horizon and 
\[
 \beta_{BH} \equiv 2\pi /\kappa.
\]
Here $\kappa$ is the surface gravity of the black hole. 
\end{lemma}

Note that $\beta_{BH}^{-1}$ is the Hawking temperature of the black
hole. This lemma states that a detailed balance condition holds 
\footnote{
It guarantees that a thermal distribution of any temperature
is mapped to a thermal distribution of some other temperature closer
to the Hawking temperature, as far as the diagonal elements are
concerned.
}. 
Summing up about all $k$'s, we expect that a thermal density matrix
$\rho_{th}(\beta_{BH},\Omega_{BH})$ in $\P$ with a temperature
$\beta_{BH}^{-1}$ and a chemical potential $\Omega_{BH}$ for azimuthal 
angular momentum quantum number will be mapped by $T$ to a thermal
density matrix $\tilde{\rho}_{th}(\beta_{BH},\Omega_{BH})$ in 
$\tilde{\P}$ with the same temperature and the same chemical
potential. To show that this expectation is true, we have to prove
that all off-diagonal elements of 
$T\left(\rho_{th}(\beta_{BH},\Omega_{BH})\right)$ are zero. For this
purpose the following lemma is proved in Appendix \ref{app:lemma}. 

%
%=====================================%
%============<  LEMMA  >==============%
%
\begin{lemma}\label{lemma:off-diagonal}
Denote a matrix element of $T$ as 
\begin{equation}
 T_{\{ n_{\!_{i}\rho}\} \{ n'_{\!_{i}\rho}\} }
	^{\{ n_{\!_{i}\gamma}\} \{ n'_{\!_{i}\gamma}\} } \equiv
 \langle\{n_{\!_{i}\rho}\}| T\left(
	|\{n_{\!_i\gamma}\}\rangle\langle\{n'_{\!_i\gamma}\}|
	\right) |\{n'_{\!_{i}\rho}\}\rangle.
\end{equation}
Then 
\begin{equation}
 T_{\{ n_{\!_{i}\rho}\} \{ n'_{\!_{i}\rho}\} }
	^{\{ n_{\!_{i}\gamma}\} \{ n'_{\!_{i}\gamma}\} } = 0, 
\end{equation}
unless
\begin{equation}
 n_{\!_{i}\gamma}-n'_{\!_{i}\gamma}= n_{\!_{i}\rho}-n'_{\!_{i}\rho}
\end{equation}
for $\forall i$.
\end{lemma}

Lemma \ref{lemma:off-diagonal} shows that all off-diagonal elements of 
$T(\rho )$ in the basis $\left\{|\{n_{\!_{i}\rho}\}\rangle\right\}$
vanish if all off-diagonal elements of $\rho$ in the basis
$\left\{|\{n_{\!_{i}\gamma}\}\rangle\right\}$ is zero. Thus, combining 
it with Lemma \ref{lemma:balance} and the well-known fact that
$|0\rangle$ is mapped to the thermal state, the following theorem is
easily proved. Note that a set of all 
$|\{n_{\!_i\gamma}\}\rangle\langle\{n'_{\!_i\gamma}\}|$ generates $\P$ 
and a set of all 
$|\{n_{\!_{i}\rho}\}\rangle\langle\{n'_{\!_{i}\rho}\}|$ generates
$\tilde{\P}$ (see the argument below (\ref{eqn:2-3:def-basis})).

%
%=====================================%
%============< THEOREM >==============%
%
\begin{theorem} \label{theorem:stability}
Consider the linear map $T$ defined by (\ref{eqn:2-3:def-T}) for a real,
massless scalar field on a background geometry which describes a
formation of a quasi-stationary black hole. Then 
\begin{equation}
 T\left(\rho_{th}(\beta_{BH},\Omega_{BH})\right)=
      \tilde{\rho}_{th}(\beta_{BH},\Omega_{BH}),
\end{equation}
where 
\begin{eqnarray}
 \rho_{th}(\beta_{BH},\Omega_{BH}) 
& \equiv &
	Z^{-1}\sum_{\{ n_{\!_{i}\gamma}\} }e^{-\beta_{BH}
	\sum_{i}n_{\!_{i}\gamma}
	\left(\omega_{i}-\Omega_{BH}m_{i}\right)}
	|\{ n_{\!_{i}\gamma}\}\rangle
	\langle\{ n_{\!_{i}\gamma}\}|,	\nonumber\\
 \tilde{\rho}_{th}(\beta_{BH},\Omega_{BH})
& \equiv &
	Z^{-1}\sum_{\{ n_{\!_{i}\rho}\} }e^{-\beta_{BH}
	\sum_{i}n_{\!_{i}\rho}
	\left(\omega_{i}-\Omega_{BH}m_{i}\right)}
	|\{ n_{\!_{i}\rho}\}\rangle
	\langle\{ n_{\!_{i}\rho}\}|,	\nonumber\\
Z
& \equiv &
	\sum_{\{ j_{i}\} }e^{-\beta_{BH}
	\sum_{i}j_{i}\left(\omega_{i}-\Omega_{BH}m_{i}\right)}.
\end{eqnarray}
\end{theorem}

$\rho_{th}(\beta_{BH},\Omega_{BH})$ and
$\tilde{\rho}_{th}(\beta_{BH},\Omega_{BH})$ can be regarded as 'grand
canonical ensemble' in $\P$ and $\tilde{\P}$ respectively, which have
a common temperature $\beta_{BH}^{-1}$ and a common chemical potential
$\Omega_{BH}$ for azimuthal angular momentum quantum number. Thus the
theorem says that the 'grand canonical ensemble' at $\I^-$ ($v>v_0$)
with special values of temperature and chemical potential evolves to a 
'grand canonical ensemble' at $\I^+$ with the same temperature and the 
same chemical potential. Note that the special values
$\beta_{BH}^{-1}$ and $\Omega_{BH}$ are determined by the background
geometry: $\beta_{BH}^{-1}$ is the Hawking temperature and
$\Omega_{BH}$ is the angular velocity of the black hole formed. This
result is used in subsection \ref{subsec:proof_GSL} to prove  the
generalized second law for the quasi-stationary black hole.

%======================================%
%<<<<<<   SUBSECTION 2-3-2    >>>>>>>>>%
%======================================%
\subsection{A proof of the generalized second law}
\label{subsec:proof_GSL}

 The generalized second law of black hole thermodynamics is 
\begin{equation}
 \Delta S_{BH}+\Delta S_{matter}\geq 0,
\end{equation}
where $\Delta$ denotes a change of quantities under an evolution of
the system, $S_{BH}$ and $S_{matter}$ are black hole entropy of the
black hole and thermodynamic entropy of the matter fields, 
respectively. For a quasi-stationary black hole, using the first law
of the black hole thermodynamics
\[
 \Delta S_{BH}=\beta_{BH}( \Delta M_{BH}-\Omega_{BH}\Delta J_{BH}),
\]
the conservation of total energy
\[
 \Delta M_{BH}+\Delta E_{matter}=0
\]
and the conservation of total angular momentum
\[
 \Delta J_{BH}+\Delta L_{matter}=0,
\]
it is easily shown that the generalized second law is equivalent
to the following inequality: 
\begin{equation}
 \Delta S_{matter}
 -\beta_{BH}(\Delta E_{matter}-\Omega_{BH}\Delta L_{matter})\geq 0,
\end{equation}
where $\beta_{BH}$, $\Omega_{BH}$, $M_{BH}$ and $J_{BH}$ are the
inverse temperature, the angular velocity, the mass and the angular
momentum of the black hole; $E_{matter}$ and $L_{matter}$ are the
energy and the azimuthal component of angular momentum of the matter
fields. Thus this is of the form 
\begin{equation}
 U[\tilde{\rho}_0 ;\beta_{BH},\Omega_{BH})\geq 
 U[\rho_0 ;\beta_{BH},\Omega_{BH}), \label{eqn:2-3:GSL'}
\end{equation}
where $U$ is a functional of a density matrix of the matter fields
defined by 
\begin{equation}
 U[\rho ;\beta_{BH},\Omega_{BH})\equiv
      -{\bm{Tr}}\left[\rho\ln\rho\right] 
      -\beta_{BH}\left({\bm{Tr}}[{\bm{E}}\rho]
      -\Omega_{BH}{\bm{Tr}}[{\bm{L}_z}\rho]\right),
\end{equation}
$\rho_0$ and $\tilde{\rho}_0$ are an initial density matrix and the
corresponding final density matrix respectively. In the expression
$\bm{E}$ and $\bm{L}_z$ are operators corresponding to the energy and
the azimuthal component of the angular momentum. Note that
(\ref{eqn:2-3:GSL'}) is an inequality between functionals of a density
matrix of matter fields 
\footnote{
Information about the background geometry appears in the inequality as
the variables $\beta_{BH}$ and $\Omega_{BH}$ which parameterize the
functional.
}. 
We will prove the generalized second law by showing that this
inequality holds. Actually we do it for a quasi-stationary black hole
which is formed by gravitational collapse, using the results of
subsection \ref{subsec:scalar_field} and a theorem given in the
following.

%<<<<<<   Non-decreasing functional    >>>>>>>>>%
\subsubsection{Non-decreasing functional}

In this subsection a theorem which makes it possible to
construct a functional which does not decrease by a physical
evolution. It is a generalization of a result of
\cite{Sorkin1986}. After that, we derive (\ref{eqn:2-3:GSL'}) for a
quasi-stationary black hole which arises from gravitational collapse,
applying the theorem to the scalar field investigated in
subsection \ref{subsec:scalar_field}.

Let us consider Hilbert spaces $\F$ and
$\tilde{\F}$. First we give some definitions needed for the theorem. 

%============< DEFINITION >==============%
%
\begin{definition}
A linear bounded operator $\rho$ on $\F$ is called a density matrix,
if it is self-adjoint, positive semi-definite and satisfies
\[
 \bm{Tr}\rho =1.
\]
\end{definition}
In the rest of this section we denote a space of all density matrices on 
$\F$ as $\P (\F )$. Evidently $\P (\F )$ is a linear convex set rather
than a linear set. 

%========================================%
%============< DEFINITION >==============%
%
\begin{definition}
A map $\T$ of $\P (\F )$ into $\P (\tilde{\F})$ is called linear, if 

\[
 \T\left( a\rho_1 +(1-a)\rho_2\right) = 
	a\T (\rho_1) + (1-a)\T (\rho_2)
\]
for $0\leq\!^\forall a\leq 1$ and $\!^\forall\rho_1$, 
$\!^\forall\rho_2$ $(\in\P (\F ))$.
\end{definition}
By this definition it is easily proved by induction that 
\begin{equation}
 \T\left(\sum_{i=1}^N a_i\rho_i\right) =
	\sum_{i=1}^N a_i\T (\rho_i ),	\label{eqn:2-3:linearity}
\end{equation}
if $a_i\geq 0$, $\sum_{i=1}^N a_i =1$ and $\rho_i\in\P (\F )$.

Now we prove the following lemma which concerns the $N\to\infty$ limit
of the left hand side of (\ref{eqn:2-3:linearity}). We use this lemma in
the proof of theorem \ref{theorem:general}.
%=====================================%
%============< LEMMA >================%
%
\begin{lemma}	\label{lemma:WOT-limit}
Consider a linear map $\T$ of $\P (\F )$ into $\P (\tilde{\F})$ and an
element $\rho_0$ of $\P (\F )$. For a diagonal decomposition
\[
 \rho_0 = \sum_{i=1}^\infty p_i|i\rangle\langle i|,
\]
define a series of density matrices of the form 
\begin{equation}
 \rho_n = \sum_{i=1}^n p_i/a_n|i\rangle\langle i|
	\qquad (n=N,N+1,\cdots ),
\end{equation}
where 
\[
 a_n \equiv \sum_{i=1}^n p_i
\]
and $N$ is large enough that $a_N >0$.
Then 
\begin{equation}
 \lim_{n\to\infty}\langle\Phi |\T(\rho_n)|\Psi\rangle =
	\langle\Phi |\T(\rho_0)|\Psi\rangle
\end{equation}
for arbitrary elements $|\Phi\rangle$ and $|\Psi\rangle$ of
$\tilde{\F}$. 
\end{lemma}

This lemma says that $\T (\rho_n)$ has a weak-operator-topology-limit
$\T (\rho_0)$.

%
%===================================%
%============< PROOF >==============%
%
\begin{proof}
By definition,
\begin{equation}
 \rho_0 = a_n\rho_n + (1-a_n)\rho'_n,
\end{equation}
where
\[
 \rho'_n = \left\{ \begin{array}{lc}
	\sum_{i=n+1}^\infty p_i/(1-a_n)|i\rangle\langle i| &
	(a_n <1)	\\
	\rho_n	&
	(a_n =1)
 \end{array}\right..
\]
Then the linearity of $\T$ shows
\[
 \langle\Phi |\T(\rho_0)|\Psi\rangle =
	a_n\langle\Phi |\T(\rho_n)|\Psi\rangle +
	(1-a_n)\langle\Phi |\T(\rho'_n)|\Psi\rangle.
\]
Thus, if $\langle\Phi |\T(\rho'_n)|\Psi\rangle$ is finite in
$n\to\infty$ limit, then the lemma is established since 
\[
 \lim_{n\to\infty}a_n=1.
\]
For the purpose of proving the finiteness of 
$\langle\Phi |\T(\rho'_n)|\Psi\rangle$, it is sufficient to show that
$|\langle\Phi |\tilde{\rho}|\Psi\rangle|$ is bounded from above by 
$\|\Phi\|\ \|\Psi\|$ for an arbitrary element $\tilde{\rho}$ of 
$\P (\tilde{\F})$. This is easy to prove as follows. 
\begin{equation}
 |\langle\Phi |\tilde{\rho}|\Psi\rangle | = 
 |\sum_i\tilde{p}_i\langle\Phi |\tilde{i}\rangle
	\langle \tilde{i}|\Psi\rangle| \leq 
 \sum_i|\langle\Phi |\tilde{i}\rangle\langle\tilde{i}|\Psi\rangle |
 \leq \|\Phi\|\ \|\Psi\| ,
\end{equation}
where we have used a diagonal decomposition
\[
 \tilde{\rho} = \sum_i\tilde{p}_i|\tilde{i}\rangle\langle\tilde{i}|.
\]
\end{proof}
\QED

%=====================================%
%============< THEOREM >==============%
%
\begin{theorem} \label{theorem:general}
Assume the following three assumptions: 
{\bm a}. $\T$ is a linear map of $\P (\F )$ into $\P (\tilde{\F})$,
$\quad$ 
{\bm b}. $f$ is a continuous function convex to below and there are
non-negative constants $c_1$, $c_2$ and $c_3$ such that 
$|f((1-\epsilon)x)-f(x)|\leq |\epsilon|(c_1|f(x)|+c_2|x|+c_3)$ for
$\!^\forall x$ $(\geq 0)$ and sufficiently small $|\epsilon |$,\quad 
{\bm c}. there are positive definite density matrices $\rho_\infty$
$(\in\P (\F ))$ and $\tilde{\rho}_\infty$ $(\in\P (\tilde{\F}))$ such
that $\T (\rho_\infty )=\tilde{\rho}_\infty$.

If $[\rho_\infty,\rho_0 ]=[\tilde{\rho}_\infty,\T (\rho_0)]=0$ and
$\bm{Tr}[\rho_\infty |f(\rho_0\rho^{-1}_\infty )|]<\infty$, then 
\begin{equation}
  \tilde{\U}\left[\T (\rho_{0})\right]\geq \U\left[\rho_{0}\right],
	\label{eqn:2-3:theorem}
\end{equation}
where
\begin{eqnarray}
 \U\left[\rho\right] & \equiv & -{\bf Tr}\left[\rho_{\infty}f\left(\rho
      \rho_{\infty}^{-1}\right)\right],\nonumber\\
 \tilde{\U}\left[\tilde{\rho}\right] & \equiv & 
      -{\bf Tr}\left[\tilde{\rho}_{\infty}f\left(\tilde{\rho}
      \tilde{\rho}_{\infty}^{-1}\right)\right].
\end{eqnarray}
\end{theorem}
As stated in the first paragraph of this subsection, theorem
\ref{theorem:general} is used in subsection \ref{subsec:proof} to
prove the generalized second law for a quasi-stationary black hole
which arises from gravitational collapse. 

%
%===================================%
%============< PROOF >==============%
%
\begin{proof}
First let us decompose the density matrices diagonally as follows:
\begin{eqnarray}
 \rho_0 
& = & \sum_{i=1}^\infty p_i|i\rangle\langle i|,\quad
 \rho_\infty
 =  \sum_{i=1}^\infty q_i|i\rangle\langle i|,	\nonumber\\
 \T (\rho_0)
& = & \sum_{i=1}^\infty\tilde{p}_i|\tilde{i}\rangle\langle\tilde{i}|,
	\quad
 \T (\rho_\infty )
 =  \sum_{i=1}^\infty\tilde{q}_i|\tilde{i}\rangle\langle\tilde{i}|.
\end{eqnarray}
Then by lemma \ref{lemma:WOT-limit} and (\ref{eqn:2-3:linearity}),
\begin{equation}
 \tilde{p}_i 
 = \langle\tilde{i}|\T (\rho_0)|\tilde{i}\rangle
 = \lim_{n\to\infty}\sum_{j=1}^n A_{ij}p_j/a_n,	\label{eqn:2-3:tilde-p}
\end{equation}
where $a_n\equiv\sum_{i=1}^np_i$ and 
$A_{ij}\equiv\langle\tilde{i}|\T 
	(|j\rangle\langle j|)|\tilde{i}\rangle$. $A_{ij}$ has the
following properties:
\[
 \sum_{i=1}^\infty A_{ij} =1,\quad 0\leq A_{ij}\leq 1.
\]
Similarly it is shown that
\[
 \tilde{q}_i
= \lim_{n\to\infty}\sum_{j=1}^n A_{ij}q_j/b_n,
\]
where $b_n\equiv\sum_{i=1}^n q_i$. By (\ref{eqn:2-3:tilde-p}) and the
continuity of $f$, it is shown that 
\begin{equation}
 f(\tilde{p}_i/\tilde{q}_i) = \lim_{n\to\infty}
	f\left(\sum_{j=1}^n A_{ij}\frac{p_j/a_n}{\tilde{q}_i}\right).
\end{equation}

Next define $C^n_i$ and $\tilde{C}^n_i$ by 
\begin{equation}
 C^n_i \equiv \sum_{j=1}^n A_{ij}q_j/\tilde{q}_i,\quad
 \tilde{C^n_i} \equiv C^n_i/a_n,
\end{equation}
then the convex property of $f$ means
\[
 f(\tilde{p}_i/\tilde{q}_i) \leq
	\lim_{n\to\infty}\sum_{j=1}^n\frac{A_{ij}q_j}{C^n_i\tilde{q}_i}
	f(\tilde{C}^n_ip_j/q_j)
\]
since
\[
 \sum_{j=1}^n\frac{A_{ij}q_j}{C^n_i\tilde{q}_i} = 1,
	\frac{A_{ij}q_j}{C^n_i\tilde{q}_i} \geq 0.
\]
Hence
\begin{equation}
 -\tilde{\U}[\T (\rho_0)] 
 = \sum_{i=1}^\infty\tilde{q}_if(\tilde{p}_i/\tilde{q}_i)
 \leq \sum_{i=1}^\infty\lim_{n\to\infty}\sum_{j=1}^n
	\frac{A_{ij}q_j}{C^n_i}
	f(\tilde{C}^n_ip_j/q_j).
\end{equation}

Since $C^n_i$ and $\tilde{C^n_i}$ satisfy 
\[
 \lim_{n\to\infty}C^n_i =\lim_{n\to\infty}\tilde{C^n_i} = 1,
\]
it is implied by the assumption about $f$ that 
\[
 \left| f\left(\frac{\tilde{C}^n_ip_j}{q_j}\right) -
	f\left(\frac{p_j}{q_j}\right)\right|
 \leq |1-\tilde{C}^n_i|\left( c_1|f(p_j/q_j)|+c_2p_j/q_j+c_3\right)
\]
for sufficiently large $n$. Therefore
\begin{eqnarray*}
 & & \left|\sum_{j=1}^n\frac{A_{ij}q_j}{C^n_i}
	\left( f(\tilde{C}^n_ip_j/q_j) - f(p_j/q_j)\right)\right|
	\nonumber\\
& & \leq
 \frac{|1-\tilde{C}^n_i|}{C^n_i}
	\left(c_1\sum_{j=1}^nA_{ij}q_j|f(p_j/q_j)| + 
	c_2\sum_{j=1}^nA_{ij}p_j + 
	c_3\sum_{j=1}^nA_{ij}q_j\right)	\\
& & \leq
 \frac{|1-\tilde{C}^n_i|}{C^n_i}
	\left(c_1\sum_{j=1}^nq_j|f(p_j/q_j)| + 
	c_2\sum_{j=1}^np_j +
	c_3\sum_{j=1}^nq_j\right),
\end{eqnarray*}
where we have used $0\leq A_{ij}\leq 1$ to obtain the last
inequality. Since the first term in the brace in the last expression
is finite in $n\to\infty$ limit by the assumption of the absolute
convergence of $\U\left[\rho_0\right]$ and all the other terms in the
brace are finite,
\[
 \lim_{n\to\infty}\left|\sum_{j=1}^n\frac{A_{ij}q_j}{C^n_i}
	\left( f(\tilde{C}^n_ip_j/q_j) - f(p_j/q_j)\right)\right|
 = 0.
\]
Moreover, by the absolute convergence of $\U\left[\rho_0\right]$, it
is easily shown that
\[
 \lim_{n\to\infty}\left|\left(\frac{1}{C^n_i}-1\right)
	\sum_{j=1}^n A_{ij}q_jf(p_j/q_j)\right|=0.
\]
Thus
\begin{equation}
 -\tilde{\U}[\T (\rho_0 )]\leq
	\sum_{i=1}^\infty\sum_{j=1}^\infty A_{ij}q_jf(p_j/q_j).
		\label{eqn:2-3:theorem'}
\end{equation}
We can interchange sum over $i$ and sum over $j$ in the right hand
side of (\ref{eqn:2-3:theorem'}) since it converges absolutely by the
absolute convergence of $\U\left[\rho_0\right]$. Hence 
\[
 -\tilde{\U}[\T (\rho_0 )]\leq
	\sum_{j=1}^\infty q_jf(p_j/q_j) = -\U [\rho_0].
\]
\end{proof}
\QED

\subsubsection{Proof of the generalized second law}
	\label{subsec:proof}

Let us combine theorem \ref{theorem:stability} with theorem
\ref{theorem:general} to prove the generalized second law. In theorem
\ref{theorem:general} set the linear map $\T$, the convex function
$f(x)$ and the density matrices $\rho_{\infty}$ and
$\tilde{\rho}_{\infty}$ as follows. 
\begin{eqnarray}
 \T & = & T,\nonumber\\
 f(x) & = & x\ln x,\nonumber\\
 \rho_{\infty} & = & \rho_{th}(\beta_{BH},\Omega_{BH}),\nonumber\\
 \tilde{\rho}_{\infty} & = & 
	\tilde{\rho}_{th}(\beta_{BH},\Omega_{BH}).
\end{eqnarray}
Note that it is theorem \ref{theorem:stability} that makes such a
setting possible. Hence, if an initial state $\rho_{0}$ and the
corresponding final state $T(\rho_{0})$ satisfy 
\begin{equation}
 \left[\rho_{0},\rho_{th}(\beta_{BH},\Omega_{BH})\right]=
 \left[T(\rho_{0}),
	\tilde{\rho}_{th}(\beta_{BH},\Omega_{BH})\right]=0
   \label{eqn:2-3:initial-cond}
\end{equation}
and $\U [\rho_0]$ converges absolutely, theorem 
\ref{theorem:general} can be applied to the system of the 
quasi-stationary black hole and the scalar field around it. Now
\begin{eqnarray}
 \U [\rho_0] & = & -\bm{Tr}\left[\rho_0\ln\rho_0\right]
	-\beta_{BH}\left(\bm{Tr}\left[\bm{E}\rho_0\right]
		-\Omega_{BH}\bm{Tr}\left[\bm{L}_z\rho_0\right]
			\right) -\ln Z	\nonumber\\
 & = & U\left[\rho_{0};\beta_{BH},\Omega_{BH}\right) -\ln Z,
					\nonumber\\
 \tilde{\U}[\T (\tilde{\rho}_0 )]& = & 
-\bm{Tr}\left[\tilde{\rho}_0\ln\tilde{\rho}_0\right]
-\beta_{BH}\left(\bm{Tr}\left[\tilde{\bm{E}}\tilde{\rho}_0\right]
-\Omega_{BH}\bm{Tr}\left[\tilde{\bm{L}}_z\tilde{\rho}_0\right]
			\right) -\ln Z	\nonumber\\
 & = & U\left[ T(\tilde{\rho}_{0});\beta_{BH},\Omega_{BH}\right) 
	-\ln Z,
	\label{eqn:2-3:U-lnZ}
\end{eqnarray}
where
\begin{eqnarray*}
 \bm{E} & \equiv & \sum_{\{n_{\!_{i}\gamma}\} }
	\left(\sum_i n_{\!_{i}\gamma}\omega_i\right)
	|\{n_{\!_{i}\gamma}\}\rangle\langle\{n_{\!_{i}\gamma}\}|,\\ 
 \bm{L}_z & \equiv & \sum_{\{n_{\!_{i}\gamma}\} }
	\left(\sum_i n_{\!_{i}\gamma}m_i\right)
	|\{n_{\!_{i}\gamma}\}\rangle\langle\{n_{\!_{i}\gamma}\}|,
\end{eqnarray*}
and 
\begin{eqnarray*}
 \tilde{\bm{E}} & \equiv & \sum_{\{n_{\!_{i}\rho}\} }
	\left(\sum_i n_{\!_{i}\rho}\omega_i\right)
	|\{n_{\!_{i}\rho}\}\rangle\langle\{n_{\!_{i}\rho}\}|,\\ 
 \tilde{\bm{L}}_z & \equiv & \sum_{\{n_{\!_{i}\rho}\} }
	\left(\sum_i n_{\!_{i}\rho}m_i\right)
	|\{n_{\!_{i}\rho}\}\rangle\langle\{n_{\!_{i}\rho}\}|.
\end{eqnarray*}
Thus the inequality (\ref{eqn:2-3:theorem}) in this case is
(\ref{eqn:2-3:GSL'}) itself, which in turn is equivalent to the
generalized second law. Finally, theorem \ref{theorem:general} proves
the generalized second law for a quasi-stationary black hole which
is formed by gravitational collapse, provided that an initial density 
matrix $\rho_0$ of the scalar field satisfies the above
assumptions. For example, it is guaranteed by lemma
\ref{lemma:off-diagonal} that if $\rho_{0}$ is diagonal in the
basis $\{ |\{n_{\!_{i}\gamma}\}\rangle\}$ then $T(\rho_{0})$ is also
diagonal in the basis $\{ |\{n_{\!_{i}\rho}\}\rangle\}$ and
(\ref{eqn:2-3:initial-cond}) is satisfied. The assumption of the absolute 
convergence of $U[\rho_{0};\beta_{BH},\Omega_{BH})$ holds whenever
initial state $\rho_0$ at $\I^-$ contains at most finite number of 
excitations. 
Therefore the assumptions are satisfied when $\rho_0$ is
diagonal in the basis $\{ |\{n_{\!_{i}\gamma}\}\rangle\}$ and contains
at most finite number of excitations.

Here we have to admit that $\ln Z$ in Eqs.~(\ref{eqn:2-3:U-lnZ})
diverges formally due to the infinite volume of the system. 
However, it should be possible to avoid this divergence by a suitable
regularization. 
It will be valuable to analyze what kind of regularization does well.

%<<<<<<<<  Concluding remark  >>>>>>>>%
\subsection{Concluding remark}
\label{subsec:remark_GSL}

Now we make a comments on Frolov and Page's statement that their proof 
of the generalized second law may be applied to the case of the black
hole formed by gravitational collapse \cite{Frolov&Page1993}. Their
proof for a quasi-stationary eternal black hole is based on the
following two assumptions: (1) a state of matter fields on the past
horizon is thermal one; (2) a set of radiation modes on the past
horizon and a set of modes on the past null infinity are quantum
mechanically uncorrelated. These two assumptions are reasonable for
the eternal case since a black hole emit a thermal radiation. In the
case of a black hole formed by gravitational collapse, we might expect
that things would go 
well by simply replacing the past horizon with a null surface at a
moment of a formation of a horizon ($v=v_0$ surface in {\it
Figure \ref{fig:modes}}). However, a state of the matter fields on
the past horizon is completely determined by a state of the fields
before the horizon formation ($v<v_0$ in {\it Figure}
\ref{fig:modes}), in which there is no causal effect of the
existence of future horizon. Since the essential origin of the thermal
radiation from a black hole is the existence of the horizon, the state
of the fields on the null surface has not to be a thermal one. Hence
the assumption (1) does not hold in this case. Although the above
replacement may be the most extreme one, a replacement of the past
horizon by an intermediate null hypersurface causes an intermediate
violation of the assumption (1) and (2) due to the 
correlation between modes on the future null infinity and modes on the
past null infinity located after the horizon formation. The
correlation can be seen in (\ref{eqn:2-3:multi}) explicitly in the
case of the replacement by the future null infinity. Thus we
conclude that their proof can not be applied to the case of the black
hole formed by a gravitational collapse.

Finally we discuss a generalization of our proof to a dynamical
background. For the case of a dynamical background, $\beta_{BH}$ and 
$\Omega_{BH}$ are changed from time to time by a possible
backreaction. Thus, to prove the generalized second law for the
dynamical background, we have to generalize theorem
\ref{theorem:stability} to the dynamical case consistently with the 
backreaction. Once this can be achieved, theorem
\ref{theorem:general}, combined with the quasi-local first law of
black hole dynamics derived in subsection \ref{subsec:quasilocal_1st}, 
seems useful to prove the generalized second law for the dynamical
background.

Here we have to admit that, in the dynamical situation, entropy
$S_{matter}$ of matter fields might lose its physical meaning, while
dynamical black hole entropy can be defined as a quarter of trapping
horizon area (see subsection \ref{subsec:dynamic-S}). 
However, if the dynamical version of the generalized second law would
be proved for some definition of $S_{matter}$ then, by integrating it
from an initial stationary state to a final stationary state, we would 
be able to obtain the generalized second law for finite changes of
black hole parameters. 
(Note that in the proof given in this section, it has been assumed
implicitly by the quasi-stationarity that $\Delta S_{BH}\ll S_{BH}$,
$\Delta M_{BH}\ll M_{BH}$, etc.)
For example, if we take a black hole as the initial state and a flat
spacetime as the final one, then the finite version of the generalized 
second law insists that matter entropy is increased by evaporation of 
the initial black hole and that the produced entropy is greater than
the initial black hole entropy. 
Thus, the generalization is necessary for a detailed investigation of
the information loss problem.

%%%%%%%%%%%%%%%%%%%%%%%%%%%%%%%%%%%%%%%%%%%%%%%%%%%%%%%%%%%%%%%%%%%%
%%%%%%%%%%%%%%%%%%%%%%%%%%%%%%%%%%%%%%%%%%%%%%%%%%%%%%%%%%%%%%%%%%%%
% CHAPTER 3
%%%%%%%%%%%%%%%%%%%%%%%%%%%%%%%%%%%%%%%%%%%%%%%%%%%%%%%%%%%%%%%%%%%%
%%%%%%%%%%%%%%%%%%%%%%%%%%%%%%%%%%%%%%%%%%%%%%%%%%%%%%%%%%%%%%%%%%%%
\chapter{Black hole entropy}
	\label{chap:BHentropy}

%%%%%%%%%%%%%%%%%%%%%%%%%%%%%%%%%%%%%%%%
%%%%%%%%%%%% SECTION 3-1 %%%%%%%%%%%%%%%
%%%%%%%%%%%%%%%%%%%%%%%%%%%%%%%%%%%%%%%%
\section{D-brane statistical-mechanics}
	\label{sec:Dbrane}

%======================================%
%<<<<<<   SUBSECTION 3-1-1    >>>>>>>>>%
%======================================%
\subsection{Black brane solution in the type IIB superstring}

The low-energy effective theory of type IIB superstring contains, as
its bosonic part, a metric field, a dilaton field, a R-R field and
NS-NS fields. By setting the NS-NS fields to be 
zero, we obtain the following low-energy effective action in the
$10$-dimensional Einstein frame. 
%============< EQUATION >==============%
%
\begin{equation}
 \frac{1}{16\pi G_{10}}\int d^{10}x\sqrt{-g}
	[R-\frac{1}{2}(\nabla\phi)^2-\frac{1}{12}e^{\phi}H^2],
	\label{eqn:3-1:action}
\end{equation}
%======================================%
where $\phi$ and $H$ are the dilaton field and the R-R
three-form field strength, respectively. (The $10$-dimensional
Newton's constant $G_{10}$ is defined so that the dilaton $\phi$
vanishes asymptotically.)

For this effective theory we consider toroidal compactification to 
five dimensions with an $S^1$ of length $2\pi R$, a $T^4$ of 
four-volume $(2\pi)^4V$, and momentum along the $S^1$: we assume the
$10$-dimensional metric of the form 
%============< EQUATION >==============%
%
\begin{equation}
 ds_{(10)}^2 = e^{-2(4\chi+\psi)/3}g^{(5)}_{\mu\nu}dx^{\mu}dx^{\nu}
	+ e^{2\psi}(dX_5+A_{\mu}dx^{\mu})^2
	+ e^{2\chi}\sum_{i=6}^9(dX_i)^2,
	\label{eqn:3-1:compactification}
\end{equation}
%======================================%
where $\mu=0,1,\cdots,4$, $i=6,\cdots,9$, and all fields depend only
on $x^{\mu}$. Here we also assume that $X_5$ is periodically
identified with period $2\pi R$ and that each of $X_i$ is identified
with period $2\pi V^{1/4}$. 
Note that the conformal factor $e^{-2(4\chi+\psi)/3}$ makes the metric
$g^{(5)}_{\mu\nu}dx^{\mu}dx^{\nu}$ be in the $5$-dimensional
Einstein frame, and that the field $A_{\mu}$ becomes a $U(1)$ gauge
field in $5$-dimensions.

In Ref.~\cite{HMS1996} a six parameter family of black brane solutions
of the equation of motion following from the action
(\ref{eqn:3-1:action}) was given. The metric is of the form
(\ref{eqn:3-1:compactification}) with 
%============< EQUATION >==============%
%
\begin{eqnarray}
 g^{(5)}_{\mu\nu}dx^{\mu}dx^{\nu} & = & 
	-f^{-2/3}\left(1-\frac{r_0^2}{r^2}\right)dt^2
	+ f^{1/3}\left[\left(1-\frac{r_0^2}{r^2}\right)^{-1}dr^2
	+ r^2 d\Omega_{(3)}^2\right],\nonumber\\
 A_{\mu}dx^{\mu} & = & 
	f_3^{-1}\frac{r_0^2\sinh{2\sigma}}{2r^2} dt,\nonumber\\
 e^{2\chi}  & = & f_1^{1/4} f_2^{-1/4},\nonumber\\ 
 e^{2\psi} & = & f_1^{-3/4} f_2^{-1/4} f_3,
\end{eqnarray}
%======================================%
where
%============< EQUATION >==============%
%
\begin{eqnarray}
 f_1 & = & 1+\frac{r_0^2\sinh^2\alpha}{r^2},\nonumber\\
 f_2 & = & 1+\frac{r_0^2\sinh^2\gamma}{r^2},\nonumber\\
 f_3 & = & 1+\frac{r_0^2\sinh^2\sigma}{r^2},\nonumber\\
 f & = & f_1 f_2 f_3.
\end{eqnarray}
%======================================%
The solution is parameterized by the six independent quantities 
$\alpha$, $\gamma$, $\sigma$, $r_0$, $R$ and $V$.

As seen as a black hole in $5$-dimensions, this black brane solution
has two charges $Q_1$ and $Q_2$ associated with the R-R field $H$ and
the charge $N$ associated with the Kaluza-Klein gauge field
$A_{\mu}$. These charges are written in terms of the six parameters as 
%============< EQUATION >==============%
%
\begin{eqnarray}
 Q_1 & = &\frac{Vr_0^2}{2g}\sinh{2\alpha},\nonumber\\
 Q_5 & = &\frac{r_0^2}{2g}\sinh{2\gamma},\nonumber\\
 N & = &\frac{R^2Vr_0^2}{2g^2}\sinh{2\sigma},
\end{eqnarray}
%======================================%
where $g$ is the $10$-dimensional string coupling, by which the
$10$-dimensional Newton's constant is written as $G_{10}=8\pi^6g^2$. 
Note that, from the general argument of Kaluza-Klein compactification,
the charge $N$ is related to the momentum $P$ around the $S^1$ as
$P=N/R$. The $5$-dimensional black hole also has charges $P_{\psi}$
and $P_{\chi}$ related to the asymptotic fall-off of $\psi$ and
$\chi$, which are given by
%============< EQUATION >==============%
%
\begin{eqnarray}
 P_{\psi} & = & \frac{RVr_0^2}{2g^2}
	[\cosh{2\sigma}-\frac{1}{2}
	(\cosh{2\alpha}+\cosh{2\gamma})],\nonumber\\
 P_{\chi} & = & \frac{RVr_0^2}{2g^2}
	(\cosh{2\alpha}-\cosh{2\gamma}).
\end{eqnarray}
%======================================%
In $10$-dimensions $P_{\psi}$ and $P_{\chi}$ are pressures which
describe how the energy changes for isentropic variations in $R$ and
$V$. 
The ADM energy associated with this $5$-dimensional black
hole is 
%============< EQUATION >==============%
%
\begin{equation}
 E = \frac{RVr_0^2}{2g^2}(\cosh{2\alpha}+\cosh{2\gamma}
	+\cosh{2\sigma}). 
\end{equation}
%======================================%
Hence the black brane solution can be parameterized by the five charges 
($Q_1$, $Q_5$, $N$, $P_{\psi}$, $P_{\chi}$) and the ADM energy $E$.

For this solution as a $5$-dimensional black hole the
Bekenstein-Hawking entropy $S_{BH}$ and the Hawking temperature
$T_{BH}$ are given by
%============< EQUATION >==============%
%
\begin{eqnarray}
 S_{BH} & \equiv & \frac{A}{4G_5} 
	= 2\pi (\sqrt{n_L}+\sqrt{n_R})
	(\sqrt{N_1}+\sqrt{N_{\bar{1}}})
	(\sqrt{N_5}+\sqrt{N_{\bar{5}}}),
	\nonumber\\
 \frac{1}{T_{BH}} & = & \frac{\pi R}{2}
	\left(\frac{1}{\sqrt{n_L}}+\frac{1}{\sqrt{n_R}}\right)
	(\sqrt{N_1}+\sqrt{N_{\bar{1}}})
	(\sqrt{N_5}+\sqrt{N_{\bar{5}}}),
\end{eqnarray}
%======================================%
where $A$ is the area of the horizon and $G_5$ is the $5$-dimensional
Newton's constant given by $G_{10}=G_5(2\pi)^5RV$.
Here $N_1$, $N_{\bar{1}}$, $N_5$, $N_{\bar{5}}$, $n_L$ and $n_R$ are
defined by 
%============< EQUATION >==============%
%
\begin{eqnarray}
 Q_1 & = & N_1 - N_{\bar{1}},\nonumber\\
 Q_5 & = & N_5 - N_{\bar{5}},\nonumber\\
 N & = & n_L - n_R,\nonumber\\
 P_{\psi} & = & -\frac{R}{2g}(N_1 + N_{\bar{1}})
	- \frac{RV}{2g}(N_5 + N_{\bar{5}}) + 
	\frac{1}{R}(n_L+n_R),\nonumber\\
 P_{\chi} & = & \frac{R}{g}(N_1 + N_{\bar{1}})
	- \frac{RV}{g}(N_5 + N_{\bar{5}}),\nonumber\\
 E & = & \frac{R}{g}(N_1 + N_{\bar{1}}) + 
	\frac{RV}{g}(N_5 + N_{\bar{5}}) + 
	\frac{1}{R}(n_L+n_R).
	\label{eqn:3-1:relations-charges}
\end{eqnarray}
%======================================%
The scales of compactification are written in terms of these
quantities as
%============< EQUATION >==============%
%
\begin{eqnarray}
 R & = & \left(\frac{g^2n_Ln_R}{N_1N_{\bar{1}}}\right)^{1/4},
	\nonumber\\
 V & = & \left(\frac{N_1N_{\bar{1}}}{N_5N_{\bar{5}}}\right)^{1/2}.
\end{eqnarray}
%======================================%

The black brane solution characterized by ($N_1$, $N_{\bar{1}}$, $N_5$,
$N_{\bar{5}}$, $n_L$, $n_R$) can be interpreted as a configuration of
strings and solitons in the type IIB superstring
theory~\cite{HMS1996}, provided that interactions among string and
soliton can be neglected. 
The configuration is composed of 
$N_1$ ``constituent'' D-strings wrapped on the $S^1$, 
$N_{\bar{1}}$ ``constituent'' anti-D-strings wrapped on the $S^1$, 
$N_5$ ``constituent'' D-fivebranes wrapped on the $T^5=T^4\times S^1$, 
and $N_{\bar{5}}$ ``constituent'' anti-D-fivebranes wrapped on the
$T^5$. 
Open strings on the D-branes have momentum along the $S^1$ so that 
$P_L\equiv n_L/R$ is the momentum along the $S^1$ which is a sum over
left-moving massless modes of open strings and that 
$P_R\equiv n_R/R$ is the momentum which is a sum over right-moving
massless modes. 
This interpretation is based on the following three facts: 
\begin{itemize}
\item[(i)]
A single-wound D-string or anti-D-string has
%============< EQUATION >==============%
%
\begin{eqnarray}
 Q_1 = \pm 1, Q_5 = 0, N = 0,\nonumber\\
 P_{\psi} = -\frac{R}{2g}, P_{\chi} = \frac{R}{g}, 
 E = \frac{R}{g},
\end{eqnarray}
%======================================%
where plus sign is for the D-string and minus sign is for the
anti-D-string. 
\item[(ii)]
A single-wound D-fivebrane or anti-D-fivebrane has
%============< EQUATION >==============%
%
\begin{eqnarray}
 Q_1 = 0, Q_5 = \pm 1, N = 0,\nonumber\\
 P_{\psi} = -\frac{RV}{2g}, P_{\chi} = -\frac{RV}{g},
 E = \frac{RV}{g},
\end{eqnarray}
%======================================%
where plus sign is for the D-fivebrane and minus sign is for the
anti-D-fivebrane.
\item[(iii)]
A left- or right- moving string with momentum $P=\pm n/R$ along the 
$S^1$ has
%============< EQUATION >==============%
%
\begin{eqnarray}
 Q_1 = 0, Q_5 = 0, N = \pm n,\nonumber\\
 P_{\psi} = \frac{n}{R}, P_{\chi} = 0,
 E = \frac{n}{R},
\end{eqnarray}
%======================================%
where plus sign is for the left-mover and minus sign is for the
right-mover. 
\end{itemize}

In the following arguments, we consider the case that there are no
anti-D-branes: $N_{\bar{1}}=N_{\bar{5}}=0$, and thus $Q_1=N_1$,
$Q_5=N_5$. In this case the Bekenstein-Hawking entropy and the Hawking
temperature are given by 
%============< EQUATION >==============%
%
\begin{eqnarray}
 S_{BH} & = & 2\pi (\sqrt{n_L}+\sqrt{n_R})\sqrt{Q_1Q_5},\nonumber\\ 
 \frac{1}{T_{BH}} & = & \frac{\pi R}{2}
	\left(\frac{1}{\sqrt{n_L}}+\frac{1}{\sqrt{n_R}}\right)
	\sqrt{Q_1Q_5}.\label{eqn:3-1:SBH&TBH}
\end{eqnarray}
%======================================%
The purpose of the remaining part of this section is to explain these
expressions of the entropy and the temperature by using the D-brane
technology.

%======================================%
%<<<<<<   SUBSECTION 3-1-2    >>>>>>>>>%
%======================================%
\subsection{Number of microscopic states}
	\label{subsec:number}

Consider a set of $Q_1$ single-wound D-strings wrapped on the $S^1$
and a set of $Q_5$ single-wound D-fivebranes wrapped on 
$T^5=T^4\times S^1$. 
The D-strings may be connected up to form a set of multiply-wound
D-strings, which is composed of $N_{q_1}^{(1)}$ D-strings of length
$2\pi Rq_1$ ($q_1=1,2,\cdots$).  
The D-fivebranes may also be connected up to form a set of
multiply-wound D-fivebranes, which is composed of $N_{q_5}^{(5)}$
D-fivebranes of length $2\pi Rq_5$ ($q_5=1,2,\cdots$) along the $S^1$.
This general configuration of D-strings and D-fivebranes was first
considered in Ref.~\cite{Mukohyama1996}. 
Note that, by the conservation of charges $Q_1$ and $Q_5$, the
integers $N_{q_1}^{(1)}$ and $N_{q_5}^{(5)}$ must satisfy the
following constraints:
\begin{eqnarray}
 Q_1 & = & \sum_{q_1} q_1N_{q_1}^{(1)},	\nonumber	\\
 Q_5 & = & \sum_{q_5} q_5N_{q_5}^{(5)},
\end{eqnarray}
since a D-string which winds $q_1$ times has charge ($Q_1$,
$Q_5$)$=$($q_1$, $0$) and a D-fivebrane which winds $q_5$ times has
charge ($Q_1$, $Q_5$)$=$($0$, $q_5$).

\begin{figure}
 \begin{center}
  \epsfile{file=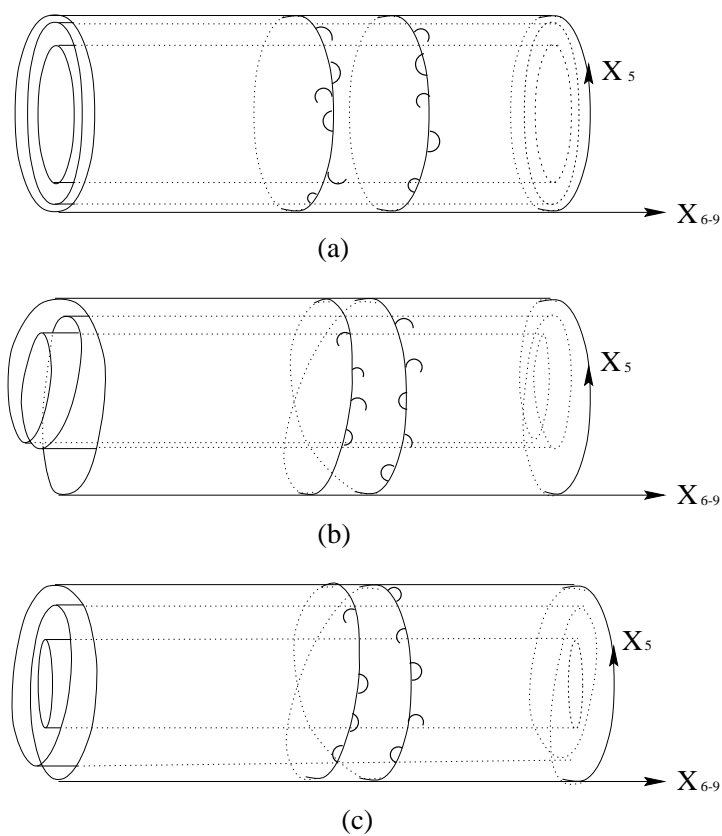,scale=0.5}
 \end{center}
 \caption{
(a) The situation considered by Callan and Maldacena is shown for
$Q_1=2$, $Q_5=3$. There are three D-fivebranes wrapped on $T^5$
($X_{5-9}$ are coordinates for the torus.) and two D-strings wrapped
on $S^1$ of the torus. ($X_5$ is a coordinate for the $S^1$.) 
(b) The situation considered by Maldacena and Susskind is shown for
$Q_1=2$, $Q_5=3$. There are a triple-wound D-fivebrane and a
double-wound D-string.
(c) Another example of the generalized configurations of D-branes for
$Q_1=2$, $Q_5=3$. 
}
\label{fig:D-brane}
\end{figure}

For example, the configuration considered by Callan and
Maldacena~\cite{Callan&Maldacena1996} is the case that
$N_1^{(1)}=Q_1$, $N_1^{(5)}=Q_5$ and all other $N_{q_1}^{(1)}$ and
$N_{q_5}^{(5)}$ are zero: no D-branes are connected up. 
({\it Figure} \ref{fig:D-brane}(a) is a schematic picture of this
situation for $Q_1=2$, $Q_5=3$.) 
On the contrary Maldacena and Susskind~\cite{Maldacena&Susskind1996}
considered the configuration such that $N_{Q_1}^{(1)}=N_{Q_5}^{(5)}=1$ 
and all other $N_{q_1}^{(1)}$ and $N_{q_5}^{(5)}$ are zero: one long
D-string and one long D-fivebrane. 
({\it Figure} \ref{fig:D-brane}(b) is a schematic picture of this
situation for $Q_1=2$, $Q_5=3$.) 
The general class of configurations introduced above includes more
abundant situations. For example, the situation in {\it Figure} 
\ref{fig:D-brane}(c) is an example for $Q_1=2$, $Q_5=3$: 
$N_2^{(1)}=$ $N_1^{(5)}=$ $N_2^{(5)}=1$ and others are zero.

%============< FIGURE >==============%
%              D-brane.ps
%======================================%

As explained in the previous subsection, the charges $n_L$ and $n_R$
are implemented in the D-brane picture so that 
$P_L\equiv n_R/R$ is the momentum along the $S^1$ which is a sum over 
left-moving massless modes of open strings on the D-branes and that 
$P_R\equiv n_L/R$ is the momentum which is a sum over right-moving
massless modes.

There are several types of open strings on the D-brane configuration:
those with both boundaries being on a common D-string; 
those with one boundary being on a D-string and another boundary being
on a different D-string; 
those with one boundary being on a D-string and another boundary being
on a D-fivebrane; etc. 
Among them we only consider a set of strings each of which connects a
D-string and a D-fivebrane, since if many such strings are excited
then all other strings become massive~\cite{BVS&Sen1996}. 
As in the situation considered in Ref.~\cite{Callan&Maldacena1996},
each string connecting a D-string and a D-fivebrane has $4$ bosonic
and $4$ fermionic degrees of freedom. 
Hence we have $4N_{q_1}^{(1)}N_{q_5}^{(5)}$ bosonic and
$4N_{q_1}^{(1)}N_{q_5}^{(5)}$  fermionic degrees of freedom for
strings with one boundary on one of $N_{q_1}^{(1)}$ $q_1$-wound
D-strings and another boundary on one of $N_{q_5}^{(5)}$
$q_5$-wound D-fivebranes.

For each of fermionic or bosonic degrees of freedom of the string 
connecting a $q_1$-wound D-string and a $q_5$-wound D-fivebrane, the
spectrum of the momentum along the $S^1$ is 
\begin{eqnarray}
 p_n(q_1,q_5) & = & \frac{\pm n}{R(q_1,q_5)_{LCM}}, 
	\quad (n=1,2,\cdots )	\label{eqn:3-1:spectrum}\\
 ('+' \mbox{ for left moving} & ; & '-' \mbox{ for right moving}),
\end{eqnarray}
where $(q_1,q_5)_{LCM}$ is the least common multiple of $q_1$ and
$q_5$, since the boundary condition of the strings is 
\begin{equation}
 X^5 \sim X^5 + 2\pi R (q_1,q_5)_{LCM}.
\end{equation}

We regard the momentum $P_L\equiv n_L/R$ ($P_R\equiv -n_R/R$)
along the $S^1$ as a sum over all left (right) moving massless modes
of the strings. It is evident from the assumption of neglecting
interactions between strings that the number $d(n_L,n_R)$ of string
states consistent with given values of $n_L$ and $n_R$ is a product of
the numbers $d(n_L)$ and $d(n_R)$ of left- and right-moving string 
states:
\begin{equation}
 d(n_L,n_R) = d(n_L)d(n_R). \label{eqn:3-1:d=dd}
\end{equation}
At this point it is easy to show that $d(n_L)$ satisfies the following
relation for an arbitrary value of $w$ satisfying $0<w<1$. 
\begin{equation}
 \sum_{l=0}^{\infty} d(n_L)w^{n_L} = 
	\prod_{q_1}\prod_{q_5}\prod_{n=1}^{\infty}
	\left[\frac{1+w^{n/(q_1,q_5)_{LCM}}}{1-w^{n/(q_1,q_5)_{LCM}}}
	\right]^{4N_{q_1}^{(1)}N_{q_5}^{(5)}},
	\label{eqn:3-1:sum-d(nL)}
\end{equation}
where $n_L=l/Q_1!Q_5!$. It is evident that $d(0)=1$. 
Note that the denominator in the r.h.s. represents a partition
function for the bosonic modes of open strings and the numerator
represents a partition function for the fermionic modes.

Note that $d(n_L)$ can be considered as a residue of the r.h.s. of
Eq.~(\ref{eqn:3-1:sum-d(nL)}) divided by $x^{l+1}$, where
$x=w^{1/Q_1!Q_5!}$. Hence, $d(n_L)$ can be written as a complex
integral: 
\begin{eqnarray}
 d(n_L) & = & \frac{1}{2\pi i}\oint dx x^{-(l+1)}
	\prod_{q_1}\prod_{q_5}\prod_{n=1}^{\infty}
	\left[\frac{1+w^{n/(q_1,q_5)_{LCM}}}{1-w^{n/(q_1,q_5)_{LCM}}}
	\right]^{4N_{q_1}^{(1)}N_{q_5}^{(5)}}\nonumber\\
	& = & \frac{1}{2\pi i}\oint dw\exp [h(w)],
	\label{eqn:3-1:d-integ}
\end{eqnarray}
where 
\begin{equation}
 h(w) = 4\sum_{q_1,q_5}\sum_n
	N_{q_1}^{(1)}N_{q_5}^{(5)}\ln
	\left[\frac{1+w^{n/(q_1,q_5)_{LCM}}}{1-w^{n/(q_1,q_5)_{LCM}}}
	\right] - (n_L+1)\ln w.
\end{equation}
The contour in the integration w.r.t. $x$ is a closed curve
surrounding $x=0$ in complex $x$-plane and, the contour in the
integration w.r.t. $w$ is a closed curve surrounding $w=0$ in complex
$w$-plane.

Now we give an asymptotic formula for $d(n_{L})$ in the limit 
\begin{equation}
 \sqrt{\frac{Q_1Q_5}{n_{L}}}\ll \min_{(q_1,q_5)\in I}(q_1,q_5)_{LCM},
	\label{eqn:3-1:limit_L}
\end{equation}
where $I \equiv \{ (q_1,q_5)|N_{q_1}^{(1)}N_{q_5}^{(5)}\neq 0\}$. 
For this purpose we use an asymptotic expression of $h(w)$.
Let us consider a function $g(x)$ defined by
\begin{equation}
 g(x) = \prod_{n=1}^{\infty}\left[\frac{1+x^n}{1-x^n}\right],
\end{equation}
where $0<x<1$ is assumed. For $x\sim 1$, we can estimate $\ln g(x)$ as 
\begin{eqnarray}
 \ln g(x) & = & \sum_{n=1}^{\infty}[\ln (1+x^n)-\ln (1-x^n)]
	\nonumber\\
	& \approx & \frac{1}{-\ln x}
	\int_0^{\infty}dy[\ln (1+e^{-y})-\ln (1-e^{-y})]
	\nonumber\\
	& = & \frac{1}{-\ln x}\cdot\frac{\pi^2}{4},
	\label{eqn:3-1:g-formula}
\end{eqnarray}
where we have estimated the summation w.r.t. $n$ by the integration
w.r.t. $y=-n\ln x$. 
Hence, we obtain the following asymptotic expression of $h(w)$ for
$w^{1/m} \sim 1$, where $m=\min_{(q_1,q_5)\in I}(q_1,q_5)_{LCM}$. 
\begin{equation}
 h(w) \approx \pi^2\sum_{q_1,q_5}
	\frac{N_{q_1}^{(1)}N_{q_5}^{(5)}}{-\ln w^{1/(q_1,q_5)_{LCM}}} 
	- (n_L+1)\ln w.
\end{equation}
We shall estimate the integration w.r.t. $w$ in
(\ref{eqn:3-1:d-integ}) by using this expression and the suddle-point
method. 
When the condition (\ref{eqn:3-1:limit_L}) is satisfied, it is shown
by using this asymptotic expression of $h(w)$ that $exp[h(w)]$ has a 
saddle point given by 
\begin{equation}
 -\ln w^{1/m} \approx 
	\frac{\pi}{m}\sqrt{\frac{
	\sum_{q_1,q_5}N_{q_1}^{(1)}N_{q_5}^{(5)}(q_1,q_5)_{LCM}}
	{n_L+1}}\qquad
	\left(\le\frac{\pi}{m}\sqrt{\frac{Q_1Q_5}{n_L+1}}\right).
\end{equation}
Note that $w^{1/m} \sim 1$ is guaranteed for this saddle point by the
condition (\ref{eqn:3-1:limit_L}). 
Thus, we obtain the following asymptotic formula of $d(n_L)$ by using
the saddle point method. 
\begin{equation}
 d(n_{L}) \approx 
	\exp\left[ 2\pi\sqrt{n_{L}\sum_{q_1}\sum_{q_5}
	N_{q_1}^{(1)}N_{q_5}^{(5)}(q_1,q_5)_{LCM}}\right]. 
\end{equation}
From this expression it is easily shown that 
\begin{equation}
 d(n_{L}) \leq \exp (2\pi\sqrt{n_{L}Q_1Q_5}),
\end{equation}
where the bound is saturated if and only if all $(q_1,q_5)$ $(\in I)$
are relatively prime.

Therefore, for a configuration satisfying 
\begin{equation}
 \sqrt{\frac{Q_1Q_5}{\min{(n_L,n_R)}}}\ll 
	\min_{(q_1,q_5)\in I}(q_1,q_5)_{LCM},
	\label{eqn:3-1:limit}
\end{equation}
$\ln{d(n_L,n_R)}$ is bounded from above by the Bekenstein-Hawking
entropy given by (\ref{eqn:3-1:SBH&TBH}): 
\begin{equation}
 \ln{d(n_L,n_R)} \leq S_{BH},\label{eqn:3-1:lndN}
\end{equation}
where the bound is saturated if and only if all
$(q_1,q_5)$ $(\in I)$ are relatively prime. 
This is one of the main results in Ref.~\cite{Mukohyama1996} and is a 
generalization of the results in
Refs.~\cite{Callan&Maldacena1996,Maldacena&Susskind1996} to the more
abundant configurations of D-strings and D-fivebranes.

When $n_R=0$, which corresponds to an extremal black hole, the
condition (\ref{eqn:3-1:limit}) does not hold since the l.h.s diverges. 
Nonetheless, even in this case, the bound (\ref{eqn:3-1:lndN})
does hold if the condition (\ref{eqn:3-1:limit_L}) is satisfied,
since $d(0)=1$. The bound is saturated if and only if all $(q_1,q_5)$
$(\in I)$ are relatively prime.

%======================================%
%<<<<<<   SUBSECTION 3-1-3    >>>>>>>>>%
%======================================%
\subsection{Canonical ensemble of open strings}
	\label{subsec:canonical}

The Hawking process can be described in the D-brane picture as a decay 
process of non-BPS excitations of
D-branes~\cite{Callan&Maldacena1996,Das&Mathur1996,Maldacena&Strominger1996}.
A collision of a right-moving open string with a left-moving one on
the D-branes results in an emission of a closed string leaving away
from the D-branes, which is interpreted as Hawking radiation. 
The spectrum of the emission of the closed string can, in principle,
be obtained from decay rates of the string, provided that the initial
state of the open strings on the D-branes is given. On the contrary,
when we do not know anything about the initial state of the open
strings, it seems that the best way to evaluate the spectrum is
summing up decay rates over all consistent initial states of open
strings on the D-branes. Formally, the summation should be done by
using a microcanonical ensemble with fixed $n_L$ and $n_R$. Since this
ideal summation is not easy, we adopt an approximation. Our strategy
of the approximation to perform the summation is to replace the 
microcanonical ensemble by the canonical ensemble, in which each
expectation value of $n_L$ and $n_R$ coincides with the given fixed
value in the original microcanonical ensemble. It is easy to see that
the corresponding spectrum of closed strings is approximately the
thermal one with temperature being the same as the temperature of this 
canonical ensemble.

Thus, in this subsection we consider a canonical ensemble of open
strings on the D-brane configuration introduced in the beginning of 
subsection~\ref{subsec:number}. From arguments in the derivation of 
Eqs.~(\ref{eqn:3-1:d=dd}) and (\ref{eqn:3-1:sum-d(nL)}), we can obtain 
the partition function of the canonical ensemble as 
\begin{eqnarray}
 Z(\beta,\alpha ) & = & Z_L(\beta,\alpha )Z_R(\beta,\alpha ),
					\nonumber\\
 Z_{L,R}(\beta,\alpha ) & = & 
	\prod_{q_1}\prod_{q_5}\prod_{n=1}^{\infty}
	\left[\frac{1+e^{-\beta e_n(q_1,q_5)-\alpha p_n(q_1,q_5)}}
	{1-e^{-\beta e_n(q_1,q_5)-\alpha p_n(q_1,q_5)}}
	\right]^{4N_{q_1}^{(1)}N_{q_5}^{(5)}},
	\label{eqn:3-1:def-Z}
\end{eqnarray}
where $\beta$ is inverse temperature and $\alpha$ is chemical
potential w.r.t. the total momentum along the $S^1$. 
Here $p_n$ is given by (\ref{eqn:3-1:spectrum}) (the plus sign is for 
$Z_L$ and the minus sign is for $Z_R$) and $e_n=|p_n|$. 
The quantities $\beta$ and $\alpha$ should be determined by requiring 
\begin{eqnarray}
 \frac{1}{R}(n_L+n_R) & = &  
	-\frac{\partial\ln Z(\beta,\alpha )}{\partial\beta},
	\nonumber\\
 \frac{1}{R}(n_L-n_R) & = &  
	-\frac{\partial\ln Z(\beta,\alpha )}{\partial\alpha}.
	\label{eqn:3-1:fix-beta-alpha}
\end{eqnarray}
Note that these requirements originate from the last and the third
equalities in (\ref{eqn:3-1:relations-charges}), respectively.

By using the formula (\ref{eqn:3-1:g-formula}), we can obtain an 
asymptotic expression of $Z_{L,R}(\beta,\alpha )$ in the limit
(\ref{eqn:3-1:limit}). The result is 
\begin{eqnarray}
 \ln Z_{L,R}(\beta,\alpha ) &  \approx & \frac{\pi^2R}{\beta\pm\alpha}
	\sum_{q_1}\sum_{q_5}N_{q_1}^{(1)}N_{q_5}^{(5)}(q_1,q_5)_{LCM},
						\nonumber	\\
 ('+' \mbox{ for } 'L' & ; & '-' \mbox{ for }'R').
\end{eqnarray}
From the condition (\ref{eqn:3-1:fix-beta-alpha}), $\beta$ and
$\alpha$ are determined as 
\begin{eqnarray}
 \beta +\alpha & \approx & \pi R\sqrt{\frac{\sum_{q_1}\sum_{q_5}
	N_{q_1}^{(1)}N_{q_5}^{(5)}(q_1,q_5)_{LCM}}{n_R}},\nonumber\\
 \beta -\alpha & \approx & \pi R\sqrt{\frac{\sum_{q_1}\sum_{q_5}
	N_{q_1}^{(1)}N_{q_5}^{(5)}(q_1,q_5)_{LCM}}{n_L}}.
\end{eqnarray}
Thus, for a configuration satisfying (\ref{eqn:3-1:limit}), the
inverse temperature $\beta$ and the entropy $S$ of the effective
canonical ensemble are 
\begin{eqnarray}
 \beta & \approx & \frac{\pi R}{2}
	\left(\frac{1}{\sqrt{n_L}}+\frac{1}{\sqrt{n_R}}\right) 
	\sqrt{\sum_{q_1}\sum_{q_5}
	N_{q_1}^{(1)}N_{q_5}^{(5)}(q_1,q_5)_{LCM}},\nonumber\\
 S & = & \ln Z(\beta,\alpha ) 
	- \beta \frac{\partial\ln Z(\beta,\alpha )}{\partial\beta}
	- \alpha \frac{\partial\ln Z(\beta,\alpha )}{\partial\alpha}
	\nonumber\\
   & \approx & 2\pi\left(\sqrt{n_L}+\sqrt{n_R}\right)
	\sqrt{\sum_{q_1}\sum_{q_5}
	N_{q_1}^{(1)}N_{q_5}^{(5)}(q_1,q_5)_{LCM}}\nonumber\\
   & \approx & \ln d(n_L,n_R),
\end{eqnarray}
where $d(n_L,n_R)$ is the number of micro-states evaluated by using
the microcanonical ensemble in the previous subsection. In fact, the
last equality is trivial from the general arguments in statistical 
mechanics: entropy in microcanonical ensemble can be calculated
approximately by the corresponding canonical ensemble. 
The temperature $\beta^{-1}$ is bounded from below by the Hawking
temperature given by (\ref{eqn:3-1:SBH&TBH}): 
\begin{equation}
 \beta^{-1} \geq T_{BH},\label{eqn:3-1:Tdecay}
\end{equation}
where the the bound is saturated if and only if all $(q_1,q_5)$
$(\in I)$ are relatively prime. Note that this condition for the
equality is the same as that for the equality in 
(\ref{eqn:3-1:lndN}). Here remember that the temperature
$\beta^{-1}$ coincides with the temperature of the thermal
spectrum of the closed string emission from the D-branes.

%======================================%
%<<<<<<   SUBSECTION 3-1-4    >>>>>>>>>%
%======================================%
\subsection{Summary and speculations}
	\label{subsec:speculations_Dbrane}

In this section the D-brane statistical-mechanics has
been reviewed by using a configuration of D-strings and D-fivebranes 
introduced in Ref.~\cite{Mukohyama1996}. 
We have considered a set of multiply-wound D-strings, which is
composed of $N_{q_1}^{(1)}$ D-strings of length $2\pi Rq_1$
($q_1=1,2,\cdots$) and a set of multiply-wound D-fivebranes, which is
composed of $N_{q_5}^{(5)}$ D-fivebranes of length $2\pi Rq_5$
($q_5=1,2,\cdots$) along the $S^1$. 
For configurations satisfying (\ref{eqn:3-1:limit}), the 
number of microscopic states $d(n_L,n_R)$ of open strings on the
D-branes is bounded from above by exponential of the Bekenstein-Hawking
entropy $S_{BH}$ of the corresponding black hole, and the temperature
$\beta^{-1}$ of a decay of D-brane excitations to closed strings is
bounded from below by the Hawking temperature $T_{BH}$ of the
corresponding black hole: Eqs.(\ref{eqn:3-1:lndN}) and
(\ref{eqn:3-1:Tdecay}). 
Note that $d(n_L,n_R)$ and $\beta^{-1}$ are evaluated microscopically
while $S_{BH}$ and $T_{BH}$ are defined in terms of macroscopic
quantities (area and surface gravity of the horizon of the
corresponding black hole).
The bounds (\ref{eqn:3-1:lndN}) and (\ref{eqn:3-1:Tdecay}) are
saturated if and only if all $(q_1,q_5)$ $(\in I)$ are relatively
prime.

Thus several speculations may be possible.
\begin{itemize}
 \item
Inside a black hole characterized by the four parameters ($Q_1$,
$Q_5$, $n_L$, $n_R$), some dynamical processes may occur. The
processes may be described in the D-brane picture: D-branes
repeat fission and fusion to settle down to one of the states for
which all $(q_1,q_5)$ $(\in I)$ are relatively prime.
 \item 
During the process the microscopic entropy increases to reach the
Bekenstein-Hawking entropy of the corresponding black hole. Moreover
the temperature of closed string radiation from the D-branes decreases
to reach the Hawking temperature of the black hole. 
\end{itemize}

These speculations may be significant to investigate the microstates
of dynamical black holes. For example let us consider a merger of two
black holes $B_1$ and $B_2$ and suppose that $B_1$ corresponds to a
D-brane configuration $\{ N_{q_1}^{(11)}, N_{q_5}^{(15)}\}$ and $B_2$
corresponds to a configuration 
$\{ N_{q_1}^{(21)}, N_{q_5}^{(25)}\}$. Just after merging, the large
black hole $B$ formed by the merger corresponds to a configuration 
$\{ N_{q_1}^{(1)}=N_{q_1}^{(11)}+N_{q_1}^{(21)}, 
N_{q_5}^{(5)}=N_{q_5}^{(15)}+N_{q_5}^{(25)} \}$, provided that
directions of the D-strings are the same for $B_1$ and $B_2$. In general
the last configuration does not saturate the bounds (\ref{eqn:3-1:lndN})
and (\ref{eqn:3-1:Tdecay}) even if the configurations for $B_1$ and
$B_2$ saturate the bounds. Thus, in general just after the merger the
microscopic entropy of $B$ does not agree with the corresponding
Bekenstein-Hawking entropy and the temperature of the closed string
radiation does not agree with the corresponding Hawking
temperature. However, after a sufficiently long time the D-branes'
fission and fusion settle the entropy and the temperature to the
Bekenstein-Hawking entropy and the Hawking temperature.

%%%%%%%%%%%%%%%%%%%%%%%%%%%%%%%%%%%%%%%%
%%%%%%%%%%%% SECTION 3-2 %%%%%%%%%%%%%%%
%%%%%%%%%%%%%%%%%%%%%%%%%%%%%%%%%%%%%%%%
\section{Brick wall model}
	\label{sec:brick_wall}

The entanglement interpretation, which will be investigated in detail
in section \ref{sec:entanglement}, seems to be implicit in, and is 
certainly closely related to a pioneering calculation done by Gerard 
`tHooft~\cite{tHooft1985} in 1985. He considered the statistical 
thermodynamics of quantum fields in the Hartle-Hawking state (i.e. 
having the Hawking temperature $T_{BH}$ at large radii) propagating on 
a fixed Schwarzschild background of mass $M$. To control divergences, 
`tHooft introduced a ``brick wall''---actually a static spherical 
mirror at which the fields are required to satisfy Dirichlet or 
Neumann boundary conditions---with radius a little larger than the 
gravitational radius $2M$. He found, in addition to the expected 
volume-dependent thermodynamical quantities describing hot fields in a 
nearly flat space, additional wall contributions proportional to the area. 
These contributions are, however, also proportional to $\alpha^{-2}$, 
where $\alpha$ is the proper altitude of the wall above the 
gravitational radius, and thus diverge in the limit $\alpha\to 0$. For 
a specific choice of $\alpha$ (which depends on the number of fields, 
etc., but is generally of order $l_{pl}$), `tHooft was able to 
recover the Bekenstein-Hawking formula with the correct coefficient.

However, this calculation raises a number of questions which have 
caused many, including `tHooft himself, to have reservations about 
its validity and consistency. 

\begin{enumerate}
 \renewcommand{\labelenumi}{(\alph{enumi})}
 \item $S_{BH}$ is here obtained as a one-loop effect, originating 
 from thermal excitations of the quantum fields. Does this material 
 contribution to $S_{BH}$ have to be {\it added} to the zero-loop 
 Gibbons-Hawking contribution which arises from the gravitational 
 part of the action and already by itself accounts for the full value 
 of $S_{BH}$?~\cite{Liberati}
 \item The ambient quantum fields were assumed to be in the 
 Hartle-Hawking state. Their stress-energy should therefore be 
 bounded (of order $M^{-4}$ in Planck units) near the gravitational 
 radius, and negligibly small for large masses. However, `tHooft's 
 calculation assigns to them enormous (Planck-level) energy densities 
 near the wall. 
 \item The integrated field energy gives a wall contribution to the
mass 
 %============< EQUATION >==============%
 %
 \begin{equation}
  \Delta M = \frac{3}{8}M \label{eqn:3-2:DeltaM=3/8M}
 \end{equation}
 %======================================%
 when $\alpha$ is adjusted to give the correct value of $S_{BH}$. This 
 suggests a substantial gravitational back-reaction~\cite{tHooft1985}
 and  that the assumption of a fixed geometrical background may be 
 inconsistent~\cite{Liberati,Belgiorno&Martellini,Susskind&Uglum1994}. 
\end{enumerate}

Our main purpose in this section is to point out that these 
difficulties are only apparent and easy to
resolve~\cite{Mukohyama&Israel1998}. The basic remark 
is that the {\it brick-wall model strictly interpreted does not 
represent a black hole}. It represents the exterior of a starlike 
object with a reflecting surface, compressed to nearly (but not 
quite) its gravitational radius. The ground state for quantum fields 
propagating around this star is not the Hartle-Hawking
state~\cite{Hartle&Hawking1976} but the Boulware
state~\cite{Boulware}, corresponding to zero 
temperature, which has a quite different behavior near the
gravitational radius.

In subsection \ref{subsec:BandHHstate} we summarize essential
properties of the Boulware and Hartle-Hawking states that play a role
in our arguments.
In subsection \ref{subsec:BWmodel1} we sketch the physical essence of
the brick-wall model by using a particle description of quantum
fields. A systematic treatment of the model is deferred to
subsection \ref{subsec:BWmodel2}, in which the results in
subsection \ref{subsec:BWmodel1} are rigorously derived from the
quantum field theory in curved spacetimes. 
In subsection \ref{subsec:complementarity} we propose a complementary
principle between the brick wall model and the Gibbons-Hawking
instanton. 
In Appendix~\ref{app:on-shell}, for completeness, we apply the
so-called on-shell method to the brick wall model and show that in the 
on-shell method we might miss some physical degrees of freedom. Hence, 
we do not adopt the on-shell method in the main body of this thesis.

%======================================%
%<<<<<<   SUBSECTION 3-2-1    >>>>>>>>>%
%======================================%
\subsection{The Boulware and Hartle-Hawking states}
	\label{subsec:BandHHstate}

It is useful to begin by summarizing briefly the essential properties
of the quantum states that will play a role in our discussion.

In a curved spacetime there is no unique choice of time
coordinate. Different choices lead to different definitions of
positive-frequency modes and different ground states.

In any static spacetime with static (Killing) time parameter $t$, the
Boulware state $|B\rangle$ is the one annuled by the annihilation
operators $a_{Kill}$ associated with ``Killing modes''
(positive-frequency in $t$). In an asymptotically flat space,
$|B\rangle$ approaches the Minkowski vacuum at infinity.

In the spacetime of a stationary eternal black hole, the
Hartle-Hawking state $|HH\rangle$ is the one annuled by $a_{Krus}$,
the annihilation operators associated with ``Kruskal modes''
(positive-frequency in the Kruskal lightlike coordinates $U$,
$V$). This state appears empty of ``particles'' to free falling
observers at the horizon, and its stress-energy is bounded there (not
quite zero, because of irremovable vacuum polarization effects).

If, just for illustrative purposes, we consider a $(1+1)$-dimensional
spacetime, it is easy to give concrete form to these remarks. We
consider a spacetime with metric 
%============< EQUATION >==============%
%
\begin{equation}
 ds^2 = -f(r)dt^2+\frac{dr^2}{f(r)},
	\label{eqn:3-2:2d-metric}
\end{equation}
%======================================%
and denote by $\kappa(r)$ the redshifted gravitational force, i.e.,
the upward acceleration $a(r)$ of a stationary test-particle reduced
by the redshift factor $f^{1/2}(r)$, so that
$\kappa(r)=\frac{1}{2}f'(r)$. A horizon is characterized by $r=r_0$,
$f(r_0)=0$, and its surface gravity defined by
$\kappa_0=\frac{1}{2}f'(r_0)$.

Quantum effects induce an effective quantum stress-energy $T_{ab}$
($a,b,\cdots =r,t$) in the background geometry
(\ref{eqn:3-2:2d-metric}). If we assume no net energy flux 
($T^r_t=0$)---thus excluding the Unruh state---$T_{ab}$ is completely
specified by a quantum energy density $\rho=-T^t_t$ and pressure
$P=T^r_r$. These are completely determined (up to a boundary
condition) by the conservation law $T^b_{a;b}=0$ and the trace
anomaly, which is 
%============< EQUATION >==============%
%
\begin{equation}
 T^a_a = \frac{\hbar}{24\pi}R
\end{equation}
%======================================%
for a massless scalar field, with $R=-f''(r)$ for the metric
(\ref{eqn:3-2:2d-metric}). Integration gives 
%============< EQUATION >==============%
%
\begin{equation}
 f(r)P(r) = -\frac{\hbar}{24\pi}(\kappa^2(r) + const.).
	\label{eqn:3-2:2d-fP}
\end{equation}
%======================================%
Different choices of the constant of integration correspond to
different boundary conditions, i.e., to different quantum states.

For the Hartle-Hawking state, we require $P$ and $\rho$ to be bounded
at the horizon $r=r_0$, giving 
%============< EQUATION >==============%
%
\begin{eqnarray}
 P_{HH} & = & \frac{\hbar}{24\pi}\frac{\kappa_0^2-\kappa^2(r)}{f(r)},
	\nonumber\\
 \rho_{HH} & = & P_{HH} + \frac{\hbar}{24\pi}f''(r).
	\label{eqn:3-2:HH-P-rho}
\end{eqnarray}
%======================================%
When $r\to\infty$ this reduces to (setting $f(r)\to 1$)
%============< EQUATION >==============%
%
\begin{eqnarray}
 \rho_{HH} & \simeq & P_{HH} = \frac{\pi}{6\hbar}T_{BH}^2,
	\nonumber\\
 T_{BH} & = & \hbar\kappa_0/2\pi,
\end{eqnarray}
%======================================%
which is appropriate for one-dimensional scalar radiation at the
Hawking temperature $T_{BH}$.

For the Boulware state, the boundary condition is $P=\rho=0$ when
$r=\infty$. The integration constant in (\ref{eqn:3-2:2d-fP}) must vanish, 
and we find 
%============< EQUATION >==============%
%
\begin{eqnarray}
 P_{B} & = & -\frac{\hbar}{24\pi}\frac{\kappa^2(r)}{f(r)},
	\nonumber\\
 \rho_{B} & = & P_{B} + \frac{\hbar}{24\pi}f''(r).
	\label{eqn:3-2:B-P-rho}
\end{eqnarray}
%======================================%
If a horizon were present, $\rho_B$ and $P_B$ would diverge there to
$-\infty$.

For the difference of these two stress tensors,
%============< EQUATION >==============%
%
\begin{eqnarray}
 \Delta T_a^b = (T_a^b)_{HH} - (T_a^b)_{B},\label{eqn:3-2:DeltaTab}
\end{eqnarray}
%======================================%
(\ref{eqn:3-2:HH-P-rho}) and (\ref{eqn:3-2:B-P-rho}) give the exactly thermal
form 
%============< EQUATION >==============%
%
\begin{eqnarray}
 \Delta P = \Delta\rho = \frac{\pi}{6\hbar}T^2(r),
	\label{eqn:3-2:2d-DP-Drho}
\end{eqnarray}
%======================================%
where $T(r)=T_{BH}/\sqrt{f(r)}$ is the local temperature in the
Hartle-Hawking state. We recall that thermal equilibrium in any static 
gravitational field requires the local temperature $T$ to rise with
depth in accordance with Tolman's law~\cite{Tolman}
%============< EQUATION >==============%
%
\begin{eqnarray}
 T\sqrt{-g_{00}} = const. \label{eqn:3-2:Tolman}
\end{eqnarray}
%======================================%

We have found, for this $(1+1)$-dimensional example, that the
Hartle-Hawking state is thermally excited above the zero-temperature
(Boulware) ground state to a local temperature $T(r)$ which grows
without bound near the horizon. Nevertheless, it is the Hartle-Hawking 
state which best approximates what a gravitational theorist would call 
a ``vacuum'' at the horizon.

These remarks remain at least qualitatively valid in
$(3+1)$-dimensions, with obvious changes arising from the 
dimensionality. In particular, the $(3+1)$-dimensional analogue of
(\ref{eqn:3-2:2d-DP-Drho}) for a massless scalar field, 
%============< EQUATION >==============%
%
\begin{eqnarray}
 3\Delta P \simeq \Delta\rho 
	\simeq \frac{\pi^2}{30\hbar^3}T^4(r),\label{eqn:3-2:DeltaP-Deltarho}
\end{eqnarray}
%======================================%
holds to a very good approximation, both far from the black hole and
near the horizon. In the intermediate zone there are deviations, 
but they always remain bounded~\cite{DP-Drho}, and will not affect our 
considerations.

%======================================%
%<<<<<<   SUBSECTION 3-2-2    >>>>>>>>>%
%======================================%
\subsection{A brief sketch of the brick wall model}
	\label{subsec:BWmodel1}
	
We shall briefly sketch the physical essence of the brick-wall 
model. (A systematic treatment is deferred to 
subsection \ref{subsec:BWmodel2})

We wish to study the thermodynamics of hot quantum fields confined to 
the outside of a spherical star with a perfectly reflecting surface 
whose radius $r_{1}$ is a little larger than its gravitational radius 
$r_{0}$. To keep the total field energy bounded, we suppose the 
system enclosed in a spherical container of radius $L\gg r_{1}$.

It will be sufficiently general to assume for the geometry outside 
the star a spherical background metric of the form 
%============< EQUATION >==============%
%
\begin{equation}
 ds^2 = -f(r)dt^2+\frac{dr^2}{f(r)}+r^2d\Omega^2.
 	\label{eqn:3-2:4d-metric}
\end{equation}
%======================================%
This includes as special cases the Schwarzschild, 
Reissner-Nordstr{\"o}m and de Sitter geometries, or any combination of 
these.

Into this space we introduce a collection of quantum fields, raised to 
some temperature $T_{\infty}$ at large distances, and in thermal 
equilibrium. The local temperature $T(r)$ is then given by Tolman's 
law (\ref{eqn:3-2:Tolman}), 
%============< EQUATION >==============%
%
\begin{equation}
 T(r) = T_{\infty}f^{-1/2}\label{eqn:3-2:T(r)-Tinfty}
\end{equation}
%======================================%
and becomes very large when $r\to r_{1}=r_{0}+\Delta r$. We shall 
presently identify $T_{\infty}$ with the Hawking temperature $T_{BH}$ 
of the horizon $r=r_{0}$ of the exterior metric 
(\ref{eqn:3-2:4d-metric}), continued (illegitimately) into the internal
domain $r<r_{1}$.

Characteristic wavelengths $\lambda$ of this radiation are small 
compared to other relevant length-scales (curvature, size of
container) in the regions of interest to us. Near the star's surface, 
%============< EQUATION >==============%
%
\begin{equation}
 \lambda \sim \hbar/T = f^{1/2}\hbar/T_{\infty}\ll r_{0}.\nonumber
\end{equation}
%======================================%
Elsewhere in the large container, at large distances from the star, 
%============< EQUATION >==============%
%
\begin{equation}
 f\simeq 1, \quad
 \lambda\simeq\hbar/T_{\infty}\sim r_{0}\ll L.\nonumber
\end{equation}
%======================================%
Therefore, a particle description should be a good approximation to 
the statistical thermodynamics of the fields (Equivalently, one can 
arrive at this conclusion by considering the WKB solution to the wave 
equation, cf. `tHooft~\cite{tHooft1985} and 
subsection \ref{subsec:BWmodel2}.)

For particles of rest-mass $m$, energy $E$, $3$-momentum $p$ and 
$3$-velocity $v$ as viewed by a local stationary observer, the energy 
density $\rho$, pressure $P$ and entropy density $s$ are given by the 
standard expressions 
%============< EQUATION >==============%
%
\begin{eqnarray}
 \rho & = & {\cal N}\int_{0}^{\infty}
 	\frac{E}{e^{\beta E}-\epsilon}\frac{4\pi p^2dp}{h^3},
 	\nonumber\\
 P & = & \frac{{\cal N}}{3}\int_{0}^{\infty}
 	\frac{vp}{e^{\beta E}-\epsilon}\frac{4\pi p^2dp}{h^3},
 	\nonumber\\
 s & = & \beta (\rho +P).\label{eqn:3-2:S-P-rho}
\end{eqnarray}
%======================================%
Here, as usual, 
%============< EQUATION >==============%
%
\begin{equation}
 E^2-p^2 =m^2, \quad v=p/E, \quad \beta =T^{-1};\nonumber
\end{equation}
%======================================%
$\epsilon$ is $+1$ for bosons and $-1$ for fermions and the factor 
${\cal N}$ takes care of helicities and the number of particle 
species. The total entropy is given by the integral 
%============< EQUATION >==============%
%
\begin{equation}
 S = \int_{r_{1}}^{L}s(r)4\pi r^2dr/\sqrt{f},
 	\label{eqn:3-2:Stot}
\end{equation}
%======================================%
where we have taken account of the proper volume element as given by 
the metric (\ref{eqn:3-2:4d-metric}). The factor $f^{-1/2}$ does not,
however, appear in the integral for the gravitational mass of the
thermal excitations~\cite{MTW} (It is canceled, roughly speaking, by 
negative gravitational potential energy):
%============< EQUATION >==============%
%
\begin{equation}
 \Delta M_{therm} = \int_{r_{1}}^{L}\rho(r)4\pi r^2dr.
	\label{eqn:3-2:Mtherm}
\end{equation}
%======================================%

The integrals (\ref{eqn:3-2:Stot}) and (\ref{eqn:3-2:Mtherm}) are dominated by 
two contributions for large container radius $L$ and for small 
$\Delta r=r_{1}-r_{0}$:
\begin{enumerate}
 \renewcommand{\labelenumi}{(\alph{enumi})}
 \item A volume term, proportional to $\frac{4}{3}\pi L^3$, 
 representing the entropy and mass-energy of a homogeneous quantum 
 gas in a flat space (since $f\simeq 1$ almost everywhere in the 
 container if $L/r_{0}\to\infty$) at a uniform temperature 
 $T_{\infty}$. This is the result that would have been expected, and 
 we do not need to consider it in detail. 
 \item Of more interest is the contribution of gas near the inner 
 wall $r=r_{1}$, which we now proceed to study further. We shall find 
 that it is proportional to the wall area, and diverging like 
 $(\Delta r)^{-1}$ when $\Delta r\to 0$.
\end{enumerate}

Because of the high local temperatures $T$ near the wall for small 
$\Delta r$, we may insert the ultrarelativistic approximations 
%============< EQUATION >==============%
%
\[
 E\gg m,\quad p\simeq E,\quad v\simeq 1
\]
%======================================%
into the integrals (\ref{eqn:3-2:S-P-rho}). This gives
%============< EQUATION >==============%
%
\begin{equation}
 P\simeq\frac{1}{3}\rho
 	\simeq\frac{{\cal N}}{6\pi^2}T^4\int_{0}^{\infty}
 	\frac{x^3dx}{e^x-\epsilon}
\end{equation}
%======================================%
in Planck units ($h=2\pi\hbar =2\pi$). The purely numerical integral 
has the value $3!$ multiplied by $1$, $\pi^4/90$ and 
$\frac{7}{8}\pi^4/90$ for $\epsilon=0,1$ and $-1$ respectively, and we 
shall adopt $3!$, absorbing any small discrepancy into ${\cal N}$. 
Then, from (\ref{eqn:3-2:S-P-rho}), 
%============< EQUATION >==============%
%
\begin{equation}
 \rho=\frac{3{\cal N}}{\pi^2}T^4,\quad s=\frac{4{\cal N}}{\pi^2}T^3
 	\label{eqn:3-2:rho-T4}
\end{equation}
%======================================%
in terms of the local temperature $T$ given by (\ref{eqn:3-2:T(r)-Tinfty}).

Substituting (\ref{eqn:3-2:rho-T4}) into (\ref{eqn:3-2:Stot}) gives for the 
wall contribution to the total entropy
%============< EQUATION >==============%
%
\begin{equation}
 S_{wall} = \frac{4{\cal N}}{\pi^2}4\pi r_{1}^2 T_{\infty}^3
 	\int_{r_{1}}^{r_{1}+\delta}\frac{dr}{f^2(r)},
 	\label{eqn:3-2:Swall}
\end{equation}
%======================================%
where $\delta$ is an arbitrary small length subject to 
$\Delta r\ll\delta\ll r_{1}$. It is useful to express this result in 
terms of the proper altitude $\alpha$ of the inner wall above the 
horizon $r=r_{0}$ of the exterior geometry (\ref{eqn:3-2:4d-metric}). 
(Since (\ref{eqn:3-2:4d-metric}) only applies for $r>r_{1}$, the physical 
space does not, of course, contain any horizon.) We assume that $f(r)$ 
has a (simple) zero for $r=r_{0}$, so we can write
%============< EQUATION >==============%
%
\begin{equation}
 f(r) \simeq 2\kappa_{0} (r-r_{0}),\quad
 \kappa_{0}=\frac{1}{2}f'(r_{0})\ne 0 \quad
 (r\to r_{0}),
\end{equation}
%======================================%
where $\kappa_{0}$ is the surface gravity. Then 
%============< EQUATION >==============%
%
\begin{equation}
 \alpha = \int_{r_{0}}^{r_{1}}f^{-1/2}dr\quad\Rightarrow\quad
 	\Delta r = \frac{1}{2}\kappa_0\alpha^2,
\end{equation}
%======================================%
and (\ref{eqn:3-2:Swall}) can be written 
%============< EQUATION >==============%
%
\begin{equation}
 S_{wall} = \frac{{\cal N}}{90\pi\alpha^2}
 	\left(\frac{T_{\infty}}{\kappa_0/2\pi}\right)^3
	\frac{1}{4}A \label{eqn:3-2:Swall2}
\end{equation}
%======================================%
in Planck units, where $A=4\pi r_{1}^2$ is the wall area.

Similarly, we find from (\ref{eqn:3-2:Mtherm}) and (\ref{eqn:3-2:rho-T4}) that 
thermal excitations near the wall contribute 
%============< EQUATION >==============%
%
\begin{equation}
 \Delta M_{them,wall} = \frac{{\cal N}}{480\pi\alpha^2}
 	\left(\frac{T_{\infty}}{\kappa_0/2\pi}\right)^3AT_{\infty}
 	\label{eqn:3-2:M-thermw-wall}
\end{equation}
%======================================%
to the gravitational mass of the system.

The wall contribution to the free energy 
%============< EQUATION >==============%
%
\begin{equation}
 F=\Delta M-T_{\infty}S \label{eqn:3-2:Gibbs-Duhem}
\end{equation}
%======================================%
is 
%============< EQUATION >==============%
%
\begin{equation}
 F_{wall}= -\frac{{\cal N}}{1440\pi\alpha^2}
 	\left(\frac{T_{\infty}}{\kappa_0/2\pi}\right)^3AT_{\infty}.
 	\label{eqn:3-2:Fwall}
\end{equation}
%======================================%
The entropy is recoverable from the free energy by the standard 
prescription
%============< EQUATION >==============%
%
\begin{equation}
 S_{wall} = -\partial F_{wall}/\partial T_{\infty}.\label{eqn:3-2:Gibbs}
\end{equation}
%======================================%
(Observe that this is an ``off-shell''
prescription~\cite{Frolov1995,FFZ1996a}:  
the geometrical quantities $A$, $\alpha$ and, in particular, the 
surface gravity $\kappa_{0}$ are kept fixed when the temperature is 
varied in (\ref{eqn:3-2:Fwall}).)

Following `tHooft~\cite{tHooft1985}, we now introduce a crude cutoff to 
allow for quantum-gravity fluctuations by fixing the wall altitude 
$\alpha$ so that 
%============< EQUATION >==============%
%
\begin{equation}
 S_{wall} = S_{BH},\quad \mbox{when}\quad T_{\infty} = T_{BH},
 	\label{eqn:3-2:Swall=SBH}
\end{equation}
%======================================%
where the Bekenstein-Hawking entropy $S_{BH}$ and Hawking temperature 
$T_{BH}$ are defined to be the {\it purely geometrical} quantities 
in terms of the wall's area $A$ and redshifted acceleration ($=$
surface gravity) $\kappa_0$.  
From (\ref{eqn:3-2:Swall=SBH}) and (\ref{eqn:3-2:Swall2}), restoring 
conventional units for a moment, we find 
%============< EQUATION >==============%
%
\begin{equation}
 \alpha = l_{pl}\sqrt{{\cal N}/90\pi},\label{eqn:3-2:normalization}
\end{equation}
%======================================%
so that $\alpha$ is very near the Planck length if the effective 
number ${\cal N}$ of basic quantum fields in nature is on the order 
of $300$.

It is significant and crucial that the normalization 
(\ref{eqn:3-2:normalization}) is {\it universal}, depending only on 
fundamental physics, and independent of the mechanical and 
geometrical characteristics of the system.

With $\alpha$ fixed by (\ref{eqn:3-2:normalization}), the wall's free 
energy (\ref{eqn:3-2:Gibbs-Duhem}) becomes 
%============< EQUATION >==============%
%
\begin{equation}
 F_{wall} = -\frac{1}{16}\left(\frac{T_{\infty}}{T_{BH}}\right)^3
 	AT_{\infty}.\label{eqn:3-2:Fwall2}
\end{equation}
%======================================%
This ``off-shell'' formula expresses $F_{wall}$ in terms of three 
independent variables: the temperature $T_{\infty}$ and the 
geometrical characteristics $A$ and $T_{BH}$. From (\ref{eqn:3-2:Fwall2}) 
we can obtain the wall entropy either from the thermodynamical Gibbs 
relation (\ref{eqn:3-2:Gibbs}) (with $T_{\infty}$ set equal to $T_{BH}$ 
{\it after} differentiation), or from the Gibbs-Duhem formula 
(\ref{eqn:3-2:Gibbs-Duhem}) which is equivalent to the 
statistical-mechanical definition $S=-{\bf Tr}(\rho\ln\rho)$. Thus the 
distinction~\cite{Frolov1995,FFZ1996a} between ``thermodynamical'' and 
``statistical'' entropies disappears in this formulation, because the 
geometrical and thermal variables are kept independent.

The wall's thermal mass-energy is given ``on-shell'' 
($T_{\infty}=T_{BH}$) by 
%============< EQUATION >==============%
%
\begin{equation}
 \Delta M_{therm,wall} = \frac{3}{16}AT_{BH}
\end{equation}
%======================================%
according to (\ref{eqn:3-2:M-thermw-wall}) and (\ref{eqn:3-2:normalization}). 
For a wall skirting a Schwarzschild horizon, so that 
$T_{BH}=(8\pi M)^{-1}$, this reduces to `tHooft's result 
(\ref{eqn:3-2:DeltaM=3/8M}).

As already noted, thermal energy is not the only source of the wall's 
mass. Quantum fields outside the wall have as their ground state the 
Boulware state, which has a negative energy density growing to Planck 
levels near the wall. On shell, this very nearly cancels the thermal 
energy density (\ref{eqn:3-2:rho-T4}); their sum is, in fact, the 
Hartle-Hawking value (cf. (\ref{eqn:3-2:DeltaTab}) and 
(\ref{eqn:3-2:DeltaP-Deltarho})):  
%============< EQUATION >==============%
%
\begin{equation}
 (T_{\mu}^{\nu})_{therm,T_{\infty}=T_{BH}} + (T_{\mu}^{\nu})_{B}
 = (T_{\mu}^{\nu})_{HH},
\end{equation}
%======================================%
which remains bounded near horizons, and integrates virtually to zero 
for a very thin layer near the wall. The total gravitational mass of 
the wall is thus, from (\ref{eqn:3-2:M-thermw-wall}) and 
(\ref{eqn:3-2:normalization}),
%============< EQUATION >==============%
%
\begin{eqnarray}
 (\Delta M)_{wall} & = & (\Delta M)_{therm,wall} + (\Delta M)_{B,wall}
 	\nonumber\\
 	& = & \frac{3}{16}AT_{BH}\left( (T_{\infty}/T_{BH})^4-1\right),
\end{eqnarray}
%======================================%
which vanishes on shell. For a central mass which is large in Planck 
units, there is no appreciable back-reaction of material near the wall 
on the background geometry (\ref{eqn:3-2:4d-metric}).

We may conclude that many earlier 
concerns~\cite{tHooft1985,Liberati,Belgiorno&Martellini} were 
unnecessary: `tHooft's brick wall model does provide a perfectly 
self-consistent description of a configuration which is 
indistinguishable from a black hole to outside observers, and which 
accounts for the Bekenstein-Hawking entropy purely as thermal entropy 
of quantum fields at the Hawking temperature (i.e. in the 
Hartle-Hawking state), providing one accepts the ad hoc but plausible 
ansatz (\ref{eqn:3-2:normalization}) for a Planck-length cutoff near the 
horizon.

The model does, however, present us with a feature which is 
theoretically possible but appears strange and counterintuitive from a 
gravitational theorist's point of view. Although the wall is 
insubstantial (just like a horizon)---i.e., space there is 
practically a vacuum and the local curvature low---it is 
nevertheless the repository of all of the Bekenstein-Hawking entropy 
in the model.

It has been argued~\cite{Frolov&Novikov1993} that this is just what
might be expected of black hole entropy in the entanglement picture. 
Entanglement will arise from virtual pair-creation in which one 
partner is ``invisible'' and the other ``visible'' (although only 
temporarily---nearly all get reflected back off the potential barrier). 
Such virtual pairs are all created very near the horizon. Thus, on
this picture, the entanglement entropy (and its divergence) arises
almost entirely from the strong correlation between nearby field
variables on the two sides of the partition, an effect already present 
in flat space~\cite{Callan&Wilczek1994}.

An alternative (but not necessarily incompatible) possibility is that
the concentration of entropy at the wall is an artifact of the model
or of the choice of Fock representation (based on a static observer's
definition of positive frequency). The boundary condition of perfect
reflectivity at the wall has no black hole counterpart. Moreover, one
may well suspect that localization of entanglement entropy is not an
entirely well-defined concept~\cite{Mukohyama1998a} or invariant under 
changes of the Fock representation.

%======================================%
%<<<<<<   SUBSECTION 3-2-3    >>>>>>>>>%
%======================================%
\subsection{The brick wall model reexamined}
	\label{subsec:BWmodel2}

In the previous section, we have investigated the statistical 
mechanics of quantum fields in the region $r_{1}<r<L$ of the spherical 
background (\ref{eqn:3-2:4d-metric}) with the Dirichlet boundary condition 
at the boundaries. By using the particle description with the local 
temperature given by the Tolman's law, we have obtained the inner-wall 
contributions of the fields to entropy and thermal energy. When the 
former is set to be equal to the black hole entropy by 
fixing the cutoff $\alpha$ as (\ref{eqn:3-2:normalization}), the 
later becomes comparable with the mass of the background geometry. 
After that, it has been shown that at the Hawking temperature the wall 
contribution to the thermal energy is exactly canceled by the negative 
energy of the Boulware state, assuming implicitly that the ground state 
of the model is the Boulware state and that the  gravitational energy 
appearing in the Einstein equation is a sum of the renormalized energy 
of the Boulware state and the thermal energy of the fields.

In this section we shall show that these implicit assumptions do 
hold. In the following arguments it will also become clear how the 
local description used in the previous section is derived from the 
quantum field theory in curved spacetime, which is globally defined.

For simplicity, we consider a real scalar field described by the 
action 
%============< EQUATION >==============%
%
\begin{equation}
 I = -\frac{1}{2}\int d^4x\sqrt{-g}\left[
 	g^{\mu\nu}\partial_{\mu}\phi\partial_{\nu}\phi
 	+ m_{\phi}^2\phi^2\right]. 
\end{equation}
%======================================%
On the background given by (\ref{eqn:3-2:4d-metric}), the action is reduced 
to 
%============< EQUATION >==============%
%
\begin{equation}
 I =  \int dt L,
\end{equation}
%======================================%
with the Lagrangian $L$ given by
%============< EQUATION >==============%
%
\begin{equation}
 L = - \frac{1}{2}\int d^3x r^2\sqrt{\Omega}\left[
 	- \frac{1}{f}(\partial_{t}\phi)^2 + f(\partial_{r}\phi)^2
 	+ \frac{1}{r^2}\Omega^{ij}\partial_{i}\phi\partial_{j}\phi
 	+ m_{\phi}^2\phi^2\right]. 
\end{equation}
%======================================%
Here $x^i$ ($i=1,2$) are coordinates on the $2$-sphere. 
In order to make our system finite let us suppose that two 
mirror-like boundaries are placed at $r=r_1$ and $r=L$ ($r_1<L$), 
respectively, and investigate the scalar field in the region between the 
two boundaries. In the following arguments we quantize the 
scalar field with respect to the Killing time $t$. Hence, the ground 
state obtained below is the Boulware state.
After the quantization, we investigate the statistical mechanics of the 
scalar field in the Boulware state. It will be shown that the resulting 
statistical mechanics is equivalent to the brick wall model.

Now let us proceed to the quantization procedure. First, the momentum 
conjugate to $\phi(r,x^i)$ is 
%============< EQUATION >==============%
%
\begin{equation}
 \pi(r,x^i) = \frac{r^2\sqrt{\Omega}}{f}\partial_{t}\phi,
\end{equation}
%======================================%
and the Hamiltonian is given by
%============< EQUATION >==============%
%
\begin{equation}
 H = \frac{1}{2}\int d^3x\left[\frac{f}{r^2\sqrt{\Omega}}\pi^2
 	+ r^2\sqrt{\Omega}f\left(\partial_{r}\phi\right)^2
 	+ \sqrt{\Omega}\Omega^{ij}\partial_{i}\phi\partial_{j}\phi
 	+ r^2\sqrt{\Omega}m_{\phi}^2\phi^2\right].
 	\label{eqn:3-2:Hamiltonian}
\end{equation}
%======================================%
Next, promote the field $\phi$ to an operator and expand it as
%============< EQUATION >==============%
%
\begin{equation}
 \phi(r,x^i) = \sum_{nlm}\frac{1}{\sqrt{2\omega_{nl}}}\left[
 	a_{nlm}\varphi_{nl}(r)Y_{lm}(x^i)e^{-i\omega_{nl}t}
 	+ a^{\dagger}_{nlm}\varphi_{nl}(r)Y_{lm}(x^i)e^{i\omega_{nl}t}
 	\right],
\end{equation}
%======================================%
where $Y_{lm}(x^i)$ are real spherical harmonics defined by
%============< EQUATION >==============%
%
\begin{eqnarray*}
 \frac{1}{\sqrt{\Omega}}\partial_{i}\left(\sqrt{\Omega}\Omega^{ij}
 	\partial_{j}Y_{lm}\right) + l(l+1)Y_{lm} & = & 0,\\
 \int Y_{lm}(x^i)Y_{l'm'}(x^i)\sqrt{\Omega(x^i)}d^2x
 	& = & \delta_{ll'}\delta_{mm'},
\end{eqnarray*}
%======================================%
and $\{\varphi_{nl}(r)\}$ ($n=1,2,\cdots$) is a set of real functions 
defined below, which is complete with respect to the space of 
$L_{2}$-functions on the interval $r_{1}\leq r\leq L$ for 
each $l$. The positive constant $\omega_{nl}$ is defined as the 
corresponding eigenvalue. 
%============< EQUATION >==============%
%
\begin{eqnarray}
 \frac{1}{r^2}\partial_{r}\left( r^2f\partial_{r}\varphi_{nl}\right) 
 	- \left[ \frac{l(l+1)}{r^2}+m_{\phi}^2\right]\varphi_{nl}
 	+ \frac{\omega_{nl}^2}{f}\varphi_{nl} & = & 0,
 	\label{eqn:3-2:field-eq}\\
 \varphi_{nl}(r_{1})=\varphi_{nl}(L) & = & 0,\nonumber\\
 \int_{r_{1}}^{L}\varphi_{nl}(r)\varphi_{n'l}(r)\frac{r^2}{f(r)}dr
 	& = & \delta_{nn'}.\nonumber
\end{eqnarray}
%======================================%
The corresponding expansion of the operator $\pi(r,x^i)$ is then:
%============< EQUATION >==============%
%
\begin{eqnarray}
 \pi(r,x^i) & = & -i\frac{r^2\sqrt{\Omega(x^i)}}{f(r)}\sum_{nlm}
 	\sqrt{\frac{\omega_{nl}}{2}}
	\nonumber\\
 & & \times \left[
 	a_{nlm}\varphi_{nl}(r)Y_{lm}(x^i)e^{-i\omega_{n}t}
 	- a^{\dagger}_{nlm}\varphi_{nl}(r)Y_{lm}(x^i)e^{i\omega_{nl}t}
 	\right]. 
\end{eqnarray}
%======================================%
Hence, the usual equal-time commutation relation 
%============< EQUATION >==============%
%
\begin{eqnarray}
 \left[\phi(r,x^i),\pi(r',{x'}^i)\right] & = &
 	i\delta(r-r')\delta^2(x^i-{x'}^i), \nonumber\\
 \left[\phi(r,x^i),\phi(r',{x'}^i)\right] & = &
 \left[\pi(r,x^i),\pi(r',{x'}^i)\right] = 0
\end{eqnarray}
%======================================%
becomes 
%============< EQUATION >==============%
%
\begin{eqnarray}
 \left[ a_{nlm},a^{\dagger}_{n'l'm'}\right] & = & 
 	\delta_{nn'}\delta_{ll'}\delta_{mm'},\nonumber\\
 \left[ a_{nlm},a_{n'l'm'}\right] & = & 0,\nonumber\\
 \left[ a^{\dagger}_{nlm},a^{\dagger}_{n'l'm'}\right] & = & 0.
\end{eqnarray}
%======================================%
The normal-ordered Hamiltonian is given by
%============< EQUATION >==============%
%
\begin{equation}
 :H: = \sum_{nlm}\omega_{nl}a^{\dagger}_{nlm}a_{nlm}.\label{eqn:3-2::H:}
\end{equation}
%======================================%
Thus, the Boulware state $|B\rangle$, which is defined by
%============< EQUATION >==============%
%
\begin{equation}
 a_{nlm}|B\rangle = 0
\end{equation}
%======================================%
for ${}^{\forall}(n,l,m)$, is an eigenstate of the normal-ordered 
Hamiltonian with the eigenvalue zero. The Hilbert space of all quantum 
states of the scalar field is constructed as a symmetric Fock space 
on the Boulware state, and the complete basis 
$\{|\{N_{nlm}\}\rangle\}$ $(N_{nlm}=0,1,2,\cdots)$ is defined by
%============< EQUATION >==============%
%
\begin{equation}
 |\{N_{nlm}\}\rangle = \prod_{nlm}\frac{1}{\sqrt{N_{nlm}!}}
 	\left( a^{\dagger}_{nlm}\right)^{N_{nlm}}|B\rangle,
\end{equation}
%======================================%
and each member of the basis is an eigenstate of the normal-ordered 
Hamiltonian:
%============< EQUATION >==============%
%
\begin{equation}
 :H: |\{N_{nlm}\}\rangle = \left(\sum_{nlm}\omega_{nl}N_{nlm}\right) 
 	|\{N_{nlm}\}\rangle.
\end{equation}
%======================================%

Now we shall investigate the statistical mechanics of the quantized 
scalar field. The free energy $F$ is given by
%============< EQUATION >==============%
%
\begin{equation}
 e^{-\beta_{\infty} F} \equiv
 	{\bf Tr}\left[ e^{-\beta_{\infty} :H:}\right]
	 = \prod_{nlm}\frac{1}{1-e^{-\beta_{\infty}\omega_{nl}}},
\end{equation}
%======================================%
where $\beta_{\infty}=T_{\infty}^{-1}$ is inverse temperature. 
For explicit calculation of the free energy we adopt the WKB 
approximation. First, we rewrite the mode function $\varphi_{nl}(r)$ 
as 
%============< EQUATION >==============%
%
\begin{equation}
 \varphi_{nl}(r) = \psi_{nl}(r)e^{-ikr},
\end{equation}
%======================================%
and suppose that the prefactor $\psi_{nl}(r)$ varies very slowly: 
%============< EQUATION >==============%
%
\begin{equation}
 \left|\frac{\partial_{r}\psi_{nl}}{\psi_{nl}}\right| \ll |k|,\qquad
 \left|\frac{\partial_{r}^2\psi_{nl}}{\psi_{nl}}\right| \ll |k|^2.
 	\label{eqn:3-2:WKB-cond1}
\end{equation}
%======================================%
Thence, assuming that
%============< EQUATION >==============%
%
\begin{equation}
 \left|\frac{\partial_{r}(r^2f)}{r^2f}\right| \ll |k|,
 	\label{eqn:3-2:WKB-cond2}
\end{equation}
%======================================%
the field equation (\ref{eqn:3-2:field-eq}) of the mode function is 
reduced to 
%============< EQUATION >==============%
%
\begin{equation}
 k^2 = k^2(l,\omega_{nl}) \equiv
 	\frac{1}{f}\left[\frac{\omega_{nl}^2}{f} 
 	- \frac{l(l+1)}{r^2} - m_{\phi}^2\right]. 
\end{equation}
%======================================%
Here we mention that the slowly varying condition 
(\ref{eqn:3-2:WKB-cond1}) can be derived from the condition 
(\ref{eqn:3-2:WKB-cond2}) and viceversa.
The number of modes with frequency less than $\omega$ is given
approximately by 
%============< EQUATION >==============%
%
\begin{equation}
 \tilde{g}(\omega) = \int \nu({l,\omega}) (2l+1) dl,
\end{equation}
%======================================%
where $\nu(l,\omega)$ is the number of nodes in the mode with 
$(l,\omega)$:
%============< EQUATION >==============%
%
\begin{equation}
 \nu(l,\omega) = \frac{1}{\pi}\int_{r_{1}}^{L} 
 	\sqrt{k^2(l,\omega)}dr.
\end{equation}
%======================================%
Here it is understood that the integration with respect to $r$ and 
$l$ is taken over those values which satisfy $r_{1}\le r\le L$ 
and $k^2(l,\omega)\geq 0$. Thus, when 
%============< EQUATION >==============%
%
\[
 \left|\frac{\partial_{r}(r^2f)}{r^2f}\right| \ll 
 \frac{1}{f\beta_{\infty}}
\]
%======================================%
is satisfied, the free energy is given approximately by
%============< EQUATION >==============%
%
\begin{equation}
 F \simeq \frac{1}{\beta_{\infty}}\int_{0}^{\infty} 
 	\ln\left(1-e^{-\beta_{\infty}\omega}\right)
 	\frac{d\tilde{g}(\omega)}{d\omega} d\omega 
 	= \int_{r_{1}}^{L}\tilde{F}(r) 4\pi r^2dr,
\end{equation}
%======================================%
where the `free energy density' $\tilde{F}(r)$ is defined by
%============< EQUATION >==============%
%
\begin{equation}
 \tilde{F}(r) \equiv 
 	\frac{1}{\beta (r)}\int_{0}^{\infty}
 	\ln\left( 1-e^{-\beta (r) E}\right)
 	\frac{4\pi p^2dp}{(2\pi )^3}.
\end{equation}
%======================================%
Here the `local inverse temperature' $\beta (r)$ is defined by the 
Tolman's law
%============< EQUATION >==============%
%
\begin{equation}
 \beta (r) = f^{1/2}(r)\beta_{\infty},
\end{equation}
%======================================%
and $E$ is defined by $E=\sqrt{p^2+m_{\phi}^2}$. Hence the total energy 
$U$ (equal to $\Delta M_{therm}$ given by (\ref{eqn:3-2:Mtherm})) 
and entropy $S$ are calculated as 
%============< EQUATION >==============%
%
\begin{eqnarray}
 U & \equiv & {\bf Tr}\left[e^{\beta_{\infty}(F-:H:)}:H:\right]
 	= \frac{\partial}{\partial\beta_{\infty}}(\beta_{\infty} F) 
 	= \int_{r_{1}}^{L}\rho(r) 4\pi r^2dr,
 	\label{eqn:3-2:energy}\\
 S & \equiv & -{\bf Tr}\left[ e^{\beta_{\infty}(F-:H:)}
 	\ln e^{\beta_{\infty}(F-:H:)}\right]
 	= \beta_{\infty}^2\frac{\partial}{\partial\beta_{\infty}}F
	\nonumber\\
 & = & \int_{r_{1}}^{L}s(r) 4\pi r^2dr/\sqrt{f(r)},
 	\label{eqn:3-2:entropy}
\end{eqnarray}
%======================================%
where the `density' $\rho (r)$ and the `entropy density' $s(r)$ are 
defined by
%============< EQUATION >==============%
%
\begin{eqnarray}
 \rho(r) & \equiv & \frac{\partial}{\partial\beta(r)}
 	(\beta(r) \tilde{F}(r)) 
 	= \int_{0}^{\infty}\frac{E}{e^{\beta(r)E}-1}
 	\frac{4\pi p^2dp}{(2\pi )^3}, \nonumber\\
 s(r) & \equiv & \beta^2(r)\frac{\partial}{\partial\beta(r)}
 	\tilde{F}(r)
  	= \beta(r)\left(\rho(r)+P(r)\right),\nonumber\\
\end{eqnarray}
%======================================%  	
where the `pressure' $P(r)$ is defined by~\footnote{
To obtain the last expression of $P(r)$ we performed an integration by 
parts. 
}
%============< EQUATION >==============%
%
\begin{equation}
 P(r) \equiv -\tilde{F}(r)
 	= \frac{1}{3}\int_{0}^{\infty}\frac{p^2/E}{e^{\beta(r)E}-1}
 	\frac{4\pi p^2dp}{(2\pi )^3}.
\end{equation}
%======================================%
These expressions are exactly same as expressions (\ref{eqn:3-2:S-P-rho}) 
for the local quantities in the statistical mechanics of gas of 
particles.

Thus, we have shown that the local description of the statistical 
mechanics used in subsection \ref{subsec:BWmodel1} is equivalent to
that of the quantized field in the curved background, which is defined
globally, and whose ground state is the Boulware state.

The stress energy tensor of the minimally coupled scalar field is given 
by 
%============< EQUATION >==============%
%
\begin{equation}
 T_{\mu\nu} = 
 	-\frac{2}{\sqrt{-g}}\frac{\delta I}{\delta g^{\mu\nu}}
 = \partial_{\mu}\phi\partial_{\nu}\phi 
	-\frac{1}{2}g_{\mu\nu}\left(g^{\rho\sigma}
	\partial_{\rho}\phi\partial_{\sigma}\phi
	+m_{\phi}^2\phi^2\right).
\end{equation}
%======================================%
In particular, the $(tt)$-component is 
%============< EQUATION >==============%
%
\begin{equation}
 T^t_t = -\frac{1}{2}\left[\frac{1}{f}(\partial_{t}\phi)^2
	+ f(\partial_{r}\phi)^2 
	+ \frac{1}{r^2}\Omega^{ij}\partial_{i}\phi\partial_{j}\phi
	+ m_{\phi}^2\phi^2\right].
\end{equation}
%======================================%
Hence, the contribution $\Delta M$ of the scalar field to the mass of
the background geometry is equal to the Hamiltonian of the field:
%============< EQUATION >==============%
%
\begin{equation}
 \Delta M \equiv -\int_{r_{1}}^{L}T^t_t 4\pi r^2dr = H,
 	\label{eqn:3-2:DeltaM}
\end{equation}
%======================================%
where $H$ is given by (\ref{eqn:3-2:Hamiltonian}). 
When we consider the statistical mechanics of the hot quantized system, 
contributions of both vacuum polarization and thermal excitations must 
be taken into account. Thus, the contribution to the mass is given by
%============< EQUATION >==============%
%
\begin{equation}
 \langle\Delta M\rangle = 
 	{\bf Tr}\left[e^{\beta_{\infty}(F-:H:)}\Delta M^{(ren)}\right],
\end{equation}
%======================================%
where $\Delta M^{(ren)}$ is an operator defined by the expression 
(\ref{eqn:3-2:DeltaM}) with $T^t_t$ replaced by the renormalized stress 
energy tensor $T_{\qquad t}^{(ren) t}$. From (\ref{eqn:3-2:DeltaM}), it is 
easy to show that 
%============< EQUATION >==============%
%
\begin{equation}
 \Delta M^{(ren)} = :H: + \Delta M_{B},
\end{equation}
%======================================%
where $:H:$ is the normal-ordered Hamiltonian given by (\ref{eqn:3-2::H:}) 
and $\Delta M_{B}$ is the zero-point energy of the Boulware state defined 
by 
%============< EQUATION >==============%
%
\begin{equation}
 \Delta M_{B} = -\int_{r_{1}}^{L}
 	\langle B|T_{\qquad t}^{(ren) t}|B\rangle 4\pi r^2dr.
\end{equation}
%======================================%
Hence, $\langle\Delta M\rangle$ can be decomposed into the
contribution of the thermal excitations and the contribution from the
zero-point energy:
%============< EQUATION >==============%
%
\begin{equation}
 \langle\Delta M\rangle = U + \Delta M_{B},
 	\label{eqn:3-2:dM=dMB+U}
\end{equation}
%======================================%
where $U$ is given by (\ref{eqn:3-2:energy}) and equal to $\Delta M_{therm}$ 
defined in (\ref{eqn:3-2:Mtherm}).

Finally, we have shown that the gravitational mass appearing in the 
Einstein equation is the sum of the energy of the thermal excitation 
and the mass-energy of the Boulware state. Therefore, as shown in 
subsection \ref{subsec:BWmodel1}, the wall contribution to the total
gravitational mass is zero on shell ($T_{\infty}=T_{BH}$) and the
backreaction can be neglected. Here, we mention that the 
corresponding thermal state on shell is called a topped-up Boulware 
state~\cite{PVI1998}, and can be considered as a generalization to
spacetimes not necessarily containing a black hole of the
Hartle-Hawking state~\cite{Hartle&Hawking1976}.

%======================================%
%<<<<<<   SUBSECTION 3-2-4    >>>>>>>>>%
%======================================%
\subsection{Complementarity}
	\label{subsec:complementarity}

Attempts to provide a microscopic explanation of the
Bekenstein-Hawking entropy $S_{BH}$ initially stemmed from two quite
different directions.

Gibbons and Hawking~\cite{Gibbons&Hawking1977} took the view that
$S_{BH}$ is of topological origin, depending crucially on the presence
of a horizon. They showed that $S_{BH}$ emerges as a boundary
contribution to the geometrical part of the Euclidean action. (A
non-extremal horizon is represented by a regular point in the
Euclidean sector, so the presence of a horizon corresponds to the {\it
absence} of an inner boundary in this sector.)

'tHooft~\cite{tHooft1985} sought the origin of $S_{BH}$ in the thermal
entropy of ambient quantum fields raised to the Hawking
temperature. He derived an expression which is indeed proportional to
the area, but with a diverging coefficient which has to be regulated
by interposing a ``brick wall'' just above the gravitational radius
and adjusting its altitude by hand to reproduce $S_{BH}$ with the
correct coefficient.

In addition, the brick wall model appears to have several problematical 
features---large thermal energy densities near the wall, producing a
substantial mass correction from thermal excitations---which have
raised questions about its self-consistency as a model in which
gravitational back-reaction is neglected.

We have shown that such caveats are seen to be unfounded once the
ground state of the model is identified correctly. Since there are no
horizons above the brick wall, the ground state is the Boulware state,
whose negative energy almost exactly neutralizes the positive energy
of the thermal excitations. 'tHooft's model is thus a perfectly
self-consistent description of a configuration which to outside
observers appears as a black hole but does not actually contain
horizons.

It is a fairly widely held opinion
(e.g. \cite{Callan&Wilczek1994,one-loop}) that the entropy contributed by
thermal excitations or entanglement is a one-loop correction to the
zero-loop (or ``classical'') Gibbons-Hawking contribution. The
viewpoint advocated in this section appears (at least superficially)
quite different. We view these two entropy sources---(a) brick wall,
no horizon, strong thermal excitations near the wall, Boulware ground
state; and (b) black hole, horizon, weak (Hartle-Hawking) stress
-energy near the horizon, Hartle-Hawking ground state---as ultimately
equivalent but mutually exclusive (complementary in the sense of Bohr) 
descriptions of what is externally virtually the same physical
situation. The near-vacuum experienced by free-falling observers near
the horizon is eccentrically but defensibly explainable, in terms of
the description (a), as a delicate cancellation between a large
thermal energy and an equally large and negative ground-state
energy---just as the Minkowski vacuum is explainable to a uniformly
accelerated observer as a thermal excitation above his negative-energy 
(Rindler) ground state. (This corresponds to setting $f(r)=r$ in the
$(1+1)$-dimensional example treated in subsection
\ref{subsec:BandHHstate}.)

That the entropy of thermal excitations can single-handedly account for 
$S_{BH}$ without cutoffs or other {\it ad hoc} adjustments can be
shown by a thermodynamical argument~\cite{PVI1998}. One considers the 
reversible quasi-static contraction of a massive thin spherical shell
toward its gravitational radius. The exterior ground state is the
Boulware state, whose stress-energy diverges to large negative
values in the limit. To neutralize the resulting backreaction, the
exterior is filled with thermal radiation to produce a ``topped-up''
Boulware state (TUB) whose temperature equals the acceleration
temperature at the shell's radius. To maintain thermal equilibrium
(and hence applicability of the first law) the shell itself must be
raised to the same temperature. The first law of thermodynamics then
shows that the shell's entropy approaches $S_{BH}$ (in the
non-extremal case) for essentially arbitrary equations of state. Thus, 
the (shell $+$ TUB) configuration passes smoothly to a black hole $+$
Hartle-Hawking state in the limit.

It thus appears that one has two complementary descriptions, (a) and
(b), of physics near an event horizon, corresponding to different Fock 
representations, i.e., different definitions of positive frequency and 
ground state. The Bogoliubov transformation that links these
representations is known~\cite{Israel1976}. However, because of the 
infinite number of field modes, the two ground states are unitarily
inequivalent~\cite{Umezawa}. This signals some kind 
of phase transition (formation of a condensate) in the passage
between description (a), which explains $S_{BH}$ as a thermal effect,
and description (b), which explains it as geometry. 
We know that a condensation actually does occur at this point; it is
more usually called gravitational collapse.

It will be interesting to explore the deeper implications of these
connections.

%%%%%%%%%%%%%%%%%%%%%%%%%%%%%%%%%%%%%%%%
%%%%%%%%%%%% SECTION 3-3 %%%%%%%%%%%%%%%
%%%%%%%%%%%%%%%%%%%%%%%%%%%%%%%%%%%%%%%%
\section{Entanglement entropy and thermodynamics}
	\label{sec:entanglement}

As explained several times,
entanglement entropy is often speculated as a strong candidate for the
origin of the black-hole entropy. To judge whether this speculation is 
true or not, it is effective to investigate the whole structure of 
thermodynamics obtained from the entanglement entropy, rather than
just to examine the apparent structure of the entropy alone or to
compare it with that of the black hole entropy. It is because entropy
acquires a physical significance only when it is related to the energy
and the temperature of a system. From this point of view, we construct
a `entanglement thermodynamics' by introducing an entanglement
energy and compare it with the black-hole thermodynamics.

Our strategy of the 'construction of entanglement thermodynamics' is
as follows. (See chapter \ref{chap:intro} for the 'construction of
black hole thermodynamics' for Schwarzschild black holes.) 
\begin{enumerate}
\item	First we give concepts and definitions of 'entanglement
	energy'. 
	(We follow the usual definition for entanglement entropy.) 
\item	Next we calculate entanglement entropy and entanglement energy
	for tractable models.
\item	Finally, we obtain 'entanglement temperature' $T_{ent}$ by
	assuming the following relation analogous to the first law of
	thermodynamics.
	\begin{equation}
	 \delta E_{ent} = T_{ent}\delta S_{ent},
		\label{eqn:3-3:1st-law-ent}
	\end{equation}
	where $S_{ent}$ and $E_{ent}$ denote entanglement entropy and
	energy, respectively. We call this relation the first law of
	entanglement thermodynamics.
\end{enumerate}

This section is organized as follows. In subsection
\ref{subsec:Sent} we review the concept of the entanglement entropy. In
subsection \ref{subsec:Eent} we propose four definitions of
entanglement energy and present general formulas for calculating the
energy. In subsection \ref{subsec:evaluation} explicit evaluations of 
entanglement entropy and energy are performed for some tractable
models with the help of the formulas prepared in subsection
\ref{subsec:Eent}. In subsection \ref{subsec:comparison} we construct
entanglement thermodynamics and compare it with the black hole
thermodynamics.

%======================================%
%<<<<<<   SUBSECTION 3-4-1    >>>>>>>>>%
%======================================%
\subsection{Entanglement entropy}	
	\label{subsec:Sent}

In this subsection we review the definition and basic properties of
the entanglement entropy.

%<<<<<<   Definition    >>>>>>>>>%
\subsubsection{Definition of entanglement entropy}
\label{subsection:Senta}

Let ${\cal F}$ be a Hilbert space constructed from two Hilbert spaces
${\cal F}_1$ and ${\cal F}$ as 
%============< EQUATION >==============%
%
\begin{equation}
{\cal F} = {\cal F}_1 \bar{\otimes} {\cal F}_2,
\label{eqn:3-3:F=F1*F2}
\end{equation}
%======================================%
where $\bar{\otimes}$ denotes a tensor product followed by a suitable
completion. We call an element $u \in {\cal F}$  {\it prime} if $u$
can be written as $u = v \otimes w$ with $v \in {\cal F}_1$ and 
$w\in {\cal F}_2$. For example, 
$u= v_1 \otimes w_1 + 2 v_1 \otimes w_2 
+ v_2 \otimes w_1 + 2 v_2 \otimes w_2$ 
is prime since $u$ can be represented as 
$u = (v_1 + v_2) \otimes (w_1 + 2w_2) $.  
On the other hand 
$u= v_1 \otimes w_1 + v_2 \otimes w_2 $ 
is not prime if neither $v_1$ and $v_2$ nor $w_1$ and $w_2$ are
linearly dependent. The entanglement entropy 
$S_{ent} : {\cal F} \rightarrow {\bf R}_+ 
=\{ \mbox{non-negative real numbers} \}$ 
defined below can be regarded as a measure of the non-prime nature of
an element of ${\cal F} = {\cal F}_1 \bar{\otimes} {\cal F}_2$. 
 
First of all, from an element $u$ of ${\cal F}$ with unit norm we
construct an operator $\rho$ 
(`density operator') by
%============< EQUATION >==============%
%
\begin{equation}
 \rho v = (u,v)u \quad {}^\forall v\in{\cal F},
\label{eqn:3-3:rho}
\end{equation}
%======================================%
where $(u,v)$ is the inner product which is antilinear with respect to 
$u$. In this context $\rho$ represents a `pure state'.
 
From $\rho$ we define another operator (`reduced density operator') 
$\rho_2$ by
%============< EQUATION >==============%
%
\begin{equation}
 \rho_2 y = \sum_{i,j}f_j(e_i\otimes f_j,\rho e_i\otimes y)
	\quad {}^\forall y\in{\cal F}_2,
\label{eqn:3-3:reduced}
\end{equation}
%======================================%
where $\{ e_i\}$ and $\{ f_j\}$ are orthonormal bases of ${\cal F}_1$
and ${\cal F}_2$ respectively. Note that 
%============< EQUATION >==============%
%
\begin{equation}
 {\rm Tr}_2\left(\rho_2 A\right) =
	{\rm Tr}\left[\rho (  1\otimes A) \right]
\end{equation}
%======================================%
for an arbitrary bounded operator $A$ on ${\cal F}_2$.
  
Finally we define the entanglement entropy with respect to $\rho$ as 
%============< EQUATION >==============%
%
\begin{equation}
  S_{ent} \left[ \rho \right] 
     \equiv   -k_B{\rm Tr}_2 \left[ \rho_2 
                      \ln \rho_2 \right].
\label{eqn:3-3:entropy}
\end{equation}
%======================================%
We can totally exchange the roles played by ${\cal F}_1$ and 
${\cal F}_2$ in 
Eq.(\ref{eqn:3-3:reduced}) and Eq.(\ref{eqn:3-3:entropy}). The entanglement
entropy is so defined as to be invariant under the exchange of 
${\cal F}_1$ and ${\cal F}_2$ when $\rho$ corresponds to a pure state,
i.e., when $\rho$ is given by Eq.(\ref{eqn:3-3:rho}). 
(See {\it Appendix} \ref{app:Sent=Sent'} for the proof of this
property.)

%<<<<<<   example    >>>>>>>>>%
\subsubsection{A simple example}

As a simple example, let us consider spin states for a system
consisting of an electron and a proton. We take 
${\cal F}_1= \{ |\uparrow\rangle_e,|\downarrow\rangle_e\}$ 
for an electron and 
${\cal F}_2= \{ |\uparrow\rangle_p,|\downarrow\rangle_p\}$ 
for a proton, where `$\uparrow$' is for `up', while `$\downarrow$' is
for `down'. Then ${\cal F} = {\cal F}_1 \otimes {\cal F}_2$ is spanned
by 
%============< EQUATION >==============% 
%
\[
 \{ |\uparrow\rangle_e \otimes |\uparrow\rangle_p,
    |\uparrow\rangle_e \otimes |\downarrow\rangle_p,
    |\downarrow\rangle_e \otimes |\uparrow\rangle_p,
    |\downarrow\rangle_e \otimes |\downarrow\rangle_p \}.
\]
%======================================%
Now let us consider a state 
%============< EQUATION >==============%
%
\begin{eqnarray*}
   | \phi \rangle &=& (\alpha |\uparrow\rangle_e 
	+ \beta |\downarrow\rangle_e)
	\otimes (\gamma |\uparrow\rangle_p 
	+ \delta |\downarrow\rangle_p), \\
     &{}& |\alpha|^2 +|\beta|^2 = |\gamma|^2 + |\delta|^2 =1,
\end{eqnarray*}
%======================================%
which is clearly a prime state. 
According to Eq.(\ref{eqn:3-3:reduced}), we then get
%============< EQUATION >==============%
%
\[
 \rho_e = 
       \left(
           \begin{array}{cc}
              |\alpha|^2      &  \alpha \beta^* \\
	      \alpha^* \beta  &  |\beta|^2     
           \end{array}	
      \right).
\]
%======================================%
Here `$e$' is for `electron'.
By a suitable diagonalization of this matrix, it is easy to see that
$S_{ent} =0$. We can exchange the roles between `electron' and
`proton': we then get
%============< EQUATION >==============%
%
\[
 \rho_p = 
       \left(
           \begin{array}{cc}
              |\gamma|^2      &  \gamma\delta^* \\
	      \gamma^*\delta  &  |\delta|^2     
           \end{array}	
      \right),
\]
%======================================%
(`$p$' is for `proton') which again leads to $S_{ent} =0$. 

On the contrary, an $s$-state 
%============< EQUATION >==============%
%
\begin{eqnarray*}
   | \phi' \rangle &=& 
	\alpha |\uparrow\rangle_e \otimes |\downarrow\rangle_p
	+ \beta |\downarrow\rangle_e \otimes |\uparrow\rangle_p,\\
     &{}& |\alpha|^2 +|\beta|^2 =1, \alpha \beta \neq 0 
\end{eqnarray*}
%======================================%
is not a prime state. For this state the reduced density operators are 
given by
%============< EQUATION >==============%
%
\[
 \rho_e = 
       \left(
           \begin{array}{cc}
              |\alpha|^2      &  0  \\
	         0  &  |\beta|^2     
           \end{array}	
      \right), 
 \rho_p = 
       \left(
           \begin{array}{cc}
              |\beta|^2      &  0  \\
	         0  &  |\alpha|^2     
           \end{array}	
      \right).
 \]
%======================================%
Therefore we get 
$S_{ent} = -k_B (|\alpha|^2 \ln |\alpha|^2 
                   + |\beta|^2 \ln |\beta|^2) > 0$.

%<<<<<<<<<<  Formula  >>>>>>>>>>>>%
\subsubsection{Formula of entanglement entropy}

Let us consider a system of coupled harmonic oscillators  
$\left\{ q^A\right\}$ $(A=1,\cdots,n_{tot})$ described by the
Lagrangian, 
%============< EQUATION >==============%
%
\begin{equation}
 L = \frac{a}{2}\delta_{AB}\dot{q}^A\dot{q}^B - 
     \frac{1}{2}V_{AB}q^A q^B.
\label{eqn:3-3:Lagrange}
\end{equation}
%======================================%
Here $\delta_{AB}$ is Kronecker's delta symbol\footnote{
From now on, we choose the units $\hbar =c=1$ and apply
Einstein's summation convention unless otherwise stated.}; 
$V$ is a real-symmetric, positive-definite matrix which does not
depend on $\left\{ q^A\right\}$. We have introduced $a$$(>0)$ as a
fundamental length characterizing the system.\footnote{
Thus $\left\{ q^A\right\}$ are treated as dimension-free quantities in
the present units.}
The corresponding Hamiltonian becomes 
%============< EQUATION >==============%
%
\begin{equation}
 H_{tot} = \frac{1}{2a}\delta^{AB}p_A p_B + 
           \frac{1}{2}V_{AB}q^A q^B,
\label{eqn:3-3:Htot}
\end{equation}
%======================================%
where $p_A=a\delta_{AB}\dot{q}^B$ is the canonical momentum conjugate
to $q^A$. 

Firstly we calculate the wave function 
$\langle\left\{q^A\right\} |0\rangle$ of the ground state
$|0\rangle$. Note that Eq.(\ref{eqn:3-3:Htot}) can be written as
%============< EQUATION >==============%
%
\begin{equation}
 H_{tot} = \frac{1}{2a}\delta^{AB} 
           \left(p_A+iW_{AC}q^C\right)\left(p_B-iW_{BD}q^D\right) +
           \frac{1}{2a}{\rm Tr} W
\end{equation}
%======================================%
by using the commutation relation 
$\left[ q^A,p_B\right] =i\delta^A_B$. Here $W$ is a symmetric matrix
satisfying $(W^2)_{AB} = a V_{AB}$. 
The ambiguity in sign is fixed by
requiring $W$ to be positive definite. Thus, 
%============< EQUATION >==============%
%
\begin{equation}
 W = \sqrt{aV}.
\end{equation}
%======================================%
Now $\langle\left\{q^A\right\} | 0\rangle$ is given as a solution to 
%============< EQUATION >==============%
%
\begin{equation}
 \left(\frac{\partial}{\partial q^A}+W_{AB}q^B\right)
	\langle\left\{q^A\right\} | 0\rangle = 0,
\end{equation}
%======================================%
since $p_A$ is expressed as $-i\frac{\partial}{\partial q^A}$.  The
solution is 
%============< EQUATION >==============%
%
\begin{equation}
 \langle\left\{q^A\right\} | 0\rangle =
      \left(\det\frac{W}{\pi}\right)^{1/4} 
      \exp\left( -\frac{1}{2}W_{AB}q^A q^B \right),
\end{equation}
%======================================%
which is normalized with respect to the standard Lebesgue measure 
$dq^1 \cdots dq^{n_{tot}}$. The density matrix $\rho_0$ 
corresponding to this ground state is represented as 
%============< EQUATION >==============%
%
\begin{eqnarray}
 \langle\left\{q^A\right\}| \rho_0 |\left\{q'^B\right\}\rangle
& = &
      \langle\left\{q^A\right\}|0\rangle	
      \langle 0|\left\{q'^B\right\}\rangle \nonumber	\\
& = & 
      \left( \det \left(\frac{W}{\pi}\right) \right)^{1/2}\nonumber\\
 & & \times \exp\left[ -\frac{1}{2}W_{AB}
       \left(q^A q^B + {q'}^A q'^B\right)\right].
 \label{eqn:3-3:ex-rho0}
\end{eqnarray}
%======================================%

Now we split $\left\{ q^A\right\}$ into two
subsystems, $\left\{ q^a\right\}$ $(a=1,\cdots,n_B)$ and 
$\left\{ q^{\alpha}\right\}$ $(\alpha =n_B+1,\cdots,n_{tot})$. (We
assign the labels `1' and `2' to the former and the latter subsystems, 
respectively.) Then we obtain the reduced density matrix associated
with the subsystem 2 (the subsystem 1),  by taking the  partial trace
of $\rho_0$ w.r.t. the subsystem 1 (the subsystem 2): 
%============< EQUATION >==============%
%
\begin{eqnarray}
 \langle\left\{q^{\alpha}\right\}| \rho_2 
                |\left\{q'^{\beta}\right\}\rangle
& = & 
      \int \prod_{c=1}^{n}dq^{c}
      \langle\left\{q^a,q^{\alpha}\right\}| \rho_0
            | \left\{q^b,q'^{\beta}\right\}\rangle  \nonumber \\
& = & 
      \left(\det\frac{D'}{\pi}\right)^{1/2}
      \exp\left[ -\frac{1}{2}D'_{\alpha\beta}
          \left( q^{\alpha}q^{\beta}+q'^{\alpha}q'^{\beta}\right)
               \right]	\nonumber  \\
& & 
\times\exp\left[ -\frac{1}{4}\left(B^{T}A^{-1}B\right)_{\alpha\beta}
		(q-q')^{\alpha}(q-q')^{\beta}
               \right]		\label{eqn:3-3:formula-rho2}
\end{eqnarray}
%======================================%
and
%============< EQUATION >==============%
%
\begin{eqnarray}
 \langle\left\{q^a\right\}| \rho_1 |\left\{q'^b\right\}\rangle
& = & 
      \int\prod_{\gamma=n+1}^{N}dq^{\gamma}
      \langle\left\{q^a,q^{\alpha}\right\}| \rho_0
             | \left\{q'^b,q^{\beta}\right\}\rangle  \nonumber \\
& = & 
      \left(\det\frac{A'}{\pi}\right)^{1/2}
      \exp\left[ -\frac{1}{2}A'_{ab}
          \left( q^a q^b + q'^a q'^b \right)
               \right]	\nonumber	\\
& & 
\times\exp\left[ -\frac{1}{4}\left(B D^{-1}{B^T}\right)_{ab}
	          (q-q')^a(q-q')^b
               \right],		\label{eqn:3-3:formula-rho1}
\end{eqnarray}
%======================================%
where $A$, $B$, $D$, $A'$ and $D'$ are defined by
%============< EQUATION >==============%
%
\begin{eqnarray}
 \left( W_{AB}\right) & = & \left(
 \begin{array}{cc}
	A_{ab}			& B_{a \beta} \\
	(B^{T})_{\alpha b}	& D_{\alpha \beta}	
 \end{array}	\right)\ \ \ ,		\nonumber \\
 A'     & = & A - BD^{-1}B^{T}\ \ \ ,	\nonumber \\
 D'     & = & D - B^{T}A^{-1}B\ \ \ .	
 \label{eqn:3-3:ABD}  
\end{eqnarray}
%======================================%
(The superscript $T$ denotes transposition.) Note that $A^T=A$
and $D^T=D$. Here we mention that, if we define $\tilde{A}$,
$\tilde{B}$ and $\tilde{D}$ by 
%============< EQUATION >==============%
%
\begin{equation}
 W^{-1}  =  
 \left(
 \begin{array}{cc}
	\tilde{A}^{ab}	& \tilde{B}^{a\beta}	\\
	(\tilde{B}^{T})^{\alpha b} & \tilde{D}^{\alpha\beta}
 \end{array}	
 \right),	\label{eqn:3-3:def-tilde-ABD}
 \end{equation}
%======================================%
then
%============< EQUATION >==============%
%
\begin{eqnarray}
 \tilde{A} & = & {A'}^{-1},\nonumber\\
 \tilde{D} & = & {D'}^{-1},\nonumber\\
 \tilde{B} & = &  - {A'}^{-1} B D^{-1}
	= - {A}^{-1} B {D'}^{-1}.
\end{eqnarray}
%======================================%

Now  the entanglement entropy $S_{ent}:=-\Tr\rho_2\ln\rho_2$ is given
as follows\cite{BKLS1986,Srednicki1993}. 
Let $\{ \Lambda_i \}$ $(i=1,\cdots,N-n_B)$ 
be the eigenvalues of a positive definite symmetric 
matrix~\footnote{
 The corresponding expression in ref.\cite{BKLS1986} 
 (``${\Lambda^a}_b:= (M^{-1})^{ac}N_{cb}$") 
 reads $\Lambda=\tilde{D} B^T A^{-1} B$ 
in the present notation. This definition 
does not give a symmetric matrix and  
should be replaced by 
``$\Lambda^{ab}:= (M^{-1/2})^{ac}N_{cd} (M^{-1/2})^{db}$",
 namely  Eq.(\ref{eqn:3-3:Lambda}).
} 
$\Lambda$,
%============< EQUATION >==============%
%
\begin{equation}
\Lambda:= \tilde{D}^{1/2} B^T A^{-1} B \tilde{D}^{1/2}.
\label{eqn:3-3:Lambda}
\end{equation}
%======================================%
Then it is easily shown that entanglement entropy is given by 
%============< EQUATION >==============%
%
\begin{eqnarray}
    S_{ent} &=& \sum_{i=1}^{N-n_B} S_i , \nonumber \\
   S_i &=& -\frac{\mu_{i}}{1-\mu_{i}}\ln{\mu_{i}} 
        - \ln (1-\mu_{i}),
\label{eqn:3-3:Sl}
\end{eqnarray}
%======================================%
where 
$\mu_i := \Lambda_i^{-1} \left( \sqrt{1+\Lambda_i} -1\right)^2$. 
(Note that $0 < \mu_i < 1$.)

%<<<<<<<<<<  Relevance  >>>>>>>>>>>>%
\subsubsection{Relevance to black hole entropy}	

In the case of black-hole physics, the presence of
the event horizon causes a natural decomposition of a Hilbert space
$\cal F$ of all states of matter fields to a tensor product of the
state spaces inside and outside a black hole as
Eq. (\ref{eqn:3-3:F=F1*F2}). 
For example, let us  take  a scalar field. 
We can suppose that its one-particle Hilbert space $\cal H$ is
decomposed as
%============< EQUATION >==============%
%
\begin{equation}
 {\cal H} = {\cal H}_1 \oplus {\cal H}_2,
	\label{eqn:3-3:H=H1+H2}
\end{equation}
%======================================%
where ${\cal H}_1$ is a space of mode functions with
supports inside the horizon and ${\cal H}_2$ is a space of mode
functions with supports outside the horizon. Then we can
construct new Hilbert spaces (`Fock spaces') ${\cal F}$, ${\cal F}_1$
and ${\cal F}_2$ from ${\cal H}$, ${\cal H}_1$ and ${\cal H}_2$,
respectively, as 
%============< EQUATION >==============%
%
\begin{eqnarray}
 {\cal F} & \equiv &
    \mbox{\boldmath C} \oplus {\cal H} \oplus 
    \left({\cal H}\bar{\otimes} {\cal H}\right)_{sym}
    \oplus \cdots,	
    \nonumber	\\
 {\cal F}_1 & \equiv &
    \mbox{\boldmath C} \oplus {\cal H}_1 \oplus 
    \left({\cal H}_1\bar{\otimes} {\cal H}_1\right)_{sym}
    \oplus \cdots,	
    \nonumber	\\
 {\cal F}_2 & \equiv &  
    \mbox{\boldmath C} \oplus {\cal H}_2 \oplus 
    \left({\cal H}_2 \bar{\otimes} {\cal H}_2 \right)_{sym} 
    \oplus \cdots,	
    \label{eqn:3-3:Fock}
\end{eqnarray}
%======================================%
where $(\cdots)_{sym}$ denotes the symmetrization. Now these three
Hilbert spaces satisfy the relation (\ref{eqn:3-3:F=F1*F2}). Hence the
entanglement entropy $S_{ent}$ is defined by the procedure given at
the beginning of this subsection (Eqs.(\ref{eqn:3-3:rho})-(\ref{eqn:3-3:entropy})) 
for each state in ${\cal F}$.

The entanglement entropy $S_{ent}$ originates from a tensor product
structure of the Hilbert space  as Eq.(\ref{eqn:3-3:F=F1*F2}), which is
caused by the existence of the  boundary between two regions (the
event horizon) through Eq.(\ref{eqn:3-3:H=H1+H2}). Furthermore the
symmetric property of $S_{ent}$ between ${\cal F}_1$ and ${\cal F}_2$
mentioned before also suggests that $S_{ent}$ is
related with a boundary between two regions. In fact $S_{ent}$ turns
out to be proportional to the area of such a boundary (a model for the
event horizon) in simple models discussed below. In view of the
Bekenstein-Hawking formula (\ref{eqn:1-0:BHformula}), thus, the
entanglement entropy has a nature similar to the black hole entropy.

The relevance of the entanglement entropy to the black hole entropy is 
also suggested by the following observation. 
Let us consider a free scalar field on a background geometry
describing a gravitational collapse to a black hole. We compare the
black hole entropy and the entanglement entropy for this system. We
begin with the black hole entropy. In the initial region of the
spacetime, there is no horizon and the entropy around this region can
be regarded as zero. In 
the final region, on the other hand, there is an event  horizon so
that the black-hole possesses non-zero entropy. As for the
entanglement entropy, the existence of the event horizon naturally
divides the Hilbert space ${\cal F}$ of all states of the scalar field
into ${\cal F}_1\bar{\otimes}{\cal F}_2$. Thus, the scalar field in
some pure state possesses non-zero entanglement entropy. In this
manner, we observe that the black-hole entropy and the entanglement
entropy come from the same origin, i.e. the existence of the event
horizon. This is the reason why the entanglement entropy is regarded
as one of the potential candidates for the origin of the black-hole
entropy.

%<<<<<<   Simple models    >>>>>>>>>%
\subsubsection{Simple models in Minkowski spacetime}

The relation between the entanglement entropy and the black hole
entropy was analyzed in terms of simple tractable models by
Bombelli, et. al.~\cite{BKLS1986} and
Srednicki~\cite{Srednicki1993}. They considered a free 
scalar field on a flat spacelike hypersurface 
embedded in a 4-dimensional Minkowski spacetime, and calculated the
entanglement entropy for a division of the hypersurface into two
regions with a common boundary $B$. 
Here the two regions and $B$ are, respectively,  the
models of the interior, the exterior of the black holes and the
horizon. Ref.\cite{BKLS1986} chooses $B$ to be a 2-dimensional flat
surface ${\bf R}^2$, and the matter state to be the ground state,
showing that the resulting entanglement entropy becomes proportional
to the area of $B$. Ref.\cite{Srednicki1993} chooses $B$ to be a
$2$-sphere $S^2$ in ${\rm R}^3$, and chooses the two regions 
to be the interior and the exterior of the sphere. The matter state is
chosen to be the ground state. Then it is shown that the resulting
entanglement entropy is again proportional to the area of $B$.

Both of the results can be expressed as 
%============< EQUATION >==============%
%
\begin{equation}
 S_{ent}[\rho_0] \simeq k_B{\cal N}_S\frac{A}{4\pi a^2},
\label{eqn:3-3:Sent-result}
\end{equation}
%======================================%
where $\rho_0$ is the  ground-state density matrix, $A$ is area
of the boundary, $a$ is a cutoff length, and ${\cal N}_S$ is a
dimensionless numerical constant of order unity. In particular, 
${\cal N}_S=0.30$ for $B=S^2$~\cite{Srednicki1993}.
This coincides with the Bekenstein-Hawking formula if the cut-off
length $a$ is chosen as 
%============< EQUATION >==============%
%
\begin{equation}
 a = \sqrt{\frac{{\cal N}_S\hbar G}{\pi c^3}}
  = \sqrt{\frac{{\cal N}_S}{\pi}}\ l_{pl},\label{eqn:3-3:a=lpl}
\end{equation}
%======================================%
where $l_{pl}$ is the Planck length. Here note that $a$ depends
only on the Planck length.

Note that the cutoff length (\ref{eqn:3-3:a=lpl}) is almost same as
the cutoff length (\ref{eqn:3-2:normalization}) for the brick wall
model.

%======================================%
%<<<<<<   SUBSECTION 3-3-2    >>>>>>>>>%
%======================================%
\subsection{Entanglement energy}	
	\label{subsec:Eent}

In this subsection we define entanglement energy to construct
entanglement thermodynamics. We give four possible
definitions of entanglement energy. The difference between them 
comes from the difference in the way to formulate the
reduction of a system (caused by, for instance, the formation of an
event horizon). 
In the first to third definitions we assume that some operators drop
out from the set of all observables, while the state of a total system
is regarded  as unchanged. 
In the fourth definition, on the contrary, we assume that the state 
undergoes a change in the course of the reduction of the system (so 
that the density matrix of the system changes actually), while
operators are regarded as unchanged. 
Since at present we cannot  judge whether and which one of these
treatments reflects the true process of reduction, the best way is to 
investigate all options. As we shall see in subsection
\ref{subsec:comparison}, the universal behavior of the entanglement
thermodynamics does not depend on the choice of the entanglement
energy.

Let us consider a
system described by a Fock space $\cal F$ constructed from a
one-particle Hilbert space $\cal H$ in the previous section. Let
$H_{tot}$ be a total  Hamiltonian acting  on $\cal F$. We assume that
the Hamiltonian $H_{tot}$ is naturally decomposed as 
%============< EQUATION >==============%
%
\begin{equation}
 H_{tot} = H_1 + H_2 + H_{int} \ \ \ ,		
\label{eqn:3-3:Htot=H1+H2+Hint}
\end{equation}
%======================================%
where $H_1$ and $H_2$ are parts acting on ${\cal F}_1$ and 
${\cal F}_2$, respectively, and $H_{int}$ is a part representing the
interaction of two regions.

%<<<<<<   Definitions a-c    >>>>>>>>>%
\subsubsection{The first to third definitions of the entanglement
energy} 

The first to third definitions of entanglement energy follow
when we regard that the operators connecting the two subsystems drop
out from the set of observables (when, for instance, an event horizon
is formed), while the state of the system is regarded as
unchanged. To be more precise, we assume that  $H_1$ and $H_2$ remain
to be observables but that $H_{int}$ is no longer an observable. In
this case it is natural to define 
the entanglement energy by one of the following three.
%============< EQUATION >==============%
%
\begin{itemize}
 \item[(a)] $E_{end}=\langle :H_1:\rangle\equiv{\rm Tr}[:H_1:\rho]$,
 \item[(b)] $E_{end}=\langle :H_2:\rangle\equiv{\rm Tr}[:H_2:\rho]$,
 \item[(c)] $E_{ent}=\langle :H_1:\rangle+\langle :H_2:\rangle$,
\end{itemize}
%======================================%
where $\rho$ is a density matrix of the total system and the two
normal orderings mean to subtract the minimum eigenvalues of $H_1$ and
$H_2$ respectively.

%<<<<<<   Definition d    >>>>>>>>>%

\subsubsection{The forth definition of the entanglement energy}

Next let us consider the case in which the total density operator
$\rho$ actually changes to the product of reduced density operators of 
each subsystems, $\rho_1$ and $\rho_2$, (when, for instance,  an event
horizon is formed), while the observables remain unchanged.

In this case $\rho$ reduces to $\rho'$ given  by 
%============< EQUATION >==============%
%
\begin{equation}
  \rho' = \rho_1 \otimes \rho_2\ \ \ .	
\label{eqn:3-3:rhoI}
\end{equation}
%======================================%
It is easy to see that the entropy associated with this density matrix
becomes 
%============< EQUATION >==============%
%
\begin{equation}
 -k_B{\rm Tr} \left[\rho'\ln\rho'\right] =
	 S_{ent}[\rho ] + S'_{ent}[\rho ],	
	 \label{eqn:3-3:Sent-tot}
\end{equation}
%======================================%
where $S_{ent}[\rho ]$ and $S'_{ent}[\rho ]$ are entanglement entropy
obtained through $\rho_1$ and $\rho_2$, respectively. $S_{ent}[\rho ]$ 
and $S'_{ent}[\rho ]$ are identical if $\rho$ is a pure state (see
the argument below Eq. (\ref{eqn:3-3:entropy})).

Since we are assuming that the observables do not change, we are led
to the following forth definition of entanglement energy:
%============< EQUATION >==============%
%
\begin{itemize}
 \item[(d)] $E_{ent}=\langle :H_{tot}:\rangle_{\rho'}\equiv 
		{\rm Tr}\left[:H_{tot}:\rho'\right]$,
\end{itemize}
%======================================%
where $:\ -\ :$ denotes the usual normal ordering (a subtraction of
the ground state energy).

%<<<<<<   Formula for the definitions a-c    >>>>>>>>>%
\subsubsection{Formula of $\langle :H_1:\rangle$ for the ground state}

What we should do next is to give formulas of entanglement energy
explicitly by choosing $\rho$ as the ground state $\rho_0$ of
$H_{tot}$.  
In the next subsection we consider a free 
scalar field and discretize it with some spatial separation for
regularization. Since the system thus obtained is 
equivalent to a set of harmonic oscillators, here, we
give a formulas of entanglement energy for the ground state
of coupled harmonic oscillators described by the Hamiltonian
(\ref{eqn:3-3:Htot}), splitting the total system as in the previous
subsection.

First, we derive formulas for the entanglement energies corresponding
to the definitions (a), (b) and (c).

Firstly we divide  the Hamiltonian (\ref{eqn:3-3:Htot}) into three terms
as Eq.(\ref{eqn:3-3:Htot=H1+H2+Hint}):
%============< EQUATION >==============%
%
\begin{eqnarray}
 H_1 
& \equiv & 
 \frac{1}{2a}\delta^{ab}p_a p_b + \frac{1}{2}{V^{(1)}}_{ab}
                 q^a q^b	\nonumber\\
& = & 
 \frac{1}{2a}\delta^{ab} 
	\left(p_a+iw^{(1)}_{ac}q^c\right)
	\left(p_b-iw^{(1)}_{bd}q^d\right) +
           \frac{1}{2a}{\rm Tr} w^{(1)},    \nonumber\\
 H_2 
& \equiv &
 \frac{1}{2a}\delta^{\alpha\beta}p_{\alpha} p_{\beta} + 
	\frac{1}{2}{V^{(2)}}_{\alpha\beta}
	     q^{\alpha} q^{\beta}          \nonumber\\
& = & 
 \frac{1}{2a}\delta^{\alpha\beta} 
	\left(p_{\alpha}+iw^{(2)}_{\alpha\gamma}q^{\gamma}\right)
	\left(p_{\beta}-iw^{(2)}_{\beta\delta}q^{\delta}\right) +
           \frac{1}{2a}{\rm Tr} w^{(2)},     \nonumber\\
 H_{int} 
 & \equiv & H_{tot} - H_1 -H_2     \nonumber\\
 &=& V_{int \  a \beta} q^a q^\beta,
\end{eqnarray}
%======================================%
where $V^{(1)}$, $V^{(2)}$ and $V_{int}$ are blocks in the matrix $V$
given by 
%============< EQUATION >==============%
%
\begin{equation}
 \left( V_{AB}\right)  =  \left(
 \begin{array}{cc}
	V^{(1)}_{ab}	&	\left({V_{int}}\right)_{a\beta}  \\
	({V_{int}}^{T})_{\alpha b} & V^{(2)}_{\alpha\beta}
 \end{array}	\right),
 \label{eqn:3-3:V}
 \end{equation}
%======================================%
and $w^{(1)}$ and $w^{(2)}$ are, respectively, the positive
square-roots of $aV^{(1)}$ and $aV^{(2)}$. Although there exists
freedom in the way of the division, the above division seems to be the 
most natural one. Here and throughout this section we adopt it.

By rescaling the variables $\left\{ q^A\right\}$ as
%============< EQUATION >==============%
%
\[
 \bar{q}^A := \delta^{AB}\left( W^{1/2}\right)_{BC}q^C,
\]
%======================================%
the expression of 
the density matrix for the vacuum state Eq.(\ref{eqn:3-3:ex-rho0}) 
gets simplified as 
%============< EQUATION >==============%
%
\[
 \langle\left\{\bar{q}^A\right\}|\rho_0|\left\{\bar{q}'^B\right\}\rangle 
	= \prod_{C=1}^N\pi^{-1/2}\exp\left[ -\frac{1}{2}\left\{
		(\bar{q}^C)^2 + (\bar{q}'^C)^2\right\}\right],
\]
%======================================%
and the normal ordered Hamiltonian $:H_1:$  is
represented as
%============< EQUATION >==============%
%
\begin{eqnarray*}
 :H_1: & = & -\frac{1}{2a}\delta^{ab}
	\left(\frac{\partial}{\partial q^a}-w_{ac}^{(1)}q^c\right)
	\left(\frac{\partial}{\partial q^b}+w_{bd}^{(1)}q^d\right)
	\nonumber\\
 & = & -\frac{1}{2a}U^{AB}
	\left(\frac{\partial}{\partial\bar{q}^A}
		-\bar{w}^{(1)}_{AC}\bar{q}^C\right)
	\left(\frac{\partial}{\partial\bar{q}^B}
		+\bar{w}^{(1)}_{BD}\bar{q}^D\right).
\end{eqnarray*}
%======================================%
Here the matrices $U$ and $\bar w^{(1)}$ are defined as
%============< EQUATION >==============%
%
\begin{eqnarray*}
 U^{AB} & := & \delta^{AC}\left( W^{1/2}\right)_{Ca}\delta^{ab}
	\left( W^{1/2}\right)_{bD}\delta^{DB},\\
 \bar{w}^{(1)}_{AB} & := & \delta_{AC}\left(W^{-1/2}\right)^{Ca}
	w^{(1)}_{ab}\left(W^{-1/2}\right)^{bD}\delta_{DB}.
\end{eqnarray*}
%======================================%
Hence the matrix elements of $:H_1:\rho$ with respect to
the basis $|{\bar q^A}\rangle$ are expressed as
%============< EQUATION >==============%
%
\begin{eqnarray*}
 & & \langle\left\{\bar{q}^A\right\}| :H_1:\rho_0
		|\left\{\bar{q}'^B\right\}\rangle \nonumber\\
 & & = 	\frac{1}{2a}\left\{\left[
		(\bar{w}^{(1)}+1)U(\bar{w}^{(1)}-1)
		\right]_{AB}\bar{q}^A\bar{q}^B
	+ {\rm Tr}\left[ U(1-\bar{w}^{(1)})\right]\right\}
			\nonumber\\
 & & \times\prod_{C=1}^N\pi^{-1/2}\exp\left[ -(\bar{q}^C)^2\right].
\end{eqnarray*}
%======================================%
From this we obtain~\footnote{
See Eqs.(\ref{eqn:3-3:ABD}) and (\ref{eqn:3-3:def-tilde-ABD}) 
for the definitions of 
the matrices $A$, $\tilde{A}$, $D$ and $\tilde{D}$.}  
%============< EQUATION >==============%
%
\begin{eqnarray}
 \langle :H_1:\rangle & = & 
	\int\left(\prod_{C=1}^N d\bar{q}^C\right)
	\langle\left\{\bar{q}^A\right\}| :H_1:\rho_0
		|\left\{\bar{q}'^B\right\}\rangle \nonumber\\
 & = & \frac{1}{4a}\left[ aV^{(1)}_{ab}(\tilde{A})^{ab}
	+ A_{ab}\delta^{ab} -2w^{(1)}_{ab}\delta^{ab}\right].
	\label{eqn:3-3:E-I-formula}
\end{eqnarray}
%======================================%
Similarly $\langle :H_2:\rangle$ is expressed as
%============< EQUATION >==============%
%
\begin{equation}
 \langle :H_2:\rangle = \frac{1}{4a}\left[ 
	aV^{(2)}_{\alpha\beta}(\tilde{D})^{\alpha\beta}
	+ D_{\alpha\beta}\delta^{\alpha\beta} 
	-2w^{(2)}_{\alpha\beta}\delta^{\alpha\beta}\right],
	\label{eqn:3-3:E-I'-formula}
\end{equation}
%======================================%
where $w^{(2)}$ is the positive square-root of $aV^{(2)}$.

%<<<<<<   Formula for the definition d    >>>>>>>>>%
\subsubsection{Formula of $\langle :H_{tot}:\rangle_{\rho'}$ for the
ground state} 
\label{subsection:Eentb}

By using the formulas (\ref{eqn:3-3:formula-rho1}) and
(\ref{eqn:3-3:formula-rho2}), $\rho'$ defined by 
Eq.(\ref{eqn:3-3:rhoI}) is represented as 
%============< EQUATION >==============%
%
\begin{eqnarray}
 \langle\left\{q^A\right\}| \rho' |\left\{q'^B\right\}\rangle
& = &
      \left(\det\frac{M}{\pi}\right)^{1/2}
      \exp\left[ -\frac{1}{2}M_{AB}
          \left( q^A q^B + q'^A q'^B \right)
               \right]	\nonumber	\\
& & 
\times\exp\left[ -\frac{1}{4}N_{AB}(q-q')^A(q-q')^B\right],
\end{eqnarray}
%======================================%
where 
%============< EQUATION >==============%
%
\begin{eqnarray}
 \left( M_{AB}\right) & = & \left(
 \begin{array}{cc}
	A'_{ab}	& 0			\\
	0	& D'_{\alpha\beta}
 \end{array}	\right),	\nonumber\\
 \left( N_{AB}\right) & = & \left(
 \begin{array}{cc}
	\left( BD^{-1}B^{T}\right)_{ab} & 0	\\
	0 & \left(B^{T}A^{-1}B\right)_{\alpha\beta}
 \end{array}	\right).
\end{eqnarray}
%======================================%

We can diagonalize $M$ and $N$ simultaneously by the following
non-orthogonal transformation: 
%============< EQUATION >==============%
%
\begin{equation}
 q^A \to \tilde{q}^A \equiv 
	\left( \tilde{U}M^{1/2}\right)^A_{\ B} q^B,
\label{eqn:3-3:change}
\end{equation}
%======================================%
where $\tilde{U}$ is a real orthogonal matrix  satisfying
%============< EQUATION >==============%
%
\begin{eqnarray}
 M^{-1/2} N M^{-1/2} 
 &=& 
 \tilde{U}^{T}\lambda \tilde{U}, 	\nonumber\\
 \lambda
& = & 
 \left( \begin{array}{cccccc}
	\lambda_1 & & \\
	& \lambda_2 & \\
	& & \ddots
               \end{array}\right).
\end{eqnarray}
%======================================%
Now in terms of  $\left\{\tilde{q}^A\right\}$,  
   $H_{tot}$ is  represented as 
%============< EQUATION >==============%
%
\begin{equation}
 H_{tot} =  
 -\frac{1}{2a}\left( \tilde{U}M\tilde{U}^{T}\right)^{AB}
	\left( \frac{\partial}{\partial \tilde{q}^A} - 
		\tilde{W}_{AC}\tilde{q}^C\right)
	\left( \frac{\partial}{\partial \tilde{q}^B} + 
		\tilde{W}_{BD}\tilde{q}^D\right)
        + \frac{1}{2a}{\rm Tr} W, 
\end{equation}
thus, 
\begin{equation}
 :H_{tot}:= 
 -\frac{1}{2a}\left( \tilde{U}M\tilde{U}^{T}\right)^{AB}
	\left( \frac{\partial}{\partial \tilde{q}^A} - 
		\tilde{W}_{AC}\tilde{q}^C\right)
	\left( \frac{\partial}{\partial \tilde{q}^B} + 
		\tilde{W}_{BD}\tilde{q}^D\right),
\end{equation}
%======================================%
where 
%============< EQUATION >==============%
%
\begin{equation}
 \tilde{W} \equiv \tilde{U}M^{-1/2}WM^{-1/2}\tilde{U}^{T}.
\end{equation}
%======================================%
Hence the density matrix $\rho'$ is expressed in terms of 
$|\left\{ \tilde{q}^A \right\} \rangle $ as\footnote{
Einstein's summation convention is not applied to
Eq.(\ref{eqn:3-3:ex-rhoI}). }
%============< EQUATION >==============%
%
\begin{equation}
 \langle\left\{\tilde{q}^A\right\}| \rho'
	|\left\{\tilde{q}'^B\right\}\rangle =  
 \prod_{C=1}^N \pi^{-1/2}
   \exp\left[ -\frac{1}{2}\left\{(\tilde{q}^C)^2 + (\tilde{q}'^C)^2 
				\right\}
                 -\frac{1}{4}
                  \lambda_C (\tilde{q}^C-\tilde{q}'^C)^2
          \right].	
\label{eqn:3-3:ex-rhoI}
\end{equation}
%======================================%
This density matrix is normalized with respect to the measure 
$d\tilde{q}^1\cdots d\tilde{q}^N$. 

Now it is easy to calculate the entanglement energy. First the matrix
components of $:H_{tot}:\rho'$ with respect to $\{\tilde{q}^A\}$ are
given by
%============< EQUATION >==============%
%
\begin{eqnarray}
 & & \langle\left\{\tilde{q}^A\right\} | :H_{tot}:\rho'|
	\left\{\tilde{q}^B\right\}\rangle	\nonumber\\
 & & =
 -\frac{1}{2a}
 \Big\{
	\left[ \tilde{U}M\tilde{U}^{T}
	-l\tilde{U}M^{-1/2}VM^{-1/2}\tilde{U}^{T}\right]_{AB}
		\tilde{q}^A \tilde{q}^B 	\nonumber\\
 & & +{\rm Tr}\left[ W- N/2 - M \right]
 \Big\}
 \prod_{C=1}^{N}\pi^{-1/2}
 \exp\left[ -(\tilde{q}^C)^2\right].
\end{eqnarray}
%======================================%
Hence the entanglement energy
$\langle :H_{tot}:\rangle_{\rho'}$ is expressed as 
%============< EQUATION >==============%
%
\begin{eqnarray}
 \langle :H_{tot}:\rangle_{\rho'} & = &
 \int(\prod_{C=1}^{N}d\tilde{q}^C)
   \langle \left\{ \tilde{q}^C  \right\} | :H_{tot}:\rho' | 
	\left\{ \tilde{q}^B \right\} \rangle \nonumber \\
& = & 
 \frac{1}{4a}{\rm Tr}\left[ aVM^{-1}+M+N-2W\right].
\end{eqnarray}
%======================================%
Here we have used the formula
$\int d{\vec x} \ {\vec x}\cdot{\cal A} {\vec x}\ 
           \exp[-{\vec x}\cdot {\vec x}] 
  = {1\over 2} \pi^{N/2} {\rm Tr} {\cal A}$, 
where $N$ is the dimension of ${\vec x}$.
With the help of the identity 
$ {\rm Tr}\left[ M+N\right] = {\rm Tr} A + {\rm Tr} D 
= {\rm Tr} W $, 
we finally arrive at the following formula for 
$\langle :H_{tot}:\rangle_{\rho'}$
%============< EQUATION >==============%
%
\begin{eqnarray}
 \langle :H_{tot}:\rangle_{\rho'} & = & 
 \frac{1}{4a}{\rm Tr}\left[ a V M^{-1} -W \right]   
 \label{eqn:3-3:Eent1'} \nonumber \\
	& = &
  \frac{1}{4}{\rm Tr}\left[ V(M^{-1}-W^{-1})\right]	\nonumber \\
	& = & 
 -\frac{1}{2}{\rm Tr}\left[ V_{int}^{T}\tilde{B}\right].	
\label{eqn:3-3:ex-Eent1}
\end{eqnarray}
%======================================%

%<<<<<<   Formula for the definition c    >>>>>>>>>%
\subsubsection{Alternative formula of 
$\langle :H_1:\rangle +\langle :H_2:\rangle$ for the ground state}
\label{subsection:Eentd}

We can calculate $\langle :H_1:\rangle +\langle :H_2:\rangle$ for the
ground state as a sum of Eqs.~(\ref{eqn:3-3:E-I-formula}) and
(\ref{eqn:3-3:E-I'-formula}). Nonetheless, for a check of numerical
calculations, it is useful to give an alternative formula of 
$\langle :H_1:\rangle +\langle :H_2:\rangle$. 

In terms of  $\left\{\bar{q}^A\right\}$, the operator $:H_1:+:H_2:$ is
written as  
%============< EQUATION >==============%
%
\begin{equation}
 :H_1: + :H_2:\  
 =  -\frac{1}{2a} \delta ^{AC} \delta ^{BD} W_{CD}
 \left(
   \frac{\partial}{\partial\bar{q}^A}-\bar{w}_{AE}\bar{q}^E
   \right)
 \left(
 \frac{\partial}{\partial\bar{q}^B}+\bar{w}_{BF}\bar{q}^F
 \right) \ \ \ ,
\end{equation}
%======================================%
where $\bar{w}$ is defined by
%============< EQUATION >==============%
%
\begin{eqnarray}
 \bar{w} & \equiv &  
       \delta_{AC} 
               \left( W^{-1/2} w W^{-1/2} \right)^{CD} 
                        \delta_{DB} \ \ \ ,    \nonumber\\
 \left( w_{AB}\right) & \equiv &  \left(
 \begin{array}{cc}
	{w^{(1)}}_{ab}		&	0	\\
	0		& 	{w^{(2)}}_{\alpha \beta}
 \end{array}	\right) \ \ \ .
\end{eqnarray}
%======================================%

From the expression we obtain
%============< EQUATION >==============%
%
\begin{eqnarray}
 & &\langle\left\{\bar{q}^A\right\}| (:H_1:+:H_2:)\rho_0
		|\left\{\bar{q}^B\right\}\rangle\nonumber\\
 & & =
 \frac{1}{2a}\left\{
	\left[(\bar{w}+1)W(\bar{w}-1)\right]_{AB}\bar{q}^A\bar{q}^B
	- {\rm Tr}\left[ W(\bar{w}-1)\right]\right\}	\nonumber\\
& &  \times \prod_{C=1}^{N}\pi^{-1/2}\exp\left[ 
	-(\bar{q}^C)^2\right].
\end{eqnarray}
%======================================%

Hence we arrive at the following expression 
$\langle :H_1:\rangle +\langle :H_2:\rangle$ for $\rho_0$. 
%============< EQUATION >==============%
%
\begin{eqnarray}
 \langle :H_1:\rangle +\langle :H_2:\rangle & = & 
 \int (\prod_{C=1}^N d\bar{q}^C)\langle\left\{\bar{q}^A\right\} |
	(:H_1:+:H_2:)\rho_0|\left\{\bar{q}^B\right\}\rangle\nonumber\\
& = & 
 \frac{1}{4a}{\rm Tr}\left[ w^2W^{-1}-W\right]
	- \frac{1}{2a}{\rm Tr}\left[ w-W\right].
\end{eqnarray}
%======================================%
With the help of the relation 
${\rm Tr} [ w^2W^{-1}] = {\rm Tr}[ aVM^{-1}]$ 
which follows from the definitions of $w$ and $M$, this formula is
simplified as
%============< EQUATION >==============%
%
\begin{eqnarray}
 \langle :H_1:\rangle +\langle :H_2:\rangle & = & 
 \frac{1}{4a}{\rm Tr}\left[ aVM^{-1} -W \right]
 -\frac{1}{2a}{\rm Tr}\left[ w-W\right] \nonumber	\\
& = & 
 \langle :H_{tot}:\rangle_{\rho'} 
	-\frac{1}{2a}{\rm Tr}\left[ w-W\right],
\label{eqn:3-3:ex-Eent2}
\end{eqnarray}
%======================================%
where Eq.(\ref{eqn:3-3:Eent1'}) has been used to obtain the last line.

%======================================%
%<<<<<<   SUBSECTION 3-3-3    >>>>>>>>>%
%======================================%
\subsection{Explicit evaluation of the entanglement entropy and energy
for a tractable model in some stationary spacetimes}
\label{subsec:evaluation}

With the help of the formulas derived in the previous two subsections, 
we now calculate entanglement entropy and energy explicitly to
construct entanglement thermodynamics for a tractable model in
Minkowski, Schwarzschild and Reissner-Nordstr{\" o}m spacetimes.

\subsubsection{Model description}

The basic idea of entanglement thermodynamics is to express
the thermodynamic quantities for a black hole in terms of 
expectation values of quantum operators dependent on the
spacetime division as in the statistical mechanics modeling
of the thermodynamics for ordinary systems. Therefore we must
specify how to divide spacetime into two regions and with respect
to what kind of state the expectation values are taken.

According to the original idea of entanglement, it is clearly most
natural to consider a dynamical spacetime describing black hole
formation from a nearly flat spacetime in the past infinity, and
divide the spacetime into the regions inside and outside the
horizon. In this situation, if we start from the asymptotic Minkowski
vacuum in the past, the entanglement entropy associated with the
division of spacetime by the horizon acquires a clear physical meaning.
However, this ideal modeling seems difficult.

Hence, we are obliged to consider a stationary spacetime, and take as
the quantum state a stationary one. 
To be specific, we consider the vacuum with respect to a Killing
time in a spherically symmetric static black hole described by a
metric of the form
%============< EQUATION >==============%
%
\begin{equation}
 ds^2 = -N(\rho)^2 dt^2 + d\rho^2 
 	+r(\rho)^2\left( d\theta + \sin^2{\theta}d\psi^2\right).
	\label{eqn:3-3:metric}
\end{equation}
%======================================%

The vacuum is well known as the Boulware state. 
As stated in subsections \ref{subsec:BandHHstate} and
\ref{subsec:BWmodel1}, the Boulware state has negative energy density
and its contribution to gravitational mass diverges for a stationary
black hole background. 
It has been shown in section \ref{sec:brick_wall} that the ground
state of the brick wall model is the Boulware state and that the
negative divergence in gravitational mass due to the Boulware state is
canceled by a positive divergence due to thermal excitations above the
Boulware state. Hence, we can safely state that the brick wall model
is (the Boulware state) $+$ (thermal excitations). Thermal characters 
of a black hole system can be explained by the latter: entropy and
temperature of a black hole are modeled by entropy and temperature of
the thermal excitations.

Our present purpose in this section is to explain the thermal
characters of the black hole system in terms of entanglement. From the
above arguments on the brick wall model, we expect that entanglement
in the Boulware state does well for this purpose: entanglement entropy 
and entanglement temperature might explain entropy and temperature of
a black hole; the negative divergence in gravitational mass due to the 
Boulware state might be canceled by entanglement energy, provided that
the sum of the vacuum energy and entanglement energy contributes to 
gravitational mass.

\begin{figure}
 \begin{center}
  \epsfile{file=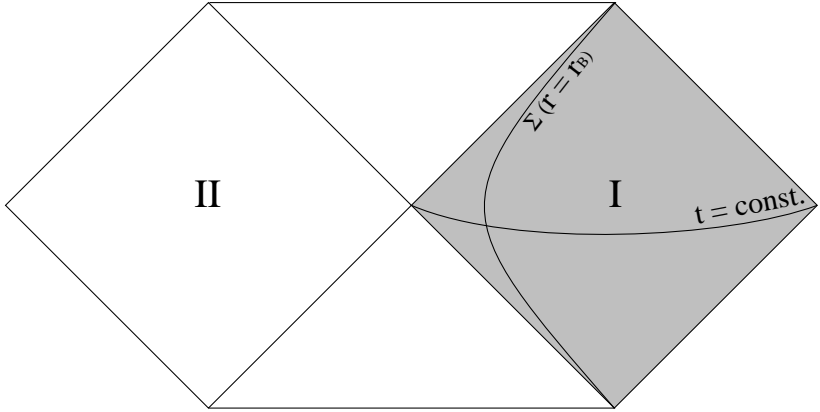,scale=0.4}
 \end{center}
\caption{
The Kruskal extension of the Schwarzschild spacetime. We
consider only the region $I$ (the shaded region). As the boundary
$\Sigma$ we take the hypersurface $r=r_B$. 
}
\label{fig:Kruskal}
\end{figure}

Here a subtlety occurs: 
in, for example, a Schwarzschild spacetime, if we require that the
quantum field is in the Boulware state on the whole extended Kruskal
spacetime and take the bifurcation surface as the boundary surface,
then the entanglement entropy vanishes because the state is expressed
as the tensor product of the Boulware states in the regions I and II 
of {\it Figure} \ref{fig:Kruskal}.  

In order to avoid this, we restrict the spacetime into the region I,
and replace the boundary by a timelike surface $\Sigma$ at a proper
distance of the order of the cut-off length of the theory to the
horizon (see {\it Figure} \ref{fig:Kruskal}). 
This prescription corresponds to taking fluctuations of the horizon
into consideration: 
quantum fluctuations of geometry near a horizon will prevent events
closer to the horizon than about the Planck length from being seen on 
the outside.
However, we should keep in mind that we have no definite criterion
regarding the exact position of the boundary. 
To minimize this ambiguity, we will also investigate the influence of
the variation of the boundary position.

As a matter content we consider a real scalar field described by
%============< EQUATION >==============%
%
\begin{equation}
  I = -\frac{1}{2}\int \sqrt{-g} dx^4
  [\partial^{\mu}\phi \partial_{\mu}\phi + m_{\phi}^2].
	\label{eqn:3-3:action_m}
\end{equation}
%======================================%
The mass $m_{\phi}$ does not play an essential role since a typical length
scale controlling the entanglement thermodynamics is much smaller than
the Compton length of an usual field.  Therefore we just set
$m_{\phi}=0$ in the following arguments.

%============< FIGURE >==============%
%              Kruskal.ps
%======================================%

%<<<<<<   Discritization    >>>>>>>>>%
\subsubsection{Discretized theory of a scalar field}

Consider a massless real scalar field described by the action 
%============< EQUATION >==============%
%
\begin{equation}
 I = -\frac{1}{2}\int dx^4\sqrt{-g}
	\partial^{\mu}\phi \partial_{\mu}\phi,
	\label{eqn:3-3:action}
\end{equation}
%======================================%
where the background geometry is fixed to be a spherically symmetric 
static spacetime with the metric (\ref{eqn:3-3:metric}).
For this system we calculate entanglement entropy and 
entanglement energy to construct entanglement thermodynamics 
using the methods developed in the previous two subsections.
Those methods are both based on a discrete system $\{ q^A\}$
$(A=1,2,\cdots,n_{tot})$ described by a Hamiltonian of the form 
(\ref{eqn:3-3:Htot}). For this discrete system it is easy to
divide the whole Hilbert space ${\cal F}$ into the form
(\ref{eqn:3-3:F=F1*F2}): 
${\cal F}$ is defined as a Fock space constructed from $\{ q^A\}$
$(A=1,2,\cdots,n_{tot})$; ${\cal F}_1$ is defined as a Fock space
constructed from $\{ q^a\}$ $(a=1,2,\cdots,n_B)$; ${\cal F}_2$ is defined as
a Fock space constructed from $\{ q^{\alpha}\}$ 
$(\alpha=n_B+1,\cdots,n_{tot})$. In order to apply this scheme to our 
problem we have to construct a discretized theory of the scalar
field whose Hamiltonian is of the form (\ref{eqn:3-3:Htot}).

First we expand the field $\phi$ in terms of the spherical harmonics as 
%============< EQUATION >==============%
%
\begin{equation}
 \phi (\rho,\theta,\psi) =
 	\sum_{l,m}\frac{N^{1/2}}{r}\phi_{lm}(\rho)Z_{lm}(\theta,\psi ),
\end{equation}
%======================================%
where $Z_{lm}=\sqrt{2}\Re Y_{lm}$ for $m>0$, $\sqrt{2}\Im Y_{lm}$ for 
$m<0$, and $Z_{l0}=Y_{l0}$. Then the Hamiltonian corresponding to the 
Killing time for the action (\ref{eqn:3-3:action}) is decomposed into a 
direct sum of contributions from each harmonics component $H_{lm}$ as
%============< EQUATION >==============%
%
\begin{equation}
 H = \sum_{lm}H_{lm}\ \ .\label{eqn:3-3:Hamiltonian2}
\end{equation}
%======================================%
Here $H_{lm}$ is given by
%============< EQUATION >==============%
%
\begin{equation}
	H_{lm} = \frac{1}{2}\int d\rho\left[
		\pi_{lm}^2 + Nr^2
		\left\{\frac{\partial}{\partial\rho}\left(
		\frac{N^{1/2}}{r}\phi_{lm}\right)\right\}^2
		+ l(l+1)\left(\frac{N\phi_{lm}}{r}\right)^2\right]
		\ \ ,\label{eqn:3-3:subHamiltonian}
\end{equation}
%======================================%
where $\pi_{lm}(\rho )$ is a momentum conjugate to 
$\phi_{lm}(\rho )$.

Note that for any $(l,m)$ the Hamiltonian
(\ref{eqn:3-3:subHamiltonian}) of the subsystem has the form 
%============< EQUATION >==============%
%
\begin{equation}
	H_{lm} = \frac{1}{2}\int d\rho \pi_{lm}^2(\rho)
		+ \frac{1}{2}\int d\rho d\rho'
		\phi_{lm}(\rho)V^{(l,m)}(\rho,\rho')\phi_{lm}(\rho'),
		\label{eqn:3-3:form-of-Hamiltonian}
\end{equation}
%======================================%
where the following algebra of Poisson brackets is understood:
%============< EQUATION >==============%
%
\begin{eqnarray}
 \{\phi_{lm}(\rho),\pi_{l'm'}(\rho')\} & = & \delta(\rho -\rho'), 
	\nonumber\\
 \{\phi_{lm}(\rho),\phi_{l'm'}(\rho')\} & = & 0, \nonumber\\
 \{\pi_{lm}(\rho),\pi_{l'm'}(\rho')\} & = & 0.
\end{eqnarray}
%======================================%
Each subsystem described by the Hamiltonian
(\ref{eqn:3-3:form-of-Hamiltonian}) can be discretized by the
following procedure: 
%============< EQUATION >==============%
%
\begin{eqnarray}
	\rho & \to & (A-1/2)a\ \ ,\nonumber\\
	\delta(\rho -\rho') & \to & \delta_{AB}/a\ \ ,
	\label{eqn:3-3:discretization}
\end{eqnarray}
%======================================%
where $A,B=1,2,\cdots$ and $a$ is a cut-off length. The corresponding 
Hamiltonian of the discretized system is of the form
(\ref{eqn:3-3:Htot}) with 
%============< EQUATION >==============%
%
\begin{eqnarray}
	\phi_{lm}(\rho) & \to & q^A\ \ ,\nonumber\\
	\pi_{lm}(\rho) & \to & p_{A}/a\ \ ,\nonumber\\
	V^{(l,m)}(\rho,\rho') & \to & V_{AB}/a^2\ \ .
\end{eqnarray}
%======================================%
In this way we obtain a discretized system with the total Hamiltonian 
(\ref{eqn:3-3:Htot}) with the matrix $V$ given by the direct sum 
%============< EQUATION >==============%
%
\begin{equation}
	V = \mathop{\oplus}_{l,m}V^{(l,m)},
\end{equation}
%======================================%
where $V^{(l,m)}$ is independent of $m$ and is explicitly expressed as 
%============< EQUATION >==============%
%
\begin{eqnarray}
 V^{(l,m)}_{AB}\phi_{lm}^A \phi_{lm}^B & = & 
	a\sum_{A=1}^{\infty}\left[
	N_{A+1/2}\left(\frac{x_{A+1/2}}{a}\right)^2\left(
	\frac{N_{A+1}^{1/2}}{x_{A+1}}\phi_{lm}^{A+1} -
	\frac{N_{A}^{1/2}}{x_{A}}\phi_{lm}^{A}\right)^2
	\right.	\nonumber\\
	& & \left.
	+ \frac{l(l+1)}{r_0^2}
	\left(\frac{N_{A}\phi_{lm}^{A}}{x_{A}}
	\right)^2		\right]\ \ .
\end{eqnarray}
%======================================%
Here
%============< EQUATION >==============%
%
\begin{eqnarray}
	x_{A} & = & r(\rho =(A-1/2)a)/r_0\ \ ,\nonumber\\
	x_{A+1/2} & = & r(\rho =Aa)/r_0\ \ ,\nonumber\\
	N_{A} & = & N(\rho =(A-1/2)a)\ \ ,\nonumber\\
	N_{A+1/2} & = & N(\rho =Aa)\ \ ,\nonumber\\
	\phi_{lm}^{A} & = & \phi_{lm}(\rho =(A-1/2)a)\ \ .
\end{eqnarray}
%======================================%
In the matrix representation $V^{(l,m)}$ is given by the 
$n_{tot}\times n_{tot}$ matrix
%============< EQUATION >==============%
%
\begin{eqnarray}
 \left( V^{(l,m)}_{AB}\right) & = & \frac{2a}{r_{0}^2}\left( 
               \begin{array}{cccccc}
                   \Sigma^{(l)}_1 & \Delta_1 & & & & \\
         \Delta_1 & \Sigma^{(l)}_2 & \Delta_2 & & & \\
       & \ddots & \ddots & \ddots & & \\
       & & \Delta_{A-1} & \Sigma^{(l)}_A & \Delta_A & \\
       & & & \ddots & \ddots & \ddots 
               \end{array}
             \right) \ \ \ , \nonumber\\
   \Sigma^{(l)}_A
          & = & \frac{1}{2}(r_{0}/a)^2N_{A}x_{A}^{-2}\left[
          	N_{A-1/2}x_{A-1/2}^2 + N_{A+1/2}x_{A+1/2}^2\right]
          	\nonumber\\
          	& & + \frac{1}{2}l(l+1)N_A^2x_A^{-2}\ \ ,\nonumber\\
   \Delta_A
          & = & -\frac{1}{2}(r_{0}/a)^2 N_{A}^{1/2}N_{A+1/2}
		N_{A+1}^{1/2}x_{A}^{-1}x_{A+1/2}^2 x_{A+1}^{-1}\ \ ,
		\label{eqn:3-3:Vlm}
\end{eqnarray}
%======================================%
where we have imposed the boundary condition 
$\phi_{lm}^{n_{tot}+1}=0$. 
In these expressions $r_0$ is an arbitrary constant, which we set to be 
area radius of a horizon~\footnote{
For the model in Minkowski spacetime we take $r_0=a$.
} 
for convenience in the following arguments.

%<<<<<<   Splitting    >>>>>>>>>%
\subsubsection{Spatial division}

We divide the total system by a stationary hypersurface $\Sigma$
defined by $r=r_B$: 
we split the system $\{\phi_{lm}^A\}$ $(A=1,\cdots,n_{tot})$ into the
two subsystems, $\{\phi_{lm}^a\}$ $(a=1,\cdots,n_{B})$ and 
$\{\phi_{lm}^{\alpha}\}$ $(\alpha =n_{B}+1,\cdots,n_{tot})$; 
the area radius $r_B$ of the boundary is given by 
$r_B=r(\rho =n_Ba)$.

Since this division preserves spherical symmetry
of the system, we can still apply the expansion by harmonics. 
Thus, entanglement quantities of the system are calculated as sums of
all contributions from those subsystems, each of which is specified by 
$(l,m)$ and described by the matrix $V^{(l,m)}$.

%<<<<<<   Convergence    >>>>>>>>>%
\subsubsection{Convergence of the summation}

Since $V^{(l,m)}$ is independent of $m=(-l,-l+1,\cdots,l-1,l)$, 
the entanglement entropy and energy are given by 
%============< EQUATION >==============%
%
\begin{eqnarray}
 S_{ent} = \sum_{l=0}^{\infty}(2l+1)S_{ent}^{(l)},\nonumber\\
 E_{ent} = \sum_{l=0}^{\infty}(2l+1)E_{ent}^{(l)},
        \label{eqn:3-3:sum-l}
\end{eqnarray}
%======================================%
where $S_{ent}^{(l)}$ and $E_{ent}^{(l)}$ are entanglement entropy and 
energy of the subsystem specified by $(l,m)$ and independent of $m$.

From Eq.(\ref{eqn:3-3:Vlm}), one can  easily show that 
%============< EQUATION >==============%
%
\begin{eqnarray}
 S_{ent}^{(l)} & \sim &O\left( (la/r_0)^{-4}\ln (la/r_0)\right)\ \ 
    {\rm as}\ \ la/r_0 \to \infty,\nonumber\\
 E_{ent}^{(l)} & \sim &O\left( (la/r_0)^{-3}\right)\ \ {\rm as}\ \ 
      la/r_0 \to \infty.
\end{eqnarray}
%======================================%
Thus the infinite sums Eq.(\ref{eqn:3-3:sum-l}) actually converge 
so that we can safely truncate them 
at some appropriate $l$, depending on the accuracy we require
and the ratio $r_0/a$ we set.

%<<<<<<   Minkowski calculation    >>>>>>>>>%
\subsubsection{Numerical calculations in Minkowski spacetime}

Before analyzing our model in black hole geometries, it is instructive 
to calculate entanglement quantities in the simple models in Minkowski 
spacetime introduced by Bombelli et.al.~\cite{BKLS1986} and
Srednicki~\cite{Srednicki1993}. They calculated entanglement entropy
for divisions of a flat hypersurface in Minkowski spacetime by ${\bf
R}^2$ and $S^2$, respectively. Their results are summarized as
Eq.~(\ref{eqn:3-3:Sent-result}).

Here, let us calculate entanglement energy in the model of Srednicki
(a division by $S^2$)~\footnote{
Entanglement energy in the model of Bombelli et.al (a division by
${\bf R}^2$) is analyzed in Appendix~\ref{app:half}.
}.
The discretized theory of a scalar field corresponding to this model
can be easily obtained by the above procedure by setting $r_0=a$, 
$r(\rho)=\rho$ and $N(\rho)=1$. 
The discretized system $\{\phi_{lm}^A\}$ $(A=1,\cdots,n_{tot})$ is
described by Hamiltonian of the form (\ref{eqn:3-3:Htot}) with the
potential given by (\ref{eqn:3-3:Vlm}).

\begin{figure}
 \begin{center}
	% GNUPLOT: LaTeX picture
\setlength{\unitlength}{0.240900pt}
\ifx\plotpoint\undefined\newsavebox{\plotpoint}\fi
\sbox{\plotpoint}{\rule[-0.200pt]{0.400pt}{0.400pt}}%
\begin{picture}(1200,900)(0,0)
\font\gnuplot=cmr10 at 10pt
\gnuplot
\sbox{\plotpoint}{\rule[-0.200pt]{0.400pt}{0.400pt}}%
\put(181.0,163.0){\rule[-0.200pt]{4.818pt}{0.400pt}}
\put(161,163){\makebox(0,0)[r]{0}}
\put(1160.0,163.0){\rule[-0.200pt]{4.818pt}{0.400pt}}
\put(181.0,302.0){\rule[-0.200pt]{4.818pt}{0.400pt}}
\put(161,302){\makebox(0,0)[r]{0.1}}
\put(1160.0,302.0){\rule[-0.200pt]{4.818pt}{0.400pt}}
\put(181.0,441.0){\rule[-0.200pt]{4.818pt}{0.400pt}}
\put(161,441){\makebox(0,0)[r]{0.2}}
\put(1160.0,441.0){\rule[-0.200pt]{4.818pt}{0.400pt}}
\put(181.0,581.0){\rule[-0.200pt]{4.818pt}{0.400pt}}
\put(161,581){\makebox(0,0)[r]{0.3}}
\put(1160.0,581.0){\rule[-0.200pt]{4.818pt}{0.400pt}}
\put(181.0,720.0){\rule[-0.200pt]{4.818pt}{0.400pt}}
\put(161,720){\makebox(0,0)[r]{0.4}}
\put(1160.0,720.0){\rule[-0.200pt]{4.818pt}{0.400pt}}
\put(181.0,163.0){\rule[-0.200pt]{0.400pt}{4.818pt}}
\put(181,122){\makebox(0,0){0}}
\put(181.0,839.0){\rule[-0.200pt]{0.400pt}{4.818pt}}
\put(466.0,163.0){\rule[-0.200pt]{0.400pt}{4.818pt}}
\put(466,122){\makebox(0,0){10}}
\put(466.0,839.0){\rule[-0.200pt]{0.400pt}{4.818pt}}
\put(752.0,163.0){\rule[-0.200pt]{0.400pt}{4.818pt}}
\put(752,122){\makebox(0,0){20}}
\put(752.0,839.0){\rule[-0.200pt]{0.400pt}{4.818pt}}
\put(1037.0,163.0){\rule[-0.200pt]{0.400pt}{4.818pt}}
\put(1037,122){\makebox(0,0){30}}
\put(1037.0,839.0){\rule[-0.200pt]{0.400pt}{4.818pt}}
\put(181.0,163.0){\rule[-0.200pt]{240.659pt}{0.400pt}}
\put(1180.0,163.0){\rule[-0.200pt]{0.400pt}{167.666pt}}
\put(181.0,859.0){\rule[-0.200pt]{240.659pt}{0.400pt}}
\put(164,911){\makebox(0,0){$E_{ent}/(r_B^2/a^3)$}}
\put(680,61){\makebox(0,0){$r_B/a$}}
\put(181.0,163.0){\rule[-0.200pt]{0.400pt}{167.666pt}}
\put(1017,831){\makebox(0,0)[r]{$\langle :H_{tot}:\rangle_{\rho'}$}}
\put(224,698){\raisebox{-.8pt}{\makebox(0,0){$\Diamond$}}}
\put(252,697){\raisebox{-.8pt}{\makebox(0,0){$\Diamond$}}}
\put(281,694){\raisebox{-.8pt}{\makebox(0,0){$\Diamond$}}}
\put(309,692){\raisebox{-.8pt}{\makebox(0,0){$\Diamond$}}}
\put(338,691){\raisebox{-.8pt}{\makebox(0,0){$\Diamond$}}}
\put(367,690){\raisebox{-.8pt}{\makebox(0,0){$\Diamond$}}}
\put(395,689){\raisebox{-.8pt}{\makebox(0,0){$\Diamond$}}}
\put(424,689){\raisebox{-.8pt}{\makebox(0,0){$\Diamond$}}}
\put(452,688){\raisebox{-.8pt}{\makebox(0,0){$\Diamond$}}}
\put(481,688){\raisebox{-.8pt}{\makebox(0,0){$\Diamond$}}}
\put(509,688){\raisebox{-.8pt}{\makebox(0,0){$\Diamond$}}}
\put(538,687){\raisebox{-.8pt}{\makebox(0,0){$\Diamond$}}}
\put(566,687){\raisebox{-.8pt}{\makebox(0,0){$\Diamond$}}}
\put(595,687){\raisebox{-.8pt}{\makebox(0,0){$\Diamond$}}}
\put(623,687){\raisebox{-.8pt}{\makebox(0,0){$\Diamond$}}}
\put(652,687){\raisebox{-.8pt}{\makebox(0,0){$\Diamond$}}}
\put(681,687){\raisebox{-.8pt}{\makebox(0,0){$\Diamond$}}}
\put(709,687){\raisebox{-.8pt}{\makebox(0,0){$\Diamond$}}}
\put(738,686){\raisebox{-.8pt}{\makebox(0,0){$\Diamond$}}}
\put(766,686){\raisebox{-.8pt}{\makebox(0,0){$\Diamond$}}}
\put(795,686){\raisebox{-.8pt}{\makebox(0,0){$\Diamond$}}}
\put(823,686){\raisebox{-.8pt}{\makebox(0,0){$\Diamond$}}}
\put(852,686){\raisebox{-.8pt}{\makebox(0,0){$\Diamond$}}}
\put(880,686){\raisebox{-.8pt}{\makebox(0,0){$\Diamond$}}}
\put(909,686){\raisebox{-.8pt}{\makebox(0,0){$\Diamond$}}}
\put(937,686){\raisebox{-.8pt}{\makebox(0,0){$\Diamond$}}}
\put(966,686){\raisebox{-.8pt}{\makebox(0,0){$\Diamond$}}}
\put(994,686){\raisebox{-.8pt}{\makebox(0,0){$\Diamond$}}}
\put(1023,686){\raisebox{-.8pt}{\makebox(0,0){$\Diamond$}}}
\put(1052,684){\raisebox{-.8pt}{\makebox(0,0){$\Diamond$}}}
\put(1087,831){\raisebox{-.8pt}{\makebox(0,0){$\Diamond$}}}
\put(1017,790){\makebox(0,0)[r]{$\langle :H_1:\rangle$}}
\put(210,522){\makebox(0,0){$+$}}
\put(238,402){\makebox(0,0){$+$}}
\put(267,366){\makebox(0,0){$+$}}
\put(295,349){\makebox(0,0){$+$}}
\put(324,339){\makebox(0,0){$+$}}
\put(352,332){\makebox(0,0){$+$}}
\put(381,327){\makebox(0,0){$+$}}
\put(409,324){\makebox(0,0){$+$}}
\put(438,321){\makebox(0,0){$+$}}
\put(466,319){\makebox(0,0){$+$}}
\put(495,318){\makebox(0,0){$+$}}
\put(524,316){\makebox(0,0){$+$}}
\put(552,315){\makebox(0,0){$+$}}
\put(581,314){\makebox(0,0){$+$}}
\put(609,313){\makebox(0,0){$+$}}
\put(638,312){\makebox(0,0){$+$}}
\put(666,312){\makebox(0,0){$+$}}
\put(695,311){\makebox(0,0){$+$}}
\put(723,310){\makebox(0,0){$+$}}
\put(752,310){\makebox(0,0){$+$}}
\put(780,309){\makebox(0,0){$+$}}
\put(809,309){\makebox(0,0){$+$}}
\put(837,309){\makebox(0,0){$+$}}
\put(866,308){\makebox(0,0){$+$}}
\put(895,308){\makebox(0,0){$+$}}
\put(923,308){\makebox(0,0){$+$}}
\put(952,308){\makebox(0,0){$+$}}
\put(980,307){\makebox(0,0){$+$}}
\put(1009,307){\makebox(0,0){$+$}}
\put(1037,307){\makebox(0,0){$+$}}
\put(1087,790){\makebox(0,0){$+$}}
\sbox{\plotpoint}{\rule[-0.500pt]{1.000pt}{1.000pt}}%
\put(1017,749){\makebox(0,0)[r]{$\langle :H_2:\rangle$}}
\put(210,385){\raisebox{-.8pt}{\makebox(0,0){$\Box$}}}
\put(238,344){\raisebox{-.8pt}{\makebox(0,0){$\Box$}}}
\put(267,329){\raisebox{-.8pt}{\makebox(0,0){$\Box$}}}
\put(295,322){\raisebox{-.8pt}{\makebox(0,0){$\Box$}}}
\put(324,318){\raisebox{-.8pt}{\makebox(0,0){$\Box$}}}
\put(352,315){\raisebox{-.8pt}{\makebox(0,0){$\Box$}}}
\put(381,313){\raisebox{-.8pt}{\makebox(0,0){$\Box$}}}
\put(409,311){\raisebox{-.8pt}{\makebox(0,0){$\Box$}}}
\put(438,310){\raisebox{-.8pt}{\makebox(0,0){$\Box$}}}
\put(466,309){\raisebox{-.8pt}{\makebox(0,0){$\Box$}}}
\put(495,308){\raisebox{-.8pt}{\makebox(0,0){$\Box$}}}
\put(524,308){\raisebox{-.8pt}{\makebox(0,0){$\Box$}}}
\put(552,307){\raisebox{-.8pt}{\makebox(0,0){$\Box$}}}
\put(581,307){\raisebox{-.8pt}{\makebox(0,0){$\Box$}}}
\put(609,306){\raisebox{-.8pt}{\makebox(0,0){$\Box$}}}
\put(638,306){\raisebox{-.8pt}{\makebox(0,0){$\Box$}}}
\put(666,306){\raisebox{-.8pt}{\makebox(0,0){$\Box$}}}
\put(695,305){\raisebox{-.8pt}{\makebox(0,0){$\Box$}}}
\put(723,305){\raisebox{-.8pt}{\makebox(0,0){$\Box$}}}
\put(752,305){\raisebox{-.8pt}{\makebox(0,0){$\Box$}}}
\put(780,305){\raisebox{-.8pt}{\makebox(0,0){$\Box$}}}
\put(809,305){\raisebox{-.8pt}{\makebox(0,0){$\Box$}}}
\put(837,304){\raisebox{-.8pt}{\makebox(0,0){$\Box$}}}
\put(866,304){\raisebox{-.8pt}{\makebox(0,0){$\Box$}}}
\put(895,304){\raisebox{-.8pt}{\makebox(0,0){$\Box$}}}
\put(923,304){\raisebox{-.8pt}{\makebox(0,0){$\Box$}}}
\put(952,304){\raisebox{-.8pt}{\makebox(0,0){$\Box$}}}
\put(980,304){\raisebox{-.8pt}{\makebox(0,0){$\Box$}}}
\put(1009,304){\raisebox{-.8pt}{\makebox(0,0){$\Box$}}}
\put(1037,304){\raisebox{-.8pt}{\makebox(0,0){$\Box$}}}
\put(1087,749){\raisebox{-.8pt}{\makebox(0,0){$\Box$}}}
\end{picture}
 \end{center}
 \caption{
The numerical evaluations of entanglement energy in Minkowski
spacetime. $E_{ent}/(r_B^2/a^3)$ is shown as functions of $r_B/a$,
where $E_{ent}$ denotes $\langle :H_1:\rangle$, $\langle :H_2:\rangle$
and $\langle :H_{tot}:\rangle_{\rho'}$, respectively, and 
$r_B\equiv n_Ba$. We have taken $n_{tot}=60$ for 
$\langle :H_1:\rangle$ and 
$\langle :H_2:\rangle$, and $n_{tot}=200$ for 
$\langle :H_{tot}:\rangle_{\rho'}$. 
}
\label{fig:Eent_Min}
\end{figure}
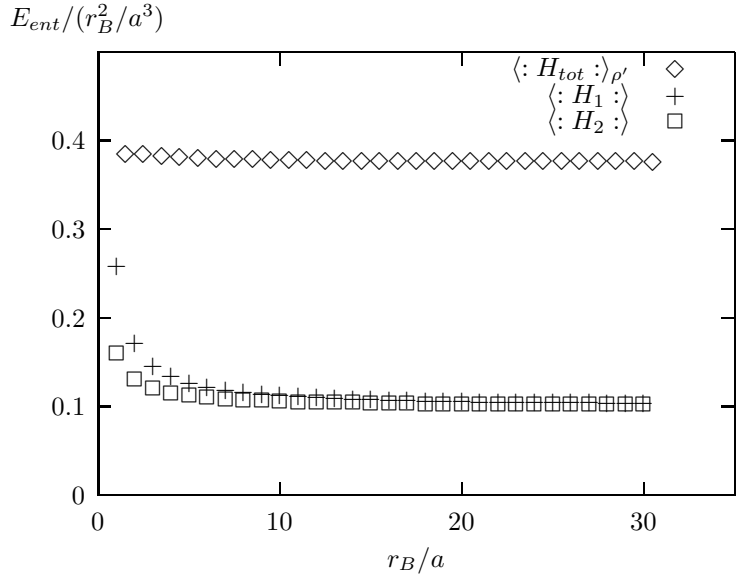

{\it Figure}~\ref{fig:Eent_Min} shows the result of numerical
calculations of $\langle :H_1:\rangle$, $\langle :H_2:\rangle$
and $\langle :H_{tot}:\rangle_{\rho'}$ in this model.

%============< FIGURE >==============%
%              Eent_Min.ps
%======================================%

From  this figure we see that $E_{ent}$ is almost
proportional to $r_B^2/a^3$:
%============< EQUATION >==============%
%
\begin{eqnarray}
 \langle :H_1:\rangle & \simeq & \langle :H_2:\rangle 
	\simeq 0.1 \frac{r_B^2}{a^3},\nonumber\\
 \langle :H_{tot}:\rangle_{\rho'}
	& \simeq & 0.4 \frac{r_B^2}{a^3}.
	\label{eqn:3-3:Eent_Min}
\end{eqnarray}
%======================================%

%<<<<<<   Schwarzschild calculations    >>>>>>>>>%
\subsubsection{Numerical calculations in Schwarzschild spacetime}

In Schwarzschild spacetime the metric is given by
%============< EQUATION >==============%
%
\begin{equation}
 ds^2 = -\left(1-\frac{r_0}{r}\right)dt^2 + 
 	\left(1-\frac{r_0}{r}\right)^{-1}dr^2 
	+ r^2 \left( d\theta + \sin^2{\theta}d\psi^2\right),
	\label{eqn:3-3:metric_Sch}
\end{equation}
%======================================%
where $r_0$ is the area radius of the horizon. As the radial 
coordinate $\rho$ we take the proper distance from the horizon:
%============< EQUATION >==============%
%
\begin{equation}
	\rho = \frac{r_0}{2}\left[\sqrt{y^2-1} + 
		\ln\left( y+\sqrt{y^2-1}\right)
		\right]	\ \ ,\label{eqn:3-3:rho-r}
\end{equation}
%======================================%
where the variable $y$ is defined by $y=2r/r_0-1$.

Using formulas given in subsection \ref{subsec:Sent}, we have
evaluated $S_{ent}$ numerically. In this calculation the outer
numerical boundary is set at $n_{tot}=100$. The summation with respect 
to $l$ in Eq.(\ref{eqn:3-3:sum-l}) is taken up to 
$l=\left[ 10r_0/a\right]$ ($\left[\ \ \right]$ is the Gauss symbol). 
From the above asymptotic behavior of $S^{(l)}_{ent}$, this guarantees
the accuracy of $10 \%$.

\begin{figure}
 \begin{center}
	% GNUPLOT: LaTeX picture
\setlength{\unitlength}{0.240900pt}
\ifx\plotpoint\undefined\newsavebox{\plotpoint}\fi
\sbox{\plotpoint}{\rule[-0.200pt]{0.400pt}{0.400pt}}%
\begin{picture}(1200,900)(0,0)
\font\gnuplot=cmr10 at 10pt
\gnuplot
\sbox{\plotpoint}{\rule[-0.200pt]{0.400pt}{0.400pt}}%
\put(181.0,163.0){\rule[-0.200pt]{4.818pt}{0.400pt}}
\put(161,163){\makebox(0,0)[r]{0.2}}
\put(1160.0,163.0){\rule[-0.200pt]{4.818pt}{0.400pt}}
\put(181.0,511.0){\rule[-0.200pt]{4.818pt}{0.400pt}}
\put(161,511){\makebox(0,0)[r]{0.3}}
\put(1160.0,511.0){\rule[-0.200pt]{4.818pt}{0.400pt}}
\put(181.0,859.0){\rule[-0.200pt]{4.818pt}{0.400pt}}
\put(161,859){\makebox(0,0)[r]{0.4}}
\put(1160.0,859.0){\rule[-0.200pt]{4.818pt}{0.400pt}}
\put(181.0,163.0){\rule[-0.200pt]{0.400pt}{4.818pt}}
\put(181,122){\makebox(0,0){0}}
\put(181.0,839.0){\rule[-0.200pt]{0.400pt}{4.818pt}}
\put(363.0,163.0){\rule[-0.200pt]{0.400pt}{4.818pt}}
\put(363,122){\makebox(0,0){20}}
\put(363.0,839.0){\rule[-0.200pt]{0.400pt}{4.818pt}}
\put(544.0,163.0){\rule[-0.200pt]{0.400pt}{4.818pt}}
\put(544,122){\makebox(0,0){40}}
\put(544.0,839.0){\rule[-0.200pt]{0.400pt}{4.818pt}}
\put(726.0,163.0){\rule[-0.200pt]{0.400pt}{4.818pt}}
\put(726,122){\makebox(0,0){60}}
\put(726.0,839.0){\rule[-0.200pt]{0.400pt}{4.818pt}}
\put(908.0,163.0){\rule[-0.200pt]{0.400pt}{4.818pt}}
\put(908,122){\makebox(0,0){80}}
\put(908.0,839.0){\rule[-0.200pt]{0.400pt}{4.818pt}}
\put(1089.0,163.0){\rule[-0.200pt]{0.400pt}{4.818pt}}
\put(1089,122){\makebox(0,0){100}}
\put(1089.0,839.0){\rule[-0.200pt]{0.400pt}{4.818pt}}
\put(181.0,163.0){\rule[-0.200pt]{240.659pt}{0.400pt}}
\put(1180.0,163.0){\rule[-0.200pt]{0.400pt}{167.666pt}}
\put(181.0,859.0){\rule[-0.200pt]{240.659pt}{0.400pt}}
\put(164,911){\makebox(0,0){$S_{ent}/(r_B/a)^2$}}
\put(680,61){\makebox(0,0){$r_B/a$}}
\put(181.0,163.0){\rule[-0.200pt]{0.400pt}{167.666pt}}
\put(1024,789){\makebox(0,0)[r]{$n_B=5$}}
\put(277,460){\raisebox{-.8pt}{\makebox(0,0){$\Diamond$}}}
\put(365,460){\raisebox{-.8pt}{\makebox(0,0){$\Diamond$}}}
\put(455,460){\raisebox{-.8pt}{\makebox(0,0){$\Diamond$}}}
\put(546,460){\raisebox{-.8pt}{\makebox(0,0){$\Diamond$}}}
\put(636,460){\raisebox{-.8pt}{\makebox(0,0){$\Diamond$}}}
\put(727,460){\raisebox{-.8pt}{\makebox(0,0){$\Diamond$}}}
\put(818,460){\raisebox{-.8pt}{\makebox(0,0){$\Diamond$}}}
\put(908,460){\raisebox{-.8pt}{\makebox(0,0){$\Diamond$}}}
\put(999,460){\raisebox{-.8pt}{\makebox(0,0){$\Diamond$}}}
\put(1090,460){\raisebox{-.8pt}{\makebox(0,0){$\Diamond$}}}
\put(1094,789){\raisebox{-.8pt}{\makebox(0,0){$\Diamond$}}}
\put(1024,748){\makebox(0,0)[r]{$n_B=2$}}
\put(273,449){\makebox(0,0){$+$}}
\put(363,446){\makebox(0,0){$+$}}
\put(454,446){\makebox(0,0){$+$}}
\put(544,446){\makebox(0,0){$+$}}
\put(635,446){\makebox(0,0){$+$}}
\put(726,446){\makebox(0,0){$+$}}
\put(817,445){\makebox(0,0){$+$}}
\put(908,445){\makebox(0,0){$+$}}
\put(998,445){\makebox(0,0){$+$}}
\put(1089,445){\makebox(0,0){$+$}}
\put(1094,748){\makebox(0,0){$+$}}
\sbox{\plotpoint}{\rule[-0.500pt]{1.000pt}{1.000pt}}%
\put(1024,707){\makebox(0,0)[r]{$n_B=1$}}
\put(272,412){\raisebox{-.8pt}{\makebox(0,0){$\Box$}}}
\put(363,410){\raisebox{-.8pt}{\makebox(0,0){$\Box$}}}
\put(454,409){\raisebox{-.8pt}{\makebox(0,0){$\Box$}}}
\put(544,409){\raisebox{-.8pt}{\makebox(0,0){$\Box$}}}
\put(635,409){\raisebox{-.8pt}{\makebox(0,0){$\Box$}}}
\put(726,409){\raisebox{-.8pt}{\makebox(0,0){$\Box$}}}
\put(817,409){\raisebox{-.8pt}{\makebox(0,0){$\Box$}}}
\put(908,409){\raisebox{-.8pt}{\makebox(0,0){$\Box$}}}
\put(998,409){\raisebox{-.8pt}{\makebox(0,0){$\Box$}}}
\put(1089,409){\raisebox{-.8pt}{\makebox(0,0){$\Box$}}}
\put(1094,707){\raisebox{-.8pt}{\makebox(0,0){$\Box$}}}
\end{picture}
 \end{center}
 \caption{
The numerical evaluations for $S_{ent}$ of the
discretized theory of the scalar field in Schwarzschild
spacetime. $S_{ent}/(r_B/a)^2$ for $n_B=1,2,5$ is shown as
functions of $r_B/a$, where $r_B\equiv r(\rho =n_Ba)$. We have
taken $n_{tot}=100$ and performed the summation over $l$ up to
$10r_0/a$. 
}
 \label{fig:Sent_Sch}
\end{figure}
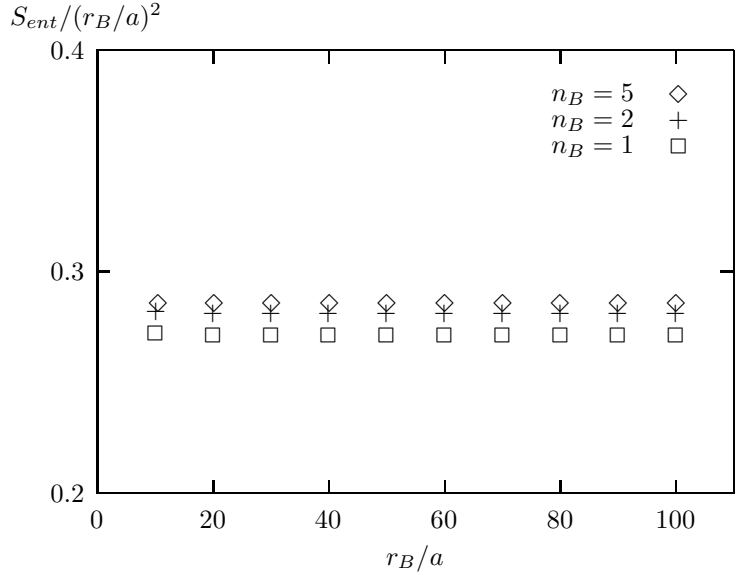

The result is shown in {\it Figure} \ref{fig:Sent_Sch}.
From this figure we see that $S_{ent}$ is proportional to $(r_B/a)^2$ 
if we change $r_0$ with $n_B$ fixed, and its coefficient has only
a weak dependence on $n_B$. Thus, we get 
\begin{equation}
     S_{ent} \simeq  0.3 \left(\frac{r_B}{a}\right)^2.
\label{eqn:3-3:resultS}
\end{equation}

%============< FIGURE >==============%
%              Sent_Sch.tex
%======================================%

This result is essentially the same as the previous result
(\ref{eqn:3-3:Sent-result}) for models in Minkowski spacetime
including the numerical coefficient. 
This can be understood in the following way.

Let us make a  coordinate change from $r$ to $x$ defined by  
\[
\frac{r}{r_0}= \frac{(x+1)^2}{4x} ,
\] 
or 
\[
x= \frac{2r}{r_0}-1 +\sqrt{\left(\frac{2r}{r_0}-1\right)^2 - 1} .
\] 
Then the metric (\ref{eqn:3-3:metric_Sch}) is rewritten as
\[
ds^2 =   -\left(\frac{x-1}{x+1}\right)^2 dt^2 +
                   r_0^2 \left(\frac{1+x}{2x}\right)^4 
                     (dx^2 + x^2 d\Omega^2) .
\]
Note that $r=0$, $r_0$ and $\infty$ correspond to $x=0$, $1$ and
$\infty$, respectively. It is easy to see that the Hamiltonian is
given in this coordinate system as
\begin{eqnarray}
 H & = & \sum_{lm}H_{lm} ,   \nonumber \\
 H_{lm} & = & \int d\xi
        \frac{64 x^4 (x-1)}{(x+1)^7}
      \biggl[ \frac{1}{2}P_{lm}^2	\nonumber\\
 & &  + \frac{1}{2} \left(\frac{(x+1)^2}{4x} \right)^4 
                     \left\{ (\partial_\xi \varphi_{lm})^2 
                     +\frac{l(l+1)}{\xi^2} \varphi_{lm}^2
                      \right\} \biggr]  ,
\label{eqn:3-3:H-note}         
\end{eqnarray}
where $\xi:=r_0 x$, and $P_{lm}$ and $\varphi_{lm}$ are expressed as
$P_{lm}:=\frac{r_0}{64}\frac{(x+1)^7}{x^4(x-1)} \dot{\phi}_{lm}$ and
$\varphi_{lm}:=r_0 \phi_{lm}$ in terms of $\phi_{lm}$.  Here note that the vacuum state is only
weakly dependent on the prefactor $ \frac{64 x^4 (x-1)}{(x+1)^7}$ in
Eq.({\ref{eqn:3-3:H-note}}).  If we neglect this prefactor, the vacuum
state is determined by the Hamiltonian which coincides with that for
the flat spacetime at $x=1$. On the other hand, $S_{ent}$ depends on
the modes in a thin layer around the boundary $\Sigma$, whose typical
thickness is a few times of $a\simeq l_{\rm Pl}$. Therefore, when
$\Sigma$ is near the horizon, the value of $S_{ent}$ should be well
approximated by the flat spacetime value.

%============< FIGURE >==============%
%              H1_Sch.tex
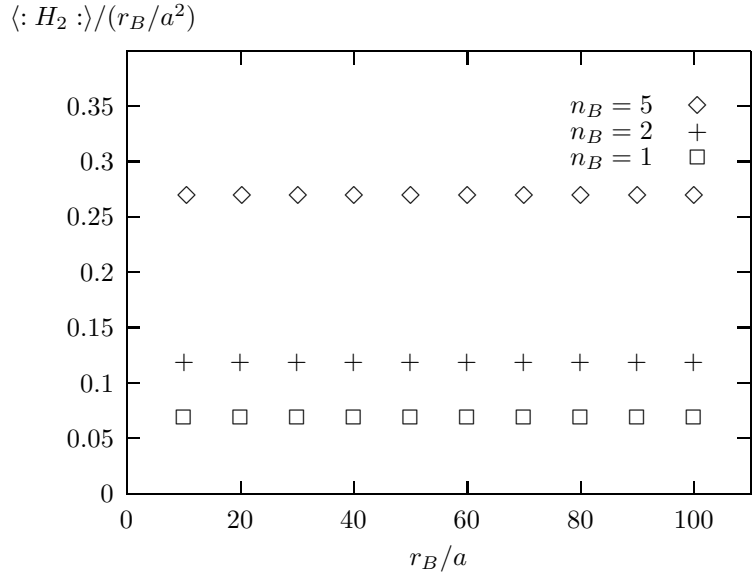
\begin{figure}
 \begin{center}
	% GNUPLOT: LaTeX picture
\setlength{\unitlength}{0.240900pt}
\ifx\plotpoint\undefined\newsavebox{\plotpoint}\fi
\sbox{\plotpoint}{\rule[-0.200pt]{0.400pt}{0.400pt}}%
\begin{picture}(1200,900)(0,0)
\font\gnuplot=cmr10 at 10pt
\gnuplot
\sbox{\plotpoint}{\rule[-0.200pt]{0.400pt}{0.400pt}}%
\put(201.0,163.0){\rule[-0.200pt]{4.818pt}{0.400pt}}
\put(181,163){\makebox(0,0)[r]{0}}
\put(1160.0,163.0){\rule[-0.200pt]{4.818pt}{0.400pt}}
\put(201.0,262.0){\rule[-0.200pt]{4.818pt}{0.400pt}}
\put(181,262){\makebox(0,0)[r]{0.05}}
\put(1160.0,262.0){\rule[-0.200pt]{4.818pt}{0.400pt}}
\put(201.0,362.0){\rule[-0.200pt]{4.818pt}{0.400pt}}
\put(181,362){\makebox(0,0)[r]{0.1}}
\put(1160.0,362.0){\rule[-0.200pt]{4.818pt}{0.400pt}}
\put(201.0,461.0){\rule[-0.200pt]{4.818pt}{0.400pt}}
\put(181,461){\makebox(0,0)[r]{0.15}}
\put(1160.0,461.0){\rule[-0.200pt]{4.818pt}{0.400pt}}
\put(201.0,561.0){\rule[-0.200pt]{4.818pt}{0.400pt}}
\put(181,561){\makebox(0,0)[r]{0.2}}
\put(1160.0,561.0){\rule[-0.200pt]{4.818pt}{0.400pt}}
\put(201.0,660.0){\rule[-0.200pt]{4.818pt}{0.400pt}}
\put(181,660){\makebox(0,0)[r]{0.25}}
\put(1160.0,660.0){\rule[-0.200pt]{4.818pt}{0.400pt}}
\put(201.0,760.0){\rule[-0.200pt]{4.818pt}{0.400pt}}
\put(181,760){\makebox(0,0)[r]{0.3}}
\put(1160.0,760.0){\rule[-0.200pt]{4.818pt}{0.400pt}}
\put(201.0,163.0){\rule[-0.200pt]{0.400pt}{4.818pt}}
\put(201,122){\makebox(0,0){0}}
\put(201.0,839.0){\rule[-0.200pt]{0.400pt}{4.818pt}}
\put(379.0,163.0){\rule[-0.200pt]{0.400pt}{4.818pt}}
\put(379,122){\makebox(0,0){20}}
\put(379.0,839.0){\rule[-0.200pt]{0.400pt}{4.818pt}}
\put(557.0,163.0){\rule[-0.200pt]{0.400pt}{4.818pt}}
\put(557,122){\makebox(0,0){40}}
\put(557.0,839.0){\rule[-0.200pt]{0.400pt}{4.818pt}}
\put(735.0,163.0){\rule[-0.200pt]{0.400pt}{4.818pt}}
\put(735,122){\makebox(0,0){60}}
\put(735.0,839.0){\rule[-0.200pt]{0.400pt}{4.818pt}}
\put(913.0,163.0){\rule[-0.200pt]{0.400pt}{4.818pt}}
\put(913,122){\makebox(0,0){80}}
\put(913.0,839.0){\rule[-0.200pt]{0.400pt}{4.818pt}}
\put(1091.0,163.0){\rule[-0.200pt]{0.400pt}{4.818pt}}
\put(1091,122){\makebox(0,0){100}}
\put(1091.0,839.0){\rule[-0.200pt]{0.400pt}{4.818pt}}
\put(201.0,163.0){\rule[-0.200pt]{235.841pt}{0.400pt}}
\put(1180.0,163.0){\rule[-0.200pt]{0.400pt}{167.666pt}}
\put(201.0,859.0){\rule[-0.200pt]{235.841pt}{0.400pt}}
\put(164,911){\makebox(0,0){$\langle :H_1:\rangle/(r_B/a^2)$}}
\put(690,61){\makebox(0,0){$r_B/a$}}
\put(201.0,163.0){\rule[-0.200pt]{0.400pt}{167.666pt}}
\put(1027,760){\makebox(0,0)[r]{$n_B=5$}}
\put(295,624){\raisebox{-.8pt}{\makebox(0,0){$\Diamond$}}}
\put(382,614){\raisebox{-.8pt}{\makebox(0,0){$\Diamond$}}}
\put(470,612){\raisebox{-.8pt}{\makebox(0,0){$\Diamond$}}}
\put(558,611){\raisebox{-.8pt}{\makebox(0,0){$\Diamond$}}}
\put(647,611){\raisebox{-.8pt}{\makebox(0,0){$\Diamond$}}}
\put(736,611){\raisebox{-.8pt}{\makebox(0,0){$\Diamond$}}}
\put(825,611){\raisebox{-.8pt}{\makebox(0,0){$\Diamond$}}}
\put(914,611){\raisebox{-.8pt}{\makebox(0,0){$\Diamond$}}}
\put(1003,610){\raisebox{-.8pt}{\makebox(0,0){$\Diamond$}}}
\put(1092,610){\raisebox{-.8pt}{\makebox(0,0){$\Diamond$}}}
\put(1097,760){\raisebox{-.8pt}{\makebox(0,0){$\Diamond$}}}
\put(1027,719){\makebox(0,0)[r]{$n_B=2$}}
\put(291,316){\makebox(0,0){$+$}}
\put(379,315){\makebox(0,0){$+$}}
\put(468,315){\makebox(0,0){$+$}}
\put(557,315){\makebox(0,0){$+$}}
\put(646,315){\makebox(0,0){$+$}}
\put(735,315){\makebox(0,0){$+$}}
\put(824,315){\makebox(0,0){$+$}}
\put(913,315){\makebox(0,0){$+$}}
\put(1002,315){\makebox(0,0){$+$}}
\put(1091,315){\makebox(0,0){$+$}}
\put(1097,719){\makebox(0,0){$+$}}
\sbox{\plotpoint}{\rule[-0.500pt]{1.000pt}{1.000pt}}%
\put(1027,678){\makebox(0,0)[r]{$n_B=1$}}
\put(290,217){\raisebox{-.8pt}{\makebox(0,0){$\Box$}}}
\put(379,217){\raisebox{-.8pt}{\makebox(0,0){$\Box$}}}
\put(468,217){\raisebox{-.8pt}{\makebox(0,0){$\Box$}}}
\put(557,217){\raisebox{-.8pt}{\makebox(0,0){$\Box$}}}
\put(646,217){\raisebox{-.8pt}{\makebox(0,0){$\Box$}}}
\put(735,217){\raisebox{-.8pt}{\makebox(0,0){$\Box$}}}
\put(824,217){\raisebox{-.8pt}{\makebox(0,0){$\Box$}}}
\put(913,217){\raisebox{-.8pt}{\makebox(0,0){$\Box$}}}
\put(1002,217){\raisebox{-.8pt}{\makebox(0,0){$\Box$}}}
\put(1091,217){\raisebox{-.8pt}{\makebox(0,0){$\Box$}}}
\put(1097,678){\raisebox{-.8pt}{\makebox(0,0){$\Box$}}}
\end{picture}
 \end{center}
 \caption{
The numerical evaluations of $\langle :H_1:\rangle$ for the
discretized theory of the scalar field in Schwarzschild
spacetime. $\langle :H_1:\rangle/(r_B/a^2)$ for $n_B=1,2,5$ is shown
as functions of $r_B/a$, where $r_B\equiv r(\rho =n_Ba)$. We have
taken $n_{tot}=100$ and performed the summation over $l$ up to
$10r_0/a$. 
}
 \label{fig:H1_Sch}
\end{figure}
%======================================%

%============< FIGURE >==============%
%              H2_Sch.tex
\begin{figure}
 \begin{center}
	% GNUPLOT: LaTeX picture
\setlength{\unitlength}{0.240900pt}
\ifx\plotpoint\undefined\newsavebox{\plotpoint}\fi
\sbox{\plotpoint}{\rule[-0.200pt]{0.400pt}{0.400pt}}%
\begin{picture}(1200,900)(0,0)
\font\gnuplot=cmr10 at 10pt
\gnuplot
\sbox{\plotpoint}{\rule[-0.200pt]{0.400pt}{0.400pt}}%
\put(201.0,163.0){\rule[-0.200pt]{4.818pt}{0.400pt}}
\put(181,163){\makebox(0,0)[r]{0}}
\put(1160.0,163.0){\rule[-0.200pt]{4.818pt}{0.400pt}}
\put(201.0,250.0){\rule[-0.200pt]{4.818pt}{0.400pt}}
\put(181,250){\makebox(0,0)[r]{0.05}}
\put(1160.0,250.0){\rule[-0.200pt]{4.818pt}{0.400pt}}
\put(201.0,337.0){\rule[-0.200pt]{4.818pt}{0.400pt}}
\put(181,337){\makebox(0,0)[r]{0.1}}
\put(1160.0,337.0){\rule[-0.200pt]{4.818pt}{0.400pt}}
\put(201.0,424.0){\rule[-0.200pt]{4.818pt}{0.400pt}}
\put(181,424){\makebox(0,0)[r]{0.15}}
\put(1160.0,424.0){\rule[-0.200pt]{4.818pt}{0.400pt}}
\put(201.0,511.0){\rule[-0.200pt]{4.818pt}{0.400pt}}
\put(181,511){\makebox(0,0)[r]{0.2}}
\put(1160.0,511.0){\rule[-0.200pt]{4.818pt}{0.400pt}}
\put(201.0,598.0){\rule[-0.200pt]{4.818pt}{0.400pt}}
\put(181,598){\makebox(0,0)[r]{0.25}}
\put(1160.0,598.0){\rule[-0.200pt]{4.818pt}{0.400pt}}
\put(201.0,685.0){\rule[-0.200pt]{4.818pt}{0.400pt}}
\put(181,685){\makebox(0,0)[r]{0.3}}
\put(1160.0,685.0){\rule[-0.200pt]{4.818pt}{0.400pt}}
\put(201.0,772.0){\rule[-0.200pt]{4.818pt}{0.400pt}}
\put(181,772){\makebox(0,0)[r]{0.35}}
\put(1160.0,772.0){\rule[-0.200pt]{4.818pt}{0.400pt}}
\put(201.0,163.0){\rule[-0.200pt]{0.400pt}{4.818pt}}
\put(201,122){\makebox(0,0){0}}
\put(201.0,839.0){\rule[-0.200pt]{0.400pt}{4.818pt}}
\put(379.0,163.0){\rule[-0.200pt]{0.400pt}{4.818pt}}
\put(379,122){\makebox(0,0){20}}
\put(379.0,839.0){\rule[-0.200pt]{0.400pt}{4.818pt}}
\put(557.0,163.0){\rule[-0.200pt]{0.400pt}{4.818pt}}
\put(557,122){\makebox(0,0){40}}
\put(557.0,839.0){\rule[-0.200pt]{0.400pt}{4.818pt}}
\put(735.0,163.0){\rule[-0.200pt]{0.400pt}{4.818pt}}
\put(735,122){\makebox(0,0){60}}
\put(735.0,839.0){\rule[-0.200pt]{0.400pt}{4.818pt}}
\put(913.0,163.0){\rule[-0.200pt]{0.400pt}{4.818pt}}
\put(913,122){\makebox(0,0){80}}
\put(913.0,839.0){\rule[-0.200pt]{0.400pt}{4.818pt}}
\put(1091.0,163.0){\rule[-0.200pt]{0.400pt}{4.818pt}}
\put(1091,122){\makebox(0,0){100}}
\put(1091.0,839.0){\rule[-0.200pt]{0.400pt}{4.818pt}}
\put(201.0,163.0){\rule[-0.200pt]{235.841pt}{0.400pt}}
\put(1180.0,163.0){\rule[-0.200pt]{0.400pt}{167.666pt}}
\put(201.0,859.0){\rule[-0.200pt]{235.841pt}{0.400pt}}
\put(164,911){\makebox(0,0){$\langle :H_2:\rangle /(r_B/a^2)$}}
\put(690,61){\makebox(0,0){$r_B/a$}}
\put(201.0,163.0){\rule[-0.200pt]{0.400pt}{167.666pt}}
\put(1027,772){\makebox(0,0)[r]{$n_B=5$}}
\put(295,631){\raisebox{-.8pt}{\makebox(0,0){$\Diamond$}}}
\put(382,631){\raisebox{-.8pt}{\makebox(0,0){$\Diamond$}}}
\put(470,631){\raisebox{-.8pt}{\makebox(0,0){$\Diamond$}}}
\put(558,631){\raisebox{-.8pt}{\makebox(0,0){$\Diamond$}}}
\put(647,631){\raisebox{-.8pt}{\makebox(0,0){$\Diamond$}}}
\put(736,631){\raisebox{-.8pt}{\makebox(0,0){$\Diamond$}}}
\put(825,631){\raisebox{-.8pt}{\makebox(0,0){$\Diamond$}}}
\put(914,631){\raisebox{-.8pt}{\makebox(0,0){$\Diamond$}}}
\put(1003,631){\raisebox{-.8pt}{\makebox(0,0){$\Diamond$}}}
\put(1092,631){\raisebox{-.8pt}{\makebox(0,0){$\Diamond$}}}
\put(1097,772){\raisebox{-.8pt}{\makebox(0,0){$\Diamond$}}}
\put(1027,731){\makebox(0,0)[r]{$n_B=2$}}
\put(291,370){\makebox(0,0){$+$}}
\put(379,370){\makebox(0,0){$+$}}
\put(468,370){\makebox(0,0){$+$}}
\put(557,370){\makebox(0,0){$+$}}
\put(646,370){\makebox(0,0){$+$}}
\put(735,370){\makebox(0,0){$+$}}
\put(824,370){\makebox(0,0){$+$}}
\put(913,370){\makebox(0,0){$+$}}
\put(1002,370){\makebox(0,0){$+$}}
\put(1091,370){\makebox(0,0){$+$}}
\put(1097,731){\makebox(0,0){$+$}}
\sbox{\plotpoint}{\rule[-0.500pt]{1.000pt}{1.000pt}}%
\put(1027,690){\makebox(0,0)[r]{$n_B=1$}}
\put(290,281){\raisebox{-.8pt}{\makebox(0,0){$\Box$}}}
\put(379,281){\raisebox{-.8pt}{\makebox(0,0){$\Box$}}}
\put(468,281){\raisebox{-.8pt}{\makebox(0,0){$\Box$}}}
\put(557,281){\raisebox{-.8pt}{\makebox(0,0){$\Box$}}}
\put(646,281){\raisebox{-.8pt}{\makebox(0,0){$\Box$}}}
\put(735,281){\raisebox{-.8pt}{\makebox(0,0){$\Box$}}}
\put(824,281){\raisebox{-.8pt}{\makebox(0,0){$\Box$}}}
\put(913,281){\raisebox{-.8pt}{\makebox(0,0){$\Box$}}}
\put(1002,281){\raisebox{-.8pt}{\makebox(0,0){$\Box$}}}
\put(1091,281){\raisebox{-.8pt}{\makebox(0,0){$\Box$}}}
\put(1097,690){\raisebox{-.8pt}{\makebox(0,0){$\Box$}}}
\end{picture}
 \end{center}
 \caption{
The numerical evaluations of $\langle :H_2:\rangle$ for the
discretized theory of the scalar field in Schwarzschild
spacetime. $\langle :H_2:\rangle/(r_B/a^2)$ for $n_B=1,2,5$ is shown
as functions of $r_B/a$, where $r_B\equiv r(\rho =n_Ba)$. We have
taken $n_{tot}=100$ and performed the summation over $l$ up to
$10r_0/a$. 
}
 \label{fig:H2_Sch}
\end{figure}
%======================================%

%============< FIGURE >==============%
%              Htot_Sch.tex
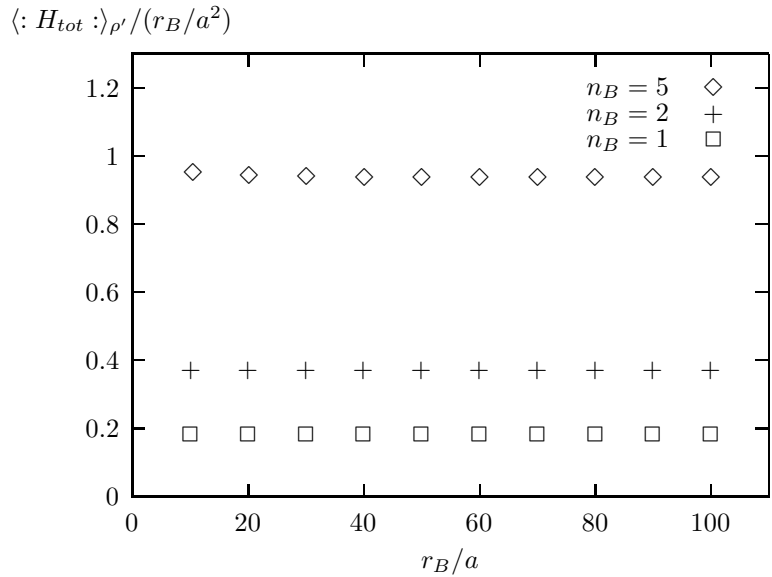
\begin{figure}
 \begin{center}
	% GNUPLOT: LaTeX picture
\setlength{\unitlength}{0.240900pt}
\ifx\plotpoint\undefined\newsavebox{\plotpoint}\fi
\sbox{\plotpoint}{\rule[-0.200pt]{0.400pt}{0.400pt}}%
\begin{picture}(1200,900)(0,0)
\font\gnuplot=cmr10 at 10pt
\gnuplot
\sbox{\plotpoint}{\rule[-0.200pt]{0.400pt}{0.400pt}}%
\put(181.0,163.0){\rule[-0.200pt]{4.818pt}{0.400pt}}
\put(161,163){\makebox(0,0)[r]{0}}
\put(1160.0,163.0){\rule[-0.200pt]{4.818pt}{0.400pt}}
\put(181.0,270.0){\rule[-0.200pt]{4.818pt}{0.400pt}}
\put(161,270){\makebox(0,0)[r]{0.2}}
\put(1160.0,270.0){\rule[-0.200pt]{4.818pt}{0.400pt}}
\put(181.0,377.0){\rule[-0.200pt]{4.818pt}{0.400pt}}
\put(161,377){\makebox(0,0)[r]{0.4}}
\put(1160.0,377.0){\rule[-0.200pt]{4.818pt}{0.400pt}}
\put(181.0,484.0){\rule[-0.200pt]{4.818pt}{0.400pt}}
\put(161,484){\makebox(0,0)[r]{0.6}}
\put(1160.0,484.0){\rule[-0.200pt]{4.818pt}{0.400pt}}
\put(181.0,591.0){\rule[-0.200pt]{4.818pt}{0.400pt}}
\put(161,591){\makebox(0,0)[r]{0.8}}
\put(1160.0,591.0){\rule[-0.200pt]{4.818pt}{0.400pt}}
\put(181.0,698.0){\rule[-0.200pt]{4.818pt}{0.400pt}}
\put(161,698){\makebox(0,0)[r]{1}}
\put(1160.0,698.0){\rule[-0.200pt]{4.818pt}{0.400pt}}
\put(181.0,805.0){\rule[-0.200pt]{4.818pt}{0.400pt}}
\put(161,805){\makebox(0,0)[r]{1.2}}
\put(1160.0,805.0){\rule[-0.200pt]{4.818pt}{0.400pt}}
\put(181.0,163.0){\rule[-0.200pt]{0.400pt}{4.818pt}}
\put(181,122){\makebox(0,0){0}}
\put(181.0,839.0){\rule[-0.200pt]{0.400pt}{4.818pt}}
\put(363.0,163.0){\rule[-0.200pt]{0.400pt}{4.818pt}}
\put(363,122){\makebox(0,0){20}}
\put(363.0,839.0){\rule[-0.200pt]{0.400pt}{4.818pt}}
\put(544.0,163.0){\rule[-0.200pt]{0.400pt}{4.818pt}}
\put(544,122){\makebox(0,0){40}}
\put(544.0,839.0){\rule[-0.200pt]{0.400pt}{4.818pt}}
\put(726.0,163.0){\rule[-0.200pt]{0.400pt}{4.818pt}}
\put(726,122){\makebox(0,0){60}}
\put(726.0,839.0){\rule[-0.200pt]{0.400pt}{4.818pt}}
\put(908.0,163.0){\rule[-0.200pt]{0.400pt}{4.818pt}}
\put(908,122){\makebox(0,0){80}}
\put(908.0,839.0){\rule[-0.200pt]{0.400pt}{4.818pt}}
\put(1089.0,163.0){\rule[-0.200pt]{0.400pt}{4.818pt}}
\put(1089,122){\makebox(0,0){100}}
\put(1089.0,839.0){\rule[-0.200pt]{0.400pt}{4.818pt}}
\put(181.0,163.0){\rule[-0.200pt]{240.659pt}{0.400pt}}
\put(1180.0,163.0){\rule[-0.200pt]{0.400pt}{167.666pt}}
\put(181.0,859.0){\rule[-0.200pt]{240.659pt}{0.400pt}}
\put(164,911){\makebox(0,0){$\langle :H_{tot}:\rangle_{\rho'}/(r_B/a^2)$}}
\put(680,61){\makebox(0,0){$r_B/a$}}
\put(181.0,163.0){\rule[-0.200pt]{0.400pt}{167.666pt}}
\put(1024,805){\makebox(0,0)[r]{$n_B=5$}}
\put(277,671){\raisebox{-.8pt}{\makebox(0,0){$\Diamond$}}}
\put(365,666){\raisebox{-.8pt}{\makebox(0,0){$\Diamond$}}}
\put(455,665){\raisebox{-.8pt}{\makebox(0,0){$\Diamond$}}}
\put(546,664){\raisebox{-.8pt}{\makebox(0,0){$\Diamond$}}}
\put(636,664){\raisebox{-.8pt}{\makebox(0,0){$\Diamond$}}}
\put(727,664){\raisebox{-.8pt}{\makebox(0,0){$\Diamond$}}}
\put(818,664){\raisebox{-.8pt}{\makebox(0,0){$\Diamond$}}}
\put(908,664){\raisebox{-.8pt}{\makebox(0,0){$\Diamond$}}}
\put(999,664){\raisebox{-.8pt}{\makebox(0,0){$\Diamond$}}}
\put(1090,664){\raisebox{-.8pt}{\makebox(0,0){$\Diamond$}}}
\put(1094,805){\raisebox{-.8pt}{\makebox(0,0){$\Diamond$}}}
\put(1024,764){\makebox(0,0)[r]{$n_B=2$}}
\put(273,362){\makebox(0,0){$+$}}
\put(363,362){\makebox(0,0){$+$}}
\put(454,362){\makebox(0,0){$+$}}
\put(544,362){\makebox(0,0){$+$}}
\put(635,362){\makebox(0,0){$+$}}
\put(726,362){\makebox(0,0){$+$}}
\put(817,362){\makebox(0,0){$+$}}
\put(908,362){\makebox(0,0){$+$}}
\put(998,362){\makebox(0,0){$+$}}
\put(1089,362){\makebox(0,0){$+$}}
\put(1094,764){\makebox(0,0){$+$}}
\sbox{\plotpoint}{\rule[-0.500pt]{1.000pt}{1.000pt}}%
\put(1024,723){\makebox(0,0)[r]{$n_B=1$}}
\put(272,260){\raisebox{-.8pt}{\makebox(0,0){$\Box$}}}
\put(363,260){\raisebox{-.8pt}{\makebox(0,0){$\Box$}}}
\put(454,260){\raisebox{-.8pt}{\makebox(0,0){$\Box$}}}
\put(544,260){\raisebox{-.8pt}{\makebox(0,0){$\Box$}}}
\put(635,260){\raisebox{-.8pt}{\makebox(0,0){$\Box$}}}
\put(726,260){\raisebox{-.8pt}{\makebox(0,0){$\Box$}}}
\put(817,260){\raisebox{-.8pt}{\makebox(0,0){$\Box$}}}
\put(908,260){\raisebox{-.8pt}{\makebox(0,0){$\Box$}}}
\put(998,260){\raisebox{-.8pt}{\makebox(0,0){$\Box$}}}
\put(1089,260){\raisebox{-.8pt}{\makebox(0,0){$\Box$}}}
\put(1094,723){\raisebox{-.8pt}{\makebox(0,0){$\Box$}}}
\end{picture}
 \end{center}
\caption{
The numerical evaluations of $\langle :H_{tot}:\rangle_{\rho'}$ for
the discretized theory of the scalar field in Schwarzschild
spacetime. $\langle :H_{tot}:\rangle_{\rho'}/(r_B/a^2)$ for
$n_B=1,2,5$ is shown as functions of $r_B/a$, where 
$r_B\equiv r(\rho =n_Ba)$. We have taken $n_{tot}=100$ and performed
the summation over $l$ up to $10r_0/a$. 
}
\label{fig:Htot_Sch}
\end{figure}
%======================================%

Next, with the helps of formulas developed in subsection
\ref{subsec:Eent}, we have numerically evaluated 
$\langle :H_1:\rangle$, $\langle :H_2:\rangle$ and 
$\langle :H_{tot}:\rangle_{\rho'}$. 
Now we have taken the numerical outer boundary at
$n_{tot}=100$. The truncation in the $l$-summation is the same as for
$S_{ent}$ (up to $l=[10r_0/a]$), which implies that the accuracy is
about $10\%$ from the above asymptotic estimate for $E^{(l)}_{ent}$.

The results of numerical calculations are shown in {\it Figure}
\ref{fig:H1_Sch}, {\it Figure} \ref{fig:H2_Sch} and {\it Figure}
\ref{fig:Htot_Sch}. In these figures, $E_{ent}/(r_B/a^2)$ is plotted
as a function of $r_B/a$ for $n_B=1,2,5$. 
All of these figures show that $E_{ent}$ is proportional
to $r_B/a^2$: 
%============< EQUATION >==============%
%
\begin{eqnarray}
 \langle :H_1:\rangle & \simeq & 0.05 (n_B-1/2)r_B/a^2,\nonumber\\
 \langle :H_2:\rangle & \simeq & 0.05 (n_B+1/2)r_B/a^2,\nonumber\\
 \langle :H_{tot}:\rangle_{\rho'} & \simeq & 0.2n_Br_B/a^2.
\end{eqnarray}
%======================================%

From these equations we immediately see that the values of
$\langle :H_1:\rangle$ and $\langle :H_2:\rangle$ coincide except for
a tiny difference independent of $n_B$. 
This difference is understood by the gravitational red-shift:
$\langle :H_1:\rangle$ comes from the modes just
inside $\Sigma$ while $\langle :H_2:\rangle$ originates from the modes
just outside $\Sigma$. In the present numerical calculations, it means
that $\langle :H_1:\rangle$ and $\langle :H_2:\rangle$ are determined
by the modes at $\rho=(n_{B}-1/2)a$ and $\rho=(n_{B}+1/2)a$,
respectively(see Eq.(\ref{eqn:3-3:Vlm})).  Hence, taking account of
the fact that the contribution of each mode to the entanglement energy
is proportional to the red-shift factor at its location, the ratio of
$\langle :H_1:\rangle$ and $\langle :H_2:\rangle$ should be
approximately given by 
%============< EQUATION >==============%
%
\begin{eqnarray}
 \langle :H_1:\rangle : \langle :H_2:\rangle
 & \sim & N^{1/2}(\rho =(n_{B}-1/2)a):N^{1/2}(\rho =(n_{B}+1/2)a)
	\nonumber\\
 & \sim & (n_{B}-1/2):(n_{B}+1/2).
\end{eqnarray}
%======================================%
This is consistent with the above numerical result.

This argument is also supported by the numerical result for the flat
spacetime model shown in {\it Figure} \ref{fig:Eent_Min}. In this
figure the values of $\langle :H_1:\rangle$ and $\langle :H_2:\rangle$ 
for a massless scalar field in the Minkowski spacetime with
$\Sigma=B\times{\bf R}=S^2\times{\bf R}$ are 
plotted. In this case there is no gravitational red-shift effect, so
we expect that $\langle :H_1:\rangle\simeq\langle :H_2:\rangle$ 
as confirmed by the numerical calculation.

%<<<<<<  R-N calculations    >>>>>>>>>%
\subsubsection{Numerical calculations in Reissner-Nordstr{\" o}m
spacetime}

In Reissner-Nordstr{\"o}m spacetime the metric is given by
%============< EQUATION >==============%
%
\begin{equation}
 ds^2 = -\left( 1-\frac{2M}{r}+\frac{Q^2}{r^2}\right)dt^2 + 
	\left(1-\frac{2M}{r}+\frac{Q^2}{r^2}\right)^{-1}dr^2 
	+ r^2 \left( d\theta^2 + \sin^2{\theta}d\psi^2\right),
\end{equation}
%======================================%
where $M$ and $Q$ are the mass and the charge of the black-hole. The area 
radius of the outer horizon $r_0$ is
%============< EQUATION >==============%
%
\begin{equation}
 r_0 = M+\sqrt{M^2-Q^2}.
\end{equation}
%======================================%
As the radial coordinate $\rho$ we take the proper distance from the 
outer horizon
%============< EQUATION >==============%
%
\begin{equation}
 \rho = \sqrt{(M^2-Q^2)(y^2-1)} + 
 	M\ln\left( y+\sqrt{y^2-1}\right),\label{eqn:3-3:rho-r'}
\end{equation}
%======================================%
where the variable $y$ is defined by 
%============< EQUATION >==============%
%
\begin{equation}
 y = \frac{r-M}{\sqrt{M^2-Q^2}}\ \ .
\end{equation}
%======================================%

Using formulas given in subsection \ref{subsec:Sent}, we have
evaluated $S_{ent}$ numerically. In this calculation the outer
numerical boundary is set at $n_{tot}=100$, and the boundary of the 
spatial division is fixed at $n_B=1$. The summation in $l$ in
Eq.(\ref{eqn:3-3:sum-l}) is taken up to $l=\left[ 10r_0/a \right]$
($\left[\ \ \right]$ is the Gauss symbol). 
From the above asymptotic behavior of $S^{(l)}_{ent}$, this guarantees
the accuracy of $10 \%$.

The result is shown in {\it Figure} \ref{fig:Sent_RN}.
From this figure we see that $S_{ent}$ is proportional to $(r_B/a)^2$ 
if we change $r_0$ with $q\equiv Q/M$ fixed, and its
coefficient has only a weak dependence on $q$. Thus, we get 
\begin{equation}
     S_{ent} \simeq  0.3 \left(\frac{r_B}{a}\right)^2.
\end{equation}

This result is essentially the same as the results in Minkowski and
Schwarzschild spacetime.

%============< FIGURE >==============%
%              Sent_Sch.tex
\begin{figure}
 \begin{center}
	% GNUPLOT: LaTeX picture
\setlength{\unitlength}{0.240900pt}
\ifx\plotpoint\undefined\newsavebox{\plotpoint}\fi
\sbox{\plotpoint}{\rule[-0.200pt]{0.400pt}{0.400pt}}%
\begin{picture}(1200,900)(0,0)
\font\gnuplot=cmr10 at 10pt
\gnuplot
\sbox{\plotpoint}{\rule[-0.200pt]{0.400pt}{0.400pt}}%
\put(181.0,163.0){\rule[-0.200pt]{4.818pt}{0.400pt}}
\put(161,163){\makebox(0,0)[r]{0.2}}
\put(1160.0,163.0){\rule[-0.200pt]{4.818pt}{0.400pt}}
\put(181.0,511.0){\rule[-0.200pt]{4.818pt}{0.400pt}}
\put(161,511){\makebox(0,0)[r]{0.3}}
\put(1160.0,511.0){\rule[-0.200pt]{4.818pt}{0.400pt}}
\put(181.0,859.0){\rule[-0.200pt]{4.818pt}{0.400pt}}
\put(161,859){\makebox(0,0)[r]{0.4}}
\put(1160.0,859.0){\rule[-0.200pt]{4.818pt}{0.400pt}}
\put(181.0,163.0){\rule[-0.200pt]{0.400pt}{4.818pt}}
\put(181,122){\makebox(0,0){0}}
\put(181.0,839.0){\rule[-0.200pt]{0.400pt}{4.818pt}}
\put(363.0,163.0){\rule[-0.200pt]{0.400pt}{4.818pt}}
\put(363,122){\makebox(0,0){20}}
\put(363.0,839.0){\rule[-0.200pt]{0.400pt}{4.818pt}}
\put(544.0,163.0){\rule[-0.200pt]{0.400pt}{4.818pt}}
\put(544,122){\makebox(0,0){40}}
\put(544.0,839.0){\rule[-0.200pt]{0.400pt}{4.818pt}}
\put(726.0,163.0){\rule[-0.200pt]{0.400pt}{4.818pt}}
\put(726,122){\makebox(0,0){60}}
\put(726.0,839.0){\rule[-0.200pt]{0.400pt}{4.818pt}}
\put(908.0,163.0){\rule[-0.200pt]{0.400pt}{4.818pt}}
\put(908,122){\makebox(0,0){80}}
\put(908.0,839.0){\rule[-0.200pt]{0.400pt}{4.818pt}}
\put(1089.0,163.0){\rule[-0.200pt]{0.400pt}{4.818pt}}
\put(1089,122){\makebox(0,0){100}}
\put(1089.0,839.0){\rule[-0.200pt]{0.400pt}{4.818pt}}
\put(181.0,163.0){\rule[-0.200pt]{240.659pt}{0.400pt}}
\put(1180.0,163.0){\rule[-0.200pt]{0.400pt}{167.666pt}}
\put(181.0,859.0){\rule[-0.200pt]{240.659pt}{0.400pt}}
\put(164,911){\makebox(0,0){$S_{ent}/(r_B/a)^2$}}
\put(680,61){\makebox(0,0){$r_B/a$}}
\put(181.0,163.0){\rule[-0.200pt]{0.400pt}{167.666pt}}
\put(1024,789){\makebox(0,0)[r]{$q=0,n_B=1$}}
\put(272,412){\raisebox{-.8pt}{\makebox(0,0){$\Diamond$}}}
\put(363,410){\raisebox{-.8pt}{\makebox(0,0){$\Diamond$}}}
\put(454,409){\raisebox{-.8pt}{\makebox(0,0){$\Diamond$}}}
\put(544,409){\raisebox{-.8pt}{\makebox(0,0){$\Diamond$}}}
\put(635,409){\raisebox{-.8pt}{\makebox(0,0){$\Diamond$}}}
\put(726,409){\raisebox{-.8pt}{\makebox(0,0){$\Diamond$}}}
\put(817,409){\raisebox{-.8pt}{\makebox(0,0){$\Diamond$}}}
\put(908,409){\raisebox{-.8pt}{\makebox(0,0){$\Diamond$}}}
\put(998,409){\raisebox{-.8pt}{\makebox(0,0){$\Diamond$}}}
\put(1089,409){\raisebox{-.8pt}{\makebox(0,0){$\Diamond$}}}
\put(1094,789){\raisebox{-.8pt}{\makebox(0,0){$\Diamond$}}}
\put(1024,748){\makebox(0,0)[r]{$q=0.9,n_B=1$}}
\put(272,410){\makebox(0,0){$+$}}
\put(363,409){\makebox(0,0){$+$}}
\put(454,409){\makebox(0,0){$+$}}
\put(544,409){\makebox(0,0){$+$}}
\put(635,409){\makebox(0,0){$+$}}
\put(726,409){\makebox(0,0){$+$}}
\put(817,409){\makebox(0,0){$+$}}
\put(908,409){\makebox(0,0){$+$}}
\put(998,409){\makebox(0,0){$+$}}
\put(1089,409){\makebox(0,0){$+$}}
\put(1094,748){\makebox(0,0){$+$}}
\sbox{\plotpoint}{\rule[-0.500pt]{1.000pt}{1.000pt}}%
\put(1024,707){\makebox(0,0)[r]{$q=0.999,n_B=1$}}
\put(272,408){\raisebox{-.8pt}{\makebox(0,0){$\Box$}}}
\put(363,409){\raisebox{-.8pt}{\makebox(0,0){$\Box$}}}
\put(453,409){\raisebox{-.8pt}{\makebox(0,0){$\Box$}}}
\put(544,409){\raisebox{-.8pt}{\makebox(0,0){$\Box$}}}
\put(635,409){\raisebox{-.8pt}{\makebox(0,0){$\Box$}}}
\put(726,409){\raisebox{-.8pt}{\makebox(0,0){$\Box$}}}
\put(817,409){\raisebox{-.8pt}{\makebox(0,0){$\Box$}}}
\put(908,409){\raisebox{-.8pt}{\makebox(0,0){$\Box$}}}
\put(998,409){\raisebox{-.8pt}{\makebox(0,0){$\Box$}}}
\put(1089,409){\raisebox{-.8pt}{\makebox(0,0){$\Box$}}}
\put(1094,707){\raisebox{-.8pt}{\makebox(0,0){$\Box$}}}
\end{picture}
 \end{center}
 \caption{
The numerical evaluations of $S_{ent}$ for the
discretized theory of the scalar field in Reissner-Nordstr{\" o}m
spacetime. $S_{ent}/(r_B/a)^2$ for $q=0,0.9,0.999$ is shown as
functions of $r_B/a$, where $r_B\equiv r(\rho =n_Ba)$. We have
taken $n_{tot}=100$ and performed the summation over $l$ up to
$10r_0/a$. 
}
 \label{fig:Sent_RN}
\end{figure}
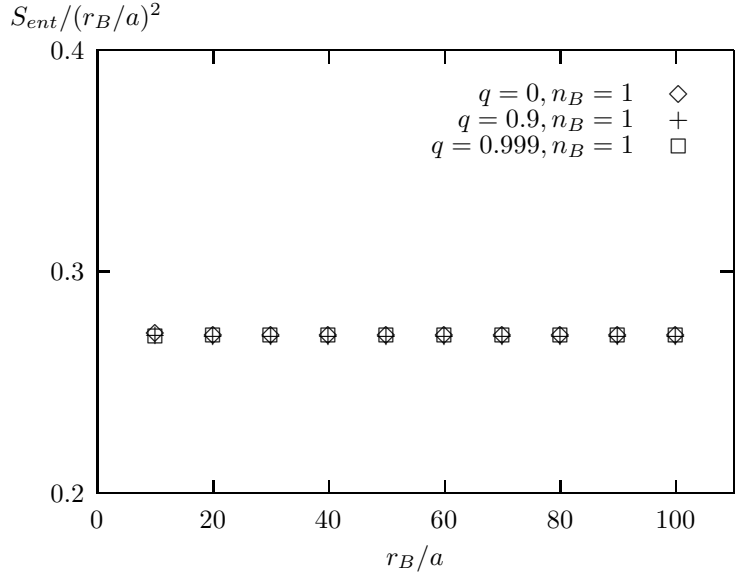
%======================================%

%============< FIGURE >==============%
%              H1_RN.tex
\begin{figure}
 \begin{center}
	% GNUPLOT: LaTeX picture
\setlength{\unitlength}{0.240900pt}
\ifx\plotpoint\undefined\newsavebox{\plotpoint}\fi
\sbox{\plotpoint}{\rule[-0.200pt]{0.400pt}{0.400pt}}%
\begin{picture}(1200,900)(0,0)
\font\gnuplot=cmr10 at 10pt
\gnuplot
\sbox{\plotpoint}{\rule[-0.200pt]{0.400pt}{0.400pt}}%
\put(221.0,163.0){\rule[-0.200pt]{4.818pt}{0.400pt}}
\put(201,163){\makebox(0,0)[r]{0}}
\put(1160.0,163.0){\rule[-0.200pt]{4.818pt}{0.400pt}}
\put(221.0,511.0){\rule[-0.200pt]{4.818pt}{0.400pt}}
\put(201,511){\makebox(0,0)[r]{0.025}}
\put(1160.0,511.0){\rule[-0.200pt]{4.818pt}{0.400pt}}
\put(221.0,859.0){\rule[-0.200pt]{4.818pt}{0.400pt}}
\put(201,859){\makebox(0,0)[r]{0.05}}
\put(1160.0,859.0){\rule[-0.200pt]{4.818pt}{0.400pt}}
\put(221.0,163.0){\rule[-0.200pt]{0.400pt}{4.818pt}}
\put(221,122){\makebox(0,0){0}}
\put(221.0,839.0){\rule[-0.200pt]{0.400pt}{4.818pt}}
\put(395.0,163.0){\rule[-0.200pt]{0.400pt}{4.818pt}}
\put(395,122){\makebox(0,0){20}}
\put(395.0,839.0){\rule[-0.200pt]{0.400pt}{4.818pt}}
\put(570.0,163.0){\rule[-0.200pt]{0.400pt}{4.818pt}}
\put(570,122){\makebox(0,0){40}}
\put(570.0,839.0){\rule[-0.200pt]{0.400pt}{4.818pt}}
\put(744.0,163.0){\rule[-0.200pt]{0.400pt}{4.818pt}}
\put(744,122){\makebox(0,0){60}}
\put(744.0,839.0){\rule[-0.200pt]{0.400pt}{4.818pt}}
\put(918.0,163.0){\rule[-0.200pt]{0.400pt}{4.818pt}}
\put(918,122){\makebox(0,0){80}}
\put(918.0,839.0){\rule[-0.200pt]{0.400pt}{4.818pt}}
\put(1093.0,163.0){\rule[-0.200pt]{0.400pt}{4.818pt}}
\put(1093,122){\makebox(0,0){100}}
\put(1093.0,839.0){\rule[-0.200pt]{0.400pt}{4.818pt}}
\put(221.0,163.0){\rule[-0.200pt]{231.023pt}{0.400pt}}
\put(1180.0,163.0){\rule[-0.200pt]{0.400pt}{167.666pt}}
\put(221.0,859.0){\rule[-0.200pt]{231.023pt}{0.400pt}}
\put(164,911){\makebox(0,0){$\langle :H_1:\rangle/(c(q)r_B/a^2)$}}
\put(700,61){\makebox(0,0){$r_B/a$}}
\put(221.0,163.0){\rule[-0.200pt]{0.400pt}{167.666pt}}
\put(1029,720){\makebox(0,0)[r]{$q=0,n_B=1$}}
\put(308,543){\raisebox{-.8pt}{\makebox(0,0){$\Diamond$}}}
\put(395,541){\raisebox{-.8pt}{\makebox(0,0){$\Diamond$}}}
\put(483,541){\raisebox{-.8pt}{\makebox(0,0){$\Diamond$}}}
\put(570,541){\raisebox{-.8pt}{\makebox(0,0){$\Diamond$}}}
\put(657,541){\raisebox{-.8pt}{\makebox(0,0){$\Diamond$}}}
\put(744,541){\raisebox{-.8pt}{\makebox(0,0){$\Diamond$}}}
\put(831,541){\raisebox{-.8pt}{\makebox(0,0){$\Diamond$}}}
\put(918,541){\raisebox{-.8pt}{\makebox(0,0){$\Diamond$}}}
\put(1006,541){\raisebox{-.8pt}{\makebox(0,0){$\Diamond$}}}
\put(1093,541){\raisebox{-.8pt}{\makebox(0,0){$\Diamond$}}}
\put(1099,720){\raisebox{-.8pt}{\makebox(0,0){$\Diamond$}}}
\put(1029,679){\makebox(0,0)[r]{$q=0.9,n_B=1$}}
\put(308,542){\makebox(0,0){$+$}}
\put(395,541){\makebox(0,0){$+$}}
\put(483,541){\makebox(0,0){$+$}}
\put(570,541){\makebox(0,0){$+$}}
\put(657,541){\makebox(0,0){$+$}}
\put(744,541){\makebox(0,0){$+$}}
\put(831,541){\makebox(0,0){$+$}}
\put(918,541){\makebox(0,0){$+$}}
\put(1006,541){\makebox(0,0){$+$}}
\put(1093,541){\makebox(0,0){$+$}}
\put(1099,679){\makebox(0,0){$+$}}
\sbox{\plotpoint}{\rule[-0.500pt]{1.000pt}{1.000pt}}%
\put(1029,638){\makebox(0,0)[r]{$q=0.999,n_B=1$}}
\put(308,541){\raisebox{-.8pt}{\makebox(0,0){$\Box$}}}
\put(395,541){\raisebox{-.8pt}{\makebox(0,0){$\Box$}}}
\put(483,541){\raisebox{-.8pt}{\makebox(0,0){$\Box$}}}
\put(570,541){\raisebox{-.8pt}{\makebox(0,0){$\Box$}}}
\put(657,541){\raisebox{-.8pt}{\makebox(0,0){$\Box$}}}
\put(744,541){\raisebox{-.8pt}{\makebox(0,0){$\Box$}}}
\put(831,541){\raisebox{-.8pt}{\makebox(0,0){$\Box$}}}
\put(918,541){\raisebox{-.8pt}{\makebox(0,0){$\Box$}}}
\put(1006,541){\raisebox{-.8pt}{\makebox(0,0){$\Box$}}}
\put(1093,541){\raisebox{-.8pt}{\makebox(0,0){$\Box$}}}
\put(1099,638){\raisebox{-.8pt}{\makebox(0,0){$\Box$}}}
\end{picture}
 \end{center}
 \caption{
The numerical evaluations of $\langle :H_1:\rangle$ for the
discretized theory of the scalar field in Reissner-Nordstr{\" o}m
spacetime. $\langle :H_1:\rangle/(c(q)r_B/a^2)$ for $q=0,0.9,0.999$ is
shown as functions of $r_B/a$, where $r_B\equiv r(\rho =n_Ba)$. We
have taken $n_{tot}=100$ and performed the summation over $l$ up to
$10r_0/a$. 
}
 \label{fig:H1_RN}
\end{figure}
%======================================%

%============< FIGURE >==============%
%              H2_RN.tex
\begin{figure}
 \begin{center}
	% GNUPLOT: LaTeX picture
\setlength{\unitlength}{0.240900pt}
\ifx\plotpoint\undefined\newsavebox{\plotpoint}\fi
\sbox{\plotpoint}{\rule[-0.200pt]{0.400pt}{0.400pt}}%
\begin{picture}(1200,900)(0,0)
\font\gnuplot=cmr10 at 10pt
\gnuplot
\sbox{\plotpoint}{\rule[-0.200pt]{0.400pt}{0.400pt}}%
\put(221.0,163.0){\rule[-0.200pt]{4.818pt}{0.400pt}}
\put(201,163){\makebox(0,0)[r]{0.05}}
\put(1160.0,163.0){\rule[-0.200pt]{4.818pt}{0.400pt}}
\put(221.0,511.0){\rule[-0.200pt]{4.818pt}{0.400pt}}
\put(201,511){\makebox(0,0)[r]{0.075}}
\put(1160.0,511.0){\rule[-0.200pt]{4.818pt}{0.400pt}}
\put(221.0,859.0){\rule[-0.200pt]{4.818pt}{0.400pt}}
\put(201,859){\makebox(0,0)[r]{0.1}}
\put(1160.0,859.0){\rule[-0.200pt]{4.818pt}{0.400pt}}
\put(221.0,163.0){\rule[-0.200pt]{0.400pt}{4.818pt}}
\put(221,122){\makebox(0,0){0}}
\put(221.0,839.0){\rule[-0.200pt]{0.400pt}{4.818pt}}
\put(395.0,163.0){\rule[-0.200pt]{0.400pt}{4.818pt}}
\put(395,122){\makebox(0,0){20}}
\put(395.0,839.0){\rule[-0.200pt]{0.400pt}{4.818pt}}
\put(570.0,163.0){\rule[-0.200pt]{0.400pt}{4.818pt}}
\put(570,122){\makebox(0,0){40}}
\put(570.0,839.0){\rule[-0.200pt]{0.400pt}{4.818pt}}
\put(744.0,163.0){\rule[-0.200pt]{0.400pt}{4.818pt}}
\put(744,122){\makebox(0,0){60}}
\put(744.0,839.0){\rule[-0.200pt]{0.400pt}{4.818pt}}
\put(918.0,163.0){\rule[-0.200pt]{0.400pt}{4.818pt}}
\put(918,122){\makebox(0,0){80}}
\put(918.0,839.0){\rule[-0.200pt]{0.400pt}{4.818pt}}
\put(1093.0,163.0){\rule[-0.200pt]{0.400pt}{4.818pt}}
\put(1093,122){\makebox(0,0){100}}
\put(1093.0,839.0){\rule[-0.200pt]{0.400pt}{4.818pt}}
\put(221.0,163.0){\rule[-0.200pt]{231.023pt}{0.400pt}}
\put(1180.0,163.0){\rule[-0.200pt]{0.400pt}{167.666pt}}
\put(221.0,859.0){\rule[-0.200pt]{231.023pt}{0.400pt}}
\put(164,911){\makebox(0,0){$\langle :H_2:\rangle/(c(q)r_B/a^2)$}}
\put(700,61){\makebox(0,0){$r_B/a$}}
\put(221.0,163.0){\rule[-0.200pt]{0.400pt}{167.666pt}}
\put(1029,720){\makebox(0,0)[r]{$q=0,n_B=1$}}
\put(308,414){\raisebox{-.8pt}{\makebox(0,0){$\Diamond$}}}
\put(395,414){\raisebox{-.8pt}{\makebox(0,0){$\Diamond$}}}
\put(483,414){\raisebox{-.8pt}{\makebox(0,0){$\Diamond$}}}
\put(570,414){\raisebox{-.8pt}{\makebox(0,0){$\Diamond$}}}
\put(657,414){\raisebox{-.8pt}{\makebox(0,0){$\Diamond$}}}
\put(744,414){\raisebox{-.8pt}{\makebox(0,0){$\Diamond$}}}
\put(831,414){\raisebox{-.8pt}{\makebox(0,0){$\Diamond$}}}
\put(918,414){\raisebox{-.8pt}{\makebox(0,0){$\Diamond$}}}
\put(1006,414){\raisebox{-.8pt}{\makebox(0,0){$\Diamond$}}}
\put(1093,414){\raisebox{-.8pt}{\makebox(0,0){$\Diamond$}}}
\put(1099,720){\raisebox{-.8pt}{\makebox(0,0){$\Diamond$}}}
\put(1029,679){\makebox(0,0)[r]{$q=0.9,n_B=1$}}
\put(308,415){\makebox(0,0){$+$}}
\put(395,414){\makebox(0,0){$+$}}
\put(483,414){\makebox(0,0){$+$}}
\put(570,414){\makebox(0,0){$+$}}
\put(657,414){\makebox(0,0){$+$}}
\put(744,414){\makebox(0,0){$+$}}
\put(831,414){\makebox(0,0){$+$}}
\put(918,414){\makebox(0,0){$+$}}
\put(1006,414){\makebox(0,0){$+$}}
\put(1093,414){\makebox(0,0){$+$}}
\put(1099,679){\makebox(0,0){$+$}}
\sbox{\plotpoint}{\rule[-0.500pt]{1.000pt}{1.000pt}}%
\put(1029,638){\makebox(0,0)[r]{$q=0.999,n_B=1$}}
\put(308,416){\raisebox{-.8pt}{\makebox(0,0){$\Box$}}}
\put(395,415){\raisebox{-.8pt}{\makebox(0,0){$\Box$}}}
\put(483,414){\raisebox{-.8pt}{\makebox(0,0){$\Box$}}}
\put(570,414){\raisebox{-.8pt}{\makebox(0,0){$\Box$}}}
\put(657,414){\raisebox{-.8pt}{\makebox(0,0){$\Box$}}}
\put(744,414){\raisebox{-.8pt}{\makebox(0,0){$\Box$}}}
\put(831,414){\raisebox{-.8pt}{\makebox(0,0){$\Box$}}}
\put(918,414){\raisebox{-.8pt}{\makebox(0,0){$\Box$}}}
\put(1006,414){\raisebox{-.8pt}{\makebox(0,0){$\Box$}}}
\put(1093,414){\raisebox{-.8pt}{\makebox(0,0){$\Box$}}}
\put(1099,638){\raisebox{-.8pt}{\makebox(0,0){$\Box$}}}
\end{picture}
 \end{center}
 \caption{
The numerical evaluations of $\langle :H_2:\rangle$ for the
discretized theory of the scalar field in Reissner-Nordstr{\" o}m
spacetime. $\langle :H_2:\rangle/(c(q)r_B/a^2)$ for $q=0,0.9,0.999$ is shown
as functions of $r_B/a$, where $r_B\equiv r(\rho =n_Ba)$. We have
taken $n_{tot}=100$ and performed the summation over $l$ up to
$10r_0/a$. 
}
 \label{fig:H2_RN}
\end{figure}
%======================================%

%============< FIGURE >==============%
%              Htot_RN.tex
\begin{figure}
 \begin{center}
	% GNUPLOT: LaTeX picture
\setlength{\unitlength}{0.240900pt}
\ifx\plotpoint\undefined\newsavebox{\plotpoint}\fi
\sbox{\plotpoint}{\rule[-0.200pt]{0.400pt}{0.400pt}}%
\begin{picture}(1200,900)(0,0)
\font\gnuplot=cmr10 at 10pt
\gnuplot
\sbox{\plotpoint}{\rule[-0.200pt]{0.400pt}{0.400pt}}%
\put(181.0,163.0){\rule[-0.200pt]{4.818pt}{0.400pt}}
\put(161,163){\makebox(0,0)[r]{0.1}}
\put(1160.0,163.0){\rule[-0.200pt]{4.818pt}{0.400pt}}
\put(181.0,511.0){\rule[-0.200pt]{4.818pt}{0.400pt}}
\put(161,511){\makebox(0,0)[r]{0.2}}
\put(1160.0,511.0){\rule[-0.200pt]{4.818pt}{0.400pt}}
\put(181.0,859.0){\rule[-0.200pt]{4.818pt}{0.400pt}}
\put(161,859){\makebox(0,0)[r]{0.3}}
\put(1160.0,859.0){\rule[-0.200pt]{4.818pt}{0.400pt}}
\put(181.0,163.0){\rule[-0.200pt]{0.400pt}{4.818pt}}
\put(181,122){\makebox(0,0){0}}
\put(181.0,839.0){\rule[-0.200pt]{0.400pt}{4.818pt}}
\put(363.0,163.0){\rule[-0.200pt]{0.400pt}{4.818pt}}
\put(363,122){\makebox(0,0){20}}
\put(363.0,839.0){\rule[-0.200pt]{0.400pt}{4.818pt}}
\put(544.0,163.0){\rule[-0.200pt]{0.400pt}{4.818pt}}
\put(544,122){\makebox(0,0){40}}
\put(544.0,839.0){\rule[-0.200pt]{0.400pt}{4.818pt}}
\put(726.0,163.0){\rule[-0.200pt]{0.400pt}{4.818pt}}
\put(726,122){\makebox(0,0){60}}
\put(726.0,839.0){\rule[-0.200pt]{0.400pt}{4.818pt}}
\put(908.0,163.0){\rule[-0.200pt]{0.400pt}{4.818pt}}
\put(908,122){\makebox(0,0){80}}
\put(908.0,839.0){\rule[-0.200pt]{0.400pt}{4.818pt}}
\put(1089.0,163.0){\rule[-0.200pt]{0.400pt}{4.818pt}}
\put(1089,122){\makebox(0,0){100}}
\put(1089.0,839.0){\rule[-0.200pt]{0.400pt}{4.818pt}}
\put(181.0,163.0){\rule[-0.200pt]{240.659pt}{0.400pt}}
\put(1180.0,163.0){\rule[-0.200pt]{0.400pt}{167.666pt}}
\put(181.0,859.0){\rule[-0.200pt]{240.659pt}{0.400pt}}
\put(164,911){\makebox(0,0){$\langle :H_{tot}:\rangle_{\rho'}/(c(q)r_B/a^2)$}}
\put(680,61){\makebox(0,0){$r_B/a$}}
\put(181.0,163.0){\rule[-0.200pt]{0.400pt}{167.666pt}}
\put(1024,720){\makebox(0,0)[r]{$q=0,n_B=1$}}
\put(272,447){\raisebox{-.8pt}{\makebox(0,0){$\Diamond$}}}
\put(363,446){\raisebox{-.8pt}{\makebox(0,0){$\Diamond$}}}
\put(454,445){\raisebox{-.8pt}{\makebox(0,0){$\Diamond$}}}
\put(544,445){\raisebox{-.8pt}{\makebox(0,0){$\Diamond$}}}
\put(635,445){\raisebox{-.8pt}{\makebox(0,0){$\Diamond$}}}
\put(726,445){\raisebox{-.8pt}{\makebox(0,0){$\Diamond$}}}
\put(817,445){\raisebox{-.8pt}{\makebox(0,0){$\Diamond$}}}
\put(908,445){\raisebox{-.8pt}{\makebox(0,0){$\Diamond$}}}
\put(998,445){\raisebox{-.8pt}{\makebox(0,0){$\Diamond$}}}
\put(1089,445){\raisebox{-.8pt}{\makebox(0,0){$\Diamond$}}}
\put(1094,720){\raisebox{-.8pt}{\makebox(0,0){$\Diamond$}}}
\put(1024,679){\makebox(0,0)[r]{$q=0.9,n_B=1$}}
\put(272,447){\makebox(0,0){$+$}}
\put(363,446){\makebox(0,0){$+$}}
\put(454,445){\makebox(0,0){$+$}}
\put(544,445){\makebox(0,0){$+$}}
\put(635,445){\makebox(0,0){$+$}}
\put(726,445){\makebox(0,0){$+$}}
\put(817,445){\makebox(0,0){$+$}}
\put(908,445){\makebox(0,0){$+$}}
\put(998,445){\makebox(0,0){$+$}}
\put(1089,445){\makebox(0,0){$+$}}
\put(1094,679){\makebox(0,0){$+$}}
\sbox{\plotpoint}{\rule[-0.500pt]{1.000pt}{1.000pt}}%
\put(1024,638){\makebox(0,0)[r]{$q=0.999,n_B=1$}}
\put(272,446){\raisebox{-.8pt}{\makebox(0,0){$\Box$}}}
\put(363,446){\raisebox{-.8pt}{\makebox(0,0){$\Box$}}}
\put(453,445){\raisebox{-.8pt}{\makebox(0,0){$\Box$}}}
\put(544,445){\raisebox{-.8pt}{\makebox(0,0){$\Box$}}}
\put(635,445){\raisebox{-.8pt}{\makebox(0,0){$\Box$}}}
\put(726,445){\raisebox{-.8pt}{\makebox(0,0){$\Box$}}}
\put(817,445){\raisebox{-.8pt}{\makebox(0,0){$\Box$}}}
\put(908,445){\raisebox{-.8pt}{\makebox(0,0){$\Box$}}}
\put(998,445){\raisebox{-.8pt}{\makebox(0,0){$\Box$}}}
\put(1089,445){\raisebox{-.8pt}{\makebox(0,0){$\Box$}}}
\put(1094,638){\raisebox{-.8pt}{\makebox(0,0){$\Box$}}}
\end{picture}
 \end{center}
\caption{
The numerical evaluations of $\langle :H_{tot}:\rangle_{\rho'}$ for
the discretized theory of the scalar field in Reissner-Nordstr{\" o}m
spacetime. $\langle :H_{tot}:\rangle_{\rho'}/(c(q)r_B/a^2)$ for
$q=0,0.9,0.999$ is shown as functions of $r_B/a$, where 
$r_B\equiv r(\rho =n_Ba)$. We have taken $n_{tot}=100$ and performed
the summation over $l$ up to $10r_0/a$. 
}
\label{fig:Htot_RN}
\end{figure}
%======================================%

Next, by using formulas in subsection \ref{subsec:Eent}, we have
numerically evaluated 
$\langle :H_1:\rangle$, $\langle :H_2:\rangle$ and 
$\langle :H_{tot}:\rangle_{\rho'}$. 
We have taken the numerical outer boundary at
$n_{tot}=100$ and the boundary of the spatial division at $n_B=1$. The
truncation in the $l$-summation is the same as for 
$S_{ent}$ (up to $l=[10r_0/a]$), which implies that the accuracy is
about $10\%$ from the above asymptotic estimate for $E^{(l)}_{ent}$.

The results of numerical calculations are shown in {\it Figure}
\ref{fig:H1_RN}, {\it Figure} \ref{fig:H2_RN} and {\it Figure}
\ref{fig:Htot_RN}. In these figures, $E_{ent}/(c(q)r_B/a^2)$ is
plotted as a function of $r_B/a$ for $n_B=1$ and $q=0,0.9,0.999$,
where $c(q)$ is defined by
%============< EQUATION >==============%
%
\begin{equation}
 c(q) = \frac{2\sqrt{1-q^2}}{1+\sqrt{1-q^2}}.
	\label{eqn:3-3:c(q)}
\end{equation}
%======================================%
All of these figures show that $E_{ent}$ is proportional
to $c(q)r_B/a^2$. The result is summarized as
%============< EQUATION >==============%
%
\begin{eqnarray}
 \langle :H_1:\rangle & \simeq & 0.025 c(q)r_B/a^2,\nonumber\\
 \langle :H_2:\rangle & \simeq & 0.075 c(q)r_B/a^2,\nonumber\\
 \langle :H_{tot}:\rangle_{\rho'} & \simeq & 0.2 c(q)r_B/a^2.
\end{eqnarray}
%======================================%

%======================================%
%<<<<<<   SUBSECTION 3-2-4    >>>>>>>>>%
%======================================%
\subsection{Comparison: entanglement thermodynamics and black-hole
thermodynamics} 
	\label{subsec:comparison}

%<<<<<<   Minkowski    >>>>>>>>>%
\subsubsection{Entanglement thermodynamics in Minkowski spacetime}

We have introduced four possible definitions of entanglement 
energy. By combining each of them  with the entanglement entropy 
$S_{ent}$, we obtain entanglement thermodynamics.

For this purpose, we consider an infinitesimal process in which 
the way of the division of the Hilbert space $\cal H$ into ${\cal
H}_1$ and ${\cal H}_2$ is changed smoothly, with the `initial' state
$\rho_0$ being fixed. (See subsection \ref{subsection:Senta}.) 
Let $\delta S_{ent}$ and $\delta E_{ent}$ be the  resultant
infinitesimal changes in  the entanglement entropy and in  the
entanglement energy, respectively. We  are dealing  with a 1-parameter
family of the infinitesimal changes for the entanglement. The
parameter is chosen to be the area radius of the boundary sphere. 
Thus, the construction of  the thermodynamics means to use the first
law of entanglement thermodynamics (\ref{eqn:3-3:1st-law-ent})
to determine $T_{ent}$, which we call entanglement temperature. 
Combining (\ref{eqn:3-3:Sent-result}) and (\ref{eqn:3-3:Eent_Min})
with Eq.(\ref{eqn:3-3:1st-law-ent}), we thus get~\footnote{
In this subsection, we recover $\hbar$ and $c$.
}  
%============< EQUATION >==============%
%
\begin{equation}
 k_B T_{ent} = \frac{{\cal N}_E}{{\cal N}_S}\cdot\frac{\hbar c}{a},
 \label{eqn:3-3:Tent}
\end{equation}
%======================================%
where ${\cal N}_E$ is a numerical factor in
Eq.~(\ref{eqn:3-3:Eent_Min}). 
Note that entanglement temperatures $T_{ent}$ obtained from the four
definitions of the entanglement energy coincides up to numerical
factors of order unity.

Let us interpret entanglement thermodynamics given by 
(\ref{eqn:3-3:Sent-result}), (\ref{eqn:3-3:Eent_Min}) and
(\ref{eqn:3-3:Tent}). 
It is helpful  to introduce the quantities
%============< EQUATION >==============%
%
\begin{eqnarray}
 n_{ent} & \equiv & \left(\frac{r_B}{a}\right)^2,\nonumber  \\
 e_{ent} & \equiv & \frac{\hbar c}{a}.
\label{eqn:3-3:unitenergy-ent}
\end{eqnarray}
%======================================%
Here $n_{ent}$ is regarded as an effective 
number of degrees of freedom of matter on the boundary $2$-surface
$B$, and $e_{ent}$ is a typical energy scale of each degree of freedom
on $B$.  

From Eqs. (\ref{eqn:3-3:Sent-result}), (\ref{eqn:3-3:Eent_Min}) and
(\ref{eqn:3-3:Tent}), we find that 
%============< EQUATION >==============%
%
\begin{eqnarray}
 S_{ent}     & \sim & k_Bn_{ent}, \nonumber   \\
 E_{ent}     & \sim & e_{ent}n_{ent}, \nonumber   \\
 k_B T_{ent} & \sim & e_{ent}.
 \label{eqn:3-3:interpret-ent}
\end{eqnarray}
%======================================%
Therefore our results can be interpreted as follows\footnote{
It is safer, however, to regard such an interpretation just as a
convenient way of representing our results. 
(See the next section for another interpretation of entanglement
entropy. This note in particular applies to the case of the black-hole
thermodynamics (see Eq.(\ref{eqn:3-3:interpret-BH})).} 
: The entanglement entropy is a measure for the number of 
microscopic degrees of  freedom on the boundary $B$; 
the entanglement energy is a measure for  the 
total energy carried by all of the degrees of freedom on $B$; 
entanglement temperature is measure for the energy carried by each 
degree of freedom on $B$.

%<<<<<<   Discrepancy    >>>>>>>>>%
\subsubsection{Discrepancy between entanglement thermodynamics in
Minkowski spacetime and black-hole thermodynamics}
\label{subsection:discrepancy}

Now we compare these results with the case of black holes.
For that purpose we express the black-hole thermodynamics in the same
form as in the previous subsection.
   
Let us introduce the  quantities  
%============< EQUATION >==============%
%
\begin{eqnarray}
 n_{BH} & \equiv & \left(\frac{r_0}{l_{pl}}\right)^2, \nonumber	\\
 e_{BH} & \equiv & \frac{\hbar c}{l_{pl}}, 
\label{eqn:3-3:unitenergy-BH}
\end{eqnarray}
%======================================%
where $r_0$ is area radius of an event horizon of a black hole. 
We can interpret that $n_{BH}$ corresponds to the effective number of
degrees of freedom on the event horizon  and $e_{BH}$ is a typical
energy scale  for each degree of freedom of matter on the horizon.

The black-hole thermodynamics can be recast in terms of 
these quantities as 
%============< EQUATION >==============%
%
\begin{eqnarray}
 S_{BH} & \sim & k_B n_{BH}, \nonumber  \\
 E_{BH} & \sim & \gamma_{BH}e_{BH} n_{BH}, \nonumber  \\
 k_B T_{BH} & \sim & \gamma_{BH}e_{BH},
\label{eqn:3-3:interpret-BH}
\end{eqnarray}
%======================================%
where $\gamma_{BH}\equiv l_{pl}/r_0$. The factor $\gamma_{BH}$ can be 
understood as a magnification of energy due to an addition of
gravitational energy or a red-shift factor of temperature since
$\sqrt{-g_{tt}}\sim l_{pl}/r_0$ at $r\sim r_0+l_{pl}^2/r_0$, which
corresponds to a stationary observer at the proper distance $l_{pl}$
away from the horizon. Thus the following interpretation is
possible\footnote{
See the footnote after Eq.(\ref{eqn:3-3:interpret-ent}).
}: 
The black-hole entropy is a measure for the number of the
microscopic degrees of freedom  on the event horizon; 
the black-hole energy is a measure at infinity for the total energy
carried by all of the degrees of freedom on the event horizon; 
the black-hole temperature is a measure at infinity for the energy
carried by each degree of freedom.

Now we compare the two types of thermodynamics characterized by
Eq.(\ref{eqn:3-3:interpret-ent}) and Eq.(\ref{eqn:3-3:interpret-BH}),
respectively. Both of them allow the interpretation that they
describe the behavior of the effective microscopic degrees of freedom
on the boundary $B$, or on the horizon. Because of the factor
$\gamma_{BH}$, however, they are hardly understood in a unified
picture. This strongly suggests that an inclusion of gravitational
effects is necessary for agreement between them.

The discrepancy is highlighted in the context of the third law of
thermodynamics.  Both types of thermodynamics fail to follow Planck's
version of the third law~\footnote{Here we mention that the black hole 
system does satisfy the third law in the sense of
Nernst~\cite{BCH1973,Israel1986}.}, 
but in quite different manners.

From Eq.(\ref{eqn:3-3:Tent}), we see that $T_{ent}$ remains constant
if $\frac{{\cal N}_E}{{\cal N}_S}$ is assumed to be
constant~\footnote{
Here we are regarding the cut-off scale $a$ as the fundamental
constant of the theory, not to be varied. 
}.
On the other hand, Eq.(\ref{eqn:3-3:Sent-result}) shows that 
$S_{ent}$ tends to zero as $A \to 0$, where $A=4\pi r_B^2$ is area of
the boundary. Therefore we obtain the
following $A$-dependence:
%============< EQUATION >==============%
%
\begin{eqnarray}
S_{ent} & \propto & A, \nonumber \\
E_{ent} & \propto & A, \nonumber \\
k_B  T_{ent}   & \propto & A^0.
\label{eqn:3-3:ent-A}
\end{eqnarray}     
%======================================%
The system  behaves  as though it is kept in touch  with 
a thermal bath with temperature $T_{ent}$.

In contrast, for the black-hole thermodynamics,
Eq.(\ref{eqn:2-0:T-BH}) and Eq.(\ref{eqn:2-0:S-BH}) along with
Eq.(\ref{eqn:2-0:constants}) give the behavior (note that 
$A \propto M_{BH}^2$) 
%============< EQUATION >==============%
%
\begin{eqnarray}
S_{BH} & \propto & A, \nonumber \\
E_{BH} & \propto & \sqrt{A}, \nonumber \\
k_B  T_{BH}   & \propto & 1/ \sqrt{A},
\label{eqn:3-3:BH-A}
\end{eqnarray}     
%======================================%
where $A=4\pi r_0^2$ is area of the horizon. 
Thus we see that $S_{BH} \to \infty$ as $T_{BH} \to 0$.

The discrepancy between Eq.(\ref{eqn:3-3:ent-A}) and 
Eq.(\ref{eqn:3-3:BH-A}) is 
quite impressive. On one hand, 
a well-known behavior (\ref{eqn:3-3:BH-A}) comes from 
the fundamental properties of the black-hole physics. On the 
other hand,  
the behavior  characterized by Eq.(\ref{eqn:3-3:ent-A}) is also 
an  universal one  in any model of the entanglement: 
the zero-point energy of the system has been subtracted in the
definitions of entanglement energy, thus 
only the degrees of freedom on the boundary $B$ contributes 
to $E_{ent}$, yielding the behavior $E_{ent} \propto A$. 
The definitions of $E_{ent}$ proposed here 
look quite reasonable though other definitions 
may  be possible. The result $E_{ent} \propto A$  also looks 
natural, being compatible with the concept of `entanglement'.
At the same time,  $S_{ent}$ also  behaves  universally as 
$S_{ent} \propto A$, which has been  the original motivation for
investigating the relation between  $S_{BH}$ and
$S_{ent}$~\cite{BKLS1986,Srednicki1993,Frolov&Novikov1993}.

%<<<<<<   Restoration    >>>>>>>>>%
\subsubsection{Restoration of the agreement by inclusion of gravity}

Let us discuss a possible restoration of the
agreement between entanglement thermodynamics and black-hole
thermodynamics by considering gravitational effects.

Although the behavior (\ref{eqn:3-3:interpret-ent}) of entanglement
thermodynamics was derived by considering models
in flat spacetime, it seems very reasonable that we regard the
quantities $S_{ent}$, $E_{ent}$ and $T_{ent}$ as those in a black-hole
background measured by a stationary observer located at the proper
distance $a$ away from the horizon \footnote{
The author thanks Professor T. Jacobson for helpful comments on this
point. 
}.
Since $S_{BH}$, $E_{BH}$ and $T_{BH}$ in (\ref{eqn:3-3:interpret-BH}) are
quantities measured at infinity, it is behavior of $S_{ent}$,
$E_{ent}$ and $T_{ent}$ at infinity that we have to compare with
(\ref{eqn:3-3:interpret-BH}).

$S_{ent}$ at infinity probably has the same
behavior as that measured by the observer near the horizon since a
number of degrees of freedom seems independent of an observer's
view-point. That is consistent with the fact that the entanglement
entropy on Schwarzschild background has the same behavior 
$S_{ent}\sim k_Bn_{ent}$ \cite{Frolov&Novikov1993,MSK1998}. On the
other hand it seems 
natural to add the gravitational energy to the entanglement energy by
replacing $E_{ent}$ with $\sqrt{-g_{tt}}E_{ent}$. Then 
entanglement temperature is determined by use of the first law
(\ref{eqn:3-3:1st-law-ent}). Thus the inclusion of gravity may alter the 
behavior (\ref{eqn:3-3:interpret-ent}) to 
%============< EQUATION >==============%
%
\begin{eqnarray}
 S_{ent}     & \sim & k_Bn_{ent} \ \ \ , \nonumber   \\
 E_{ent}     & \sim & \gamma_{ent}e_{ent}n_{ent} \ \ \ , \nonumber   \\
 k_B T_{ent} & \sim & \gamma_{ent}e_{ent} \ \ \ ,
	\label{re-interpret-ent}
\end{eqnarray}
%======================================%
where $\gamma_{ent}\equiv a/r_0$. The factor $\gamma_{ent}$ represents
the gravitational magnification of the entanglement energy due to the 
addition of gravitational energy since on the corresponding
Schwarzschild background $\sqrt{-g_{tt}}\sim a/r_0$ at $r\sim
r_0+a^2/r_0$, which corresponds to a stationary observer at the proper
distance $a$ away from the horizon (see the argument below
(\ref{eqn:3-3:interpret-BH})). Here $r_0$ is the area radius of the
horizon.

The revised behavior (\ref{re-interpret-ent}) shows a
complete agreement with (\ref{eqn:3-3:interpret-BH}), provided that
$r_B\simeq r_0$ and $a\simeq l_{pl}$. Note that the last equality in 
(\ref{re-interpret-ent}) is consistent with an interpretation that the
entanglement temperature is red-shifted by the factor
$\gamma_{ent}$. Thus the inclusion of gravitational effects restores
the agreement between the entanglement thermodynamics and the
black-hole thermodynamics at least qualitatively.

%<<<<<<   Schwarzschild    >>>>>>>>>%
\subsubsection{Entanglement thermodynamics in Schwarzschild spacetime}

From the numerical results in subsection \ref{subsec:evaluation}, 
entanglement entropy and entanglement energy in Schwarzschild
spacetime are expressed as
%============< EQUATION >==============%
%
\begin{eqnarray}
 S_{ent} & \simeq & k_B{\cal N}_S\left(\frac{r_{B}}{a}\right)^2,
	\nonumber\\
 E_{ent} & \simeq & \hbar c{\cal N}_E\frac{r_{B}}{a^2},
\end{eqnarray}
%======================================%
for all definitions of $E_{ent}$, 
where $r_{B}$ is the area radius of the boundary defined by
$r_{B}=r(\rho=n_{B}a)$. 
Here ${\cal N}_S$ and ${\cal N}_E$ are numerical factors of order $1$:
${\cal N}_S=0.3$; 
${\cal N}_E=0.05(n_B-1/2)$, $0.05(n_B+1/2)$, $0.2n_B$ for 
$\langle :H_1:\rangle$, $\langle :H_2:\rangle$ and 
$\langle :H_{tot}:\rangle_{\rho'}$, respectively.

From these expressions and the first law of entanglement
thermodynamics (\ref{eqn:3-3:1st-law-ent}),
entanglement temperature $T_{ent}$ is determined as
%============< EQUATION >==============%
%
\begin{equation}
 T_{ent} \simeq {\cal N}_T  T_{BH},
\end{equation}
%
%======================================%
where $T_{BH}$ is the Hawking temperature of the background geometry
and ${\cal N}_T$ is a constant defined by 
${\cal N}_T=2\pi{\cal N}_E/{\cal N}_S$. 
Numerical values of ${\cal N}_T$ is ${\cal N}_T=(n_B-1/2)\pi/3$,
$(n_B+1/2)\pi/3$, $4n_B\pi/3$ for $\langle :H_1:\rangle$, 
$\langle :H_2:\rangle$ and $\langle :H_{tot}:\rangle_{\rho'}$,
respectively.

These results have several interesting features. First of all we
immediately see that the entanglement thermodynamics on the
Schwarzschild spacetime shows exactly the same behavior as the black 
hole thermodynamics. This behavior is just what we expected
from the above intuitive argument on gravitational effects: 
the gravitational redshift effect modifies the area dependence of
$E_{ent}$ so as to make the entanglement thermodynamics behave just
like the black hole thermodynamics.

Second it should be noted that the temperature $T_{ent}$ becomes
independent of the cut-off scale $a$ for all definitions of
entanglement energy.

It is also suggestive that the average of entanglement temperatures
for $\langle :H_1:\rangle$ and $\langle :H_2:\rangle$ with $n_B=1$ 
gives almost the same value as $T_{BH}$. This averaging corresponds to
averaging out the difference in the red-shift factors for the one-mesh
`inside' and the one-mesh `outside' of the boundary.  Therefore such
an averaging may have some meaning.

%<<<<<<   R-N    >>>>>>>>>%
\subsubsection{Entanglement thermodynamics in Reissner-Nordstr{\" o}m
spacetime} 

From the numerical results in subsection \ref{subsec:evaluation}, 
entanglement entropy and entanglement energy in 
Reissner-Nordstr{\" o}m spacetime are expressed as
%============< EQUATION >==============%
%
\begin{eqnarray}
 S_{ent} & \simeq & 
	k_B{\cal N}_S \left(\frac{r_{B}}{a}\right)^2,
	\nonumber\\
 E_{ent} & \simeq & \hbar c{\cal N}_E c(q)\frac{r_{B}}{a^2},
\end{eqnarray}
%======================================%
for all definitions of $E_{ent}$, 
where $r_{B}$ is the area radius of the boundary defined by
$r_{B}=r(\rho=n_{B}a)$. 
Here ${\cal N}_S$ and ${\cal N}_E$ are numerical factors of order $1$, 
whose values are the same as those for Schwarzschild spacetime, 
and the coefficient $c(q)$ is a function of $q=Q/M$ given by
Eq.~(\ref{eqn:3-3:c(q)}).
The coefficient $c(q)$ approaches to zero in the limit $q\to 1$.

Combining these expressions with the first law of entanglement
thermodynamics (\ref{eqn:3-3:1st-law-ent}), we obtain entanglement
temperature $T_{ent}$ as 
%============< EQUATION >==============%
%
\begin{equation}
 T_{ent} \simeq {\cal N}_T T_{BH},
\end{equation}
%
%======================================%
where $T_{BH}$ is the Hawking temperature of the background geometry
and ${\cal N}_T$ is a constant defined by 
${\cal N}_T=2\pi{\cal N}_E/{\cal N}_S$.

Note that both $T_{BH}$ and $T_{ent}$ become zero in the extremal
limit ($q\to 1$) of the background spacetime. Therefore as in the case
of the Schwarzschild spacetime, the entanglement thermodynamics in the 
Reissner-Nordstr{\"o}m spacetime has the same structure as that of the
black-hole thermodynamics.

%======================================%
%<<<<<<   SUBSECTION 3-3-5    >>>>>>>>>%
%======================================%
\subsection{Concluding remark}
	\label{subsec:remark_entanglement}

In this section we have constructed entanglement thermodynamics for a 
massless scalar field in Minkowski, Schwarzschild and 
Reissner-Nordstr{\"o}m spacetimes. The entanglement thermodynamics in
Minkowski spacetime differs significantly from black-hole
thermodynamics. On the contrary, the entanglement thermodynamics in
Schwarzschild and  Reissner-Nordstr{\"o}m spacetimes has the same
structure as that of black-hole thermodynamics. In particular, it has
been shown that entanglement temperature in the Reissner-Nordstr{\"o}m
spacetime approaches zero in  the extremal limit.

Our model analysis strongly suggests a tight connection
between the entanglement thermodynamics and the black hole
thermodynamics. Of course, our model is too simple to give any
definite conclusion based on it. In particular, the ambiguity in
the definition of the energy comes from neglecting
backreaction of the quantum field on gravity.

Finally we comment on possible extensions of the entanglement
thermodynamics. The first is the inclusion of a charged field as matter. 
In particular, it will be valuable to analyze the entanglement 
thermodynamics for a charged field in Reissner-Nordstr{\"o}m
spacetime. In this case we may be able to define the entanglement
charge as an expectation value of the charge of the field for the
coarse-grained state. The second is a generalization
to non-spherically-symmetric spacetimes. For example it is expected
that construction of entanglement thermodynamics in Kerr spacetime
requires the introduction of a concept of an entanglement 
angular-momentum.

%%%%%%%%%%%%%%%%%%%%%%%%%%%%%%%%%%%%%%%%
%%%%%%%%%%%% SECTION 3-4 %%%%%%%%%%%%%%%
%%%%%%%%%%%%%%%%%%%%%%%%%%%%%%%%%%%%%%%%
\section{A new interpretation of entanglement entropy}
	\label{sec:interpretation}

In this section a new interpretation of entanglement entropy is
proposed:
entanglement entropy of a pure state with respect to a division of a 
Hilbert space into two subspaces $1$ and $2$ is an amount of
information, which can be transmitted through $1$ and $2$ from a
system interacting with $1$ to another system interacting with
$2$. The transmission medium is quantum entanglement between $1$
and $2$. In order to support the interpretation, suggestive arguments
are given: variational principles in entanglement thermodynamics and
quantum teleportation. It is shown that a quantum state having 
maximal entanglement entropy plays an important role in quantum
teleportation. Hence, the entanglement entropy is, in some sense, an
index of efficiency of quantum teleportation.

In subsection \ref{subsec:entropies}, based on a relation between the 
entanglement entropy and so-called conditional entropy, we propose an
interpretation of the entanglement entropy. In subsection
\ref{subsec:variational} variational principles in entanglement 
thermodynamics are used to determine quantum states. In particular, a 
state having maximal entanglement entropy is determined and is used 
in subsection \ref{subsec:QT} to transmit information about an unknown
quantum state. Subsection \ref{subsec:remark_interpretation} is
devoted to a summary of this section and to discuss implications for
the information loss problem and Hawking radiation.

%======================================%
%<<<<<<   SUBSECTION 4-3-1    >>>>>>>>>%
%======================================%
\subsection{Conditional entropy and entanglement entropy}
	\label{subsec:entropies}

Entropy plays important roles not only in statistical mechanics but
also in information theory. In the latter, 
entropy of a random experiment, each of whose outcomes has an 
attached probability, represents uncertainty about the outcome before 
performing the experiment~\cite{Jumarie}. Besides the well-known 
Shannon entropy, there exist various definitions of entropies in 
information theory. For example, the so-called conditional 
entropy of an experiment $A$ on another experiment $B$ is defined by 
$H(A|B) = -\sum_{a,b}p(a,b)\ln p(a|b)$, where $a$ and $b$ represent 
outcomes of $A$ and $B$, respectively, 
$p(a,b)$ is a joint probability of $a$ and $b$, and $p(a|b)=p(a,b)/p(b)$ 
is a conditional probability of $a$ on $b$. Here $p(b)$ is a probability 
of $b$. The conditional entropy corresponds to an uncertainty about the
outcome of $A$ after the experiment $B$ is done. In other words it can 
be regarded as the amount of information about $A$ which cannot be 
known from the experiment $B$. The quantum analogue of the conditional 
entropy was considered in references~\cite{Cerf&Adami1,Cerf&Adami2}
and is called the von Neumann conditional entropy. Consider a Hilbert
space $\cal{F}$ of the form (\ref{eqn:3-3:F=F1*F2}) and let $\rho$ be a
density matrix on $\cal{F}$. The von Neumann conditional entropy of
$\rho$ about the subsystem $1$ on the subsystem $2$ is defined by 
%============< EQUATION >==============%
%
\begin{equation}
	S_{1|2} = {\bf Tr}\left[\rho\sigma_{1|2}\right],
\end{equation}
%======================================%
where $\sigma_{1|2}={\bf 1}_{1}\otimes\ln\rho_{2}-\ln\rho$. The von 
Neumann conditional entropy $S_{2|1}$ of $\rho$ about the subsystem $2$ 
on the subsystem $1$ is defined in a similar way. It is expected that 
$S_{1|2}$ (or $S_{2|1}$) represents the amount of the information about 
the subsystem $1$ (or $2$) which cannot be known from $2$ (or 
$1$, respectively).

The von Neumann conditional entropy can be negative. In fact,
it is easy to see that 
%============< EQUATION >==============%
%
\begin{equation}
	S_{1|2} = S_{2|1} = - S_{ent}, 
\end{equation}
%======================================%
if $\rho$ is a pure state. Hence, if $\rho$ is a pure state then
the conditional entropy is zero or negative. Our question now is `what
is the meaning of the negative 
conditional entropy of a pure state?' It might be expected that
$|S_{1|2}|$ ($=S_{ent}$) is the amount of the information about $1$
(or $2$) which can be known from $2$ (or $1$, respectively). 
However, this statement is not precise. A precise statement is that it 
is an amount of information, which can be transmitted through $1$ and
$2$ from a system interacting with $1$ to another system interacting
with $2$. The transmission medium is quantum entanglement
between $1$ and $2$.

The purpose of the remaining part of this section is to give
suggestive arguments for this statement.

%======================================%
%<<<<<<   SUBSECTION 4-3-2    >>>>>>>>>%
%======================================%
\subsection{Variational principles in entanglement thermodynamics}
	\label{subsec:variational}
	
In statistical mechanics, the von Neumann entropy is used to determine 
an equilibrium state: an equilibrium state of an 
isolated system is determined by maximizing the entropy. Thus, we
expect that the entanglement entropy may be used to determine a 
quantum state.

As an illustration we consider a simple system of two 
particles, each with spin $1/2$: we consider a Hilbert space 
$\cal{F}$ of the form (\ref{eqn:3-3:F=F1*F2}) and denote an orthonormal 
basis of ${\cal{F}}_{i}$ by 
$\left\{ |\uparrow\rangle_{i},|\downarrow\rangle_{i}\right\}$ 
($i=1,2$). Let $|\phi\rangle$ be an element of $\cal{F}$ with unit 
norm and expand it as 
%============< EQUATION >==============%
%
\begin{equation}
	|\phi\rangle = a|\uparrow\rangle_{1}\otimes|\uparrow\rangle_{2}
		+ b|\uparrow\rangle_{1}\otimes|\downarrow\rangle_{2}
		+ c|\downarrow\rangle_{1}\otimes|\uparrow\rangle_{2}
		+ d|\downarrow\rangle_{1}\otimes|\downarrow\rangle_{2},
		\label{eqn:3-4:stateinF}
\end{equation}
%======================================%
where $|a|^2+|b|^2+|c|^2+|d|^2=1$ is understood. The corresponding 
reduced density matrix is given by 
%============< EQUATION >==============%
%
\begin{eqnarray}
 \rho_{2} & = &
 	(|a|^2+|c|^2)|\uparrow\rangle_{2}{}_{2}\langle\uparrow|
 	+ (ab^*+cd^*)|\uparrow\rangle_{2}{}_{2}\langle\downarrow|
 	\nonumber\\
 & & + (a^*b+c^*d)|\downarrow\rangle_{2}{}_{2}\langle\uparrow|
 	+ (|b|^2+|d|^2)|\downarrow\rangle_{2}{}_{2}\langle\downarrow|
\end{eqnarray}
%======================================%
and the entanglement entropy can be easily calculated from it. The resulting 
expression for the entanglement entropy is 
%============< EQUATION >==============%
%
\begin{equation}
	S_{ent} = -\frac{1+x}{2}\ln\left(\frac{1+x}{2}\right)
		-\frac{1-x}{2}\ln\left(\frac{1-x}{2}\right),
\end{equation}
%======================================%
where $x=\sqrt{1-4|ad-bc|^2}$. By requiring $dS_{ent}/dx=0$ we 
obtain the condition $|ad-bc|=1/2$. Thus a state maximizing the 
entanglement entropy is
%============< EQUATION >==============%
%
\begin{equation}
	|\phi\rangle = \frac{1}{\sqrt{2}}\left(
		|\uparrow\rangle_{1}\otimes|\downarrow\rangle_{2}
		-|\downarrow\rangle_{1}\otimes|\uparrow\rangle_{2}\right)
\end{equation}
%======================================%
up to a unitary transformation in ${\cal{F}}_{1}$ and the 
corresponding maximal value of the entanglement entropy is $\ln 2$. 
This state is well known as the EPR state.

It is notable that the 
corresponding reduced density matrix $\rho_{2}$
represents the microcanonical ensemble. This fact is related to the
fact that the maximum of entropy gives the microcanonical ensemble in
statistical mechanics. Thus, in general, if ${\cal{F}}_{1}$ and
${\cal{F}}_{2}$ have the same finite dimension $N$ then a state
maximizing the entanglement entropy is written as
%============< EQUATION >==============%
%
\begin{equation}
	|\phi\rangle = \frac{1}{\sqrt{N}}\sum_{n=1}^N\left(
		|n\rangle_{1}\otimes|n\rangle_{2}\right)
		\label{eqn:3-4:EPRstate}
\end{equation}
%======================================%
up to a unitary transformation in ${\cal{F}}_{1}$, 
where $|n\rangle_{1}$ and $|n\rangle_{2}$ ($n=1,2,\cdots,N$) are 
orthonormal basis of ${\cal{F}}_{1}$ and ${\cal{F}}_{2}$, respectively. 
(See Appendix \ref{app:A} for a systematic derivation. )
In the next section we use the state (\ref{eqn:3-4:EPRstate}) to transmit
information about an unknown quantum state.

In statistical mechanics, free energy $F=E-TS$ can also be used to 
determine a statistical state: its minimum corresponds to an
equilibrium state of a subsystem in contact with a heat bath of
temperature $T$, provided that $T$ is fixed. This variational
principle in statistical mechanics is based on the following three
assumptions. 
\begin{enumerate}
\item The total system (the subsystem $+$ the heat bath) obeys the
principle of maximum of entropy.
\item Total energy (energy of the subsystem $+$ energy of the heat
bath) is conserved.
\item The first law of thermodynamics holds for the heat bath. 
\end{enumerate}
Is there a corresponding variational principle in 
the quantum system in the Hilbert space $\cal{F}$ of the form 
(\ref{eqn:3-3:F=F1*F2})? The answer is yes. In section
\ref{sec:entanglement} a concept of entanglement energy 
has been introduced and a thermodynamical structure, which we call 
entanglement thermodynamics, has been constructed by using the
entanglement entropy and the entanglement energy. Thus we expect that
entanglement free energy $F_{ent}$ defined as follows plays an
important role in entanglement thermodynamics.
%============< EQUATION >==============%
%
\begin{equation}
	F_{ent} = E_{ent} - T_{ent}S_{ent},\label{eqn:3-4:Fent}
\end{equation}
%======================================%
where $E_{ent}$ is the entanglement energy and $T_{ent}$ is a 
constant. Among several options, we adopt the second definition (b) of 
the entanglement energy given in subsection \ref{subsec:Eent}:
%============< EQUATION >==============%
%
\begin{equation}
 E_{ent} = \langle :H_2:\rangle. 
	\label{eqn:3-4:Eent}
\end{equation}
%======================================%

As shown in the following arguments, by minimizing 
the entanglement free energy, we can obtain a state in $\cal{F}$ 
characterized by the constant $T_{ent}$. Before doing it, here we 
consider a physical meaning of the principle of minimum of the
entanglement free energy. Let us introduce another Hilbert space
${\cal{F}}_{bath}$, which plays the role of the heat bath in the above
statistical-mechanical consideration, and decompose it to the direct
product
${\cal{F}}_{bath}={\cal{F}}_{bath1}\otimes{\cal{F}}_{bath2}$. In this 
situation it is expected that the principle of the minimum entanglement 
free energy corresponds to the following situation.
\begin{enumerate}
\item The total system 
${\cal{F}}_{tot}\equiv{\cal{F}}\otimes{\cal{F}}_{bath}$ obeys the 
principle of  maximum of the entanglement entropy with respect to the 
decomposition 
${\cal{F}}_{tot}={\cal{F}}_{tot1}\otimes{\cal{F}}_{tot2}$, where
${\cal{F}}_{tot1}\equiv{\cal{F}}_{1}\otimes{\cal{F}}_{bath1}$ and 
${\cal{F}}_{tot2}\equiv{\cal{F}}_{2}\otimes{\cal{F}}_{bath2}$.
\item Total entanglement energy (entanglement energy for ${\cal{F}}$ 
$+$ entanglement energy for ${\cal{F}}_{bath}$) is conserved.
\item The first law of entanglement thermodynamics
(\ref{eqn:3-3:1st-law-ent}) holds for ${\cal{F}}_{bath}$. In this
situation we call the constant $T_{ent}$ the entanglement
temperature. 
\end{enumerate}
It must be mentioned here that the variational principle of minimum of
the entanglement free energy is not as fundamental as the principle of
maximum of the entanglement entropy but is an approximation to the 
latter principle for a large system. However, like the principle of
minimum free energy in statistical mechanics, the former
principle should be a very useful tool to determine a quantum state.

We now calculate $F_{ent}$ for the system of two spin-$1/2$ particles 
and minimize it. For simplicity we adopt the following Hamiltonian for 
the subsystem $2$:
%============< EQUATION >==============%
%
\begin{eqnarray}
	{}_{2}\langle\uparrow|:H_{2}:|\uparrow\rangle_{2} & = & 
		\epsilon,\nonumber\\
	{}_{2}\langle\uparrow|:H_{2}:|\downarrow\rangle_{2} & = & 
		0,\nonumber\\
	{}_{2}\langle\downarrow|:H_{2}:|\downarrow\rangle_{2} & = & 
		0,
\end{eqnarray}
%======================================%
where $\epsilon$ is a positive constant. The entanglement 
free energy $F_{ent}$ for the state (\ref{eqn:3-4:stateinF}) is given by
%============< EQUATION >==============%
%
\begin{equation}
	F_{ent} = \epsilon(|a|^2 + |c|^2)
		+ T_{ent}\left[\frac{1+x}{2}\ln\left(\frac{1+x}{2}\right)
		+\frac{1-x}{2}\ln\left(\frac{1-x}{2}\right)\right].
\end{equation}
%======================================%
By minimizing it we obtain the following expression for the state 
$|\phi\rangle$ up to a unitary transformation in ${\cal{F}}_{1}$.
%============< EQUATION >==============%
%
\begin{equation}
	|\phi\rangle = \frac{1}{\sqrt{Z}}\left[
		e^{-\epsilon/2T_{ent}}
		|\uparrow\rangle_{1}\otimes|\uparrow\rangle_{2}
		+ |\downarrow\rangle_{1}\otimes|\downarrow\rangle_{2}
		\right],
\end{equation}
%======================================%
where $Z=e^{-\epsilon/T_{ent}}+1$.

The corresponding reduced density matrix $\rho_{2}$ on ${\cal{F}}_{2}$
represents a canonical ensemble with temperature $T_{ent}$. This fact
is related to the fact that the principle of minimum of free energy
results in a canonical ensemble in statistical mechanics. Thus, in
general, if ${\cal{F}}_{1}$ and ${\cal{F}}_{2}$ have the same finite
dimension $N$ then a state minimizing the entanglement free
energy is written as 
%============< EQUATION >==============%
%
\begin{equation}
	|\phi\rangle = \frac{1}{\sqrt{Z}}\sum_{n=1}^N\left[
		e^{-E_{n}/2T_{ent}}
		|n\rangle_{1}\otimes|n\rangle_{2}\right]
		\label{eqn:3-4:TFDstate}
\end{equation}
%======================================%
up to a unitary transformation in ${\cal{F}}_{1}$, where
$Z=\sum_{n=1}^Ne^{-E_{n}/T_{ent}}$, and $E_{n}$ and $|n\rangle_{2}$
($n=1,2,\cdots,N$) are eigenvalues and orthonormalized eigenstates of
the normal-ordered Hamiltonian of the subsystem $2$. (See Appendix
\ref{app:A} for a systematic derivation.)

The state (\ref{eqn:3-4:TFDstate}) can be
obtained also from another version of the principle of maximum of the 
entanglement entropy: if we maximize $S_{ent}$ with $E_{ent}$ fixed
then the state (\ref{eqn:3-4:TFDstate}) is obtained. In this case, the
constant $T_{ent}$ is determined so that the entanglement energy
coincides with the fixed value.

Note that in Eq. (\ref{eqn:3-4:TFDstate}) the infinite-dimensional limit
$N\to\infty$ can be taken, provided that $T_{ent}$ is bounded. 
In this limit, the state
(\ref{eqn:3-4:TFDstate}) has the same form as those appearing in 
the thermo field dynamics of black holes~\cite{Israel1976} and the
quantum field theory on a collapsing star
background~\cite{Parker1975}. In fact, if we can set the value of the
entanglement temperature of ${\cal{F}}_{bath}$ to be the black hole
temperature then the state (\ref{eqn:3-4:TFDstate}) in the limit
completely coincides with those in
Refs.~\cite{Israel1976,Parker1975}. 
In section \ref{sec:entanglement} it has been shown numerically that
the entanglement temperature for a real massless scalar field in
Schwarzschild and Reissner-Nordstr{\"o}m spacetimes is finite and
equal to the black hole temperature of the background geometry up to a 
numerical constant of order $1$. The finiteness of the entanglement
temperature in the black hole spacetimes is a result of 
cancellation of divergences in entanglement entropy and entanglement
energy~\cite{MSK1998}. Thus, the finiteness is preserved even in the
limit of zero cutoff length ($a\to 0$).

%======================================%
%<<<<<<   SUBSECTION 4-3-3    >>>>>>>>>%
%======================================%
\subsection{Quantum teleportation}
	\label{subsec:QT}

In Ref.~\cite{Bennett-etal} Bennet et al. proposed a method of 
teleportation of an unknown quantum state from one place to another. It 
is called quantum teleportation. In their method the information
about the quantum state is separated into a `quantum channel' and a
`classical channel', and each channel is sent separately from a sender
``Alice'' to a receiver ``Bob''. What is important is that the quantum
channel is sent in a superluminal way by using a quantum correlation
or entanglement, while the classical channel is transmitted at most in
the speed of light. Here we mention that causality is not violated in
an informational sense since Bob cannot obtain any useful information
about the unknown state before the arrival of the classical
channel. Hence Alice has to deliver the classical channel to Bob
without fail. On the contrary she
does not need to worry about whether the information in the quantum
channel arrives at Bob's hand since the arrival is guaranteed by the
quantum mechanics. It is notable that recently quantum
teleportation was confirmed by
experiments~\cite{Bouwrneester-etal,Boschi-etal}.

In this section we generalize the arguments in
Ref.~\cite{Bennett-etal} to more abundant situations and try to
reformulate it in terms of the entanglement entropy.

Let us consider a Hilbert space ${\cal{F}}$ of the form 
(\ref{eqn:3-3:F=F1*F2}) with ${\cal{F}}_{i}$ constructed from Hilbert 
spaces ${\cal{F}}_{i\pm}$ as
%============< EQUATION >==============%
%
\begin{equation}
	{\cal{F}}_{i} = {\cal{F}}_{i+}\otimes{\cal{F}}_{i-}. 
	\label{Fi=Fi+*Fi-}
\end{equation}
%======================================%
For example, consider matter fields in a black hole spacetime formed by 
gravitational collapse. In this situation, let ${\cal{H}}_{1}$ be a space 
of all wave packets on the future event horizon and ${\cal{H}}_{2}$ be 
a space of all wave packets on the future null infinity, and decompose 
each ${\cal{H}}_{i}$ into a high frequency part ${\cal{H}}_{i+}$ and a 
low frequency part ${\cal{H}}_{i-}$. Typically, we suppose the 
decomposition at an energy scale of Planck order. If we define 
${\cal{F}}_{i\pm}$ as Fock spaces constructed from ${\cal{H}}_{i\pm}$, 
respectively, then the space $\cal{F}$ of all quantum states of the matter 
fields is given by (\ref{eqn:3-3:F=F1*F2}) with (\ref{Fi=Fi+*Fi-}). 
Although the following arguments do not depend on the construction 
of the Hilbert space $\cal{F}$, this example should be helpful for us
to understand the physical meaning of the results obtained.

For simplicity we consider the case 
that all ${\cal{F}}_{i\pm}$ have the same finite dimension $N$ 
although in the above example of field theory the dimensions of the
Hilbert spaces are infinite~\footnote{
In applying the results for finite dimensions to field theory,
we have to introduce a regularization scheme to make the system
finite. For example, we can discretize the system by introducing a
cutoff length. After that, we can consider a finite dimensional
subspace of the total Hilbert space of the discretized theory, for
example, by restricting total energy to be less than the mass of the
background geometry. After performing all calculations, we have to
confirm that the infinite-dimensional limit can be taken. See, for
example, the final paragraph of the previous subsection.
}. 
The main purpose of this
section is to see general properties of the entanglement
entropy by using a finite system. Anyway, the above example of field
theory may be helpful in understanding the following arguments. 
In the finite dimensional case we assume the following three physical
principles. 
\begin{enumerate}
 \renewcommand{\labelenumi}{(\alph{enumi})}
\item A quantum state $|\phi\rangle$ in $\cal{F}$ is a direct product 
state given by $|\phi_{+}\rangle_{+}\otimes|\phi_{-}\rangle_{-}$, 
where $|\phi_{\pm}\rangle_{\pm}$ are elements of 
${\cal{F}}_{\pm}={\cal{F}}_{1\pm}\otimes{\cal{F}}_{2\pm}$, respectively.
\item $|\phi_{+}\rangle_{+}$ is determined by the principle of
maximum of the entanglement entropy with respect to the decomposition 
${\cal{F}}_{+}={\cal{F}}_{1+}\otimes{\cal{F}}_{2+}$.
\item A complete measurement of the von Neumann type on the joint 
system ${\cal{F}}_{1}$ is performed by a sender (Alice) in the
orthonormal basis $\{|\psi_{nm}\rangle_{1}\}$, each of 
which maximizes the entanglement entropy with respect to the 
decomposition ${\cal{F}}_{1}={\cal{F}}_{1+}\otimes{\cal{F}}_{1-}$. 
\end{enumerate}
In other words the assumption (c) is stated as follows: 
the state $|\phi\rangle$ is projected by one of the basis 
$|\psi_{nm}\rangle_{1}$.

In the following arguments, under these 
assumptions, we show a possibility of quantum teleportation of the 
state $|\phi_{-}\rangle_{-}$ in ${\cal{F}}_{-}$ to ${\cal{F}}_{2}$: we 
make a clone of $|\phi_{-}\rangle_{-}$ by using the quantum
entanglement which the state $|\phi_{+}\rangle_{+}$ has. 
Therefore a receiver (Bob), who cannot contact with ${\cal{F}}_{1}$,
may be able to get all information about the state
$|\phi_{-}\rangle_{-}$ in ${\cal{F}}_{-}$, provided that he can manage
to get the classical channel.

Now let us show that explicitly. By the assumption (b) and the
arguments in subsection \ref{subsec:variational} (see
Eq.~(\ref{eqn:3-4:EPRstate})), the state $|\phi_{+}\rangle_{+}$ can be
written as 
%============< EQUATION >==============%
%
\begin{equation}
	|\phi_{+}\rangle_{+} = \frac{1}{\sqrt{N}}\sum_{n=1}^N
		|n\rangle_{1+}\otimes|n\rangle_{2+},\label{eqn:3-4:phi+}
\end{equation}
%======================================%
where $\{|n\rangle_{i+}\}$ ($n=1,2,\cdots,N$) are an orthonormal basis 
of ${\cal{F}}_{i+}$. Next, expand $|\phi_{-}\rangle_{-}$ as
%============< EQUATION >==============%
%
\begin{equation}
	|\phi_{-}\rangle_{-} = \sum_{nm}C_{nm}
		|n\rangle_{1-}\otimes|m\rangle_{2-},\label{eqn:3-4:phi-}
\end{equation}
%======================================%
where $\{|n\rangle_{i-}\}$ ($n=1,2,\cdots,N$) are an orthonormal basis 
of ${\cal{F}}_{i-}$, and $\sum_{nm}|C_{nm}|^2=1$ is understood. 
To impose the assumption (c), we adopt the following 
basis $\{|\psi_{nm}\rangle_{1}\}$ ($n,m=1,2,\cdots,N$), each of which 
maximizes the entanglement entropy.
%============< EQUATION >==============%
%
\begin{equation}
 |\psi_{nm}\rangle_{1} = \frac{1}{\sqrt{N}}\sum_{j=1}^N 
 e^{2\pi ijn/N}|(j+m)modN\rangle_{1+}\otimes|j\rangle_{1-}.
 	\label{eqn:3-4:BELLstates}
 \end{equation}
%======================================%
In Appendix \ref{app:B} it is proved that (\ref{eqn:3-4:BELLstates}) is
unique up to a unitary transformation in ${\cal{F}}_{1+}$. 
Hence, $|\phi\rangle=|\phi_{+}\rangle_{+}\otimes|\phi_{-}\rangle_{-}$ is 
written as 
%============< EQUATION >==============%
%
\begin{equation}
 |\phi\rangle = \frac{1}{N}\sum_{nm}|\psi_{nm}\rangle_{1}\otimes
		U^{(2+)}_{nm}|\tilde{\phi}_{2}\rangle_{2},
		\label{eqn:3-4:phi-decomp}
\end{equation}
%======================================%
where $|\tilde{\phi}_{2}\rangle_{2}$ is a state in ${\cal{F}}_{2}$ 
given by 
%============< EQUATION >==============%
%
\begin{equation}
 |\tilde{\phi}_{2}\rangle_{2} = 
	\sum_{n'm'}C_{n'm'}|n'\rangle_{2+}\otimes|m'\rangle_{2-},
	\label{eqn:3-4:tilde-phi2}
\end{equation}
%======================================%
and $U^{(2+)}_{nm}$ $(n,m=1,2,\cdots,N$) are unitary transformations 
in ${\cal{F}}_{2+}$ defined by 
%============< EQUATION >==============%
%
\begin{equation}
 U^{(2+)}_{nm} = \sum_{k=1}^N 
	e^{-2\pi ikn/N}|(k+m)modN\rangle_{2+}{}_{2+}\langle k|.
	\label{eqn:3-4:Unitary-tr}
 \end{equation}
%======================================%
(See Appendix \ref{app:B} for an explicit derivation of
(\ref{eqn:3-4:phi-decomp}).)

Thus, after the measurements in the basis $\{|\psi_{nm}\rangle_{1}\}$ 
by the sender (Alice), the original state $|\phi\rangle$ jumps to one
of the states $|\tilde{\phi}_{nm}\rangle$ defined by 
%============< EQUATION >==============%
%
\begin{equation}
 |\tilde{\phi}_{nm}\rangle = |\psi_{nm}\rangle_{1}\otimes
		U^{(2+)}_{nm}|\tilde{\phi}_{2}\rangle_{2}. 
		\label{eqn:3-4:tild-phi}
\end{equation}
%======================================%
This state can be seen by the receiver (Bob), who cannot contact 
with ${\cal{F}}_{1}$, as the state 
$U^{(2+)}_{nm}|\tilde{\phi}_{2}\rangle_{2}$ in ${\cal{F}}_{2}$. Here 
note that the unitary transformation $U^{(2+)}_{nm}$ in 
${\cal{F}}_{2+}$ is completely determined by a pair of integers $n$ and 
$m$ (outcome of the experiment by Alice). Thus, if the two integers 
are sent to the receiver (Bob) in the classical channel, then by
operating the inverse transformation of the corresponding unitary
transformation in ${\cal{F}}_{2+}$ the receiver (Bob) can obtain the
`clone' state $|\tilde{\phi}_{2}\rangle_{2}$ ($\in {\cal F}_{2}$) of
$|\phi_{-}\rangle$ ($\in {\cal F}_{-}$). It is evident that
$|\tilde{\phi}_{2}\rangle_{2}$ has all information about the original 
state $|\phi_{-}\rangle$.

It is remarkable that information to be 
sent to the receiver (Bob) in the classical channel is only two 
integers $n$ and $m$, while information included in the unknown state
$|\phi_-\rangle_-$ is a set of complex constants $\{C_{nm}\}$
($n,m=1,2,\cdots,N$) with a constraint $\sum_{nm}|C_{nm}|^2=1$. Thus a 
large amount of information is sent in the quantum channel. 
Here we mention that tracing out ${\cal{F}}_{2+}$ from the state 
$U^{(2+)}_{nm}|\tilde{\phi}_{2}\rangle_{2}$ or
$|\tilde{\phi}_{2}\rangle_{2}$ results in the following density matrix
$\rho_{2-}$ on ${\cal{F}}_{2-}$: 
%============< EQUATION >==============%
%
\begin{equation}
 \rho_{2-} = \sum_{nm}\left(\sum_{j}C_{jn}C_{jm}^*\right)
 	|n\rangle_{2-}{}_{2-}\langle m|, \label{eqn:3-4:rho2-}
\end{equation}
%======================================%
which is equivalent to the density matrix obtained by tracing out 
${\cal{F}}_{1-}$ from the original unknown state
$|\phi_{-}\rangle_{-}$. Hence, if the receiver (Bob) cannot contact
with ${\cal F}_{2+}$, he does not obtain any information from the
sender (Alice).

Finally it must be mentioned that the success of quantum 
teleportation is due to quantum entanglement in the state
$|\phi_{+}\rangle_{+}$ which has maximal entanglement entropy. If we
took $|\phi_{+}\rangle_{+}$ 
with less entanglement entropy then the teleportation would be less 
successful. Therefore, the entanglement entropy can be regarded as an 
index of efficiency of quantum teleportation. This consideration 
supports the interpretation of the entanglement entropy proposed in 
subsection \ref{subsec:entropies}.

%======================================%
%<<<<<<   SUBSECTION 4-3-4    >>>>>>>>>%
%======================================%
\subsection{Concluding remark and physical implications}
	\label{subsec:remark_interpretation}

In this section a new interpretation of entanglement entropy has been
proposed based on its relation to the so-called conditional entropy and 
a well-known meaning of the latter. It is conjectured that 
entanglement entropy of a pure state with respect to a division of a 
Hilbert space into two subspaces $1$ and $2$ is an amount of
information, which can be transmitted through $1$ and $2$ from a
system interacting with $1$ to another system interacting with
$2$. The medium of the transmission is quantum entanglement between
$1$ and $2$.

To support the interpretation we have given the following two suggestive
arguments: variational principles in entanglement thermodynamics and
quantum teleportation. The most important variational principle we
considered is the principle of maximum of entanglement entropy. This
principle determines a state uniquely up to a unitary transformation
in one of the two Hilbert subspaces (not in the whole Hilbert
space). From the proposed conjecture it is expected that information
can be transmitted most effectively through the two subspaces by using 
the maximal entanglement of the state. 
In fact, reformulating the quantum teleportation in terms of the
entanglement entropy, we have shown that the state having maximal
entanglement entropy plays an important role in quantum
teleportation. This consideration gives strong support to our
interpretation.

As a by-product we have shown that the variational principle of 
minimum of entanglement free energy is useful to determine a quantum 
state. The resulting quantum state has exactly the same form as those 
appearing in the thermo field dynamics of black holes~\cite{Israel1976}
and the quantum field theory on a collapsing black hole
background~\cite{Parker1975}, provided that the entanglement temperature
$T_{ent}$ is set to be the black 
hole temperature. It is remarkable that, as shown in section 
\ref{sec:entanglement}, $T_{ent}$ for a real massless scalar 
field in Schwarzschild and Reissner-Nordstr{\"o}m spacetimes is equal
to the black hole temperature of the background geometry up to a
numerical constant of order $1$. Thus we can say that
the variational principle of minimum of entanglement free energy gives
a new derivation of the Hawking radiation. 
Finally, we mention that with this variational principle the
entanglement thermodynamics is equivalent to 'tHooft's brick wall
model~\cite{tHooft1985}.

It will be valuable to analyze how to generalize arguments in this
section to the situation that divergences in entropy and energy are
absorbed by renormalization~\cite{Susskind&Uglum1994,FFZ1996b}. 
If the generalization is achieved, the physical meaning of the 
entanglement entropy in black hole physics will become clearer. 
It is noteworthy that in the brick wall model, as shown in section 
\ref{sec:brick_wall}, the divergence in thermal energy is exactly
canceled by divergence in negative
energy~\cite{Mukohyama&Israel1998}.

Now the final comment is in order. It is worthwhile to clarify in what 
physical situations the variational principles can be applicable. (In 
thermodynamics the second law supports the principle of maximum 
entropy.) In other words, in what situations does the entanglement 
entropy increase? 
In what situations does the entanglement free energy decrease? 
To answer these questions, theorem \ref{theorem:general} in subsection
\ref{subsec:proof_GSL} or its generalization may be useful.

%%%%%%%%%%%%%%%%%%%%%%%%%%%%%%%%%%%%%%%%%%%%%%%%%%%%%%%%%%%%%%%%%%%%
%%%%%%%%%%%%%%%%%%%%%%%%%%%%%%%%%%%%%%%%%%%%%%%%%%%%%%%%%%%%%%%%%%%%
% CHAPTER 5
%%%%%%%%%%%%%%%%%%%%%%%%%%%%%%%%%%%%%%%%%%%%%%%%%%%%%%%%%%%%%%%%%%%%
%%%%%%%%%%%%%%%%%%%%%%%%%%%%%%%%%%%%%%%%%%%%%%%%%%%%%%%%%%%%%%%%%%%%
\chapter{Discussions}
	\label{chap:summary}

In this thesis we have analyzed properties and the origin of the black
hole entropy in detail from various points of view. 

% SUMMARY of CHAPTER 2

First, in chapter~\ref{chap:BHlaws} laws of black hole thermodynamics
have been reviewed. In particular, the first and generalized second
laws have been investigated in detail. It is in these laws that the
black hole entropy plays key roles.

%%%%%%%%%%%% SUMMARY of SECTION 2-1 %%%%%%%%%%%%%%%

In section~\ref{sec:1st_statics} we have re-analyzed Wald and
Iyer-Wald derivation of the first law of black hole mechanics in a
general covariant theory of gravity, following
Ref.~\cite{Mukohyama1998b}. 
In particular, two issues listed in the beginning of
section~\ref{sec:1st_statics} have been discussed in detail: 
(a) gauge conditions and (b) near-stationary black hole entropy. 
It has been shown that the minimal set of gauge conditions
necessary for the derivation of the first law for stationary black
holes is that $t^a$ and $\varphi^a$ are fixed at spatial infinity,
where $t^a$ is the stationary Killing field with unit norm at
infinity, and $\varphi^a$ denotes axial Killing fields. 
It has also been shown that for non-stationary perturbations about a
stationary perturbation the first law does hold to first order in
perturbation.

%%%%%%%%%%%% SUMMARY of SECTION 2-2 %%%%%%%%%%%%%%%

However, this first law cannot be applied to a purely dynamical
situation. In this sence we have called it the first law of black hole 
statics. 
The purpose of section~\ref{sec:1st_dynamics} has been to
consider dynamical definition of black hole entropy and to derive a
dynamical version of the first law of black hole, which we call the
quasi-local first law of black hole dynamics. For simplicity, we have
considered the general relativity only. Extension to a general
covariant theory of gravity will be valuable.

%%%%%%%%%%%%%%%

In subsection~\ref{subsec:dynamic-S} we have considered two
non-statistical definitions of entropy for dynamic (non-stationary)
black holes in spherical symmetry. 
The first is analogous to the original Clausius definition
of thermodynamic entropy:
there is a first law containing an energy-supply term
which equals surface gravity times a total differential.
The second is Wald's Noether-charge method,
adapted to dynamic black holes by using the Kodama flow.
It has been shown that both definitions give the same answer for
Einstein gravity: one-quarter the area of the trapping
horizon~\cite{HMA1998}.

%%%%%%%%%%%%%

In subsection~\ref{subsec:dynamic-S} the quasi-local first law of
black hole dynamics has been derived without assuming any symmetry and 
any asymptotic condition~\cite{Mukohyama&Hayward1998}.  In the
derivation we have given a new definition of dynamical surface
gravity. In spherical symmetry it reduces to that defined in
Ref.~\cite{Hayward1998a}.

%%%%%%%%%%%% SUMMARY of SECTION 2-3 %%%%%%%%%%%%%%%

In section~\ref{sec:GSL} we have proved the generalized second law for
a quasi-stationary black hole which is formed by gravitational
collapse~\cite{Mukohyama1997a}.
After that, in subsection~\ref{subsec:remark_GSL}, we have discussed a 
generalization of our proof to a dynamical background. It has been
suggested that the generalization may be possible by using the
quasi-local first law derived in subsection~\ref{subsec:dynamic-S}.

% SUMMARY of CHAPTER 3

Next, in chapter~\ref{chap:BHentropy} three candidates for the origin
of the black hole entropy have been analyzed: the D-brane
statistical-mechanics, the brick wall model, and the entanglement
thermodynamics.

%%%%%%%%%%%% SUMMARY of SECTION 3-1 %%%%%%%%%%%%%%%

In section~\ref{sec:Dbrane} the D-brane statistical-mechanics has
been reviewed by using a configuration of D-strings and D-fivebranes
wrapped on $T^5=T^4\times S^1$, which was introduced in
Ref.~\cite{Mukohyama1996}. 
We have consider a set of multiply-wound D-strings, which is composed
of $N_{q_1}^{(1)}$ D-strings of length $2\pi Rq_1$ ($q_1=1,2,\cdots$)
along the $S^1$ and a set of multiply-wound D-fivebranes, which is
composed of $N_{q_5}^{(5)}$ D-fivebranes of length $2\pi Rq_5$
($q_5=1,2,\cdots$) along the $S^1$. 
Here $\{ N_{q_1}^{(1)}\}$ and $\{ N_{q_5}^{(5)}\}$ are sets of
arbitrary non-negative integers, and $R$ is radius of the $S^1$. 
It has been shown that the number of microscopic states 
of open strings on the D-branes is bounded from above by exponential
of the Bekenstein-Hawking entropy of the corresponding black
hole, and that the temperature of a decay of D-brane
excitations to closed strings is bounded from below by the Hawking
temperature of the corresponding black hole. This result has
been summarized as Eqs.(\ref{eqn:3-1:lndN}) and
(\ref{eqn:3-1:Tdecay}). 
The necessary and sufficient condition for these bounds to be
saturated has been shown explicitly and some speculations has been
given.

%%%%%%%%%%%% SUMMARY of SECTION 3-2 %%%%%%%%%%%%%%%

In section~\ref{sec:brick_wall} we have re-examined the brick wall
model to solve problems concerning this model. 
In particular, it has been shown that the wall contribution to the
total gravitational mass is zero if and only if temperature of thermal 
gas measured at infinity is set to be the Hawking temperature, and
that the backreaction can be neglected~\cite{Mukohyama&Israel1998}.

%%%%%%%%%%%% SUMMARY of SECTION 3-3 %%%%%%%%%%%%%%%

In section~\ref{sec:entanglement} we have constructed entanglement 
thermodynamics for a massless scalar field in
Minkowski~\cite{MSK1997}, Schwarzschild~\cite{MSK1998} and
Reissner-Nordstr{\"o}m spacetimes. The entanglement thermodynamics 
in Minkowski spacetime differs significantly from black-hole
thermodynamics. On the contrary, the entanglement thermodynamics in
Schwarzschild and  Reissner-Nordstr{\"o}m spacetimes has the same
structure as that of black-hole thermodynamics. In particular, it has
been shown that entanglement temperature in the Reissner-Nordstr{\"o}m
spacetime approaches zero in the extremal limit.

%%%%%%%%%%%% SUMMARY of SECTION 3-4 %%%%%%%%%%%%%%%

In section~\ref{sec:interpretation} a new interpretation of
entanglement entropy has been proposed based on its relation to the
so-called conditional entropy and a well-known meaning of the
latter~\cite{Mukohyama1998a}. It has been conjectured that
entanglement entropy of a pure state with respect to a division of a
Hilbert space into two subspaces $1$ and $2$ is an amount of
information, which can be transmitted through $1$ and $2$ from a
system interacting with $1$ to another system interacting with
$2$. The medium of the transmission is quantum entanglement between
$1$ and $2$. 
To support the interpretation we have given the following two
suggestive arguments: variational principles in entanglement
thermodynamics and quantum teleportation. 
It has been shown that the state having maximal entanglement entropy
plays an important role in quantum teleportation. This consideration
gives strong support to our interpretation.

%%%%%%%%%%%% SEMICLASSICAL CONSISTENCIES %%%%%%%%%%%%%%%

Now let us discuss about semiclassical consistencies of the brick wall 
model and the entanglement thermodynamics.

It has been shown that the brick wall model is a consistent
semiclassical description of black hole entropy: thermal excitations
raised to the Hawking temperature above the Boulware state explains
the black hole entropy; the positive divergence in thermal energy is
canceled by the negative divergence in the vacuum energy of the
Boulware state. 
Namely, the following simultaneous equations have a solution:
%============< EQUATION >==============%
%
\begin{eqnarray}
 S_{wall} & = &S_{BH},\nonumber\\
 T_{\infty} & = & T_{BH},\nonumber\\
 (\Delta M)_{therm,wall}+(\Delta M)_{B,wall} & = & 0.
\end{eqnarray}
%======================================%
It has been found that the last equation is equivalent to the second
one. Thus, the solution is 
%============< EQUATION >==============%
%
\begin{eqnarray}
 \alpha & = & \sqrt{\frac{{\cal N}}{90\pi}}l_{pl},\nonumber\\
 T_{\infty} & = &T_{BH}.
\end{eqnarray}
%======================================%

On the other hand, entanglement thermodynamics has the following
properties common to all definitions of entanglement energy. 
%============< EQUATION >==============%
%
\begin{eqnarray}
 S_{ent} & \simeq & N{\cal N}_S 
		\left(\frac{r_0}{a}\right)^2,\nonumber\\
 E_{ent} & \simeq &  N{\cal N}_E \frac{r_0}{a^2},\nonumber\\ 
 T_{ent} & \simeq & {\cal N}_T T_{BH},
\end{eqnarray}
%======================================%
where ${\cal N}_S$ and ${\cal N}_E$ are numerical factors of order
unity, and ${\cal N}_T=2\pi{\cal N}_E/{\cal N}_S$. Here, we multiplied
entanglement entropy $S_{ent}$ and energy $E_{ent}$ by the number $N$
of fields~\footnote{
Note that $N=1$ corresponds to one bosonic field. Hence, 
${\cal N}=N\pi^4/90$.
}. 
These properties are expected to hold also for states different from
the Boulware state, provided that the definition of entanglement
quantities are properly modified. For other states, of course, the
numerical factors ${\cal N}_S$ and ${\cal N}_E$ will be different from
those for the Boulware state.

Our question now is whether the entanglement thermodynamics is 
a consistent semiclassical description of black hole
thermodynamics or not. 
Here let us assume that the sum of the vacuum energy and the
entanglement energy contributes to gravitational energy. 
Hence, the question is restated as whether the following
simultaneous equations hold or not.
%============< EQUATION >==============%
%
\begin{eqnarray}
 S_{ent} & = & S_{BH},\nonumber\\
 T_{ent} & = & T_{BH},\nonumber\\
 E_{ent} + (\Delta M)_{B,wall} & = & 0.
	\label{eqn:4-0:consistency_ent}
\end{eqnarray}
%======================================%
Since the cutoff length $\alpha$ in section \ref{sec:brick_wall} seems
to be related to the cutoff length $a$ in section
\ref{sec:entanglement} as 
%============< EQUATION >==============%
%
\begin{equation}
 \alpha \simeq n_B a,
\end{equation}
%======================================%
a solution of Eqs.~(\ref{eqn:4-0:consistency_ent}) is given by
%============< EQUATION >==============%
%
\begin{eqnarray}
 a & \simeq & \sqrt{\frac{N{\cal N}_S}{\pi}}l_{pl},\nonumber\\
 {\cal N}_E & \simeq & \frac{{\cal N}_S}{2\pi},\nonumber\\
 n_B & \simeq & \sqrt{\frac{\pi^4}{90}\cdot\frac{1}{240{\cal N}_S}}.
	\label{eqn:4-0:sol_ent}
\end{eqnarray}
%======================================%
Unfortunately, since ${\cal N}_S=0.3$, the r.h.s of the last equation
is less then unity. Hence the last equation does not hold for positive 
integer value of $n_B$. Even if $n_B$ would be allowed to be less than
unity, the value of $n_B$ given by the last equation would not be
consistent with the second equation for all our definitions of
entanglement energy~\footnote{
Note that ${\cal N}_E$ depends on $n_B$.
}. 
This discrepancy suggests that our
model of entanglement thermodynamics suffers from a strong backreaction
near horizon. This might mean that our choice of the pure state (the 
Boulware state) would be wrong. 
However, we strongly expect that there is a state for which the
consistency condition (\ref{eqn:4-0:sol_ent}) holds. 
It will be worthwhile to analyze whether Eqs.~(\ref{eqn:4-0:sol_ent})
are satisfied for the Hartle-Hawking state or not.

%%%%%%%%%%%% VARIATIONAL PRINCIPLE %%%%%%%%%%%%%%%

Next, we shall combine a variational principle in entanglement 
thermodynamics with the third equation in the semiclassical
consistency conditions. 
As explained in subsection~\ref{subsec:variational}, the
principle of maximum of entanglement entropy introduced in
Ref.~\cite{Mukohyama1998a} determines a quantum state to be 
(\ref{eqn:3-4:EPRstate}) up to a unitary transformation in one of two
Hilbert subspaces. 
Moreover, the principle of minimum of entanglement free energy
determines a quantum state to be (\ref{eqn:3-4:TFDstate}) up to a
unitary transformation in one of two subspaces, too. 
It has also been mentioned in the second-to-last paragraph of
subsection~\ref{subsec:variational} that, if we maximize entanglement
entropy with entanglement energy fixed, the state
(\ref{eqn:3-4:TFDstate}) is obtained. In this case, entanglement
temperature should be determined so that the entanglement energy
coincides with the fixed value. 
On the other hand, the third equation in the semiclassical
consistency conditions (\ref{eqn:4-0:consistency_ent}) does fix
entanglement energy, provided that we do not change 
$(\Delta M)_{B,wall}$. We shall call this condition the small
backreaction condition. Hence, the principle of maximum of
entanglement entropy combined with the small backreaction condition
determines a state of a quantum field near horizon to be
(\ref{eqn:3-4:TFDstate}) up to a unitary transformation in one of two
Hilbert subspaces, provided that a suitable regularization scheme is
introduced. This state has exactly the same form as those appearing in
the thermo field dynamics of black holes~\cite{Israel1976} and the
quantum field theory on a collapsing black hole
background~\cite{Parker1975}, provided that the entanglement
temperature $T_{ent}$ can be set to be the Hawking temperature. 
In this case, since the entanglement energy is the same as thermal
energy in the brick wall model because of the small backreaction
condition, and since the entanglement entropy is expected to have the
almost same value as the Bekenstein-Hawking entropy, the entanglement
temperature also seems to coincide with the Hawking temperature up to
a numerical factor of order unity. 
Therefore, it seems that the principle of maximum of entanglement
entropy combined with the small backreaction condition may provide a
new derivation of the Hawking radiation. 
Further investigations on this point will be valuable.

%%%%%%%%%%%% INFORMATION LOSS PROBLEM %%%%%%%%%%%%%%%

Finally, let us discuss the information loss problem.

We have proposed a new interpretation of entanglement entropy: 
entanglement entropy of a pure state with respect to a division of a 
Hilbert space into two subspaces $1$ and $2$ is an amount of
information, which can be transmitted through $1$ and $2$ from a
system interacting with $1$ to another system interacting with $2$. 
On the other hand, it has been confirmed in section
\ref{sec:entanglement} and in many
references~\cite{BKLS1986,Srednicki1993,Frolov&Novikov1993,MSK1998} 
that the entanglement entropy has the same value as the black hole
entropy up to a numerical constant of order unity, provided that a
cutoff length of Planck order is introduced in the theory. Hence we
have a large amount of entanglement entropy to transmit information
from inside to outside of a black hole by using quantum entanglement.

Hence, in our interpretation, it seems that the
entanglement entropy is a quantity which cancels the black hole
entropy to restore information loss, provided that the black hole
entropy represents the amount of the information loss. 
For example, suppose that a black hole is formed from 
an initial state with zero entropy ($S=0$). In this case, 
non-zero black hole entropy is generated ($S_{BH}>0$) from the zero
entropy state. At the same time, entanglement entropy and negative
conditional entropy are also generated and their absolute values are as
large as the black hole entropy ($S_{ent}=|S_{cond}|\simeq S_{BH}$). 
After that, the black hole evolves by emitting Hawking 
radiation, changing the value of $S_{BH}$ and $S_{ent}$
($=|S_{cond}|$) with $S_{BH}\simeq S_{ent}$ kept. Finally, when the
black hole evaporates, the entanglement entropy cancels the
black hole entropy to settle the final entropy to be zero
($S=0$). To summarize, the black hole entropy is an amount of
temporarily missing information and the entanglement entropy is a
quantity which cancels the black hole entropy. Both entropies appear
and disappear together from the sea of zero entropy state.

%%%%%%%%%%%% SPECULATIONS %%%%%%%%%%%%%%%

We conclude this thesis by giving some speculations. 
\begin{itemize}
 \item[(i)]
The black hole entropy should be related to a number of microscopic 
states in the corresponding black hole. The microscopic description
should be possible by using a quantum theory of gravity
(eg. superstring theory, loop quantum gravity, etc.). 
 \item[(ii)]
The existence of the horizon prevents the microscopic states in the
black hole from being seen on the outside. Hence, information about
the microscopic states is lost at least temporarily. 
(See Ref.~\cite{Hawking1998} for Hawking's objection to
string-theorist's point of view.) 
At the same time, the horizon generates entanglement entropy of matter
fields which has the same value as the black hole entropy.  
 \item[(iii)]
The entanglement entropy is an amount of information, which can be
transmitted through the matter fields from a
system interacting with the matter fields in the black hole to another
system interacting with the matter fields outside the black hole,
provided that the classical channel in the sense of the quantum
teleportation can be transmitted properly. The former system carries
the temporarily missing information corresponding to the black
hole entropy. Hence the quantum channel of all temporarily missing
information about the microscopic states can be transmitted to
the outside of the black hole before the black hole evaporates
completely. It is the classical channel that is necessary for
restoring the original information from the quantum channel. 
 \item[(iv)]
We propose the conjecture that one of the following two should be
realized. 
	\begin{itemize}
	 \item[(a)]
	The black hole evaporates completely, and the classical
	channel is transmitted at the final stage of the black hole
	evaporation.  
	 \item[(b)]
	There remains a remnant at the end of the Hawking radiation, 
	and all or a part of the classical channel is carried by
	the remnant forever. 
	\end{itemize}
If (a) is correct, then it is in principle possible to restore all
information about the microscopic states of the black hole after the
evaporation. If (b) is correct, then all or a part of the information 
about the microscopic states cannot be restored although all the
information remain to exist: the information cannot be decoded since
the classical 
and quantum channels are located separately. 
\end{itemize}

%%%%%%%%%%%%%%%%%%%%%%%%%%%%%%%%%%%%%%%%%%%%%%%%%%%%%%%%%%%%%%%%%%%%
%%%%%%%%%%%%%%%%%%%%%%%%%%%%%%%%%%%%%%%%%%%%%%%%%%%%%%%%%%%%%%%%%%%%
% ACKNOWLEDGMENTS 
%%%%%%%%%%%%%%%%%%%%%%%%%%%%%%%%%%%%%%%%%%%%%%%%%%%%%%%%%%%%%%%%%%%%
%%%%%%%%%%%%%%%%%%%%%%%%%%%%%%%%%%%%%%%%%%%%%%%%%%%%%%%%%%%%%%%%%%%%

\chapter*{Acknowledgments}
\pagestyle{myheadings}
\markboth{Acknowledgments}{Acknowledgments}
\addcontentsline{toc}{chapter}{Acknowledgments}

I would like to thank my advisor, Professor H. Kodama, for his letting
me know how to proceed research works and encouraging me continuously. 
I would like to express my appreciation to Professor W. Israel for
helpful discussions and continuing encouragement during and after his
stay in Kyoto as a visiting professor. 
I am very grateful to M. Seriu for many helpful discussions and
constructive suggestions. 
I would like to acknowledge stimulating discussions with M. Siino,
T. Chiba, K. Nakao and S. A. Hayward, and with other colleagues in
Yukawa Institute for Theoretical Physics and Department of Physics in
Kyoto University. 
Finally, I am grateful to my wife, Kyoko. 

\newpage

%%%%%%%%%%%%%%%%%%%%%%%%%%%%%%%%%%%%%%%%%%%%%%%%%%%%%%%%%%%%%%%%%%%%
%%%%%%%%%%%%%%%%%%%%%%%%%%%%%%%%%%%%%%%%%%%%%%%%%%%%%%%%%%%%%%%%%%%%
% Appendix
%%%%%%%%%%%%%%%%%%%%%%%%%%%%%%%%%%%%%%%%%%%%%%%%%%%%%%%%%%%%%%%%%%%%
%%%%%%%%%%%%%%%%%%%%%%%%%%%%%%%%%%%%%%%%%%%%%%%%%%%%%%%%%%%%%%%%%%%%

\appendix
\chapter*{Appendix}
\setcounter{chapter}{1}
\pagestyle{myheadings}
\markboth{Appendix}{Appendix}
\addcontentsline{toc}{chapter}{Appendix}

%%%%%%%%%%%%%%%%%%%%%%%%%%%%%%%%%%%%%%%%
%%%%%%%%%%%%% APPENDIX 1 %%%%%%%%%%%%%%%
%%%%%%%%%%%%%%%%%%%%%%%%%%%%%%%%%%%%%%%%

\section{The conditional probability}	\label{app:probability}

In this appendix we reduce (\ref{eqn:2-3:def-P}) to
(\ref{eqn:2-3:probability}). First the $S$-matrix obtained by
\cite{Wald1975&1976} is
\begin{eqnarray}
 S|0\rangle & = & 
      N\sum_{n=0}^{\infty}\frac{\sqrt{(2n)!}}{2^{n}n!}
      \left(\stackrel{n}{\otimes}\epsilon\right)_{sym},\nonumber\\
 Sa^{\dagger}(A\ _{i}\gamma)S^{-1} & = & 
      R_ia^{\dagger}(\!_{i}\rho)+T_ia^{\dagger}(\!_{i}\sigma),
\end{eqnarray}
where $\epsilon$ and $N$ are a bivector and a normalization constant
defined by 
\[
 \epsilon = 2\sum_{i}x_{i}(\!_{i}\lambda\otimes\!_{i}\tau)_{sym},
	\quad N=\prod_i\sqrt{1-x_i},
\]
where
\[
 x_{i} = \exp\left( -\pi
	(\omega_{i}-\Omega_{BH}m_{i})/\kappa\right).
\]
In the expression, $\omega_{i}$ and $m_{i}$
are a frequency and an azimuthal angular momentum quantum number of a
mode specified by integer $i$, $\Omega_{BH}$ and $\kappa$ are an
angular velocity and a surface gravity of the black
hole. $\!_{i}\gamma$, $\!_{i}\rho$, $\!_{i}\sigma$, $\!_{i}\lambda$
and $\!_{i}\tau$ are unit vectors in $\H_{\I^+}\oplus\H_{H^+}$ defined
in \cite{Wald1975&1976}, and the former four are related as follows:
\begin{eqnarray}
 \!_{i}\gamma^{a}  & = & T_{i}\ _{i}\sigma^{a}
                         +R_{i}\ _{i}\rho^{a},	\nonumber\\
 \!_{i}\lambda^{a} & = & t_{i}\ _{i}\rho^{a}
                         +r_{i}\ _{i}\sigma^{a},	
			\label{eqn:2-3:gamma-lambda}
\end{eqnarray}
where $t_{i}$, $T_{i}$ are transmission coefficients for the mode
specified by the integer $i$ on the Schwarzschild metric \cite{Wald1975&1976}
and $r_{i}$, $R_{i}$ are reflection coefficients. They satisfy 
\footnote{The last two equations are consequences of the time
reflection symmetry of the Schwarzschild metric.}
\begin{eqnarray}
 |t_{i}|^{2}+|r_{i}|^{2} & = & |T_{i}|^{2}+|R_{i}|^{2}=1, 
				\nonumber\\
 t_{i} = T_{i}, & & r_{i}=-R_{i}^{*}T_{i}/T_{i}^{*}.\label{eqn:2-3:trTR}
\end{eqnarray}

By using the S-matrix, we obtain
\begin{eqnarray}
 S|\{ n_{\!_{i}\gamma}\}\rangle 
 & = & N\left[\prod_{i}\frac{1}{\sqrt{n_{\!_{i}\gamma}!}}\left[ 
      R_{i}a^{\dagger}(\!_{i}\rho )+T_{i}a^{\dagger}(\!_{i}\sigma )
      \right]^{n_{\!_{i}\gamma}}\right]\sum_{n=0}^{\infty}
      \frac{\sqrt{(2n)!}}{2^{n}n!}
      \left(\stackrel{n}{\otimes}\epsilon\right)_{sym}\nonumber\\
 & = & N\sum_{n=0}^{\infty}{\sum}'\left[\prod_{i}
      \frac{1}{\sqrt{n_{\!_{i}\gamma}!}}\left(\begin{array}{c}
      n_{\!_{i}\gamma}\\m_{i}\end{array}\right)R_{i}^{m_{i}}
      T_{i}^{n_{\!_{i}\gamma}-m_{i}}\right]\nonumber\\
 & & \times
      \sqrt{(2n)!}\left[\prod_{i}\frac{x_{i}^{n_{i}}}{n_i!}
      \left(\begin{array}{c}n_{i}\\l_{i}\end{array}\right)t_i^{l_{i}}
      r_i^{n_{i}-l_{i}}\right]  \nonumber \\
 & &  \times\sqrt{\frac{(2n+\sum_{i}n_{\!_{i}\gamma})!}
      {(2n)!}}\left(\prod_{i}\stackrel{n_{i}}{\otimes}\!_{i}\tau
      \stackrel{l_{i}+m_{i}}{\otimes}\!_{i}\rho
      \stackrel{n_{i}-l_{i}+n_{\!_{i}\gamma}-m_{i}}{\otimes}
      \!_{i}\sigma\right)_{sim} \nonumber\\
 & = & N\sum_{n_{i}=0}^{\infty}
      \sum_{m_{i}=0}^{n_{\!_{i}\gamma}}\sum_{l_{i}=0}^{n_{i}}
      \sqrt{\left(\sum_{i}(2n_{i}+n_{\!_{i}\gamma})\right) !}
	\nonumber\\
 & & \times
      \prod_{i}\left[\frac{1}
      {\sqrt{n_{\!_{i}\gamma}!}}\cdot\frac{x_{i}^{n_{i}}}{n_{i}!}
      \left(\begin{array}{c}n_{\!_{i}\gamma}\\m_{i}\end{array}\right)
      \left(\begin{array}{c}n_{i}\\l_{i}\end{array}\right)
      R_{i}^{m_{i}}T_{i}^{n_{\!_{i}\gamma}-m_{i}}t_{i}^{l_{i}}r_{i}
      ^{n_{i}-l_{i}}\right]   \nonumber  \\
 & &  \times\left(\prod_{i}\stackrel{n_{i}}{\otimes}
      \!_{i}\tau\stackrel{l_{i}+m_{i}}{\otimes}\!_{i}\rho
      \stackrel{n_{i}-l_{i}+n_{\!_{i}\gamma}-m_{i}}{\otimes}
      \!_{i}\sigma\right)_{sym},	\label{eqn:2-3:S|n>}
\end{eqnarray}
where ${\sum}'$ denotes a summation with respect to $n_i$, $m_i$ and
$l_i$ over the following range: $\sum_in_i=n$, $n_i\geq 0$, 
$0\leq m_i\leq n_{\!_i\gamma}$, $0\leq l_i\leq n_i$. In
(\ref{eqn:2-3:S|n>}), those orthonormal basis vectors in 
$|\{ n_{\!_{i}\rho}\}\rangle\otimes\F (\H_{H^+})$ that have a
non-vanishing inner product with $S|\{ n_{\!_{i}\gamma}\}\rangle$
appear in the form 
\footnote{The number of the 'particle' $\!_i\sigma$ in
(\ref{eqn:2-3:S|n>}) is $n_{i}+n_{\!_{i}\gamma}-n_{\!_{i}\rho}$, setting
the number of the 'particle' $\!_i\rho$ to $n_{\!_{i}\rho}$.}
\begin{equation}
 \sqrt{\frac{\left(\sum_{i}(2n_{i}+n_{\!_{i}\gamma})\right) !}
      {\prod_{i}[n_{i}!n_{\!_{i}\rho}!(n_{i}+n_{\!_{i}\gamma}
      -n_{\!_{i}\rho})!]}}\left(\prod_{i}\stackrel{n_{i}}{\otimes}
      \!_{i}\tau\stackrel{n_{\!_{i}\rho}}{\otimes}\!_{i}\rho
      \stackrel{n_{i}+n_{\!_{i}\gamma}-n_{\!_{i}\rho}}
      {\otimes}\!_{i}\sigma\right)_{sym}.\label{non-zero-term}
\end{equation}
Thus, when we calculate
$\left(\langle\{ n_{\!_{i}\rho}\} |\otimes\langle H|\right)S 
	|\{ n_{\!_i\gamma}\}\rangle$, 
the summation in (\ref{eqn:2-3:S|n>}) is reduced to a summation with
respect to $n_i$ and $m_i$ over the range 
$n_i\geq\max (0,n_{\!_{i}\rho}-n_{\!_{i}\gamma})$,  
$\max (0,n_{\!_{i}\rho}-n_{i})\leq m_{i}\leq
	\min(n_{\!_{i}\rho},n_{\!_{i}\gamma})$ 
with $l_i=n_{\!_{i}\rho}-m_i$ 
\footnote{
The range is obtained by inequalities $n_i\geq 0$, 
$0\leq m_i\leq n_{\!_{i}\gamma}$, $0\leq l_i\leq n_i$, 
$l_i+m_i=n_{\!_{i}\rho}$ and 
$n_i+n_{\!_{i}\gamma}-n_{\!_{i}\rho}\geq 0$.
}.
Here $|H\rangle$ is an element of $\F(\H_{H^+})$. Paying attention to
this fact, we can obtain the following expression of the conditional
probability.
\begin{eqnarray}
 & & P(\{ n_{\!_{i}\rho}\} |\{ n_{\!_{i}\gamma}\} )\nonumber\\
 & & = |N|^{2}
	\sum_{\{ n_{i}\geq\max (0,n_{\!_{i}\rho}-n_{\!_{i}\gamma})\}}
      \left(\sum_{i}(2n_{i}+n_{\!_{i}\gamma})\right) ! \nonumber\\
 & &  \times\prod_{i}\bigg[\frac{x_{i}^{2n_{i}}}
      {n_{\!_{i}\gamma}!(n_{i}!)^{2}}
      \bigg|\sum_{m_{i}=\max (0,n_{\!_{i}\rho}-n_{i})}
      ^{\min (n_{\!_{i}\rho},n_{\!_{i}\gamma})}
      \left(\begin{array}{c}n_{\!_{i}\gamma}\\m_{i}\end{array}\right)
      \left(\begin{array}{c}n_{i}\\n_{\!_{i}\rho}-m_{i}\end{array}\right)
	\nonumber\\
 & & \times
      R_{i}^{m_{i}}T_{i}^{n_{\!_{i}\gamma}-m_{i}}
      t_{i}^{n_{\!_{i}\rho}-m_{i}}r_{i}^{n_{i}-n_{\!_{i}\rho}+m_{i}}
      \bigg|^{2}\bigg]   \nonumber\\
 & &  \times\left|\langle\sqrt{
      \frac{\left(\sum_{i}(2n_{i}+n_{\!_{i}\gamma})\right) !}
      {\prod_{i}\left[n_{i}!
      n_{\!_{i}\rho}!(n_{i}+n_{\!_{i}\gamma}-n_{\!_{i}\rho})!\right]}}
      \prod_{i}\left(\stackrel{n_{i}}{\otimes}
      \!_{i}\tau\stackrel{n_{\!_{i}\rho}}{\otimes}\!_{i}\rho
      \stackrel{n_{i}+n_{\!_{i}\gamma}-n_{\!_{i}\rho}}
      {\otimes}\!_{i}\sigma\right)_{sym},\right.\nonumber\\
 & &  \left.\hspace{2cm}
      \prod_{i}\left(\stackrel{n_{i}}{\otimes}
      \!_{i}\tau\stackrel{n_{\!_{i}\rho}}{\otimes}\!_{i}\rho
      \stackrel{n_{i}+n_{\!_{i}\gamma}-n_{\!_{i}\rho}}
      {\otimes}\!_{i}\sigma\right)_{sym}\rangle\right|^{2}.
\end{eqnarray}
The inner product in the last expression equals to 
\footnote{(\ref{non-zero-term}) is normalized to have unit norm.}
\[
 \sqrt{\frac{\prod_{i}\left[ 
      n_{i}!n_{\!_{i}\rho}!(n_{i}+n_{\!_{i}\gamma}-n_{\!_{i}\rho})!\right]}
      { \left(\sum_{i}(2n_{i}+n_{\!_{i}\gamma})\right) !}}.
\]
Finally, by using (\ref{eqn:2-3:trTR}) and exchanging the order of the
summation suitably, we can obtain
\begin{eqnarray*}
 & & P(\{ n_{\!_{i}\rho}\} |\{ n_{\!_{i}\gamma}\} )
	\nonumber\\
 & & = \prod_{i}\left[ (1-x_i)x_i^{2n_{\!_{i}\rho}}
      \left( 1-|R_{i}|^{2}\right)^{n_{\!_{i}\gamma}+n_{\!_{i}\rho}}
		\right.\nonumber	\\
 & &  \times\sum_{l_{i}=0}^{\min (n_{\!_{i}\gamma},n_{\!_{i}\rho})}
      \sum_{m_{i}=0}^{\min (n_{\!_{i}\gamma},n_{\!_{i}\rho})}
      \frac{\left[ -|R_{i}|^{2}/(1-|R_{i}|^{2})
      \right]^{l_{i}+m_{i}}n_{\!_{i}\gamma}!n_{\!_{i}\rho}!}
      {l_{i}!(n_{\!_{i}\gamma}-l_{i})!(n_{\!_{i}\rho}-l_{i})!
       m_{i}!(n_{\!_{i}\gamma}-m_{i})!(n_{\!_{i}\rho}-m_{i})!} 
		\nonumber\\
 & &  \left.\times\sum_{n_i=n_{\!_{i}\rho}-\min (l_{i},m_{i})}
      ^{\infty}\frac{n_{i}!(n_{i}-n_{\!_{i}\rho}+n_{\!_{i}\gamma})!}
      {(n_{i}-n_{\!_{i}\rho}+l_{i})!(n_{i}-n_{\!_{i}\rho}+m_{i})!}
      (x_{i}^{2}|R_{i}|^{2})^{n_{i}-n_{\!_{i}\rho}}\right].
\end{eqnarray*}
This is what we had to show.

%%%%%%%%%%%%%%%%%%%%%%%%%%%%%%%%%%%%%%%%
%%%%%%%%%%%%% APPENDIX 2 %%%%%%%%%%%%%%%
%%%%%%%%%%%%%%%%%%%%%%%%%%%%%%%%%%%%%%%%
\section{A proof of Lemma 2}	\label{app:lemma}
In this appendix we give a proof of Lemma \ref{lemma:off-diagonal}.
%
%===================================%
%============< PROOF >==============%
%
\begin{proof}
Since a set of all $\!_i\tau$ and $\!_i\sigma$ generates $\H_{H^+}$
\cite{Wald1975&1976}, the definition of 
$T_{\{ n_{\!_{i}\rho}\} \{ n'_{\!_{i}\rho}\} }
^{\{ n_{\!_{i}\gamma}\} \{ n'_{\!_{i}\gamma}\} }$ leads
\begin{equation}
  T_{\{ n_{\!_{i}\rho}\} \{ n'_{\!_{i}\rho}\} }
	^{\{ n_{\!_{i}\gamma}\} \{ n'_{\!_{i}\gamma}\} } =
 \sum_{\{ n_{\!_i\sigma}\},\{ n_{\!_i\tau}\}}
	\langle\{n_{\!_i\tau},n_{\!_i\rho},n_{\!_i\sigma}\}|
	S|\{n_{\!_i\gamma}\}\rangle
	\langle\{n'_{\!_i\gamma}\}|
	S|\{n_{\!_i\tau},n'_{\!_i\rho},n_{\!_i\sigma}\}\rangle,
\end{equation}
where
\[
 |\{n_{\!_i\tau},n_{\!_i\rho},n_{\!_i\sigma}\}\rangle	\equiv
	\prod_i\left[\frac{1}
	{\sqrt{n_{\!_i\tau}!n_{\!_i\rho}!n_{\!_i\sigma}!}}
	\left(a^{\dagger}(\!_{i}\tau)\right)^{n_{\!_{i}\tau}}
	\left(a^{\dagger}(\!_{i}\rho)\right)^{n_{\!_{i}\rho}}
	\left(a^{\dagger}(\!_{i}\sigma)\right)^{n_{\!_{i}\sigma}}
	\right] |0\rangle.
\]
In the expression, $S|\{n_{\!_i\gamma}\}\rangle$ is given by
(\ref{eqn:2-3:S|n>}) and $S|\{n'_{\!_i\gamma}\}\rangle$ is obtained by
replacing $n_{\!_i\gamma}$ with $n'_{\!_i\gamma}$ in
(\ref{eqn:2-3:S|n>}). Now, those orthonormal basis vectors of the form 
$|\{n_{\!_i\tau},n_{\!_i\rho},n_{\!_i\sigma}\}\rangle$ that have a
non-zero inner product with $S|\{n_{\!_i\gamma}\}\rangle$ must also be
of the form (\ref{non-zero-term}). Thus, 
$T_{\{ n_{\!_{i}\rho}\} \{ n'_{\!_{i}\rho}\} }
^{\{ n_{\!_{i}\gamma}\} \{ n'_{\!_{i}\gamma}\} }$ vanishes unless there 
exist such a set of integers $\{n_i, n'_i\}$ $(i=1,2,\cdots )$ that
\begin{eqnarray}
 n_{i}&=&n'_{i}\nonumber\\
 n_{i}+n_{\!_{i}\gamma}-n_{\!_{i}\rho}
      &=&n'_{i}+n'_{\!_{i}\gamma}-n'_{\!_{i}\rho}
\end{eqnarray}
for $\forall i$. The existence of $\{n_i\}$ and $\{n'_i\}$ is
equivalent to the condition 
$n_{\!_{i}\gamma}-n'_{\!_{i}\gamma}=n_{\!_{i}\rho}-n'_{\!_{i}\rho}$ 
for $\forall i$.
\end{proof}
\QED

%%%%%%%%%%%%%%%%%%%%%%%%%%%%%%%%%%%%%%%%
%%%%%%%%%%%%% APPENDIX 3 %%%%%%%%%%%%%%%
%%%%%%%%%%%%%%%%%%%%%%%%%%%%%%%%%%%%%%%%
\section{On-shell brick wall model}
	\label{app:on-shell}

When we performed the differentiation with respect to 
$\beta_{\infty}$ to obtain the total energy and the entropy in
section~\ref{sec:brick_wall}, the 
surface gravity $\kappa_{0}$ of the black hole and the inverse 
temperature $\beta_{\infty}$ of gas on the black hole background were
considered as independent quantities. Since in equilibrium these quantities 
are related by $\beta_{\infty}^{-1}=\kappa_0/2\pi$, we have imposed this
relation, which we call the on-shell condition,  after the differentiation. 
In fact, we have shown that the wall contribution to gravitational energy 
is zero and the backreaction can be neglected, if and only if the 
on-shell condition is satisfied.

On the other hand, in the so-called on-shell 
method~\cite{Frolov1995,FFZ1996a,Belgiorno&Martellini}, the on-shell
condition is implemented before the differentiation. 
Now let us investigate what we might call an on-shell brick wall model. 
With the on-shell condition, the wall contribution to the free energy 
of the scalar field considered in subsection \ref{subsec:BWmodel2} is 
calculated as 
%============< EQUATION >==============%
%
\begin{equation}
 F_{wall}^{(on-shell)} = 
 	-\frac{A}{4}\frac{\beta_{\infty}^{-1}}{360\pi}\frac{1}{\alpha^2}.
\end{equation}
%======================================%
If we define total energy and entropy in the on-shell method by 
%============< EQUATION >==============%
%
\begin{eqnarray}
 U_{wall}^{(on-shell)} & \equiv & 
 	\frac{\partial}{\partial\beta_{\infty}}
 	\left(\beta_{\infty} F_{wall}^{(on-shell)}\right),\nonumber\\
 S_{wall}^{(on-shell)} & \equiv & 
 	\beta_{\infty}^2\frac{\partial}{\partial\beta_{\infty}}
 	F_{wall}^{(on-shell)},
\end{eqnarray}
%======================================%
then these quantities can be calculated as 
%============< EQUATION >==============%
%
\begin{eqnarray}
 U_{wall}^{(on-shell)} & = & 0,\nonumber\\
 S_{wall}^{(on-shell)} & = & 
 	\frac{A}{4}\frac{1}{360\pi}\frac{1}{\alpha^2}
	=\frac{1}{4}S_{wall},
\end{eqnarray}
%======================================%
where $S_{wall}$ is the wall contribution (\ref{eqn:3-2:Swall2}) to 
entropy of the scalar field with $T_{\infty}=T_{BH}$.

It is notable that the total energy $U_{wall}^{(on-shell)}$ in the 
on-shell method is zero irrespective of the value of the cutoff 
$\alpha$. However, $S_{wall}^{(on-shell)}$ is always smaller than 
$S_{wall}$. It is because some physical degrees of freedom are frozen 
by imposing the on-shell condition before the differentiation. Thus, we 
might miss the physical degrees of freedom in the on-shell method.

%%%%%%%%%%%%%%%%%%%%%%%%%%%%%%%%%%%%%%%%
%%%%%%%%%%%%% APPENDIX 4 %%%%%%%%%%%%%%%
%%%%%%%%%%%%%%%%%%%%%%%%%%%%%%%%%%%%%%%%
\section{Symmetric property of the entanglement entropy for a pure state}
	\label{app:Sent=Sent'}

In this appendix we first give an abstract expression for the reduced
density operators $\rho_1$ and $\rho_2$ corresponding to 
a pure state $u$ in ${\cal F}={\cal F}_1\bar{\otimes}{\cal F}_2$,
which do not use the subtrace. Then with the help of them we prove
that $S_{ent}$ obtained from $\rho_1$ and $\rho_2$ coincide 
with each other. We follow the notations in \S\ref{subsection:Senta}. 

%
%====================================%
%=========< PROPOSITION >============%
%
\begin{prop}	\label{prop:A_u}
For an arbitrary element $u$ of 
${\cal F}={\cal F}_1\bar{\otimes}{\cal F}_2$, there are antilinear
bounded operators $A_u$ 
$\in\bar{{\cal B}}({\cal F}_1,{\cal F}_2)$ and $A^*_u$ 
$\in\bar{{\cal B}}({\cal F}_2,{\cal F}_1)$ such that 
%============< EQUATION >==============%
%
\begin{equation}
 (A_ux,y) = (A^*_uy,x) = (u,x\otimes y)	\label{eqn:3-3:A_u-A^*_u}
\end{equation}
%======================================%
for ${}^{\forall}x\in{\cal F}_1$ and ${}^{\forall}y\in{\cal F}_2$.
\end{prop}

%
%===================================%
%============< PROOF >==============%
%
{\it Proof.}
Fix an arbitrary element $x$ of ${\cal F}_1$. Then $(u,x\otimes y)$
gives a linear bounded functional of $y (\in{\cal F}_2)$ since 
%============< EQUATION >==============%
%
\[
 \left| (u,x\otimes y)\right| \leq ||u|| ||x|| ||y||.
\]
%======================================%
Hence by Riesz's theorem there is a unique element $z_{u,x}$ of 
${\cal F}_2$ such that 
%============< EQUATION >==============%
%
\begin{equation}
 (z_{u,x},y) = (u,x\otimes y)
\end{equation}
%======================================%
for ${}^{\forall}y\in{\cal F}_2$. 
Let us define $A_u$ by $A_u:x\to z_{u,x}$. It is evident that $A_u$
is an antilinear bounded operator from ${\cal F}_1$ to ${\cal F}_2$ since
%============< EQUATION >==============%
%
\[
 ||A_{u}x|| = ||z_{u,x}|| = ||u|| ||x||.
\]
%======================================%

Exchanging the roles played by ${\cal F}_1$ and ${\cal F}_2$ in the above
argument, it is shown that there is an antilinear bounded operator 
$A^*_u$ from ${\cal F}_2$ to ${\cal F}_1$ such that 
$(A^*_uy,x) = (u,x\otimes y)$.
\hfill $\Box$

Note that $A_u$ and $A^*_u$ defined above are written as 
%============< EQUATION >==============%
%
\begin{eqnarray}
 A_ux & = & \sum_jf_j(x\otimes f_j,u)\ \ \ ,	\nonumber\\
 A^*_uy & = & \sum_ie_i(e_i\otimes y,u)\ \ \ .
\end{eqnarray}
%======================================%
Using this expression, it is easily shown that 
%============< EQUATION >==============%
%
\begin{eqnarray}
 A^*_uA_ux & = & \sum_{ij}e_i\left( e_i\otimes f_j,
			(u,x\otimes f_j)u\right)\ \ \ ,	\nonumber\\
 A_uA^*_uy & = & \sum_{ij}f_j\left( e_i\otimes f_j,
			(u,e_i\otimes y)u\right)\ \ \ .
\end{eqnarray}
%======================================%
These coincide with $\rho_1$ and $\rho_2$, respectively,
if $u$ has unit norm (see Eq. (\ref{eqn:3-3:rho}) and (\ref{eqn:3-3:reduced})). 
Therefore the following proposition says that $\rho_1$ and
$\rho_2$ have the same spectrum and the same multiplicity and
that entropy of them are identical.

%
%====================================%
%=========< PROPOSITION >============%
%
\begin{prop}
$\rho_u^{(1)}$ $(\in{\cal B}({\cal F}_1))$ and 
$\rho_u^{(2)}$ $(\in{\cal B}({\cal F}_2))$ defined by
%============< EQUATION >==============%
%
\begin{eqnarray}
 \rho_u^{(1)} & = & A^*_uA_u\ \ \ ,\nonumber\\
 \rho_u^{(2)} & = & A_uA^*_u\ \ \ 
\end{eqnarray}
%======================================%
are non-negative, trace-class self-adjoint operators, where $A_u$ and
$A^*_u$ are defined in Proposition \ref{prop:A_u} for an arbitrary
element $u$ of ${\cal F}$. The spectrum and the multiplicity of
$\rho_u^{(1)}$ and $\rho_u^{(2)}$ are identical for all
non-zero eigenvalues. 
\end{prop}

%
%===================================%
%============< PROOF >==============%
%
{\it Proof.}
In general
%============< EQUATION >==============%
%
\[
 (x',\rho_u^{(1)}x) = (A_ux,A_ux')
\]
%======================================%
for ${}^{\forall}x,x'$ ($\in{\cal F}_1$) by definition. Therefore 
%============< EQUATION >==============%
%
\begin{equation}
 (x,\rho_u^{(1)}x) = ||A_ux||^2 \geq 0
\end{equation}
%======================================%
and 
%============< EQUATION >==============%
%
\begin{eqnarray}
 {\rm Tr}_{{\cal F}_1} (\rho_u^{(1)}) 
	& = & \sum_i(A_ue_i,A_ue_i)	\nonumber\\
	& = & \sum_{i,j}(A_ue_i,f_j)(f_j,A_ue_i)	\nonumber\\
	& = & \sum_{i,j}|(u,e_i\otimes f_j)|^2	\nonumber\\
	& = & ||u||^2,
\end{eqnarray}
%======================================%
i.e. $\rho_u^{(1)}$ is non-negative and trace-class. In general a
non-negative operator is self-adjoint and a trace-class operator is
compact. Thus the eigenvalue expansion theorem for a self-adjoint
compact operator says that all eigenvalues of $\rho_u^{(1)}$ are
discrete except zero and have finite multiplicity. For a later
convenience let us denote the non-zero eigenvalues and the
corresponding eigenspaces as $\lambda_i$ and ${\cal F}_{1,i}$
($i=1,2,\cdots $). 

Similarly, it is shown that $\rho_u^{(2)}$ is non-negative and
trace-class and that all eigenvalues of it are discrete except zero
and have finite multiplicity.

Now $\ker\rho_u^{(2)}=\ker A^*_u$ since 
$\rho_u^{(2)}y=A_uA^*_uy$ and 
$(y,\rho_u^{(2)}y)=||A^*_uy||^2$ for an arbitrary element $y$
of ${\cal F}_2$ by definitions. Moreover, from (\ref{eqn:3-3:A_u-A^*_u}) it
is evident that  
%============< EQUATION >==============%
%
\[
 y\perp\mbox{Ran}A_u \Leftrightarrow y\in\ker A^*_u.
\]
%======================================%
With the help of these two facts ${\cal F}_2$ is decomposed as
%============< EQUATION >==============%
%
\begin{equation}
 {\cal F}_2 = \ker \rho_u^{(2)}\oplus\overline{\mbox{Ran}A_u}.
	\label{eqn:3-3:W=ker+ran}
\end{equation}
%======================================%
where the overline means to take a closure.

Moreover, it is easily shown by definitions that 
%============< EQUATION >==============%
%
\begin{eqnarray}
 \rho_u^{(2)}A_ux
	& = & \lambda_iA_ux\ \ \ ,\nonumber\\
 \left( A_ux,A_ux'\right) & = & \lambda_i(x',x)
\end{eqnarray}
%======================================%
for ${}^{\forall}x$ ($\in{\cal F}_{1,i}$) and 
${}^{\forall}x'$ ($\in{\cal F}_{1,i'}$). Hence $A_u$ maps the eigenspace
${\cal F}_{1,i}$ to a eigenspace of $\rho_u^{(1)}$ with the 
same eigenvalue, preserving its dimension. Taking account of
Eq.(\ref{eqn:3-3:W=ker+ran}), this implies that the spectrum and the 
multiplicity of $\rho_u^{(1)}$ and $\rho_u^{(2)}$ are
identical for all non-zero eigenvalues.
\hfill $\Box$

%%%%%%%%%%%%%%%%%%%%%%%%%%%%%%%%%%%%%%%%
%%%%%%%%%%%%% APPENDIX 5 %%%%%%%%%%%%%%%
%%%%%%%%%%%%%%%%%%%%%%%%%%%%%%%%%%%%%%%%
\section{Entanglement energy for the case of $B={\rm R}^2$ in
Minkowski spacetime}
	\label{app:half}

First we take $B={\rm R}^2$. Without loss of generality the
resulting two half-spaces are represented as 
$\{(x_1,x_2,x_3):$ $x_1>0\}$ and 
$\{(x_1,x_2,x_3):$ $x_1<0\}$~\cite{BKLS1986}.

Here some comments are in order. Since all the degrees of freedom on
and across $B$, which is infinite, contribute to the entanglement
energy, a suitable cut-off length $a(>0)$
should be introduced to avoid the ultra-violet divergence. For the
same reason, the infra-red divergence is also anticipated in
advance, since $B$ is non-compact in this model. The latter is taken
care of by considering the massive case since the inverse of the mass 
characterizes a typical size of the spreading of the field. Clearly
$a$ should be taken short enough in the unit of the Compton length of
the field, $m_{\phi}^{-1}$, to obtain meaningful results. Therefore we shall
only pay attention to the leading order in the limit $m_{\phi}a\to 0$ in
the course of calculation as well as in the final results. These
remarks are valid in any model of this type, and the same remarks
apply to the case of the entanglement entropy,
too~\cite{BKLS1986,Srednicki1993}.

In order to calculate $\langle :H_{tot}:\rangle_{\rho'}$ for the
present case, we first note 
that the term $ V_{AB}q^A q^B$ in Eq.(\ref{eqn:3-3:Lagrange}) corresponds
to the expression 
$\int\left[( {\vec{\nabla}} \phi )^2 + m_{\phi}^2\phi^2  \right] d^3 x$ 
read off from Eq.(\ref{eqn:3-3:action_m}), which defines an operator 
$V(x,y)$ acting on a space 
${\cal W}=\left( \left\{ \phi (\cdot) \right\}, d^3 x \right)$. 
In order to use  the formula (\ref{eqn:3-3:ex-Eent1}), thus, we need
the inverse of the positive square-root of $aV$. For this purpose, it is
convenient to work in the momentum representation of 
$\cal W$~\cite{BKLS1986} given by 
%============< EQUATION >==============%
%
\begin{eqnarray}
 \phi (\vec{x}) &=& \int \frac{d^3 k}{(2\pi)^3} \ 
                 \phi_{\vec{k}} 
                 \exp[i \vec{k} \cdot \vec{x}], \nonumber\\
 \phi_{\vec{k}} &=& \int d^3 x \ 
                 \phi (\vec{x}) 
                 \exp[-i \vec{k} \cdot \vec{x}]. 
\end{eqnarray}
%======================================%
  The results are  
%============< EQUATION >==============%
%
\begin{eqnarray}
 V(\vec{x},\vec{y})
& = & 
	\int\frac{d^3 k}{ (2\pi )^3}(\vec{k}^2+m_{\phi}^2)
	\exp [i\vec{k} \cdot (\vec{x}-\vec{y})]\ \ \ ,\nonumber\\
 W^{-1}(\vec{x},\vec{y}) 
& = &
	\int_{{\rm R}^3} \frac{d^3 k}{(2\pi )^3}
	 (\vec{k}^2  +  m_{\phi}^2)^{-1/2}
	 \exp [i\vec{k} \cdot(\vec{x}-\vec{y})].
\end{eqnarray}
%======================================%
Note that both $V(\vec{x},\vec{y})$ and 
$W^{-1}(\vec{x},\vec{y})$ are 
symmetric under the exchange of $\vec{x}$ and $\vec{y}$. 
(The cut-off must preserve this property.) 
Now the formula (\ref{eqn:3-3:ex-Eent1}) gives 
%============< EQUATION >==============%
%
\begin{eqnarray}
 \langle :H_{tot}:\rangle_{\rho'} & = & 
 -\frac{1}{2}\int_{y_1 <-a}d^3 y \int_{x_1 >a}d^3 x
	\int_{|k_1|<a^{-1}}\frac{d^3 k}{(2\pi)^3}
	    ({\vec{k}}^2  +  m_{\phi}^2)
	\exp[i\vec{k}\cdot(\vec{y}-\vec{x})]\nonumber\\
& & \times	\int_{|k'_1|<a^{-1}}\frac{d^3 k'}{(2\pi)^3}
	    ({\vec k}^{'2}  +  m_{\phi}^2)^{-1/2}
	\exp[i\vec{k}' \cdot (\vec{x}-\vec{y})],
\end{eqnarray}
%======================================%
where, as discussed above, 
a cut-off length $a$ was introduced in the integral. 

Since the integrand is invariant under the translation along $B$, the
integral with respect to $x_2$ and $x_3$ yields a divergent factor 
$A=\int_{R^2}dx_2dx_3$. Clearly this factor should be interpreted as
the area of $B$. If this divergent integral $A$ is factored out, we
obtain the following convergent expression for $\langle :H_{tot}:\rangle_{\rho'}$:
%============< EQUATION >==============%
%
\begin{eqnarray}
 \langle :H_{tot}:\rangle_{\rho'} 
& = &  
 -\frac{A}{2} \int_{-\infty}^{-a}dy_1 
              \int_{a}^{\infty}dx_1 
	\int\frac{d^2 k_\parallel}{(2\pi)^2}
	\int_{-a^{-1}}^{a^{-1}}\frac{dk_1}{2\pi}
	\int_{-a^{-1}}^{a^{-1}}\frac{d{k'}_1}{2\pi}	\nonumber\\
& & \times
	(\vec{k_\parallel}^2 + {k_1}^2 + m_{\phi}^2)
	(\vec{k_\parallel}^2 + {{k'}_1}^2 + m_{\phi}^2)^{-1/2}
	\nonumber\\
& & \times \exp[i(k_1 - {k'}_1)(y_1 -x_1 )].
\end{eqnarray}
%======================================%
Here $\vec{k}_\parallel$ is a 2-vector lying along $B$ 
and $k_1$, ${k'}_1$ are components normal to $B$  
(if we make an obvious identification of ${\rm R}^3$ with 
its Fourier space). 
Let us change   
the  variables from $x_1$ and $y_1$ to 
$z\equiv x_1 -y_1$ and $u \equiv (x_1 +y_1 )/2$. 
Then $z$ and $u$ take values in the range 
$z \leq 2a$ and $-({z \over 2} -a) \geq u \geq ({z \over 2}-a)$, 
respectively. Hence the integration with respect to $u$ yields 
%============< EQUATION >==============%
%
\begin{eqnarray}
 \langle :H_{tot}:\rangle_{\rho'} 
& = &  
 -\frac{A}{2}\int_{2a}^{\infty}dz(z-2a)
	\int \frac{d^2 k_\parallel}{(2\pi)^2}
	\int_{-a^{-1}}^{a^{-1}}\frac{dk_1}{2\pi}
	\int_{-\infty}^{\infty}\frac{dk'_1}{2\pi}	\nonumber\\
& & \times	({\vec{k}_\parallel}^2 + {k_1}^2 + m_{\phi}^2)
	({\vec{k}_\parallel}^2 + {{k'}_1}^2 + m_{\phi}^2)^{-1/2}
	\exp[-i({k_1}-{{k'}_1})z]	             \nonumber\\
& = & 
 -\frac{A}{2}\int_{2a}^{\infty}dz(z-2a)
	\int_{-a^{-1}}^{a^{-1}}\frac{dk_1}{2\pi}   \cos({k_1} z)
	\int_m ^\infty  
	    \frac{d\kappa}{2\pi}\kappa (\kappa^2+{k_1}^2)\nonumber\\
& & \times	\int_{-\infty}^{\infty}
	   \frac{d{k'}_1}{2\pi}
	(\kappa^2 + {{k'}_1}^2)^{-1/2} \cos({k'}_1 z)
\end{eqnarray}
%======================================%
in the leading order, where $\kappa$ is defined by
$\kappa^2 = {\vec{k}_\parallel}^2  +  m_{\phi}^2$. 
Here note that in this expression, the integration with respect to 
${k'}_1$ followed by that with respect to $\kappa$ leads to an
infra-red divergence if we set $m=0$, in accordance with 
our discussion at the beginning of this subsection. 

Now let us recollect some  formulas with the 
modified Bessel functions~\cite{Bessel}:
%============< EQUATION >==============%
%
\begin{eqnarray}
 K_0(x) & = &
	\int_{0}^{\infty}dt\frac{\cos t}{\sqrt{t^2+x^2}}, 
	\nonumber\\
 \int_{x_0}^{\infty}dx\ xK_0(x) & = &
	x_0 K_1(x_0)\ \ \ , \nonumber\\
 \int_{x_0}^{\infty}dx \ x^3 K_0(x) & = &
	x_0^3 K_1(x_0)+2x_0^2 K_2(x_0).
\end{eqnarray}
%======================================%
With the help of  these formulas
 $E_{ent}^I$ is written as
%============< EQUATION >==============%
%
\begin{eqnarray}
 E_{ent}^I &=&
 -\frac{A}{2} \int_{2a}^{\infty}dz (z-2a)
    \int_{-a^{-1} }^{a^{-1}}\frac{dk_1}{2\pi}
	\cos({k_1} z) \nonumber	\\
& & \times \frac{1}{2\pi^2}\left[
         \frac{m}{z} ({k_1}^2 + m_{\phi}^2) K_1(mz)  
         +  2 \frac{m_{\phi}^2}{z^2} K_2(mz) \right], \nonumber \\
& = & 
 \frac{A}{4\pi^3 a^3}
  [ \alpha_1(ma) + \alpha_2(ma)+ \alpha_3(ma)]
 \label{eqn:3-3:resultA1}
\end{eqnarray}
%======================================%
in the leading order. Here we have introduced 
%============< EQUATION >==============%
%
\begin{eqnarray}
 \alpha_1(x)
     & \equiv &
      -x \int_2^\infty  d\xi \frac{\xi -2}{\xi^4}   K_1(x\xi )
	\left[ 2\xi \cos\xi  + (\xi^2 - 2 ) \sin\xi \right],  
	\nonumber\\
 \alpha_2(x)
     & \equiv &
     -2x^2 \int_2^\infty  d\xi \frac{\xi -2}{\xi^3} K_2(x\xi )
              \sin\xi,                     \nonumber\\
 \alpha_3(x)
     & \equiv &
     -x^3 \int_2^\infty d\xi \frac{\xi -2}{\xi^2} K_1(x\xi )
              \sin\xi.
\end{eqnarray}
%======================================%
A numerical evaluation shows
%============< EQUATION >==============%
%
\[
 \left[ \alpha_1(x) + \alpha_2(x) + \alpha_3(x) \right] \sim 0.05
   \ \ {\rm as} \ \      x \to 0.
\]
%======================================%
Therefore we get~\footnote{
If we adopt another regularization scheme with the same cut-off length
$a$, the result may change. However, the change is in sub-leading order.}
%============< EQUATION >==============%
%
\begin{equation}
  \langle :H_{tot}:\rangle_{\rho'} \approx \frac{0.05A}{4\pi^3a^3}.
\end{equation}
%======================================%

%%%%%%%%%%%%%%%%%%%%%%%%%%%%%%%%%%%%%%%%
%%%%%%%%%%%%% APPENDIX 6 %%%%%%%%%%%%%%%
%%%%%%%%%%%%%%%%%%%%%%%%%%%%%%%%%%%%%%%%
\section{States determined by variational principles}
	\label{app:A}

In this appendix we give derivations of formulas (\ref{eqn:3-4:EPRstate})
and (\ref{eqn:3-4:TFDstate}).

We consider a Hilbert space $\cal{F}$ of the form
%============< EQUATION >==============%
%
\begin{equation}
 {\cal{F}} = {\cal{F}}_{1}\otimes{\cal{F}}_{2},
\end{equation}
%======================================%
where ${\cal{F}}_{1}$ and ${\cal{F}}_{2}$ are Hilbert spaces with the
same finite dimension $N$. An arbitrary unit element $|\phi\rangle$ of 
$\cal{F}$ is decomposed as 
%============< EQUATION >==============%
%
\begin{equation}
 |\phi\rangle = \sum_{n=1}^N \sum_{m=1}^N 
	C_{nm}|n\rangle_1 \otimes |m\rangle_2,
\end{equation}
%======================================%
where $|n\rangle_{1}$ and $|n\rangle_{2}$ ($n=1,2,\cdots,N$) are 
orthonormal bases of ${\cal{F}}_{1}$ and ${\cal{F}}_{2}$,
respectively, and $\sum_{n,m}|C_{nm}|^2=1$ is
understood. Here, without loss of generality, we can choose the
orthonormal basis of ${\cal{F}}_{2}$ to be eigenstates of
the normal-ordered Hamiltonian $:H_2:$ of the sub-system as
%============< EQUATION >==============%
%
\begin{equation}
 :H_2: |n\rangle_2 = E_n |n\rangle_2. 
\end{equation}
%======================================%
Since $C^{\dagger}C$ is a non-negative hermitian matrix, we can define a
set of non-negative real numbers $\{ p_n\}$, each of which is the 
eigenvalue of the matrix $C^{\dagger}C$. Hence,
%============< EQUATION >==============%
%
\begin{equation}
 C^{\dagger}C = V^{\dagger}PV,
\end{equation}
%======================================%
where $V$ is a unitary matrix and $P$ is a diagonal matrix with
diagonal elements $\{ p_n\}$. With these definitions, the entanglement
entropy $S_{ent}$ and the entanglement free energy $F_{ent}$ are
calculated as 
%============< EQUATION >==============%
%
\begin{eqnarray}
 S_{ent} & = & -\sum_{n=1}^N p_n\ln p_n,\label{eqn:3-4:Sent-pn}\\
 F_{ent} & = & \sum_{n=1}^N\sum_{m=1}^N E_n p_m |V_{mn}|^2
		+ T_{ent}\sum_{n=1}^N p_n\ln p_n. 
		\label{eqn:3-4:Fent-pn}
\end{eqnarray}
%======================================%
The constraints $\sum_{n,m}|C_{nm}|^2=1$ and $V^{\dagger}V={\bf 1}$
are equivalent to 
%============< EQUATION >==============%
%
\begin{eqnarray}
 \sum_{n=1}^N p_n & = & 1,\nonumber\\
 \sum_{l=1}^N V^{*}_{ln}V_{lm} & = & \delta_{nm}.
		\label{eqn:3-4:constraint}
\end{eqnarray}
%======================================%
Thus, the variational principles are restated as follows: to maximize
(\ref{eqn:3-4:Sent-pn}) under the constraints (\ref{eqn:3-4:constraint}); to
minimize (\ref{eqn:3-4:Fent-pn}) under the constraints
(\ref{eqn:3-4:constraint}).

Now, we shall show that expressions (\ref{eqn:3-4:Sent-pn}) and
(\ref{eqn:3-4:Fent-pn}) are same as those for entropy and free energy
in statistical mechanics in the subspace ${\cal{F}}_2$.  Let us consider
a density operator $\bar{\rho}$ on ${\cal{F}}_2$: 
%============< EQUATION >==============%
%
\begin{equation}
 \bar{\rho} = \sum_{n=1}^N \sum_{m=1}^N \tilde{P}_{nm}
		|n\rangle_{2}{}_{2}\langle m|,
\end{equation}
%======================================%
where $\tilde{P}_{nm}$ is an $N\times N$ non-negative hermitian matrix
with unit trace.  By diagonalizing the matrix $\tilde{P}$ as
%============< EQUATION >==============%
%
\begin{equation}
 \tilde{P} = \bar{V}^{\dagger}\bar{P}\bar{V},
\end{equation}
%======================================%
we obtain the following expressions for entropy and free energy.
%============< EQUATION >==============%
%
\begin{eqnarray}
 S & = & -\sum_{n=1}^N \bar{p}_n\ln \bar{p}_n,\nonumber\\
 F & = & \sum_{n=1}^N\sum_{m=1}^N E_n \bar{p}_m |V_{mn}|^2
		+ T\sum_{n=1}^N \bar{p}_n\ln \bar{p}_n,
\end{eqnarray}
%======================================%
where $\{\bar{p}_n\}$ are diagonal elements of the matrix $\bar{P}$ 
and $T$ is temperature. The constraints $\bf{Tr}\bar{\rho}=1$  and
$\bar{V}^{\dagger}\bar{V}={\bf 1}$ are restated as 
%============< EQUATION >==============%
%
\begin{eqnarray}
 \sum_{n=1}^N \bar{p}_n & = & 1,\nonumber\\
 \sum_{l=1}^N \bar{V}^{*}_{ln}\bar{V}_{lm} & = & \delta_{nm}.
\end{eqnarray}
%======================================%

At this point, it is evident that the variational principles of
maximum of entropy are the same in entanglement thermodynamics and
statistical mechanics and that the principles of minimum of free
energy are also the same in the two schemes. Hence, the principle of
maximum of the entanglement entropy gives 
%============< EQUATION >==============%
%
\begin{equation}
 \left(C^{\dagger}C\right)_{nm} = \frac{1}{N}\delta_{nm}, 
		\label{eqn:3-4:minSent}
\end{equation}
%======================================%
as the principle of maximum of entropy gives the microcanonical
ensemble 
%============< EQUATION >==============%
%
\begin{equation}
 \tilde{P}_{nm} = \frac{1}{N}\delta_{nm}
\end{equation}
%======================================%
in statistical mechanics. Similarly, the principle of minimum of
the entanglement free energy gives
%============< EQUATION >==============%
%
\begin{equation}
 \left( C^{\dagger}C\right)_{nm} = 
		Z^{-1}e^{-E_n/T_{ent}}\delta_{nm},
			\label{eqn:3-4:maxFent}
\end{equation}
%======================================%
as the principle of minimum of the free energy gives the canonical
ensemble 
%============< EQUATION >==============%
%
\begin{equation}
 \tilde{P}_{nm} = \bar{Z}^{-1}e^{-E_n/T}\delta_{nm}
\end{equation}
%======================================%
in statistical mechanics, where $Z=\sum_ne^{-E_n/T_{ent}}$ and
$\bar{Z}=\sum_ne^{-E_n/T}$ . It is easy 
to see that (\ref{eqn:3-4:minSent}) and (\ref{eqn:3-4:maxFent}) are equivalent 
to (\ref{eqn:3-4:EPRstate}) and (\ref{eqn:3-4:TFDstate}), respectively, up to
a unitary transformation in ${\cal{F}}_1$.

Finally we comment on the generalization of the analysis when the 
Hilbert space is divided into two subspaces with different 
dimensions ($dim{\cal F}_{1}>dim{\cal F}_{2}$). 
In this case, by defining $S_{ent}$ from $\rho_{2}$, we obtain 
similar results.

%%%%%%%%%%%%%%%%%%%%%%%%%%%%%%%%%%%%%%%%
%%%%%%%%%%%%% APPENDIX 7 %%%%%%%%%%%%%%%
%%%%%%%%%%%%%%%%%%%%%%%%%%%%%%%%%%%%%%%%
\section{Bell states}
	\label{app:B}

In this appendix we show that in the finite dimensional case the
orthonormal basis $\{|\psi_{nm}\rangle_{1}\}$ defined in the physical
principle (c) is given by (\ref{eqn:3-4:BELLstates}) uniquely up to a
unitary transformation in ${\cal{F}}_{1+}$. After that, we derive the
equation (\ref{eqn:3-4:phi-decomp}).

We consider the following decomposition of the Hilbert space
${\cal{F}}_{1}$. 
%============< EQUATION >==============%
%
\begin{equation}
 {\cal{F}}_{1}={\cal{F}}_{1+}\otimes{\cal{F}}_{1-},
\end{equation}
%======================================%
where ${\cal{F}}_{1+}$ and ${\cal{F}}_{1-}$ are Hilbert spaces with
the same finite dimension $N$. From the arguments in Appendix
\ref{app:A}, each of the basis $\{|\psi_{nm}\rangle_{1}\}$ is obtained
by applying a unitary transformation in ${\cal{F}}_{1+}$ to the
following state in ${\cal{F}}_1$. 
%============< EQUATION >==============%
%
\begin{equation}
 |\phi\rangle_1 = \frac{1}{\sqrt{N}}\sum_{j=1}^N\left(
		|j\rangle_{1+}\otimes|j\rangle_{1-}\right),
\end{equation}
%======================================%
where $|j\rangle_{1+}$ and $|j\rangle_{1-}$ ($j=1,2,\cdots,N$) are 
orthonormal bases of ${\cal{F}}_{1+}$ and ${\cal{F}}_{1-}$,
respectively. Evidently, any states given by (\ref{eqn:3-4:BELLstates})
are obtained by this procedure. Moreover, it is easily confirmed as
follows that a set of all states given by (\ref{eqn:3-4:BELLstates}) is a
complete orthonormal basis in the $N\times N$ dimensional Hilbert
space ${\cal{F}}_1$. 
%============< EQUATION >==============%
%
\begin{equation}
 {}_{1}\langle\psi_{nm}|\psi_{n'm'}\rangle_{1} =
	\frac{1}{N}\sum_{j=1}^{N} e^{2\pi ij(n'-n)/N}\delta_{mm'}
	= \delta_{nn'}\delta_{mm'}.
\end{equation}
%======================================%

Let us suppose another complete orthonormal basis
$\{|\bar{\psi}_{nm}\rangle_{1}\}$ in ${\cal{F}}_1$, each of which
maximizes the entanglement entropy with respect to the decomposition
${\cal{F}}_{1}={\cal{F}}_{1+}\otimes{\cal{F}}_{1-}$.
Since both $\{|\psi_{nm}\rangle_{1}\}$ and
$\{|\bar{\psi}_{nm}\rangle_{1}\}$ are complete orthonormal basis in
${\cal{F}}_1$, they are related by a unitary transformation $U$ in
${\cal{F}}_1$. Moreover, $U$ is a unitary transformation in
${\cal{F}}_{1+}$, since any states maximizing the entanglement entropy 
are related by unitary transformations in ${\cal{F}}_{1+}$ as
shown in Appendix \ref{app:A}. Therefore, the orthonormal basis 
$\{|\psi_{nm}\rangle_{1}\}$ defined in the physical principle (c) is
unique up to a unitary transformation in ${\cal{F}}_{1-}$ and is
given by (\ref{eqn:3-4:BELLstates}).

Now let us show the equation (\ref{eqn:3-4:phi-decomp}). The right hand
side is transformed as follows.
%============< EQUATION >==============%
%
\begin{eqnarray}
 & &\frac{1}{N}\sum_{nm}|\psi_{nm}\rangle_{1}\otimes
		U^{(2+)}_{nm}|\tilde{\phi}_{2}\rangle_{2}
	\nonumber\\
 & &  = \frac{1}{\sqrt{N}}\sum_{jk}
	\left(\frac{1}{N}\sum_{n}e^{2\pi i(j-k)n/N}\right)\times
	\sum_{mm'n'}{}_{2+}\langle k|n'\rangle_{2+} C_{n'm'}
	\nonumber\\
 & & 	\times |(j+m)modN\rangle_{1+}\otimes |j\rangle_{1-}
	\otimes|(k+m)modN\rangle_{2+}\otimes |m'\rangle_{2-}
	\nonumber\\
 & & =  \sum_{mm'n'}\left(\frac{1}{\sqrt{N}}|(n'+m)modN\rangle_{1+}
	\otimes |(n'+m)modN\rangle_{2+}\right) 
	\nonumber\\
 & &	\times 	C_{n'm'}|n'\rangle_{1-}\otimes|m'\rangle_{2-}
	\nonumber\\
 & & =	\left(\frac{1}{\sqrt{N}}\sum_{m''}|m''\rangle_{1+}
	\otimes |m''\rangle_{2+}\right) 
	\nonumber\\
 & &	\times \left(\sum_{n'm'}C_{n'm'}
	|n'\rangle_{1-}\otimes|m'\rangle_{2-}\right).
\end{eqnarray}
%======================================%
The final expression is $|\phi\rangle$ itself.

\pagestyle{headings}

%%%%%%%%%%%%%%%%%%%%%%%%%%%%%%%%%%%%%%%%%%%%%%%%%%%%%%%%%%%%%%%%%%%%
%%%%%%%%%%%%%%%%%%%%%%%%%%%%%%%%%%%%%%%%%%%%%%%%%%%%%%%%%%%%%%%%%%%%
% References
%%%%%%%%%%%%%%%%%%%%%%%%%%%%%%%%%%%%%%%%%%%%%%%%%%%%%%%%%%%%%%%%%%%%
%%%%%%%%%%%%%%%%%%%%%%%%%%%%%%%%%%%%%%%%%%%%%%%%%%%%%%%%%%%%%%%%%%%%

\end{document}